\documentclass[12pt,twoside]{report}
\usepackage[english]{babel}
\usepackage[pdftex,dvips]{graphicx}
\usepackage[dvips]{graphicx}
\usepackage{color}
\usepackage{amsmath}
\usepackage{amssymb}
\usepackage{latexsym}
\usepackage{bbm}
\usepackage{chicago}
\usepackage{hangcaption,fancyheadings,ctagsplt}
\usepackage{graphicx}
\usepackage{amsthm}
\usepackage{epstopdf}
\usepackage[final]{pdfpages}
\usepackage{bm}

\newcommand{\beq}{\begin{equation}}
\newcommand{\eeq}{\end{equation}}
\newcommand{\barr}{\begin{eqnarray}}
\newcommand{\earr}{\end{eqnarray}}

\newcommand{\ket}[1]{\left\vert#1\right\rangle}
\newcommand{\bra}[1]{\left\langle#1\right\vert}
\newcommand{\braket}[2]{\langle #1 \vert #2 \rangle}

\newcommand{\Ham}{\mathcal H}

\newcommand{\Tr}{\mathrm{Tr}}

\newcommand{\avg}[1]{\left< #1 \right>}
\newcommand{\Ord}[1]{{\cal O}\left( #1\right)}

\def\cZ{{\cal Z}}

\def\cM{\mathcal{M}}
\def\cH{\mathcal{H}}
\def \cZ {\mathcal{Z}}
\def\e{\mathrm{e}}

\newcommand{\bras}[2]{{}_{#2}\langle{#1}|}

\definecolor{dgreen}{rgb}{0,0.8,0}

\renewcommand{\maketitle}{\begin{titlepage}%
    \setlength{\voffset}{0.cm}
\begin{center}
  \begin{center}
    \begin{figure}
     \begin{center}
    \includegraphics[width=3.5cm]{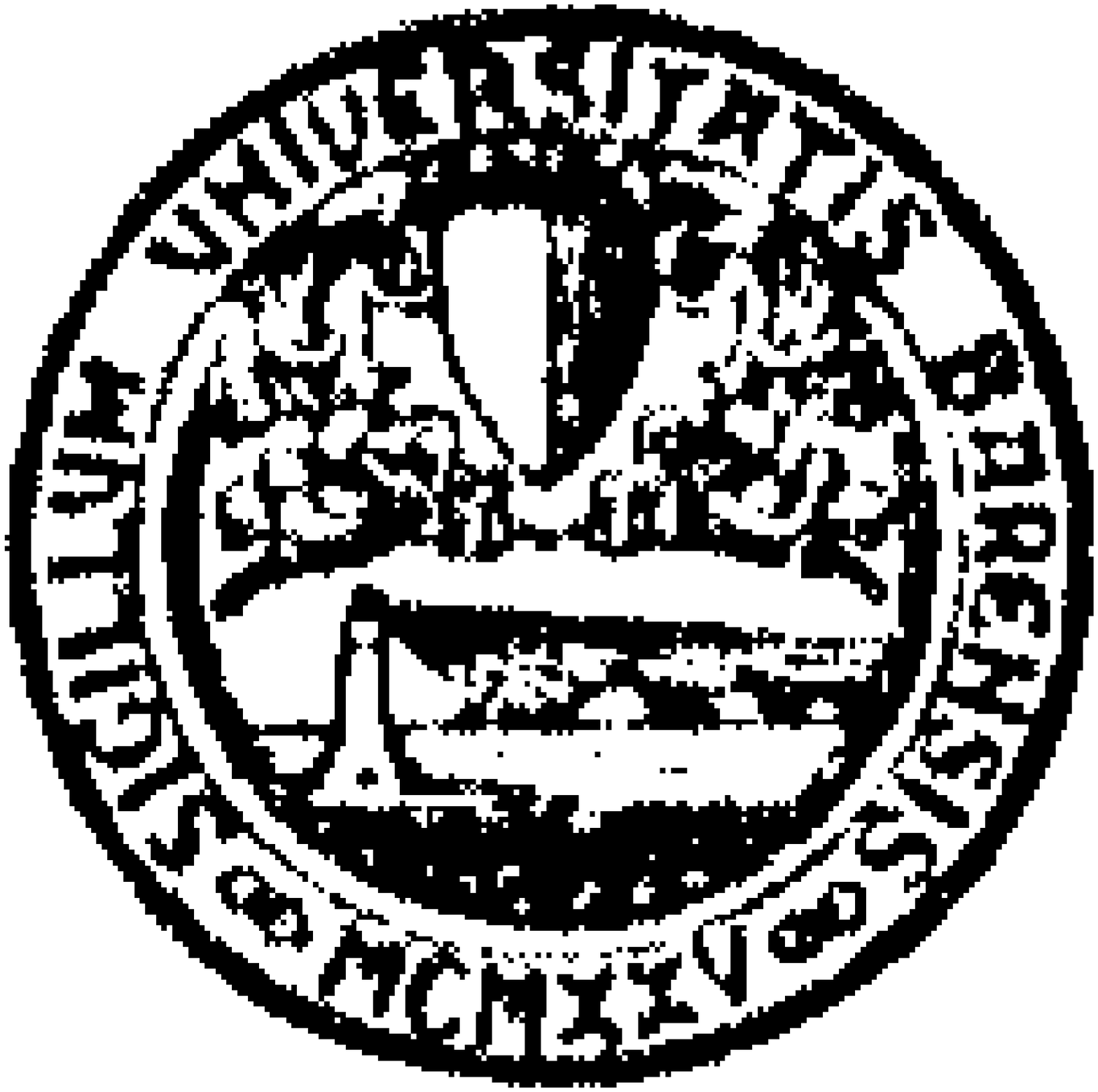}
      \end{center} 
    \end{figure}
		    {\Large {\bf UNIVERSIT\`A DEGLI STUDI DI BARI \\  \vskip 1.5 mm ALDO MORO}} \\  \vskip 2mm
		    {{\bf DIPARTIMENTO INTERATENEO DI FISICA  \\ \vskip 1mm ``MICHELANGELO MERLIN"}} \\ \vskip 1mm
		    		     {{\bf DOTTORATO DI RICERCA IN FISICA - Ciclo XXIII}} \\ \vskip 1mm  {{ \bf Settore Scientifico Disciplinare FIS/02}}
				     
		     \rule{380pt}{1pt} \end{center}

  \vspace{3.cm}
   \setlength{\textwidth}{12cm}
  \begin{minipage}{\textwidth}
   \begin{center}
       {\Large {\bf BIPARTITE ENTANGLEMENT     \\ OF LARGE QUANTUM SYSTEMS \\}}
   \end{center}
  \end{minipage}
\end{center}
      \vspace{1.5cm}
      \begin{center}  {{ \large \bf Dottoranda}}\\ \vskip 3mm{\large\sc Antonella De Pasquale}\end{center}
  \vskip 2.8cm
  \begin{center}
\large{ESAME FINALE 2011}
\end{center}
  \vspace{8em}
  \nopagebreak[4]

\newpage
\thispagestyle{empty}

  \end{titlepage}
  \if@twoside
  \thispagestyle{empty}\cleardoublepage
  \fi}

\begin{document}
\maketitle
\newpage
 \newpage
\thispagestyle{empty}
\cleardoublepage
\thispagestyle{empty}

\vspace{7cm}\hspace{9cm}
{\Large \textit{to  Mum and Dad}}

\newpage
\ \thispagestyle{empty}
\thispagestyle{empty}
\mbox{}
\pagestyle{myheadings}
\pagenumbering{Roman}
\addtocounter{page}{-1}
\markboth{Index}{Index}
\tableofcontents

\newpage
\thispagestyle{empty}
\mbox{}

\chapter*{Introduction}\label{intro}

\pagenumbering{arabic}
\markboth{Introduction}{Introduction}
\addcontentsline{toc}{chapter}{Introduction}

Entanglement measures the nonclassical correlations between the components of a quantum system. The characterization of this intimately quantum property of physical systems is one of most challenging issues of quantum information and computation theory. The reason why the profound comprehension of entanglement is considered so important is twofold: on the one hand it is spurred by an obvious interest for the foundations of quantum mechanics, on the other hand entanglement represents one of the most important resources in quantum information
processing and quantum enabled technologies~\cite{nielsen,benenticasati}. Generation of highly entangled quantum states is therefore one of the key elements for the realization of the ideas of quantum information. In this scenario a dominant role is played by the investigation of  one dimensional spin chains, as possible candidates for modeling quantum computers~\cite{Lieb1961,pfeuty1970,Takahashi}. However, experimental and theoretical difficulties, such as decoherence and imperfections in the quantum hardware, impose strong bounds on the realization of large scale systems and at the same time have boosted a high interest in finite size systems~\cite{finSize1,Schulz2001,osterloh2000,xxdiag,xydiag}.
The bipartite entanglement between systems of small dimension (such as a pair
of qubits) can be given a quantitative characterization in terms of
several physically equivalent measures, such as entropy and
concurrence~\cite{woot}, as will be briefly recalled in Chap.~\ref{chap1}. A direct investigation of quantum correlations and other basic properties of quantum systems, can be also reached through very simple tomographic techniques~\cite{paris2004}. Among them, scattering has always been considered a very powerful way to investigate many physical systems in a wide range of fields of physics, from elementary particles to condensed matter physics. Some very simple examples of state reconstruction through scattering of a probe qubit can be found in~\cite{KawabataSpinFilter,qpfesclong,spintokyo}. The relationship between entanglement and tomography is ``bidirectional'': if on the one hand tomographic techniques can be very useful in order to unveil quantum correlations between subsystems, on the other hand it has been proved that quantum effects enable an increase in precision in estimating the parameters of a given target state. The latter phenomenon is related to the fact that entangled states can evolve faster than untangled configurations, employing the same amount of resources~\cite{giovannetti2006}.
More involved is the characterization of the global features of bipartite entanglement when we deal with systems of larger dimension. This represents a very appealing issue from more then one point of view. Indeed besides its application in quantum information theory and related fields of investigation, such as complexity~\cite{parisi}, it represents an interesting problem in statistical mechanics. This will be the central theme of this thesis. We will tackle this problem by studying a random matrix model that describes the statistical properties of the purity of one of the two subsystems, namely the local purity of the global state. For the sake of clarity, we will divide this work in two parts.

In the first part, we will focus on the case of pure global states, yielding the description of isolated quantum systems. We will study the  eigenvalues distribution for the reduced density matrix of a subsystem $A$. Before the advent of the field of quantum information, this random matrix model was studied by Lubkin in 1978~\cite{Lubkin78}, who was moved by a more fundamental motivation: ``With Lucretius~\cite{lucretius} I find disorder of the universe repugnant. $\ldots$ my favorite key to understanding quantum mechanics is that subsystems cannot be isolated by tracing from an enveloping pure state without generating impurity: The probabilities associated with measurements develop because the observer must implicitly trace himself away from the observed system''. In this paper, Lubkin investigated the following problem: if an $L$ dimensional pure bipartite state is chosen at random, what probability distribution will describe the eigenvalues of one of the two subsystems? This problem results completely defined if we interpret by random pure states those vectors of the Hilbert space selected by the unique uniform Haar measure on the unitary group $\mathcal{U}(L)$, that is \textit{typical} states (with respect to this uniform measure). From the computation of the first two cumulants for $L\gg1$, he concluded that for a typical pure quantum state of a large system, the smaller subsystems are very nearly maximally mixed, that is ``impure'' showing almost no signs of the fact that the (initial) global state (the universe) is pure.  It was only almost ten years later that this topic became of central importance in quantum information theory, when Loyd and Pagel, apparently unaware of Lubkin's work, after determining the explicit expression of the distribution of the Schmidt coefficients of the reduced density matrix, computed the approximate expression for the typical entropy~\cite{Lloyd1988}, whose exact formula was conjectured by Page~\cite{page} and later proved in~\cite{foong,sanchez1995,sen1996}. For the measure on the Schmidt simplex, see also~\cite{Zyczkowski01}.  All these results confirmed the conclusions of Lubkin, which translated in quantum information language sound: if the global system is divided in two subsystems $A$ and $B$, the amount of quantum information contained in the whole system is greater then the sum of the information in the separate parts, that is typical (random) states are entangled with high probability.

Random states play a key role in quantum communication algorithms, such as quantum data hiding protocols~\cite{divincenzo2002} and superdense coding~\cite{harrow2004}. This has spoored the appearance of many algorithms for generating random states~\cite{zanardi2000}, based also on chaotic maps ~\cite{Bandyopadhyay02,Scott2004}, pseudo-integrable maps~\cite{Giraud2005} or operators~\cite{weinstein2005} and sequence of two-qubit gates~\cite{Znidaric2007}. Ideally, we would like to compute the complete expression for the probability distribution of the eigenvalues of the reduced density matrix. Alternatively, one should compute all its moments or cumulants. In particular, the exact expression for the first three cumulants have been computed in~\cite{Scott2003},  while in~\cite{Giraud} we find the exact formula for the $n$-th generic moment (and cumulant), and the explicit expression for  the first five cumulants. The probability distribution for the nonnull eigenvalues of the reduced density matrix of $A$ or $B$ have been computed for the case in which the dimensions of the corresponding Hilbert spaces $\cH_A$ and $\cH_B$ are  $\dim \cH_A=3 \leq \dim \cH_B=4$ and $\dim \cH_A= \dim \cH_B=4$, in~\cite{Giraud2007bis}. In this thesis we will introduce a different and more general approach for studying the statistics bipartite entanglement, relying on the techniques of classical statistical mechanics~\cite{paper1,metastable}. In particular, in Chap.~\ref{chap2} we will introduce a partition function for the canonical ensemble, the role of the energy being played by the purity of subsystem $A$, chosen as a measure of the quantum correlations between $A$ and $B$.  The partition function will depend on a Lagrange multiplier $\beta$ which corresponds to a fictitious inverse temperature, which fixes the value of the average purity. In other words,   being $\beta$ the conjugate variable of the purity it deforms the Haar measure, localizing it on the  corresponding manifold of states with a given average entanglement, whose associated uncertainty becomes smaller as the dimension of the quantum system increases. We will prove the equivalence, for our system, between the canonical ensemble and microcanonical description in terms of isoentangled manifolds~\cite{Kus01}.  We will therefore explore the entire Hilbert space, from typical to maximally entangled and separable states, corresponding to different regimes of temperatures, see Chaps.~\ref{chap3} and~\ref{chap5}. In Chap.~\ref{chap4} we will describe a metastable solution for the system in the region of negative temperatures. In particular, by computing the dominant contribution of the partition function thru the method of steepest descent and minimizing (or maximizing) the free energy of the system for positive (or negative temperatures), we will determine the probability distribution of the Schmidth coefficients corresponding to a given value of average entanglement, and not only for $\beta=0$ (typical states). Of course, once we know the partition function we can also compute the moments of the purity distribution and we will prove the consistency of our results with the cumulants' analysis provided in the papers cited above. We will unveil the presence of three main regimes in the system, the separable, the typical and the maximally entangled phases, separated by first and second order phase transitions. Finally in Chap.~\ref{chap6} we will overview our results abandoning the temperature and using the purity as our physical variable. One of the main results of our analysis will be the computation of the volume of the isopurity manifolds in the Hilbert space, that is the probability distribution of the purity. In particular this volume will be maximum for typical states, while the iso-entangled manifolds given by maximally entangled and separable states shrink to a vanishing volume. The latter case has been predicted in~\cite{Zyczkowski98} for the more general case of arbitrary mixed states, by proving the existence of a topological lower bound for this volume.

In the second part of the thesis, we will extend the above analysis to the case of a large quantum system in a mixed state, with a fixed value of the global purity. A statistical mixture describes the lack of complete knowledge of our physical system, when due to the interaction with the environment, it undergoes a dissipative evolution. In general, a complete microscopic description of the dynamical evolution of a system coupled to the environment (or bath) is a complex many-body problem which requires the solution of a potentially infinite number of coupled dynamical equations. According to an open system approach, this issue is tackled by retaining only basic information about the environment and describing the system dynamics in terms of a master equation~\cite{petruccione,gardiner2004}. The lack of a complete knowledge about the bath leads to master equation coefficients (MECÕs), which may be either unknown or obtained from a microscopical derivation carried out within some approximation scheme. Some tomographic approaches, based on the idea of studying the evolution of a Gaussian probe in order to provide the sought-for bridge between dynamical parameters and measurable quantities  (the probe's cumulant) have been proposed in order to retrieve the above coefficients or checking the approximation schemes for a wide class of Gaussian Shape Preserving master equations, both in the Markovian and convolution less non-Markovian evolution~\cite{openSystems,Tips,Mecs}. If we deal with isolated quantum systems, their density operator reduces to a rank one projection, and the analysis of its quantum correlations can be performed with the techniques introduced in Chaps.~\ref{chap2}-\ref{chap6}.
The general case of arbitrary mixed states is far more involved than the case of pure states studied in the first  part, due to more than one reason. The first one is related to the definition of a random sampling for the system. While for the case of pure states the only natural choice is obtained by requiring invariance under the full group of unitary transformations, for the case of mixed states, if we impose this unitary invariance we are just stating that the probability density depends only on the eigenvalues, while on the other hand we do not have a unique natural measure on the simplex of eigenvalues~\cite{woot1990}. It follows that, since no single distinguished probability measure exists, we can introduce many ensembles for random density operators. One possibility is given by adding an ancillary system of the same dimension, then determine a purification of our density operator, and consider the measure induced on it, by partial tracing, for instance from the Haar measure on the unitary larger group~\cite{Braunstein,hall1998,Zyczkowski01}. This strategy will be followed in Chap.~\ref{chap7}. Other sampling techniques are based on the Bures distance~\cite{hall1998,Slater,Bengtsson} or, alternatively on the multi-partite systems~\cite{Collins2010,Zyczkowski2010}. The absence of a unique measure for the set of density matrices is an extremely delicate point. Indeed, there are some properties which strictly depend on the chosen measure, such as the volume of separable states, while other statistical properties, such as the relation between the probability of entanglement as a function of the purity of the global state, are not very sensitive to the measure~\cite{Zyczkowski99}. Another aspect that makes the generalization of the approach introduced for the set of pure states very hard, is related to fact that from the definition of statistical mixture it follows that the purity of one subsystem (the {\it local purity} of the global system) cannot be considered a bonafide measure for the quantum correlations between subsystems $A$ and $B$, but should be substituted by its convex roof~\cite{woot,Amico08,h4}. We will discuss this point at the end of Chap.~\ref{chap1}. However, the above substitution does not allow for a simple analytic treatment of the partition function. We will thus devote the second part of the thesis to the study of the statistical distribution of the $A$-local purity, which can be considered as a lower bound for the bipartite entanglement between two parts of the global system. By generalizing the partition function introduced for the case of pure states, we will compute the canonical moments for the local purity, following different techniques. In particular in Chap.~\ref{chap8}, we will introduce the Gaussian approximation for the elements of the unitary groups on the set of density matrices describing our system and on their purification, and compute the average $A$-purity for $\beta=0$ up to order $\Ord{1/N}$, being $\dim \cH_A=N$. In Chap.~\ref{chap9} we will compute the exact expression for the first moment of the local purity at $\beta=0$, by exploiting the properties of the twirling transformations (establishing a formal connection between our statical problem and the theory of quantum channel), and then, thru the solution due to Zuber of some basic  integrals over the unitary group, we will determine the exact expression for the $n$-th generic moment, and explicitly compute the first two cumulants for the local purity at $\beta=0$, that is for typical random states. We will therefore generalize the exact expressions computed for the pure case in~\cite{Scott2003,Giraud}. Finally, we will compute the high temperature expansion of the first moment of the local purity in the canonical ensemble.

As a final remark, notice that the classical statistical mechanics approach we will follow in this thesis in order to characterize the behavior of bipartite entanglement for a large quantum system is general, and for instance can be adopted also for determining the statistical distribution of the potential of multipartite entanglement, as shown in~\cite{multipartitoClassStat}.

\newpage

 \newpage
\thispagestyle{empty}
\mbox{}
\newpage
\thispagestyle{empty}
\mbox{}
\chapter{An introduction to bipartite entanglement} \label{chap1}
\markboth{An introduction to bipartite entanglement}
{An introduction to bipartite entanglement}
\begin{figure}[h]
  \begin{center}
    \includegraphics[scale=0.3]{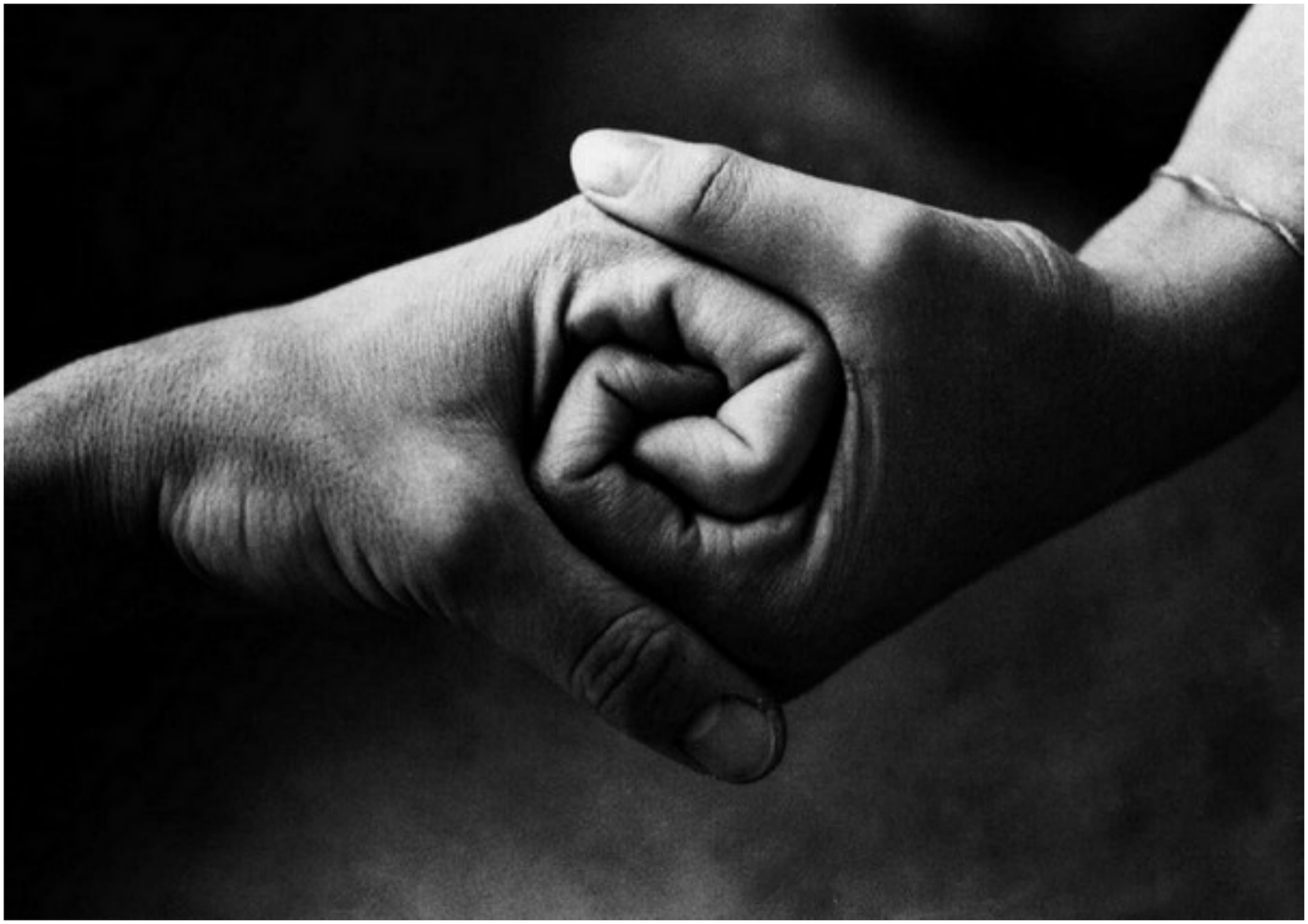}
    \caption{``Entanglement: folding the arms"}
    \label{fig:verschrankung}
  \end{center}
\end{figure}

The origin of the word ``entanglement" goes back to 1935. It is the English translation made by E. Schr\"odinger of the German word ``Vershr\"alung", which is a colloquial expression indicating the action of ``folding the arms", Fig.~\ref{fig:verschrankung}. Schr\"odinger introduced this name in the dictionary of quantum physics in order to describe one of the most striking manifestations 
of quantum phenomena, that is
``the best possible knowledge of a whole does not include the best possible knowledge of its parts, even though they may be entirely separated"~\cite{Schrodinger1935a,Schrodinger1935b,Schrodinger1936}. This is due to presence of correlations of exclusively quantum nature, namely entanglement. 
In other words, the presence of quantum correlations, with no classical counterparts, is responsible for the fact that while for a classical bipartite system the knowledge of the total system is completely equivalent to the knowledge of both its subsystems, in the quantum case this is only a necessary condition. 

The phenomenon of entanglement is profoundly rooted in the mathematical  structure of quantum mechanics, whose natural framework is given by complex Hilbert spaces. In particular, entanglement is a direct consequence of the linearity of the Hilbert space (expression of the superposition principle), when applied to composite systems, that are described by the tensor product of the Hilbert spaces associated to each subsystem~\cite{h4,facchiMultip}.

The peculiarities of this quantum phenomenon make the entanglement an extremely important resource in quantum information processing. It is believed to be the main responsible for the speed up in quantum computation and quantum communication~\cite{nielsen,benenticasati}. Furthermore, many quantum protocols could not be conceived without the existence of entangled states, an outstanding example is given by quantum teleportation~\cite{bennett}. Besides its important applications in relatively simple systems, that can be described in terms of a few effective quantum variables, entanglement has also been widely investigated in many-body systems~\cite{Amico08,h4}, from many points of view. In particular, on the one hand, the methods used in quantum information have proved to be extremely useful for the control and the manipulation of this kind of systems, on the other hand the interplay between quantum information and quantum statistical mechanics has unveiled some peculiar properties of entanglement in quantum critical models~\cite{osterloh02,osborne02,vidal03}. As a consequence of the powerful role played by entanglement in the study of fundamental quantum phenomena and its applications to innovative technologies, the task of characterizing and quantifying entanglement has emerged as a prominent theme of quantum information theory.

This chapter is devoted to the analysis of some fundamental aspects of bipartite entanglement. We will show that there are many basic open questions, revealing that we are still far from a complete comprehension of this phenomenon. We will start from the definition of entangled state in Sec.~\ref{bipartite entanglement}. Then, in Sec.~\ref {pure states}, we will focus on the only class of states for which the basic properties of bipartite entanglement, together with an appropriate measure, can be considered well understood, namely the pure quantum states. Finally, in Sec.~\ref{mixed states} we will show in what sense the study of entanglement and its quantification become extremely involved for an arbitrary quantum state, and report some fundamental results obtained in the last decades.

\section{Bipartite entanglement}\label{bipartite entanglement}
E. Sch\"odinger defined entanglement as ``not \textit{one} but rather \textit{the} characteristic trait of quantum mechanics". As already mentioned, this phenomenon is strictly related to the mathematical formulation of quantum mechanics, relying on the structure of complex Hilbert spaces.

The smallest non trivial Hilbert space is two dimensional, and describes the simplest quantum system, the \textit{qubit}. This is the fundamental unit of quantum information, as the \textit{bit} is the indivisible unit of classical information. A privileged orthonormal basis of this two dimensional Hilbert space is the computational basis, $\{\ket{0},\ket{1}\}$, which can be seen as the quantum version of the only two possible values associated to a single bit, $\{0,1\}$. The way a qubit differs from a bit is that a linear combination of the states $\ket{0}$ and $\ket{1}$ is still a possible state for the qubit. This derives from the vectorial structure of the Hilbert space, and represents the simplest expression of one of the cornerstone of quantum mechanics, the \textit{superposition principle}.

Let us now consider a composite bipartite system, and indicate with $A$ and $B$ its two parts. In classical physics, the states of the total system belong to the Cartesian product of the spaces associated to $A$ and $B$, and are always given by a convex combination of products of states describing each subsystem independently. The dimension of the global system is then the sum of the dimensions of its subspaces. On the contrary, the Hilbert space $\Ham$ where the states of the total system live is given by the  tensor product of the Hilbert subspaces $\Ham_A$ and $\Ham_B$ associated to $A$ and $B$ respectively, $\Ham=\Ham_A \otimes \Ham_B$, and the dimension of $\Ham$ is the product of the dimensions of $\Ham_A$ and $\Ham_B$. If for instance, we refer to a system of $n$ qubits, we can study the properties of two subsystems given by $n_1$ and $n_2=n-n_1$ qubits. The associated Hilbert spaces have dimensions $2^n$, $2^{n_1}$ and $2^{n_2}$, respectively ($2^{n}=2^{n_1+n_2}=2^{n_1}2^{n_2}$). Let us now come back to a generic bipartite system, and to the difference between the classical and quantum approach. It is again the superposition principle which plays the key role. Indeed an arbitrary pure state $\ket{\psi} \in \Ham$ cannot in general be expressed by the product of separate states of each subsystem, as would happen in the classical case: this is the formal definition of the entanglement. In other words, we will define the state $\ket{\psi}$ entangled, with respect of the bipartition $(A,B)$ if and only if
\begin{equation} \label{eq:entanglement}
\ket{\psi}\neq\ket{\psi}_A\otimes\ket{\psi}_B \ ,
\end{equation}
for some vectors $\ket{\psi}_A$ and $\ket{\psi}_B$ of each subsystem.  If condition (\ref {eq:entanglement}) is not satisfied, the state will be said to be separable, or unentangled. Let us remark that the separability of the state of the global system strictly depends on the bipartition we are considering, as it refers to the quantum correlations between the specific subsystems.
From Eq. (\ref{eq:entanglement}) it follows that if a pure state $\ket{\psi}$ is entangled it is given by a superposition of at least two product states:
\begin{equation} \label{eq:defmultipurent}
\ket{\psi}=\sum_{i_1, i_2} c_{i_1 i_2} \ket{\psi_{i_1}}_A\otimes\ket{\psi_{i_2}}_B.
\end{equation}
This describes interference, i.e. non classical correlation, among probability amplitudes for the two subsystems. For example, the probability amplitude of having subsystem $A$ in the state $\ket{\psi_{j_1}}_A$ and subsystem $B$ in the state $\ket{\psi_{j_2}}_B$ interferes with the probability amplitude of having subsystems $A$ and $B$ in states $\ket{\psi_{k_1}}_A$ and $\ket{\psi_{k_2}}_B$. It is due to the nonclassical properties of quantum states that many schemes of quantum information and quantum technologies (quantum computation ~\cite{nielsen}, quantum teleportation ~\cite{bennett}, dense coding ~\cite{bennett92} and quantum cryptography~\cite{crypto0,ekert,Deutsch,crypto3}) have been realized.

Until now we have only referred to \textit{pure} states, which correspond to normalized vectors of a given Hilbert space. They describe isolated quantum systems whose states are completely determined as far as the theory allows. However, in practice the state of a physical system is not often perfectly determined. For instance in a laboratory the systems we measure undergo  uncontrolled interactions with the environment. An example is given by the emission of atoms from a thermal source: we only know the distribution of the kinetic energy of the emitted particles, not the kinetic energy of each of them. In such cases, we say that our knowledge of the system is incomplete. We only know that the state of the system belongs to a given ensemble of pure states associated to a set of probabilities, $\{ p_1, \ldots p_n\}$ satisfying the condition of unit total probability, $\sum_{i} p_i =1$. The system is then defined to be in a \textit{statistical mixture}, or equivalently in a \textit{mixed} state. 
From a mathematical point of view,  an arbitrary quantum state is described by a density operator $\rho$ on $\mathcal{H}$ (a positive, self-adjoint and unit trace operator), which in case of a pure state $\ket{\psi} \in \Ham$ reduces to the projection operator $\rho=\ket{\psi}\bra{\psi}$. A generic mixed state can be expressed in an infinite variety of convex combinations of pure states all of which have exactly the same consequence for any conceivable observation of the system. It naturally follows that entangled mixed states are no longer equivalent to non-product states, as for pure states. A mixed bipartite state is separable if and only if it can be written as a convex combination of product states:
\begin{equation}\label{eq:defmultimixent}
\rho = \sum_{1 \leq i \leq n} p_i \ \rho_{A,i} \otimes   \rho_{B,i} \ ,   \quad p_{i}>0,  \quad \sum_{1 \leq i \leq n} p_i=1, \quad n \geq 1,
\end{equation}
where $ \rho_{A,i}$ and $ \rho_{B,i}$ are in general mixed states of the corresponding subsystems $A$ and $B$. If condition (\ref{eq:defmultimixent}) is not satisfied, $\rho$ is said to be an entangled state.
It can be easily proved that entanglement is responsible for those properties of composite systems which do not change under local transformations and classical communication (exchange of classical bits between the two subsystems). We say that bipartite entanglement is invariant under \textit{LOCC}, \textit{L}ocal \textit{O}peration and \textit{C}lassical \textit{C}ommunication. This has led to the introduction of some criterions in order to properly identify classes of equally entangled states~\cite{grassl1998,Linden1999,albeverio2001}, and to the complementary study of the orbits in the Hilbert space associated to a given quantum state by means of local unitary operations ~\cite{Kus01}.

Despite the simplicity of definitions (\ref{eq:defmultimixent}) and (\ref{eq:entanglement}), it is in general quite hard to check whether a state is separable or not, with respect to a given bipartition. Another open problem is the quantification of entanglement. In Sec.~\ref{pure states} we will answer these two points for the case of  pure states, and in Sec.~\ref{mixed states} we will then show how this analysis becomes far more involved for the case of mixed states.   



\section{Bipartite pure states}\label{pure states}
Bipartite pure states represent one of the few cases for which the problem of revealing and also quantifying quantum correlations has a quite exhaustive answer. In the first part of this section we will discuss some fundamental properties of bipartite entanglement, by the introduction of one important tool, the Schmidt decomposition. In the second part, we will consider the problem of quantification of bipartite entanglement through the von Neumann entropy. 

\subsection{Schmidt decomposition}\label{sec:schmidtdec}
Let us consider a pure bipartite state $\ket{\psi}$ in $\Ham=\Ham_A \otimes \Ham_B$.  We introduce the following notation for the dimension of the three Hilbert spaces considered:  $\dim \Ham_A=N$, $\dim \Ham_B=M$ and $\dim \Ham = L =N M$. Without loss of generality we also set $N \leq M$.  If we introduce an orthonormal product basis $\{\ket{e_n}_A \otimes \ket{e_m}_B\}$, $1 \leq n \leq N$ and  $1 \leq m \leq M$, we can represent the state as
\begin{equation}
\ket{\psi}=\sum_{1\leq i \leq N} \sum_{1 \leq j \leq M} C_{ij} \ket{e_i}_A \otimes \ket{e_j}_B.
\end{equation}
The coefficients $C_{i j}$ are the elements of an $N \times M$ complex rectangular
matrix $C$, whose singular values $\sqrt{\lambda_k}$, $1\leq k \leq N$ (i.e. the square root of
the eigenvalues of the matrix $C^\dagger C$) determine the 
\textit{Schmidt decomposition} of the total state $\ket{\psi}$:
\begin{equation}\label{eq:pureSchmidtDec}
\ket{\psi}=\sum_{1 \leq k \leq N} \sqrt{\lambda_k} \ket{k}_A \otimes \ket{k}_B.
\end{equation}
It is uniquely determined only if the singular values are not degenerate. Indeed, the previous expression can be derived from the singular value decomposition of the coefficient matrix, namely $C=U D V^{\dagger}$, with $U$ and $V$ unitary matrices on $\Ham_A$ and $\Ham_B$, respectively, and $D$ the diagonal matrix $D=\mathrm{diag} \{ \sqrt{\lambda_1}, \ldots, \sqrt{\lambda_{N}} \} $. The unitary matrices $U$ and $V$, 
determine the orthonormal set
\begin{equation}
 \ket{k}_A \otimes \ket{k}_B=\sum_{1 \leq i \leq N} \sum_{1 \leq j \leq M} U_{k i} \ket{e_i}_A \otimes V_{k j}\ket{e_j}_B, \quad \forall \ k \in \{1,\ldots N\}.
 \end{equation}
 If the singular values of $C$ are not degenerate its singular value decomposition is unique, up to multiplication of one or more columns of $U$ by unit phase factors and simultaneous multiplication of the corresponding columns of $V$ by the same unit phase factors. On the other hand, degenerate singular values do not have unique singular vectors. Consequently, if there are degenerate $\sqrt{\lambda_{i}}$, $1\leq i \leq N$, the singular value decomposition of $C$ is not unique.
The real numbers $\lambda_i$ are called \textit{Schmidt coefficients} and, due to the normalization condition on the state of the global system, $\braket{\psi}{\psi}=1$, obey  $\sum_{i}\lambda_i=1$ (notice that in the literature $\sqrt{\lambda_i}$ instead of $\lambda_i$, $1\leq i \leq N$, are usually defined as Schmidt coefficients). The set of all possible vectors $\vec{\lambda}=\{\lambda_1, \ldots, \lambda_{N}\}$ forms a $(N-1)$ dimensional simplex $\Lambda_A$, known as the \textit{Schmidt simplex}.  The Schmidt decomposition is of central importance for the characterization and the quantification of the entanglement associated with pure states. It enables to investigate many important properties of a quantum composite system by looking at the behavior of its subsystems, through the analysis of the corresponding reduced density operators. As it will be more clear at the end of this section, this is strongly related to the very essence of entanglement. 
The state of each subsystem, $A$ (or $B$), is described by the density operator $\rho_{A}$ (or $\rho_{B}$) defined by partial tracing over the complementary subsystem, 
$\rho_{A}=\Tr_{B} \left( \ket{\psi}\bra{\psi} \right)$ (or $\rho_{B}=\Tr_{A} \left( \ket{\psi}\bra{\psi} \right)$), where $\Tr_{S}$ indicates the partial trace over subsystem $S$ (this definition of a reduced density operator is also valid for an arbitrary state $\rho$).  From Eq. (\ref{eq:pureSchmidtDec}) we immediately get that $\rho_A$ and $\rho_B$ have the same nonzero eigenvalues, the Schmidt coefficients, and their eigenbases coincide with the orthonormal bases $\ket{i}_A$ and $\ket{i}_B$, $1\leq i \leq N$, called \textit{Schmidt bases} for $\Ham_A$ and $\Ham_B$. The number of nonvanishing $\lambda_i$ is equal to the rank of both reduced density operators. It is called \textit{Schmidt number} for the state $\ket{\psi}$, and together with the Schmidt coefficients, is invariant under local unitary operations: they will be the key ingredients for quantifying the entanglement between subsystems $A$ and $B$. Indeed, a bipartite pure state is separable with respect to the bipartition $(A,B)$ if and only if there is only one nonzero Schmidt coefficient, which must be equal to $1$. This is equivalent to saying that each subsystem is in a pure state. If it is not the case, the state $\ket{\psi}$ is entangled. In particular, a pure state $\ket{\psi} \in \Ham$ is called \textit{maximally entangled} if and only if all its Schmidt coefficients are equal to $1/{N}$, or equivalently subsystem $A$ is in a completely mixed state.
It is worth noticing that we have introduced the expression ``maximally entangled state" without referring to any measure of entanglement. Indeed these states present the maximal degree of entanglement with respect to every (reasonable) measure. They are a very peculiar class of quantum states, and can be considered the greatest manifestation of the presence of quantum correlations between subsystems. More formally, they give rise to the most striking violation of a class of inequalities, Bell inequalities, that must be satisfied by classical correlations.  If a bipartite system is in a maximally entangled state, we have that to a perfect knowledge of the state of the composite system ($\ket{\psi}$ is a pure state) it corresponds the maximal ignorance of the state of its two parts ($\rho_A$ is in a completely mixed state), since the global information is completely  shared between the two subsystems. This situation never happens in the classical case, where the complete knowledge of the state of the global system and of the state of both subsystems is equivalent. In quantum mechanics this represents just a necessary condition, that becomes sufficient only for the case of separable (unentangled) bipartite states, when both the global system and its subsystems $A$ and $B$ are in a pure state~\cite{facchiMultip}. This behavior, is the most impressive signature of quantum correlations, i.e. entanglement. 
\subsection{von Neumann entropy and purity}\label{pure state measures}
Bipartite entanglement is a very complex property of quantum states, and the task of capturing all its manifestations with an appropriate measure is in general very hard. Until now is has not been found a unique answer to this question. Nevertheless, pure bipartite states represent one of the few cases for which this intricate feature simplifies very much.  
The Schmidt decomposition is an elementary necessary and sufficient criterion for separability of pure bipartite states. In the last section we have seen that the Schmidt coefficients, and obviously their number, are invariant under local unitary operations on the system. Therefore, any reasonable measure of entanglement for pure bipartite states must be an appropriate function of these quantities. A widely accepted measure of entanglement is the von Neumann entropy of the reduced density operators $\rho_A$ and $\rho_B$ describing the two subsystems $A$ and $B$ of the global pure state $\ket{\psi}$:
\begin{equation}
S(\rho_A)=S(\rho_B)= - \Tr (\rho_A \log_{2} \rho_A)=-\sum_{1 \leq i \leq N} \lambda_i \log (\lambda_i),
\end{equation}
$\lambda_i$, $1 \leq i \leq N$, being the Schmidt coefficients of $\ket{\psi}$, and $0 \log 0$ defined to be zero. 
The von Neumann entropy goes from $0$ for separable states to $\log_2 N$ for maximally entangled states. It is the unique entanglement measure for pure states that beside being invariant under LOCC is also continuous and additive if there are several copies of the system ($S (\ket{\psi_1} \otimes \ldots \otimes\ket{\psi_n})=\sum_{j \leq n}S(\ket{\psi_j})$). These properties are some of the requirements that, according to an axiomatic approach, a good measure of entanglement has to satisfy.  We refer to Sec.~\ref{generic state measures} for a brief review about some peculiar aspects of entanglement quantification. 



Historically, the von Neumann entropy was the first entanglement measure to have an operational interpretation: it quantifies the quantum information of the system, i.e. the
deeply nonclassical interplay between the information provided by an entangled state about the whole system and the information regarding its subsystems. In 1995 Schumacher showed that the von Neumann entropy corresponds to the minimum number of qubits needed to encode a quantum state produced by a statistical source, in an ideal coding scheme~\cite{schumacher1995}.  In this sense it is exactly the quantum counterpart of the classical Shannon entropy which counts the minimum number of bits (units of classical information) one needs to encode the output of a random source using an ideal code.  As already pointed out at the end of Sec.~\ref{bipartite entanglement}, the main difference between classical and quantum information, is that if a state is entangled with respect to a given bipartition, it provides more information about the global system then about its two parts, whereas this can never happen for a classical state. Translated in terms of statistical mechanics, the subsystems of an entangled state can exhibit more entropy than global system, i.e. they can show more disorder than the system as a whole. This never happens in two cases: when we have a classical system (whose statistical disorder is quantified by the Shannon entropy) and if we deal with a quantum separable state. Once again, entanglement manifests itself as the distinctive character of quantum mechanics.  

The definition of the von Neumann entropy  as a good candidate for measuring bipartite entanglement in case of pure states is supported by the so-called ``uniqueness theorem"~\cite{vidal2000,Horodecki2000prl,Nielsen2000,Donald2002}  which states that in the asymptotic regime all other entanglement measures, introduced on the basis either of quantum information or of thermodynamical considerations, coincide on pure bipartite states and reduce to the von Neumann entropy of the corresponding reduced density matrices. 
A more convenient measure of entanglement from a computational point of view is the purity $\pi_{AB}$. It corresponds (up to a constant) to the so-called linear entropy $\mathcal{L}_{AB}$, that is the first-order term of the expansion of the von Neumann entropy:
\begin{equation}
S(\rho_{A}) = - \Tr (\rho_A \log_{2} \rho_A) \sim  - \Tr (\rho_A (\rho_A-1))=1-\Tr (\rho_A^2) = \frac{N-1}{N}\mathcal{L}_{AB}
\end{equation}
and the purity of one of the two subsystems is defined as:
\begin{equation}
\pi_{AB}=\Tr (\rho_A^2)=\Tr (\rho_B^2)=\sum_{1 \leq i \leq N} \lambda_i^2.
\end{equation}
We can easily see that the purity lies in the compact interval $[1/N, 1]$, whose boundaries from left to right refer to maximally entangled and separable states, with respect to the bipartition $(A, B)$ (the corresponding bounds of the linear entropy are $1$ and $0$). In this thesis the purity will be the fundamental tool in order to investigate the statistical behavior of bipartite entanglement of pure quantum states. 

\section{Bipartite mixed states}\label{mixed states}
Beside, and maybe ``before", the problem of properly quantifying entanglement, an open question of quantum information theory is to determine whether an arbitrary (mixed) quantum state is separable or not with respect to a given bipartition, namely the separability problem. We have already stressed that pure bipartite states represent a singularity in this complex scenario
since the Schmidt decomposition on the one hand provides a simple necessary and sufficient criterion for checking the separability of pure states, on the other hand it naturally leads to the quantification of entanglement in terms of the mixedness of its subsystems. On the contrary, the complete characterization of the convex set of separable mixed states reveals itself to be extremely involved~\cite{Zyczkowski99}. We will devote the present section to a brief overview of the main separability criteria and measures of bipartite entanglement for a generic mixed state~\cite{Amico08,h4,Bruss2002}. 
\subsection{Separability criteria}
\begin{figure}[h]
  \begin{center}
    \includegraphics[scale=0.5]{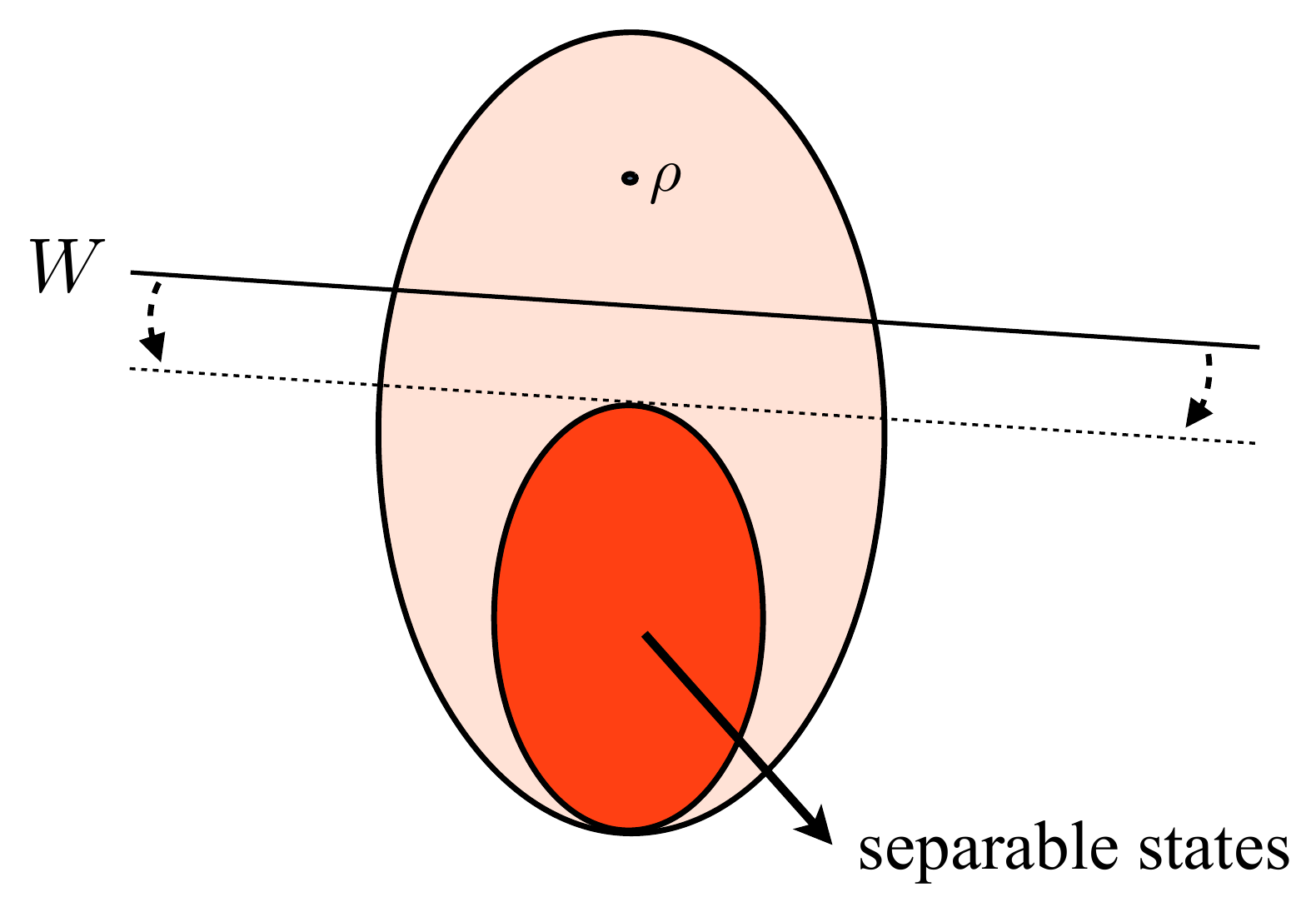}
    \caption{Intuitive picture of an entanglement witness $W$ (continuous straight line) and its optimizations (dashed lines).}
    \label{fig:w}
  \end{center}
\end{figure}

A very strong necessary separability condition for quantum bipartite states is given by \textit{Peres' positive partial transpose} (\textit{PPT}) criterion~\cite{peres1996,horodecki1996}. It says that if a quantum state $\rho$ is separable, the state obtained by partial transposition is positive. Partial transposition indicates the application of the partial transpose operator to only one of two subsystems, namely $(\rho^{T_A})_{m \mu, n \nu}=(\rho)_{n \mu, m \nu}$, where the Latin indices refer to subsystem $A$  ($m,n = 1, \ldots N$) and the Greek ones to subsystem $B$ ($\mu,\nu = 1, \ldots M$). The same holds for partial transposition with respect to subsystem $B$. This condition which is necessary for arbitrary dimensions of $\Ham_A$ and $\Ham_B$, is also sufficient for low dimensional systems, $2 \times 2$ and $2 \times 3$. The PPT criterion, is an ``operational recipe" for checking the separability of a given state. Other operational approaches are the reduction~\cite{horodecki1999} and the majorization criteria~\cite{NielsenKempe2001}. From these effective techniques, especially from the PPT condition, a more general technique in terms of linear maps has been derived. However, even if linear maps criteria provide both necessary and sufficient conditions to be satisfied by any bipartite separable system, they do not supply any simple procedure for determining if a bipartite state is separable. In this sense they can be defined ``non operational"~\cite{Bruss2002}.
The positive maps criterion says that a quantum state $\rho$ is separable with respect to a given bipartition $(A, B)$ if and only if for any positive maps $\Lambda_A$  and $\Lambda_B$ one has that $(\Lambda_A \otimes 1) \rho \geq 0$ and $(1 \otimes \Lambda_B)\rho \geq 0$. This condition is, automatically satisfied by the PPT criterion, with $\Lambda_{A/B}=T_{A/B}$. However, the positive maps criterion is nonoperational, since its application would require a complete knowledge of the set of all positive maps, and this is itself an open problem. As special case of positive maps, there is the criterion of the so-called entanglement witnesses.
It says that a quantum state $\rho$ is entangled if and only if there exists an Hermitian operator $W$ such that $\Tr(W \rho)<0$ and $\Tr(W\rho^{sep})\geq0$ for any separable state $\rho^{sep}$.
However, if the expectation value of $W$ on $\rho$ is non negative, this does not guarantee that $\rho$ is separable. This can be clarified introducing the following geometric interpretation: each entanglement witness defines an hyperplane in the Hilbert space $\Ham_{AB}$ that separates $\rho$ from the set of separable states. This geometric approach also helps to understand how to optimize $W$: one can perform a parallel transport of the hyper-planes~\cite{Lewenstein2000} or replace them by curved manifolds~\cite{GŸhne2004} both tangent to the set of separable states, see Fig.~\ref{fig:w}. Entanglement witnesses are explicitly related to positive maps by the Jamiolkowski isomorphism~\cite{Jamiolkowski}, according to which one has that given an entanglement witness $W$ there exist a positive map $\Lambda$, such that $W=(1\otimes\Lambda) P$, where $P$ is the projection operator on maximally entangled states ($P=1/N_{A} \sum_{i,j=1}^{N} \ket{i i} \bra{j j})$). Observe that notwithstanding these optimization techniques, the problem of finding a complete characterization of the convex set of separable states~\cite{Zyczkowski99} is still open, since it would need in principle an infinite number of entanglement witnesses. 

 \subsection{Entanglement measures}\label{generic state measures}
Quantifying entanglement is one of the central topics of quantum information theory. As already mentioned, entanglement represents an extremely complex property of quantum states and for arbitrary states there is not a unique definitive measure.
In Sec.~\ref{pure state measures}, we have seen that in the asymptotic regime, when we consider the tensor product states of a large number of identical copies of the system, due to the uniqueness theorem, the entanglement finds a sort of regularization, at least on pure states. Following an axiomatic approach, there exists a set of postulates that a good measure of entanglement should fulfill: monotonicity under LOCC, vanishing on separable states, continuity, additivity, subadditivity and convexitiy~\cite{h4,Amico08,Bruss2002}. However, there are contrasting believes about the necessity of all these requirements for properly quantifying entanglement, and furthermore it is still not known if there exists a measure satisfying all of them~\cite{vedral97,vidal2000}. A natural question may also be whether, despite the absence of a unique measure of entanglement, we can order all the states of a given bipartite system with respect to the degree of entanglement~\cite{Virmani2000}. The answer is negative, since there exist incomparable states even in the case of pure states. This lack of a single order can be explained, at least qualitatively, in terms of the complexity of entanglement, in the sense that there are many manifestations of entanglement and it may happen that one state shows more entanglement of one type than another state, and the situation is reversed for another type of entanglement~\cite{Miranowicz2004,Verstraete2004}. 

Two very important measures of quantum correlations are the entanglement of distillation $E_D$, defined as the maximal number of singlets that can be produced from a given bipartite quantum state by means of any LOCC operation, and the entanglement cost $E_C$, which is complementary to $E_D$ and measures the minimal number of singlets needed to produce a given quantum state by  LOCC operations. We recall that the singlet state is a particular maximally entangled state for a qubit, whose expression in the computational basis is $1/\sqrt{2}(\ket{0 1}-\ket{1 0})$ (for simplicity, we indicate the tensor product $\ket{\alpha}_A \otimes \ket{\beta}_B$ as $\ket{\alpha \beta}$). It has been proved that the entanglement of distillation and the entanglement cost are, respectively, the lower and upper bounds of any entanglement measure satisfying appropriate postulates in the asymptotic regime~\cite{Horodecki2000prl}. Both $E_D$ and $E_C$ fulfill the additivity axiom, but their continuity has not been completely proved. There are also indications that the entanglement of distillation is not convex~\cite{Shor2001}.  

A technique which enables to extend entanglement measures defined on pure states to arbitrary states is the convex-roof method. It says that if $E$ is a measure defined on pure states, its natural  extension to a mixed state $\rho$ is $E(\rho)= \min \sum_{i}p_i E(\ket{\psi_i})$, where the minimum is taken over all  possible convex combinations of pure states, ${\rho=\sum_k p_k\ket{\psi_k}\bra{\psi_k}}$, being $E$ an increasing function of bipartite entanglement. In this way it is defined another important measure of entanglement, the entanglement of formation:
\begin{equation}
E_{F}(\rho)=\min_{\rho=\sum_k p_k
\ket{\psi_k}\bra{\psi_k}} \sum_k p_k 
S(\rho_{A,k}), \label{eq:entFormation}
\end{equation}
being $\rho_{A,k}=\Tr_B({\ket{\psi_k}\bra{\psi_k}})$. It is the convex-roof extension of the von Neumann entropy~\cite{woot}.
This measure is continuous and convex but its full additivity for bipartite systems has not been established yet~\cite{Vidal2002}.
Finally, by analogy with pure states, a measure of entanglement more convenient for analytical treatment is the convex-roof extension of the purity of one of the two subsystems:
\begin{equation}
E_{P}(\rho)=\max_{\rho=\sum_k p_k
\ket{\psi_k}\bra{\psi_k}} \sum_k p_k 
\pi_{AB}  (\rho_{A,k}). \label{eq:convexpurity}
\end{equation}
Note that in this case we have to maximize over the mixture of pure states corresponding to $\rho$, since the purity is a decreasing function of the bipartite entanglement.

The main topic of this thesis will be the study of the statistical properties of bipartite entanglement, shared by large bipartite quantum systems.  A complete overview will be discussed for the case of bipartite pure states, as anticipated at the end of Sec.~\ref{pure state measures}, by considering the purity as the tool for measuring entanglement. The generalization of this analysis to the case of bipartite mixed state, as might be expected, will reveal far more complicated then the previous one.

 \newpage
\thispagestyle{empty}
\mbox{}
\part{PURE STATES}

 \newpage

\label{intropart1}
\newpage
\thispagestyle{empty}
\mbox{}

\chapter*{Summary}
\addcontentsline{toc}{chapter}{Summary}

In the first part of this thesis we will characterize the statistical properties of bipartite entanglement for a large quantum system in a pure state. This problem will be tackled by studying a random matrix model that describes the
distribution of an entanglement measure,
the purity of one of the two subsystems~\cite{paper1,metastable}. In particular, we will introduce a partition function for the canonical ensemble as a
function of a fictitious temperature. The purity will play the role of the energy of the system, that is different temperatures will correspond to
different degrees of entanglement.  The outcome of this statistical approach to bipartite entanglement will be twofold. On the one hand  
we will distinguish three main entanglement phases (or regions): maximally entangled, typical and separable phase. These regions are
separated by a first and a second order phase transition. On the other hand, by the explicit computation of the entropy of the system, we will evaluate the volume of
the manifolds in the Hilbert space with constant purity (isopurity manifolds), answering to the question of determining how probable is finding an entangled state, for a large quantum system. 

This part of the thesis is organized as follows. In Chap.~\ref{chap2} we will introduce the notation and set the bases of our classical statistical mechanics approach to the problem.
In Chap.~\ref{chap3} we will study the case of positive temperatures, where at very low temperatures we will find very entangled states.
Negative temperatures will be investigated in Chaps.~\ref{chap4} and~\ref{chap5} where we will show the existence of two branches, a stable one associated to a partial factorization of the state, and a metastable branch which contains the 2D quantum gravity point. We will also investigate finite size corrections. Finally, in Chap.~\ref{chap6} we will overview the results of the previous chapters, by reinterpreting them directly from the point of view of quantum information.



 \newpage
\thispagestyle{empty}
\mbox{}
\chapter {A statistical approach to bipartite
entanglement} \label{chap2}
\markboth{A statistical approach to the study of bipartite
entanglement}
{A statistical approach to the study of bipartite
entanglement}

In this chapter we will introduce the bases of our approach to the study of bipartite entanglement for a large quantum system. 
The first requirement will be setting the statistical ensemble of pure states, and this will led to the introduction of a random matrix model for the system in Sec.~\ref{Haar measure}. The next step will be the definition, in Sec.~\ref{model}, of the fundamental tool for this statistical approach: the partition function of the system for the canonical ensemble, where the role of the energy will be played by the entanglement measure we will choose, the purity of one of the two subsystems. Since the exact computation of the partition function results very complicated due to the positivity constraint on the eigenvalues, we will recur to the method of steepest descent and impose a set of saddle point equations for the Schmidt coefficients. 

\section{Random matrix model and Haar measure} \label{Haar measure}

\begin{figure}[h]
  \begin{center}
    \includegraphics[scale=0.6]{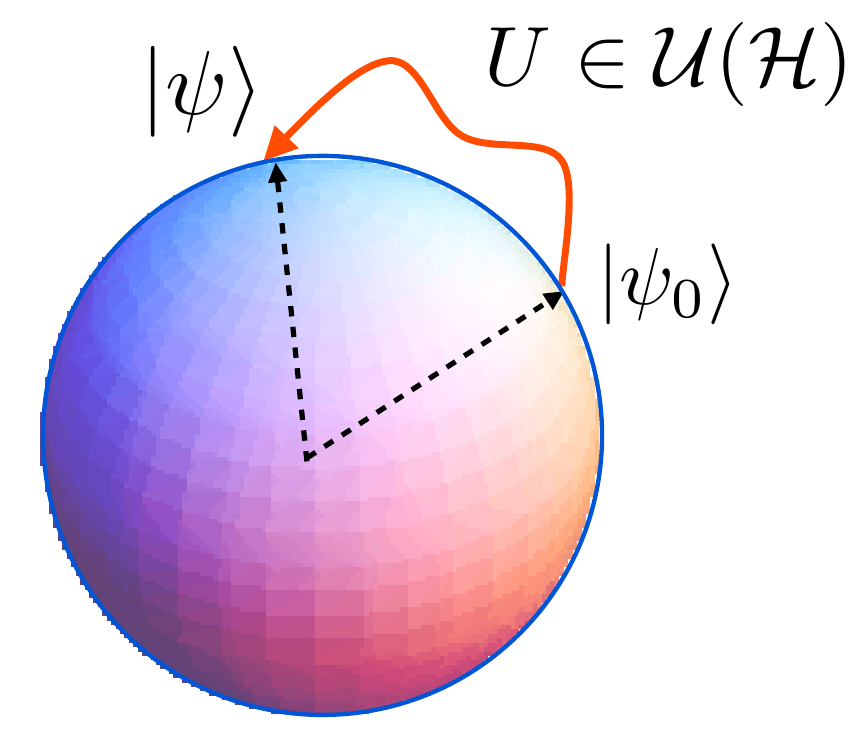}
    \caption{Typical states $\ket{\psi}$ sampled according to the unique left and right invariant Haar measure on the unitary group $\mathcal{U}(\Ham)$, from an arbitrary state $\ket{\psi_0}$: $\ket{\psi}=U\ket{\psi_0}$, with $U \in \mathcal{U}(\Ham).$}
    \label{fig:Haar}
  \end{center}
\end{figure}

Consider a bipartite system ($A$,$B$). In Sec.~\ref{bipartite entanglement} we have seen that the states of the total system live in the tensor product Hilbert space $\mathcal{H}=\mathcal{H}_A\otimes\mathcal{H}_B$. Without loss of generality we set
$\dim\mathcal{H}_A=N \leq \dim\mathcal{H}_B=M$. We also assume that the system is in a pure state $\ket{\psi}\in\mathcal{H}$.
A statistical analysis of bipartite entanglement is equivalent to answer the following question: what is the probability of finding a state with a given value of entanglement in the Hilbert space $\Ham$? In order to answer this question, we have to solve two preliminary points:
\begin{itemize}
\item introducing a proper measure of entanglement 
\item defining a probability measure on $\Ham$, according to which the states are sampled.
\end{itemize} 
The answer to the first point is almost straightforward. In Chap.~\ref{chap1} we have seen that for the case of pure states the von Neumann entropy of one of the two subsystems, $S(\rho_A)=S(\rho_B)=-\rho_{A}\log{\rho_A}$, is  a good measure of bipartite entanglement. However, at the end of Sec.~\ref{pure state measures} we have also noticed that the von Neumann entropy can be replaced for computational convenience by a more simple function, the purity, which derives from its linearization:
\begin{equation}
\pi_{AB}=\Tr(\rho_A^2)=\Tr(\rho_B^2)
=\sum_{j=1}^N \lambda_j^2  \qquad \pi_{AB}\in [1/N,1] .
\label{eq:purityN}
\end{equation}
The minimum is attained when all the eigenvalues $\lambda_j$, the Schmidt coefficients of the global state $\ket{\psi}$, are equal to
$1/N$  (subsystem $A$ in a completely mixed state and maximal entanglement between the two bipartitions), while the maximum is attained when one eigenvalue is 1 and all others are 0 (this
detects a factorized (unentangled) state). 
Let us now switch to the second point. The idea of taking a quantum state randomly is equivalent to assuming the minimal knowledge about the system. Random states can be considered \textit{typical} in the sense that they represent the states to which an arbitrary evolving quantum state can be compared. For the case of pure states there is a unique unbiased measure, as it is  enough to identify the minimal knowledge with the maximal symmetry for the system. Thus the sampling criterion has to be invariant under the full group of unitary transformations~\cite{hall1998}, and corresponds to the unique (left and right) invariant Haar (probability) measure $d\mu_{H}(U)$ on the unitary group $U\in\mathcal{U}(\Ham)\simeq\mathcal{U}(L)$:
\begin{equation}
d\mu_{H}(U)=d\mu_{H}(V), \qquad \forall \ U,V \in \mathcal{U}(\Ham).
\end{equation}
We will define the typical vector states $\ket{\psi}$~\cite{Lubkin78,page} as pure random states given by the action of a random unitary matrix $U\in\mathcal{U}(\Ham)$ on an arbitrary reference state $\ket{\psi_0}\in\Ham$, $\ket{\psi}=U\ket{\psi_0}$, see Fig.~\ref{fig:Haar}. The final state $\ket{\psi}$ will therefore be
independent on
$\ket{\psi_0}$.  Henceforth, all measures will be tacitly assumed to be probability measures (normalized to $1$), if not stated otherwise. 
For example, $\mu_H(\mathcal{U}(\Ham))=1$. 
Nevertheless, it is important to add that the Harr measure is not the only possible sampling choice. Other distributions can also be considered but they encode additional information on the system (in this sense, Haar is the most neutral). These alternative distributions could be treated in our approach by constraining the system by means of Lagrange multiplier. For an approximate realization of this Haar measure by means of short quantum circuits see~\cite{Arrow}, where it is proved that one can extract Haar-distributed random states by applying only a polynomial number of random gates. 

\subsection{Induced measure on subsystems}\label{sec:Induced measure on subsystemsPURE}
In order to describe the statistical properties of the purity of a bipartite system we need to determine the induced measure on the reduced density operator describing the state of one of the two subsystems, for instance $\rho_A$. Representing $\rho_A$ in terms of its spectral decomposition $\rho_A= \sum_{i}^{N} \lambda_i P_i$, with $\{P_n\}_{n=1, \ldots, N}$ a complete set of orthogonal projections defined up to a unitary rotation, 
we have that the set of the Hermitian reduced density matrices $\mathcal{S}_A$ can be written as a Cartesian product~\cite{Zyczkowski98,Zyczkowski99}:
\begin{equation}
\mathcal{S}_A= P_A \times \Lambda_A,
\end{equation}
where $P_A$ is the family of the complete sets of orthonormal projections on $\Ham$ and $\Lambda_A$ is the Schmidt simplex. 
It follows that the measure on the reduced density operators $\rho_A$ is a product measure
\begin{equation}\label{eq:measure on A}
d\mu(\mathcal{S}_A)\equiv d\mu(\rho_A)=d\mu(U_A) \times d\sigma({\Lambda_A}),
\end{equation}
where $d\mu(U_A)$ is defined on the unitary group $\mathcal{U}(\Ham_A)\simeq \mathcal{U}(N)$, i.e. on the eigenvectors of $\rho_A$, and $d\sigma({\Lambda_A})$ on its eigenvalues. By construction the two measures are both induced by the Haar measure over the larger unitary group $\mathcal{U}(\Ham)$ and are given by the Haar measure on $\mathcal{U}(\Ham_A)$, $d\mu(U_A)=d\mu_H(U_A)$ and, for the eigenvalues, by~\cite{Lubkin78,page}: 
\begin{eqnarray}
d\mu(\Lambda_A) = C_{N, M} \prod_{1 \leq i<j \leq N}(\lambda_i-\lambda_j)^2
 \prod_{1 \leq \ell \leq N} \lambda_\ell^{(M-N)} \delta \left(1-\sum_{1 \leq k \leq N} \lambda_k \right)d^{N} \lambda.
\label{eq:measure}
\end{eqnarray}
The square of the Vandermonde determinant ($\prod_{1 \leq i<j \leq N}(\lambda_i-\lambda_j)$) derives from integrating out the eigenvectors, and the normalization constant can be written in terms of the Euler gamma function $\Gamma(x)$:
\begin{equation}\label{eq:normcoeffpure}
C_{N,M}=\frac{\Gamma(L)}{\prod_{0 \leq j \leq N -1} \Gamma(N-j+1) \Gamma(M-j)}.
\end{equation}
It is worth observing that the measure over the eigenvalues in not uniform. In particular the Vandermonde determinant introduces a repulsion between pairs of Schmidt coefficients. 
\section{The model}\label{model}
In Sec.~\ref{bipartite entanglement} we mentioned some studies related to the idea of determining iso-entanglement manifolds, through the introduction of invariants under LOCC 
\cite{grassl1998,Linden1999,albeverio2001}, and by means of the complementary study of the orbits in the Hilbert space associated to a given quantum state ~\cite{Kus01}.
The statistical approach presented in this thesis is slightly different, in the sense that we will describe manifolds with constant average purity. However, at the end of this chapter (Sec.~\ref{equivalntensembles}) we will prove the equivalence between this approach and the one based on the microcanonical ensemble, in which the purity of distinct manifolds is fixed.  
Let us start with the introduction of a partition function~\cite{paper1} from which all the thermodynamic quantities, for example the entropy or
the free energy, can be computed:
\begin{equation}
\label{eq:partitionfunction}
\cZ_{AB}(\beta)=\int d\mu(\rho_A)  \exp\left(-\beta {{N}^{\alpha}}
\pi_{AB}\right).
\end{equation}
Recall that the measure $d\mu(\rho_A)$ is a product measure, see Eq.~(\ref{eq:measure on A}).
The parameters $\alpha$ and $\beta$ in definition (\ref{eq:partitionfunction}) are extremely relevant for our analysis.  
The coefficient $\beta$ can be interpreted as the inverse of a (fictitious) temperature $T$ selecting different regions of entanglement, $\beta=1/T$. More formally, it is a Lagrange multiplier which enlightens different regions of the Hilbert space with a given value of average purity, $\langle\pi_{AB}\rangle$.  In particular  we have that:
\begin{itemize}
\item for  $\beta=0$ we obtain typical states (sampled by $d\mu(\rho_A))$, 
\item for $\beta >0$ we expect to find more entangled states, which in the limit $\beta\rightarrow\infty$ should become maximally entangled ($\langle\pi_{AB}\rangle=1/N$),
\item for $\beta<0$  we are going towards the set of separable states, reached when $\beta \rightarrow -\infty$ ($\langle\pi_{AB}\rangle=1$).
\end{itemize}
In this scenario, the purity corresponds to the energy of the system, and varies in the interval $[1/N, 1]$ according to the different profiles of the distribution of the Schmidt coefficients. The parameter $\alpha$ is a \textit{scaling coefficient}. It is a positive integer (either 2 or 3, as we shall see) and refers to the scaling properties of the system. Its value needs to be chosen in
order to yield the correct thermodynamic limit, that is in order to keep the exponent of the partition function extensive. Since the number of degrees of freedom of $\rho_A$ is ${N}^2-1\simeq {N}^2$ (for a large system), $\alpha$ needs to satisfy the constraint:
\begin{equation}
\label{eq:piscale}
{N}^\alpha\langle\pi_{AB}\rangle=\Ord{{N}^2}.
\end{equation}
Around maximally entangled states (for $\beta>0$) we have $\langle\pi_{AB}\rangle=\Ord{1/N}$ so $\alpha=3$, while
around separable states (for $\beta<0$) we have
$\langle\pi_{AB}\rangle=\Ord{1}$ and hence $\alpha=2$. 

In the following we will assume $N=M$, since this does not change the qualitative picture, the extension to $N \neq M$ being straightforward but computationally cumbersome. \subsection{Saddle point equations}\label{The saddle point equations}
From the cyclic property of the trace, we have that the purity depends only on the eigenvalues of  $\rho_A$, thus the integral over the unitary group deriving from the measure on $\rho_A$ factorizes in the partition function, and we have only to integrate on the (positive) Schmidt coefficients, modulo the constant factor $C_{N,N}$:
\begin{eqnarray}\label{eq: partition function explicit}
\cZ_{AB}=\int_{\lambda_i\geq 0}
d^N\lambda\prod_{1 \leq i<j \leq N}(\lambda_i-\lambda_j)^2\delta\left(1-\sum_{1\leq i \leq N}\lambda_i\right)e^{-\beta
N^\alpha \sum_{1\leq i \leq N}\lambda_i^2}\; . 
\end{eqnarray}
Recall they we are assuming all the measures to be probability measures, thus $\mu_{H}(\mathcal{U}(\Ham_A))=1$.
By introducing a Lagrange multiplier for the delta function in Eq. (\ref{eq: partition function explicit}), we get:
\begin{eqnarray}\label{eq: partition function alpha}
\cZ_{AB}&=&N^2
\int_{-\infty}^\infty\frac{d\xi}{2\pi}\int_{\lambda_i\geq 0}
d^N\lambda \ e^{i N^2 \xi(1-\sum_{1 \leq i \leq N} \lambda_i)-\beta N^\alpha
\sum_{1 \leq i \leq N}\lambda_i^2+2\sum_{1 \leq i<j \leq N}\ln|\lambda_i-\lambda_j|}\; .\nonumber\\ 
\end{eqnarray}
The argument of the partition function can be interpreted as the Boltzmann factor of a gas of $N$
point charges (Coulomb gas) at positions $\lambda_i$'s on the positive half-line~\cite{Dyson1962}. The potential energy of the gas of eigenvalues is given by a harmonic potential and a two dimensional electrostatic repulsion between pairs of charges. The analogous of the integral in Eq.~(\ref{eq: partition function alpha}) is known for the case in which the integration limits
are $-\infty<\lambda_i<+\infty$, as Selberg's integral~\cite{Mehta2004}.
The constraint of the positivity of the eigenvalues makes the
computation of this integral far more complicated.  The exact solution can always be found by means of the
orthogonal polynomials method, but the expressions for $\cZ_{AB}$ grows
enormously in complexity with increasing $N$, see~\cite{Giraud,paper1} for the first few moments.
In the limit of large $N$, we will overcome this problem by looking for the stationary point of the exponent of the partition function $\cZ_{AB}$:
\begin{equation}
\beta V= \beta N^\alpha
\sum_{1 \leq i \leq N}\lambda_i^2 - 2 \sum_{1\leq i<j \leq N}\ln|\lambda_i-\lambda_j| - i N^2 \xi \left( 1- \sum_{1 \leq i \leq N} \lambda_i\right) \, .
 \label{eq:freeF}
 \end{equation}
The function $V$ represents the potential associated to the Coulomb gas of eigenvalues. By deriving $\beta V$
with respect to both the $\lambda_i$'s and
$\xi$ we get $N+1$ \textit{saddle point} equations for the system:
\begin{eqnarray}
\label{eq:stat1}-2\beta N^\alpha \lambda_i+2\sum_{1\leq j\neq i \leq N}\frac{1}{\lambda_i-
\lambda_j}-iN^2 \xi&=&0,\quad \forall \ i \in \{1,\ldots,N\},
\label{eq:normal0}
\\
\label{eq:normal}\sum_{1\leq i \leq N}\lambda_i&=&1.
\end{eqnarray}
In particular, the domain of integration for $\xi$ lies on
the real axis, but we will see that the saddle point for $\xi$
lies on the imaginary $\xi$ axis. Thus the contour needs to be deformed to pass through this point along the
line of steepest descent, see appendix~\ref{sec:Method of steepest descent}. The minimum or the maximum of $V$, for positive or negative temperatures respectively, in the thermodynamic limit, will correspond to the free energy of the system. 
In the following chapters we will separately analyze the range of
positive and negative temperatures, and unveil the presence of some critical temperatures  for the system, 
associated to first and second order phase transitions.

\section{Equivalence between canonical and microcanonical ensemble}\label{equivalntensembles}
Before considering different ranges of temperatures, we will devote this section to the proof of the equivalence, for our system, between  the microcanonical and the canonical ensembles. In the first case we represent the total Hilbert space $\Ham$ as given by isopurity manifolds, that is with a fixed value of the purity. On the other hand in the canonical ensemble we fix only the average purity for each submanifold. What follows will be independent on the specific value of the temperature $1/\beta$. 

\begin{figure}[h]
\centering
\includegraphics[height=0.4\columnwidth]{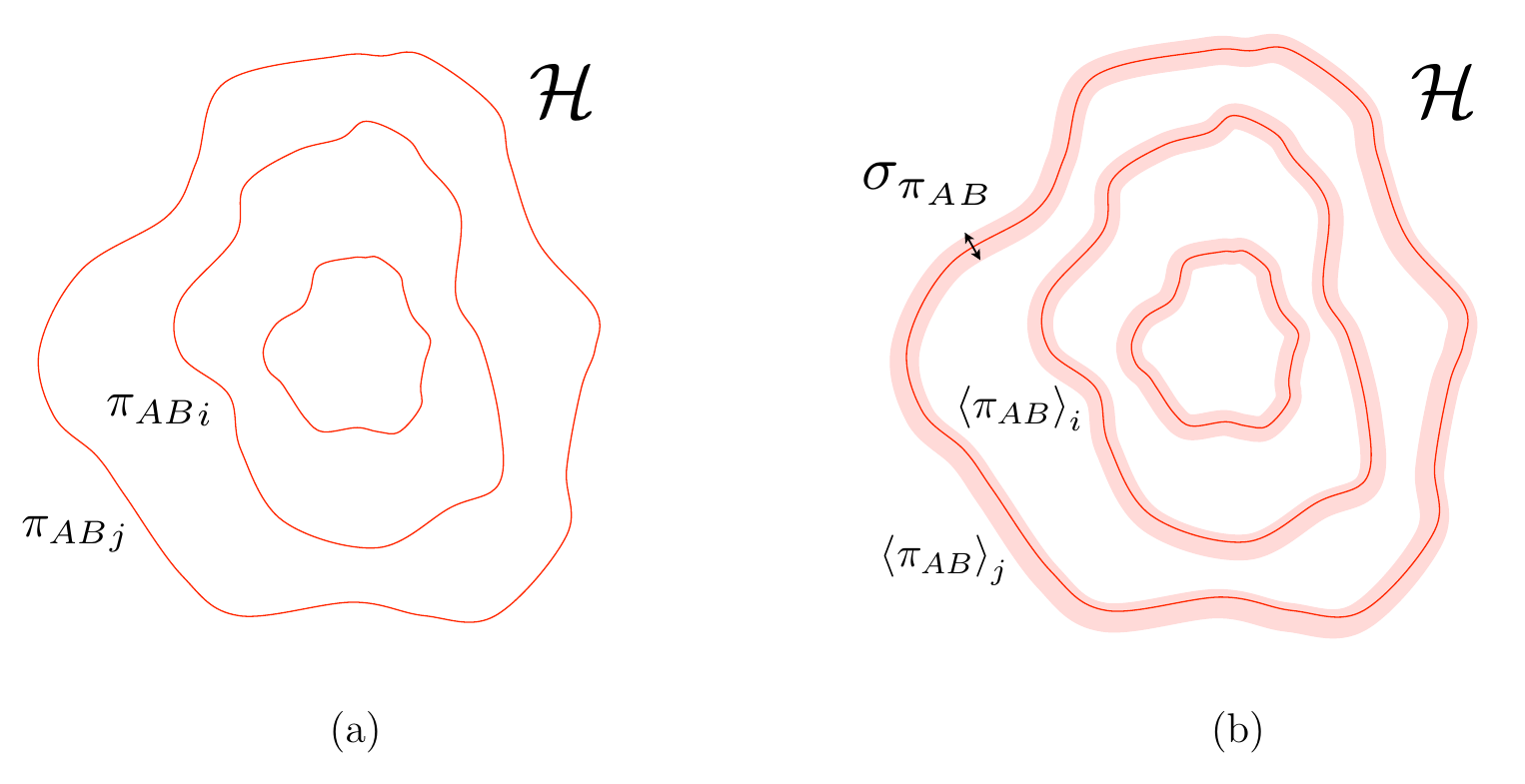}
\caption{Microcanonical (a) and canonical (b) ensembles. In the microcanonical ensemble the Hibert space $\Ham$ of a bipartite system is represented in terms of iso-purity manifolds, that is  manifolds with a fixed value of the purity ${\pi_{AB}}_i$, the index $i$ identifying the corresponding submanifold. With the canonical ensemble we release the assumption of fixing the exact value of the purity on each submanifold, and fix only its average value ${\langle \pi_{AB} \rangle}_i$ by taking into account its fluctuations $\sigma_{\pi_{AB}}$.}
 \label{fig:microCan}
\end{figure}

In order to compare the two approaches, we compute the relative fluctuations of the rescaled average purity:
\begin{equation}
\mathcal{F}_{\pi'_{AB}}= \frac{\sigma_{\pi'_{AB}}}{\langle \pi'_{AB} \rangle}, 
\end{equation}
where the average purity $\langle {\pi'_{AB}} \rangle=N^{\alpha} \langle \pi_{AB} \rangle$ (${\pi'_{AB}} =N^{\alpha} \pi_{AB}$)and its mean square fluctuation  $\sigma^2_{\pi'_{AB}}$ in the canonical ensemble are given by~\cite{huang}:
\begin{eqnarray}
\langle \pi'_{AB} \rangle &=& \frac{\int d\mu(\rho_A) \; \pi'_{AB} \;e^{-\beta
\pi'_{AB}}}{\cZ_{AB}} = - \frac{1}{\beta} \frac{\partial\ln{\cZ_{AB}} }{\partial \beta} \label{eq:avPI}\\ 
\sigma^2_{\pi'_{AB}}&=& \langle {\pi'_{AB}}^2 \rangle - \langle {\pi'_{AB}} \rangle^2 = -\frac{\partial\langle {\pi'_{AB}} \rangle}{\partial \beta}.\label{eq:varPI}
\end{eqnarray}
We recall that the coefficient $\alpha$ has to be chosen in order to yield an extensive quantity at the exponent of the partition function, namely of $\mathcal{O}({N}^2)$ (see Eq. (\ref{eq:piscale})). It can be easily shown that $\mathcal{F}_{\pi_{AB}}=\mathcal{F}_{\pi_{AB}'}$. See Fig. \ref{fig:microCan}: the indices $i$ and $j$, labeling the purity and its average, identify different isopurity manifolds in $\Ham$.
\\
The proof of the equivalence of the two ensembles is thus straightforward:
\begin{equation}
\mathcal{F}_{\pi_{AB}}=\mathcal{F}_{\pi_{AB}'} = \frac{(-\partial\langle \pi_{AB}' \rangle/\partial \beta \, )^{1/2}}{\langle \pi_{AB}' \rangle}= \mathcal{O}\left(\frac{1}{N}\right),
\end{equation}
namely the fluctuations of the average purity can be neglected for a large quantum system, $N >>1$.

\chapter{Positive temperatures}\label{chap3}
\markboth{Positive temperatures}{Positive temperatures}

\label{sec:positivetemp}

In this chapter we will study how bipartite entanglement between $A$ and $B$ can affect the probability distribution of eigenvalues of one of the two subsystems, in the range of positive temperatures. In particular, we will show the emergence of two classes of symmetries. In Sec.~\ref{natural scaling}, by the introduction of a natural scaling for the eigenvalues of subsystem $A$ in the limit of a large quantum system, we will translate the problem of finding the most probable set of eigenvalues $\lambda_i$, into the task of getting the corresponding density distribution. In Sec.~\ref{tricomi} the saddle point equations will be then reduced to an integral equation for this density function, know as Tricomi equation. In Sec.~\ref{solution} we will find a two parameter continuous family of solutions, among which we will select the one yielding the higher Boltzmann factor for the Coulomb gas of eigenvalues, i.e. the most probable density distribution for  every fixed value of the temperature. We will see that there exist two classes of symmetries for the density distribution, lying in nearby intervals of positive temperatures, that can be interpreted as two phases for the system: the \textit{maximally entangled} and the \textit{typical} phases. The two regimes are separated by a second order phase transition for the system, whose presence will be unveiled in Sec.~\ref{secondOrderPhTR} by studying the entropy of the system. We will conclude with a reinterpretation, in Sec.~\ref{EntropyVSentanglement}, of the results obtained in this chapter, directly from the point of view of entanglement. 

\section{Natural scaling and density of eigenvalues}\label{natural scaling}
\begin{figure}[h]
\centering
\includegraphics[height=0.30\columnwidth]{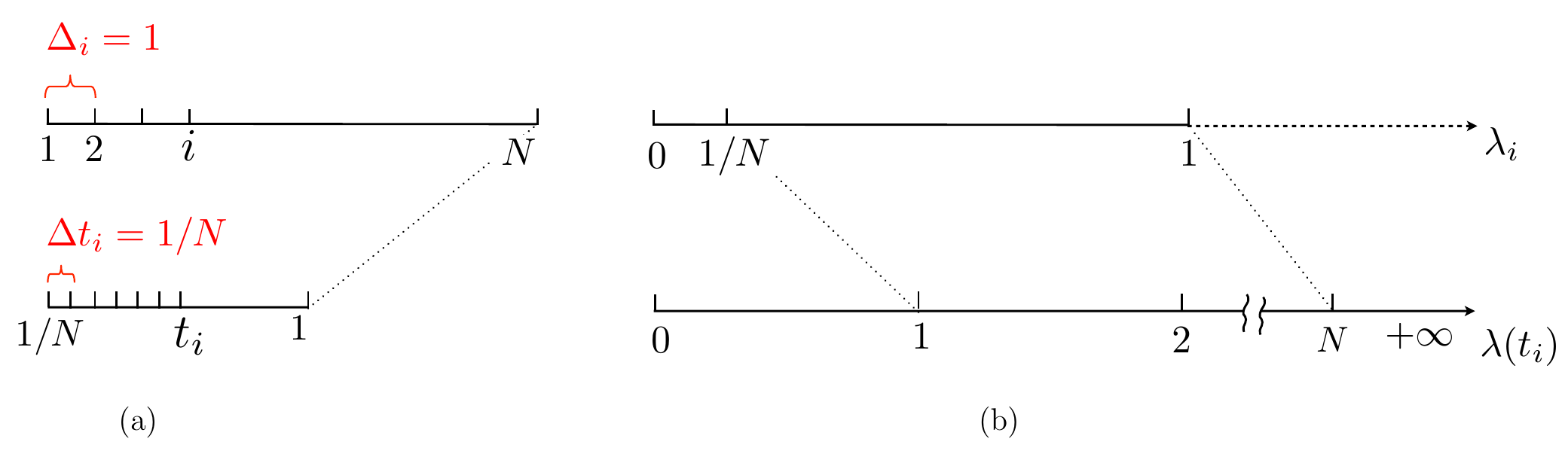} 
\caption{Natural scaling. (a) Contraction of indices $i$  from $\{1, \ldots N\}$ to the interval $[1/N,\ldots,1]$ associated to $t_i$. (b) Expansion of the eigenvalues $\lambda_i$ from the interval $[0,1]$ to the interval $[0,N]$ for $\lambda(t_i)$.}
\label{fig:natscaling1}
\end{figure}

In Sec.~\ref{model} we argued that, in order to yield the correct thermodynamic limit, the exponent of the partition function
\begin{equation}
\cZ_{AB}=\int d\mu(\rho_A)  \exp\left(-\beta {{N}^{\alpha}}
\pi_{AB}\right),
\end{equation}
has to be an extensive quantity, that is of the same order $N^2$ of the volume of states $\rho_A$. See Eq.~(\ref{eq:piscale}).
As far as we increase $\beta$ typical states enhance their degree of entanglement and become maximally entangled in the limit $\beta \rightarrow + \infty$. It then follows that for this range of temperatures the correct scaling exponent  is $\alpha=3$.
In order to determine the thermodynamic properties of the system, we
solve the saddle point equations (\ref{eq:stat1})-(\ref{eq:normal})
in the continuous limit by introducing the natural scaling
\begin{equation}
\lambda_i=\frac{1}{N} \lambda (t_i), \qquad 0< t_i=\frac{i}{N}\leq 1\quad \mbox{with} \quad \Delta t_i = t_{i+1}-t_i=\frac{1}{N}.
\label{eq:scalinglambda}
\end{equation}
Observe that $\lambda_i \Delta_i = \lambda(t_i) \Delta t_i $, where $\Delta_i=(i+1)-i=1$. In Fig.~\ref{fig:natscaling1} we show the effects of this scaling from the point of view of the eigenvalues and their labeling indices: they become continuous functions in the limit $N \rightarrow \infty$.
\\In this scaled setup, the properties of the system are determined by the \textit{density distribution} $\rho(\lambda)$ of the Schmidt coefficients. Obviously, it will depend on $\beta$, that is on the level of entanglement between $A$ and $B$.
In the limit $N\rightarrow\infty$ we define in general: 
\begin{equation}
\rho(\lambda)=\int_0^1 dt \; \delta(\lambda-\lambda(t)).
\label{eq:rhodef}
\end{equation}
The probability distribution of the Schmidt coefficients needs to satisfy two consistency conditions:
\begin{itemize}
\item nonnegativity: $\rho(\lambda) \geq 0$, for all $\lambda$
\item normalization:
\begin{equation}
\int d\lambda \; \rho(\lambda) = 1.
\label{eq:norm}
\end{equation}
\end{itemize}
As will be discussed in the next section, the above constraints on $\rho(\lambda)$ will select the physical solutions of the saddle point equations, for different values of $\beta$. Notice that the standard definition (\ref{eq:rhodef}) for $\rho(\lambda)$ naturally satisfies the above constraints. 
\section{Tricomi's equation}\label{tricomi}
The set of $N+1$ saddle point equations (\ref{eq:stat1})-(\ref{eq:normal}) for the Schmidt coefficients $\lambda_i$ and the Lagrange multiplier $\xi$, reduces to a couple of integral equations for the density $\rho(\lambda)$ and $\xi$. In both cases, they depend on the inverse temperature $\beta$. 

\begin{figure}[h]
\centering
\includegraphics[width=1\columnwidth]{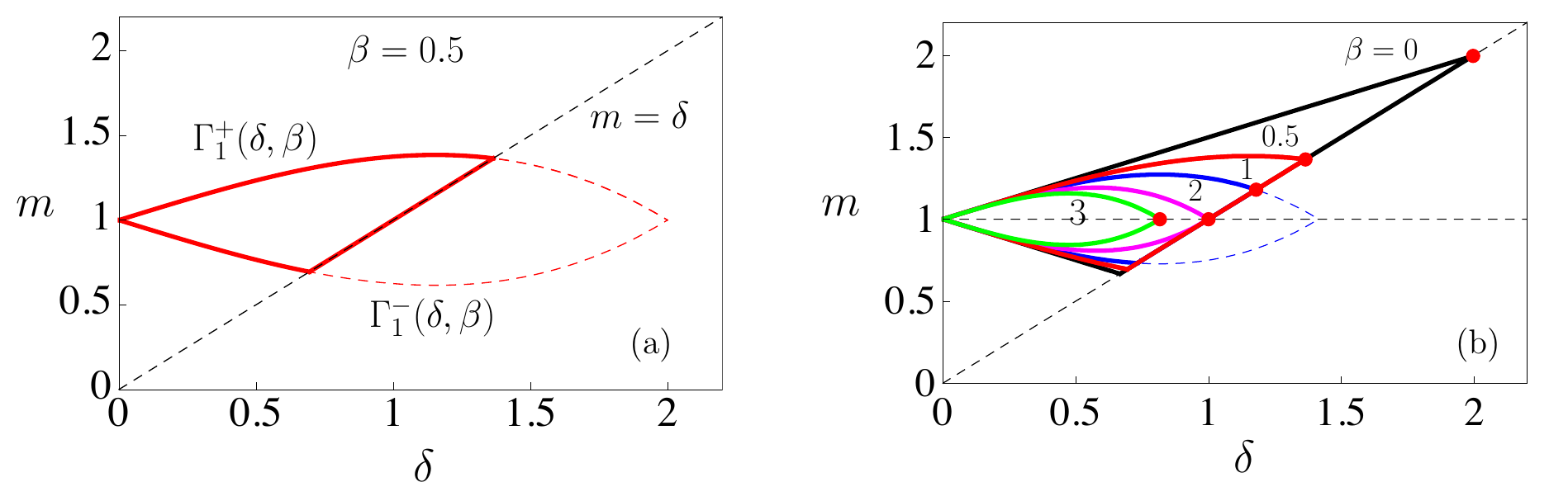}
\caption{(a) Solution domain for $\beta=0.5$. The domain is the region with full color, whereas the curved and straight dashed lines are $\Gamma_1^{\pm} (\delta,\beta)$ and $m=\delta$, respectively. The line $m=\delta$ corresponds to the positive eigenvalues condition. (b) Solution domains for different
temperatures: for each value of $\beta$ (indicated) the relative full line encloses the region of the parameter space where the
eigenvalue density is positive. For each temperature, the red dot indicates the right corner of the eye-shaped domain.}
\label{fig:positive_eye}
\end{figure}

Let us consider the sum over index $j$ in Eq. (\ref{eq:stat1}). According to the natural scaling it becomes:
\begin{eqnarray}\label{eq:tecnica}
 \sum_{j\neq i} \frac {N \Delta t_j}{\frac{\lambda(t_i)}{N}-\frac{\lambda(t_j)}{N}}&\sim& N^2 P \int_{0}^{1}  \frac{d y}{\lambda(x)-\lambda(y)}\nonumber\\&=&N^2 P \int_{0}^{1} \frac{ d y}{\lambda(x)-\lambda(y)} \int_{0}^{\infty}  d \lambda' \delta(\lambda'  -\lambda(y))\nonumber\\&=&N^2 P  \int_{0}^{\infty}  d \lambda'\frac{\rho(\lambda')} {\lambda(x)-\lambda'},
\end{eqnarray}
where $P$ indicates the Chauchy principal value:
\begin{equation}
P  \int_{0}^{\infty}  d \lambda'\frac{\rho(\lambda')} {\lambda-\lambda'}=\lim_{\; \varepsilon \to 0^+}\Bigg[\int_{0}^{\lambda-\varepsilon}  d \lambda'\frac{\rho(\lambda')} {\lambda-\lambda'}+\int_{\lambda+\varepsilon}^{\infty}  d \lambda'\frac{\rho(\lambda')} {\lambda-\lambda'}\Bigg].
\end{equation}
In Eq. (\ref{eq:tecnica}) we have also introduced the identity $ \int_{0}^{\infty}  d\lambda' \delta(\lambda'  -\lambda)=1$.
With this technique, we get that the saddle point equations, with $\alpha=3$ and $\beta\geq0$, become:
\begin{eqnarray}
-\beta\lambda + P\int_0^\infty
d\lambda'\frac{\rho(\lambda')}{\lambda-\lambda'}- i\frac{\xi}{2}&=&0 \label{eq:sadpt2},\\
\int_{0}^{\infty} d\lambda \; \rho(\lambda)  \lambda&=&1.
\label{eq:normeigv}
\end{eqnarray}
Eq. (\ref{eq:sadpt2}) is an integral equation for the density of eigenvalues $\rho(\lambda)$. It is a singular Fredholm equation of the first kind, known as Tricomi's
equation~\cite{tricomi}. According to Tricomi~\cite{tricomi} the
solution, $\rho(\lambda)$ lies in a compact interval
$[a,b]$, ($0\leq a \leq b$). We expect that as this integral equation depends on the temperature, also its solution will depend on $\beta$. 
Let us set
\begin{equation}
\label{eq:new variable}
\lambda= m+ x \delta,\qquad
\phi(x)= \rho(\lambda ) \delta,
\end{equation}
where
\begin{equation}
\label{eq:mdeltadef}
m=\frac{a+b}{2}, \qquad \delta=\frac{b-a}{2}, \qquad 0\leq\delta\leq m.
\end{equation}
The last expression maps the domain $[a,b]$ of $\rho(\lambda)$ into the interval $[-1, 1]$ associated to $\phi(x)$. Observe that by definition $\phi(x)$ fulfills the normalization condition (\ref{eq:norm})
\begin{equation}
\int_{-1}^{1} d x \;\phi(x)=1.
\end{equation}
In the new rescaling Eq. (\ref{eq:sadpt2}) becomes
\begin{equation} \frac{1}{\pi} P\int_{-1}^1 d y\;\frac{\phi(y)}{y-x} =
g(x),
\label{eq:Tricomi}
\end{equation}
with
\begin{equation}\label{gx}
g(x)=-\frac{1}{\pi}\left(i\xi \frac{\delta}{2} +\beta \delta m +\beta\delta^2 x\right).
\end{equation}
We recall that $\xi$ is a Lagrange multiplier introduced in the partition function due to the unit trace condition on the reduced density operator $\rho_A$, see Eq. (\ref{eq: partition function alpha}).  According to Tricomi's theorem, the normalized solution of this integral equation is
\begin{equation}\label{phix}
\phi(x)=-\frac{1}{\pi}P\int_{-1}^1 d y\;
\sqrt{\frac{1-y^2}{1-x^2}}\frac{g(y)}{y-x}+\frac{1}
{\pi\sqrt{1-x^2}}\ .
\end{equation}
The constraint (\ref{eq:normeigv})
\begin{equation}
\int_{-1}^{1} \lambda\, \phi(x)\, dx=1
\end{equation}
fixes the Lagrange multiplier $\xi$
\begin{equation}
\xi= i \left(\frac{4}{\delta^2}+m\left(2 \beta -\frac{4}{\delta^2}\right)\right).
\end{equation} 
Notice that $\xi$ is an imaginary number, as anticipated at the end of Sec.~\ref{The saddle point equations}. Thus the contour of integration for $\xi$ in the partition function needs to be deformed with respect to real axes, in order to pass through this point along the line of steepest descent. Finally, by substituting the last expression in Eq.~(\ref{gx}) and performing the principal value integral in (\ref{phix}) we get
\begin{equation}\label{eq:phi(x)}
\phi(x)=
\frac{1}{\pi \sqrt{1-x^2}}\left[ 1+\frac{\beta \delta^2}{2}
+\frac{2(1-m)}{\delta}x -\beta \delta^2 x^2\right].
\end{equation}
The solution of the Tricomi equation depends on the fictitious (inverse) temperature $\beta$.
The physical solutions must have a density $\phi(x)$ that is nonnegative for all $x\in[-1,1]$.
The numerator of (\ref{eq:phi(x)}) is non negative for $x\in [x_-,x_+]$, with 
\begin{equation}
\label{eq:roots}
x_{\pm}=\frac{1}{\beta\delta^2} \left(\frac{1-m}{\delta}\pm\sqrt{\Delta}\right),
\end{equation}
where
\begin{equation}
\label{eq:discriminant}
\Delta=\left(\frac{1-m}{\delta}\right)^2+\beta\delta^2 \left(1+\frac{\beta \delta^2}{2}\right). 
\end{equation}
Notice that $\Delta\geq0$ for every $m$ and $\delta$. 
The condition $[-1,1]\subseteq [x_{-}, x_{+}]$, imposes that the points $(\delta,m)$ should be restricted to a (possibly cut) \textit{eye-shape} domain given by:
\begin{equation}\label{eq:positive_temp_eyesShapeDomain}
\max\left\{ \delta,\Gamma_1^{-} (\delta,\beta)\right\} \leq m \leq \Gamma_1^{+} (\delta,\beta),
\end{equation}
where  (\ref{eq:mdeltadef}) expresses the positivity of eigenvalues and $m=\Gamma_1^{\pm} (\delta,\beta)$ are the level curves corresponding to $x_{\pm}=\pm1$:
\begin{eqnarray}
\label{Gamma1}
\Gamma_1^{\pm} (\delta,\beta) = 1\pm\frac{\delta}{2}\left(1-\frac{\beta\delta^2}{2}\right).
\end{eqnarray}
See Fig.~\ref{fig:positive_eye}(a).
\\
Observe that the level curves $\Gamma_1^{\pm} (\delta,\beta)$ are symmetric with respect to the line $m=1$ and intersect at $\delta=0$ and at $\delta=\sqrt{2/\beta}$. The right corner of the eye is at
\begin{equation}
(\delta,m)=\left(\sqrt{\frac{2}{\beta}},1\right),
\label{eq:beta>2}
\end{equation}
and belongs to the boundary as long as $\beta\geq2$. For $\beta<2$ the eye is cut by the line $m=\delta$, as shown in Fig.~\ref{fig:positive_eye}(b). 
In other words the domain is symmetric as long as $\beta \geq 2$, whereas for lower values of $\beta$ this symmetry is lost. Let  us remark that all points inside each domain correspond to solutions of the saddle point equations, that is, we have a \emph{two parameter  continuous family} of solutions.

In the next section we will show how to select the density of eigenvalues which
maximizes the partition function. It  corresponds to the most probable distribution of the Schmidt coefficients, for a given temperature. In particular, we will distinguish two classes of distributions. They will correspond to different ranges of $\beta$, associated to the two symmetries we have found for the physical domains of the solution of the Tricomi equation: $0 \leq \beta < 2$ and $\beta \geq 2$. 
\section {The solution}\label{solution}
We will now compute the density of eigenvalues which gives the maximum contribution to the partition function. It is the solution corresponding to the maximum Boltzmann factor for the gas of eigenvalues, or equivalently to the minimum of $\beta V$ (or $V$ since $\beta\geq 0$).
The potential computed for this minimizing solution will be the free energy of the system. 

\subsection {Thermodynamics of the system and minimization of the free energy}
\begin{figure}[h]
\centering
\includegraphics[width=0.78\columnwidth]{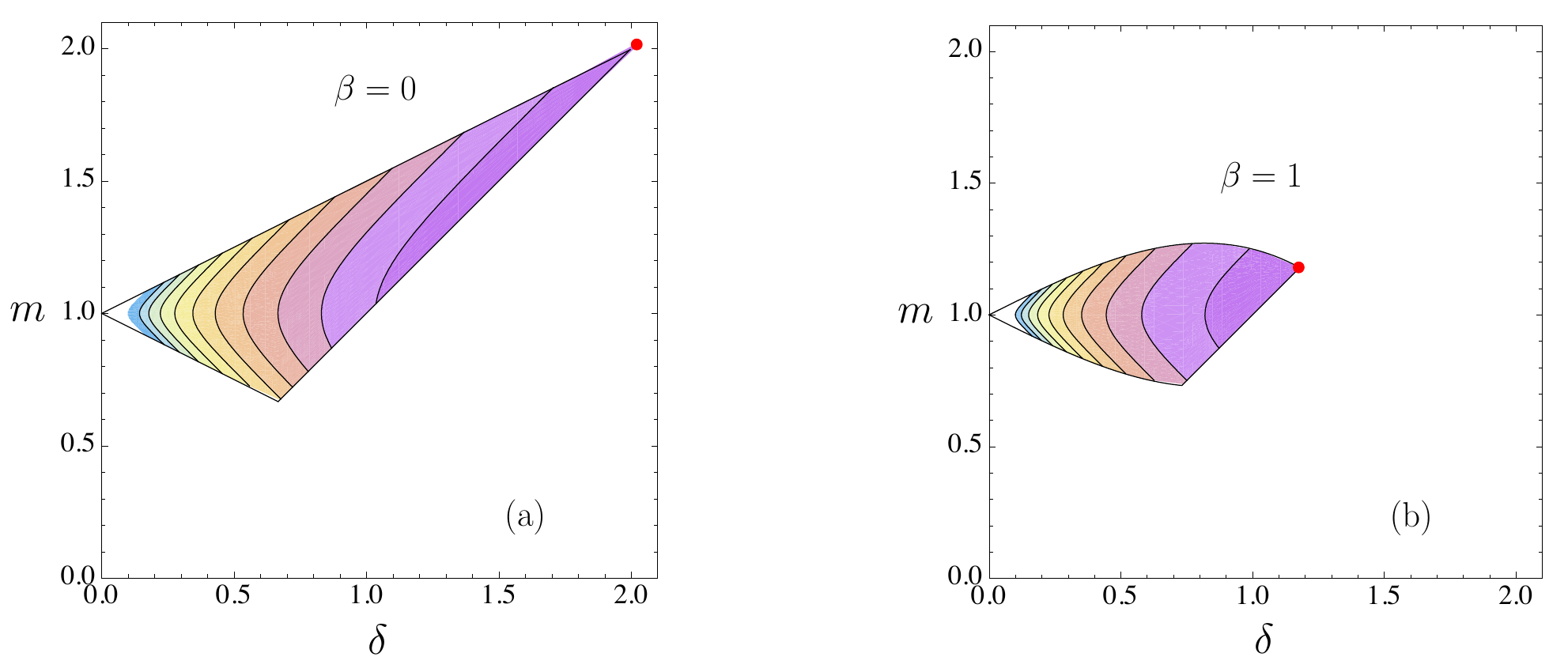}
\\  \vspace{0.3cm}
\includegraphics[height=0.35\columnwidth]{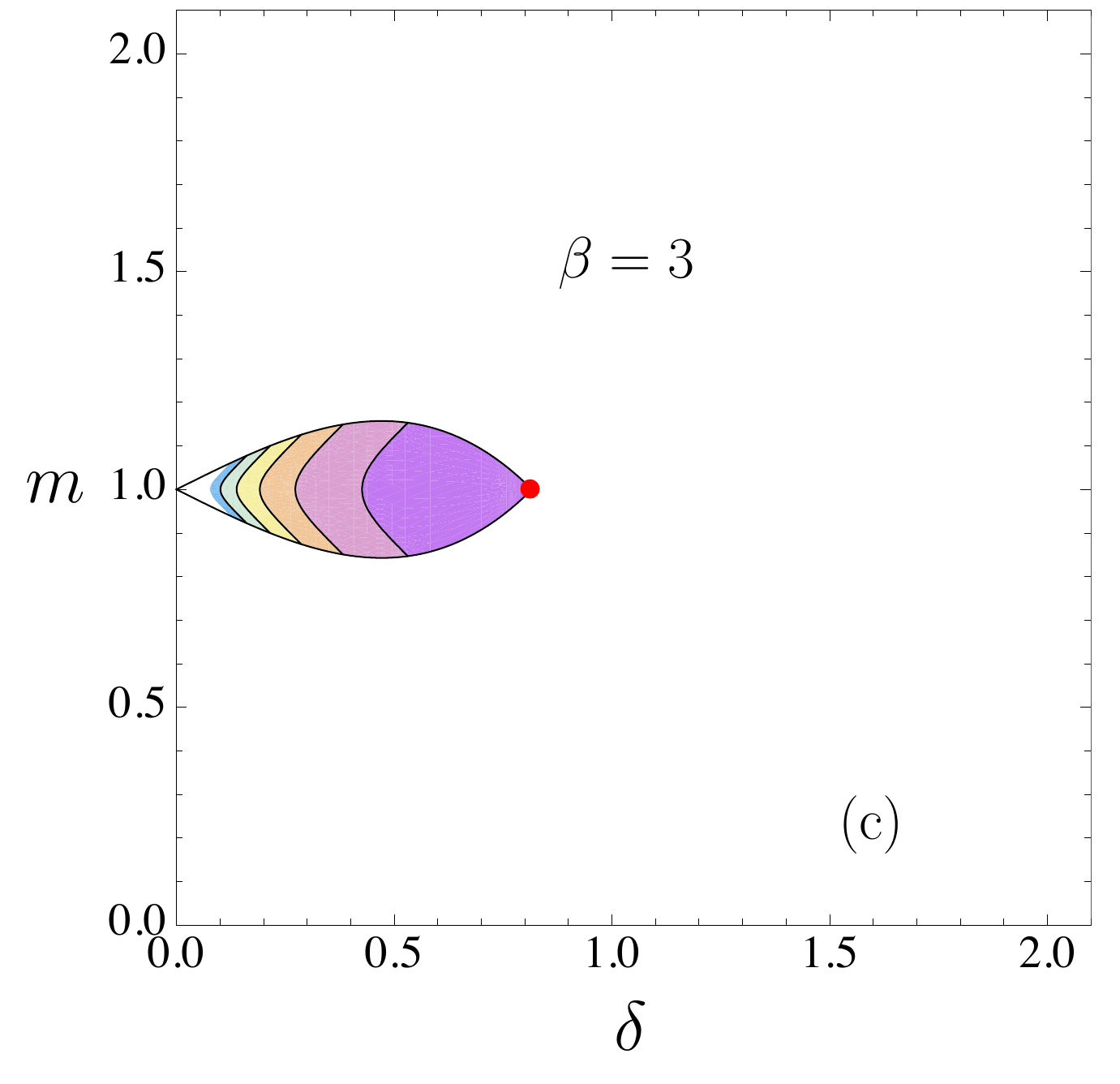}
\caption{Contour plots of $\beta f$ in regions of the parameter
space such that $\phi(x)\geq0$, for (a) $\beta=0$, (b) $\beta=1$ and (c) $\beta=3$. Darker regions have lower free energy.}
\label{fig:contourspos}
\end{figure}

\begin{figure}[h]
\centering
\includegraphics[width=0.9\columnwidth]{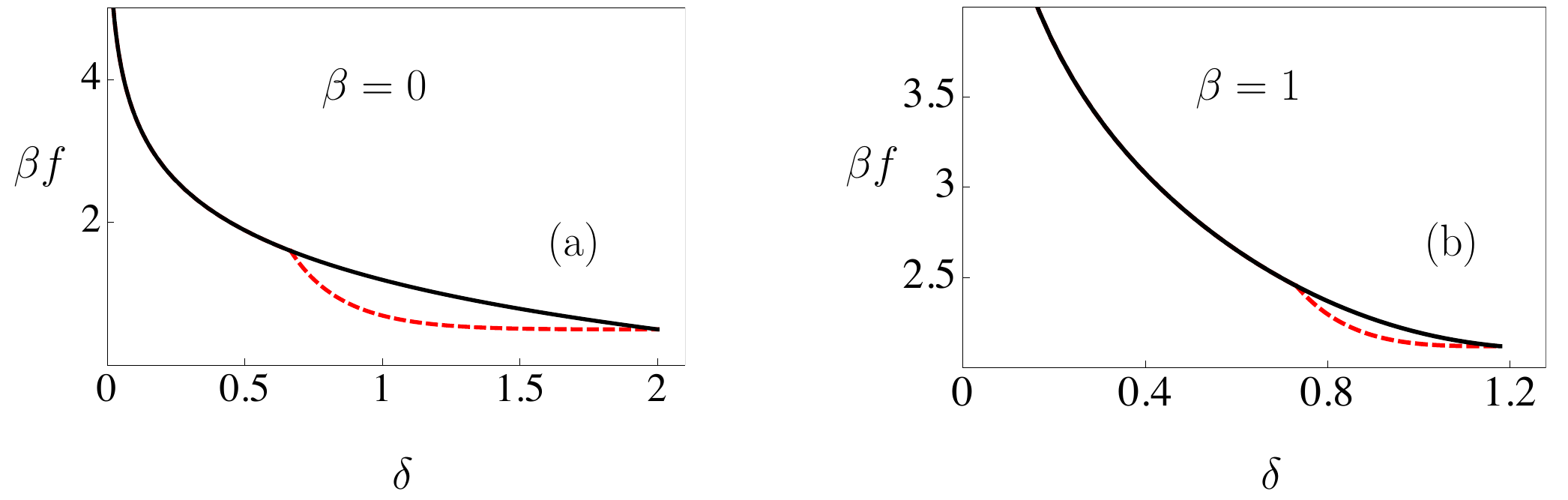}\\ \vspace{0.2cm}
\includegraphics[width=0.45\columnwidth]{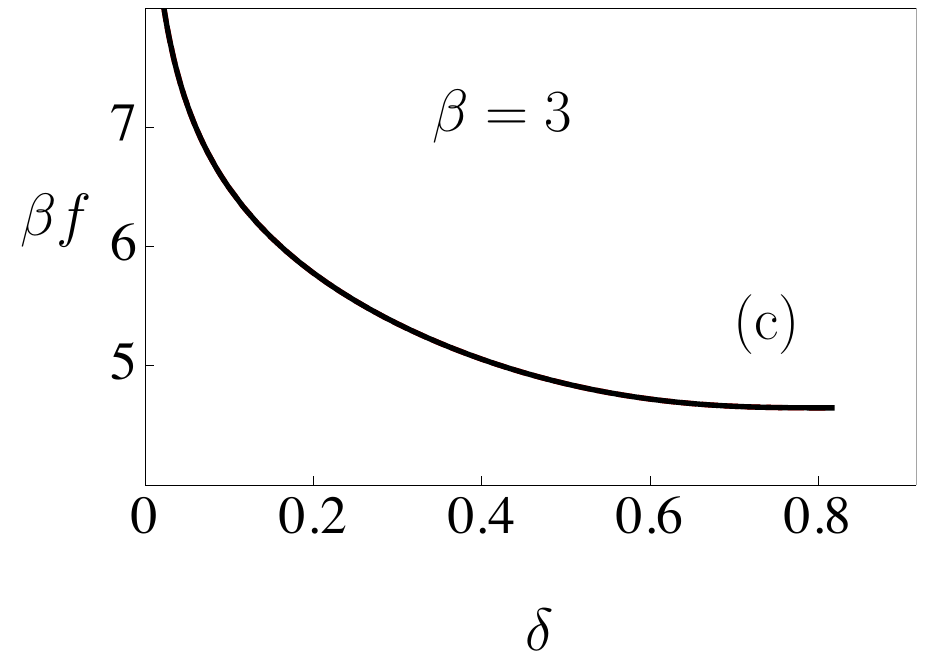}
\caption{Free energy density $\beta f$ on the boundary of the region of the domain
where $\phi(x)\geq0$, for different temperatures (indicated).
Dashed lines: $\beta f$ on the lower boundary of the eye-shaped domain.
Full lines: $\beta f$ on the upper boundary. The sought minima of the free energy can be inferred from the graphs and coincide with the dots in Figs.~\ref{fig:positive_eye} and~\ref{fig:contourspos}.}
\label{fig:3free}
\end{figure}

By applying the natural scaling (\ref{eq:piscale}), the potential for the gas of eigenvalues (\ref{eq:freeF}) with $\alpha=3$ becomes
\begin{eqnarray}
f_N &=& \frac{V}{N^2} =
 \frac{1}{N}
\sum_{1\leq i \leq N} \lambda(t_i)^2 - \frac{2}{N^2\beta} \sum_{1\leq i \leq N}\sum_{i < j \leq N}\ln|\lambda(t_i)-\lambda(t_j)| 
\nonumber\\&&+ \frac{2}{N^2 \beta} \sum_{1\leq i \leq N}\sum_{i < j \leq N}\ln N
\nonumber\\
&=&  u - \frac{1}{\beta} s + \frac{1}{\beta} \ln N + \Ord{\frac{\ln N }{N}}
\nonumber\\
&=& f +  \frac{1}{\beta} \ln N + \Ord{\frac{\ln N }{N}}.
\end{eqnarray}
where
\begin{eqnarray}
u&=& \frac{1}{N}\sum_{1 \leq i \leq N}\lambda(t_i)^2\\
s&=&\frac{2}{N^2}\sum_{1 \leq i<j \leq N}\ln|\lambda(t_i)-\lambda(t_j)| \\
f&=& \lim_{N \to \infty}\left(f_N-\frac{1}{\beta}\ln N\right).
\end{eqnarray}
In the limit $N\rightarrow \infty$ we have:
\begin{eqnarray}
u&=&\int_{-1}^{1}d x \;  \lambda^2 \phi(x),\label{eq:vu}
 \\
s&=&  \int_{-1}^{1} d x \; \int_{-1}^{1} d y \  \phi(x)\phi(y)\ln(\delta | x - y|) .\label{eq:vs}
\end{eqnarray}
Observe that $f$, $u$ and $s$ satisfy the thermodynamic relation
 \begin{equation}\label{eq:formal free energy from energy density and entropy density}
 \beta f= \beta u - s,
\end{equation}
and they will be the free energy, the internal energy and the entropy densities of the Coulomb gas of eigenvalues when computed on their most probable distribution.
In order to compute $s$  we
integrate Eq. (\ref{eq:Tricomi}) 
\begin{eqnarray}
\int_{-1}^{z} d x \; P  \! \int_{-1}^{1} d y \; \frac{\phi(x)}{y-x} = \pi \int_{-1}^{z} dz\; g(z)
\end{eqnarray}
from which it follows
\begin{eqnarray}
\int_{-1}^{1} dx\; \phi(x) \int_{-1}^{1} dy\; \phi(y) \ln |y-x| &=&
\int_{-1}^{1} d x \; \phi(x)\ln (x+1)\nonumber\\
&&- \pi \int_{-1}^{1} d x \;
\phi(x)\int_{-1}^{x} dy\;g(y) .
\end{eqnarray}
Thus by inserting in Eqs. (\ref{eq:vu})-(\ref{eq:formal free energy from energy density and entropy density}) the physical solution (\ref{eq:phi(x)}) of the Tricomi equation 
and observing that
\begin{eqnarray}
\int_{-1}^{1}dy \ \frac{\ln (y+1)}{\pi \sqrt{1-y^2}}=-\ln 2,  \qquad\quad \int_{-1}^{1} dy \ \frac{y \ln (y+1)}{\pi \sqrt{1-y^2}}= 1,\\ \nonumber \\
\int_{-1}^{1} dy \ \frac{y^2 \ln (y+1)}{\pi \sqrt{1-y^2}}=-\frac{1}{4}-\frac{1}{2}\ln 2 \qquad  \qquad
\end{eqnarray}
we get   
\begin{eqnarray}\label{eq:energy density1}
u (\delta,m,\beta) &=& 1-(1-m)^2+ \frac{\delta^2}{2}-\frac{\beta \delta^4}{8},\\
\label{eq:entrophy density1}
s (\delta,m,\beta) &=&-\frac{2(1-m)^2}{\delta^2}-\frac{\beta^2 \delta^4 }{16} +\ln
{\frac{\delta}{2}},\\
\beta f(\delta,m,\beta)&=&\beta -\beta (1-m)^2 +  \frac{2(1-m)^2}{\delta^2}+
\frac{\beta\delta^2}{2}- \frac{\beta^2 \delta^4}{16}-\ln {\frac{\delta}{2}}.
\label{eq:freeenergy} 
\end{eqnarray}
In Sec.~\ref{tricomi} we have determined the domain of $\rho(\lambda)$ in the space $(m, \delta)$, for different ranges of the inverse temperature $\beta$. As already observed all the points of this domain yield a solution of the saddle point equations (\ref{eq:normal0})-(\ref{eq:normal}). In order to find the most probable distribution of the Schmidt coefficients we now have to minimize $\beta f$ (or $f$) in the physical domain (\ref{eq:positive_temp_eyesShapeDomain}).
The contour plots of the free energy density are shown in Fig.~\ref{fig:contourspos}.
\\
Notice that $f$, as well as $u$ and $s$, is symmetric with respect to the line $m=1$. This $\mathbb{Z}_2$ symmetry will play a major role in the following.
The only stationary point (a saddle point) of $\beta f$ is at  the right corner of the eye  (\ref{eq:beta>2}), see Figs.~\ref{fig:positive_eye}
and~\ref{fig:contourspos}. Thus, the absolute minimum is on the boundary.
The behavior of the free energy at the boundaries of the allowed domain is shown in Fig.~\ref{fig:3free} for different temperatures.
\\
We deal with two regimes of solutions corresponding to different ranges of temperature. Recall that the temperature in our model is the Lagrange multiplier associated to isopurity manifolds of the Hilbert space. Different symmetries of the physical domain will correspond to different symmetries in the distribution of the Schmidt coefficients. More precisely we have:
\begin{itemize}
\item for $\beta\geq\beta_+$, where $\beta_+=2$, when the domain of the physical solution of the Tricomi equation is eye-shaped, the absolute minimum
is at the corner of the eye, Eq. (\ref{eq:beta>2})
\begin{equation}\label{eq:mdeltasemicircle}
m=1, \quad \delta=\sqrt{\frac{2}{\beta}}
\end{equation}
\item for $0 \leq \beta<\beta_+$,  the absolute minimum is at the right upper corner of the cut eye-shaped allowed region,
$\delta=\Gamma_1^+(\delta,\beta)$,
namely at
\begin{equation}
m=\delta, \quad \beta=\frac{4}{\delta^3}-\frac{2}{\delta^2}.
\label{eq:beta<2}
\end{equation}
\end{itemize}
See the red dots in Figs.~\ref{fig:positive_eye}(b) and~\ref{fig:contourspos}. For the distribution of eigenvalues corresponding to these minima, $f$ coincides with the free energy of the system, while $u$ and $s$ can be identified with the internal energy density and the entropy density, respectively. 
\subsection{Distribution of the Schmidt coefficients}\label{SemicircleEWishart}
We will study the behavior of our system starting from high values of $\beta$, that is low values of internal energy $u$ (purity). 
In the previous sections we unveiled the presence of two regimes. In this section we will show that they correspond to a different symmetry in the distribution of the eigenvalues.
\subsubsection{Wigner semicircle distribution: $\beta \geq \beta_{+}$ - towards maximally entangled states}
\begin{figure}
\centering
\includegraphics[width=1\columnwidth]{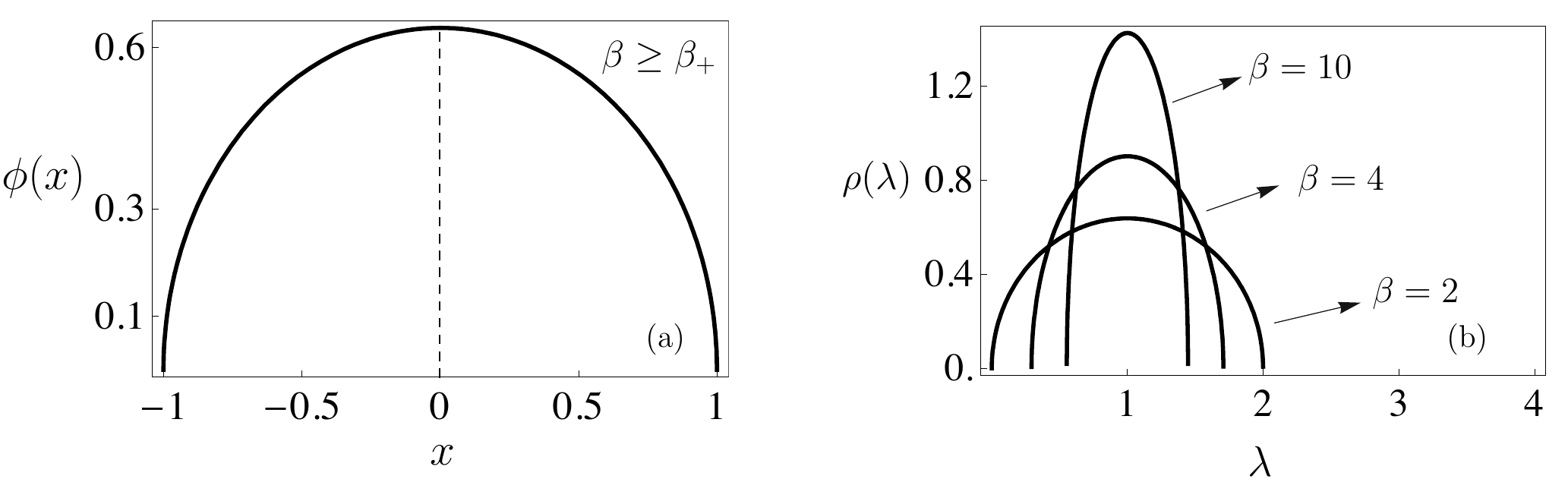}
\caption{(a) Density of the eigenvalues for $\beta\geq\beta_+=2$. (b) Density of the eigenvalues for $\beta=2, 4$ and
$10$. In the temperature range $\beta \in [\beta_{+},\infty[$ the solution is given by the semicircle
law.}\label{fig:semicircle}
\end{figure}

For $\beta\geq \beta_{+}=2$, we have found that the minimum of the free energy for the gas of eigenvalues corresponds to the corner of the eye (Eq. (\ref{eq:mdeltasemicircle})) 
From the expression (\ref{eq:phi(x)}) for the physical solution of the saddle 
equations, we get the semicircle law (see Fig.~\ref{fig:semicircle})
\begin{equation}
\phi(x)= \frac{2}{\pi}\sqrt{1-x^2},
\label{eq:semicircle0}
\end{equation}
whence, by (\ref{eq:new variable}),
\begin{equation}
\label{eq:semicircle}
\rho(\lambda)=\frac{\beta}{\pi}\sqrt{\lambda-a}\sqrt{b-\lambda},
\end{equation}
where
\begin{equation}
a=1-\delta = 1 -\sqrt{\frac{\beta_+}{\beta}}, \qquad b=1+\delta=
1+\sqrt{\frac{\beta_+}{\beta}}.
\end{equation}
This distribution is displayed in Fig.~\ref{fig:semicircle}. 
Observe that as $\beta$ becomes larger the distribution of the Schmidt coefficients becomes increasingly peaked around 1.
This means that all the eigenvalues tend to $1/N$. See the natural scaling (\ref{eq:scalinglambda}): for temperatures $T =1/\beta$ tending to zero quantum states become maximally entangled.
The internal energy and the entropy of the system for this solution can be easily obtained by plugging the condition on the parameters $m$ and $\delta$ (\ref{eq:mdeltasemicircle}) into the expressions (\ref{eq:energy
density1})-(\ref{eq:entrophy density1}) for $u$ and $s$ respectively:
\begin{eqnarray}
u&=&1+ \frac{\delta^2}{4} = 1+\frac{1}{2\beta} ,
\label{eq:puri2}\\
\label{eq:entropy density for semicircle}
s&=&-\frac{1}{4} + \ln \frac{\delta}{2}= -\frac{1}{4} -\frac{1}{2} \ln (2\beta),
\end{eqnarray}
and thus
\begin{equation}
\beta f= \frac{2}{\delta^2}+\frac{3}{4}- \ln \frac{\delta}{2}=
\beta+\frac{3}{4} +\frac{1}{2} \ln(2\beta).
\end{equation}
\subsubsection{Wishart distribution: $0\leq \beta < \beta_+$ - typical states}
\begin{figure}[h]
\centering
\includegraphics[width=1\columnwidth]{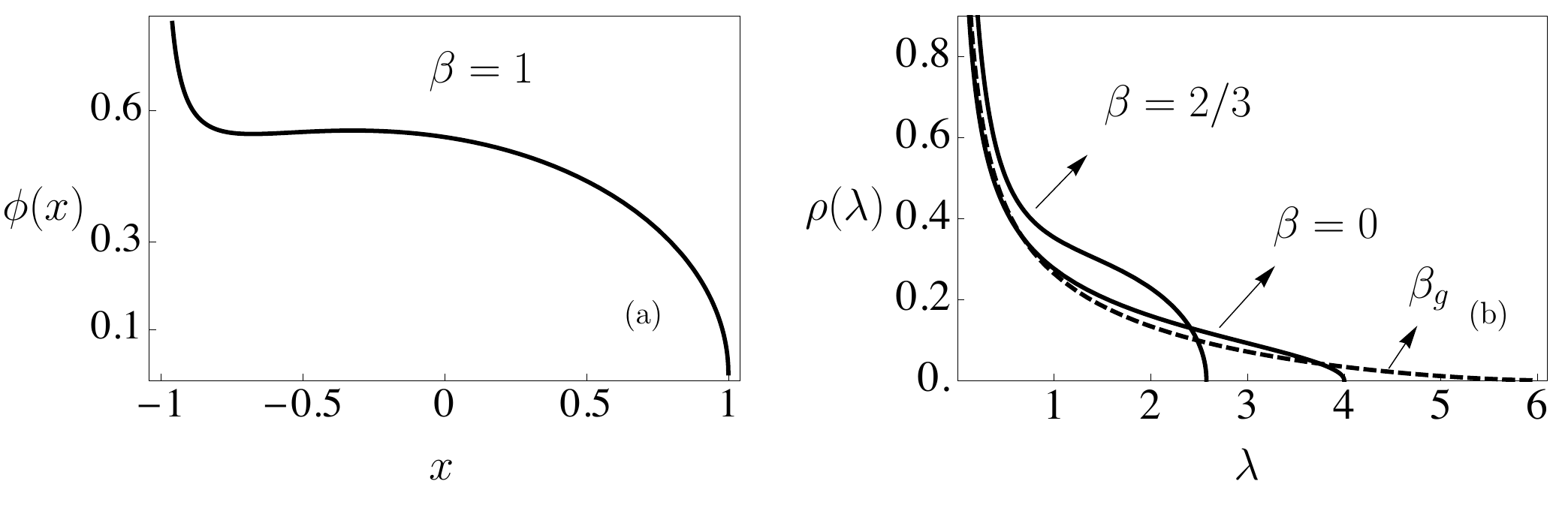}
\caption{(a) Density of the eigenvalues for $\beta=1$. (b) Density of eigenvalues for $\beta=0$, $\beta=2/3$, 
and $\beta=\beta_g=-2/27$ (dashed). In the range of temperatures $\beta \in (\beta_g,\beta_+)$, with $\beta_+=2$, the solution is given by the Wishart distribution. We refer to the next chapter for the range $\beta_g \leq \beta <0$.}
\label{fig:Wisharts}
\end{figure}

\begin{figure}[h]
\centering
\includegraphics[width=0.5\columnwidth]{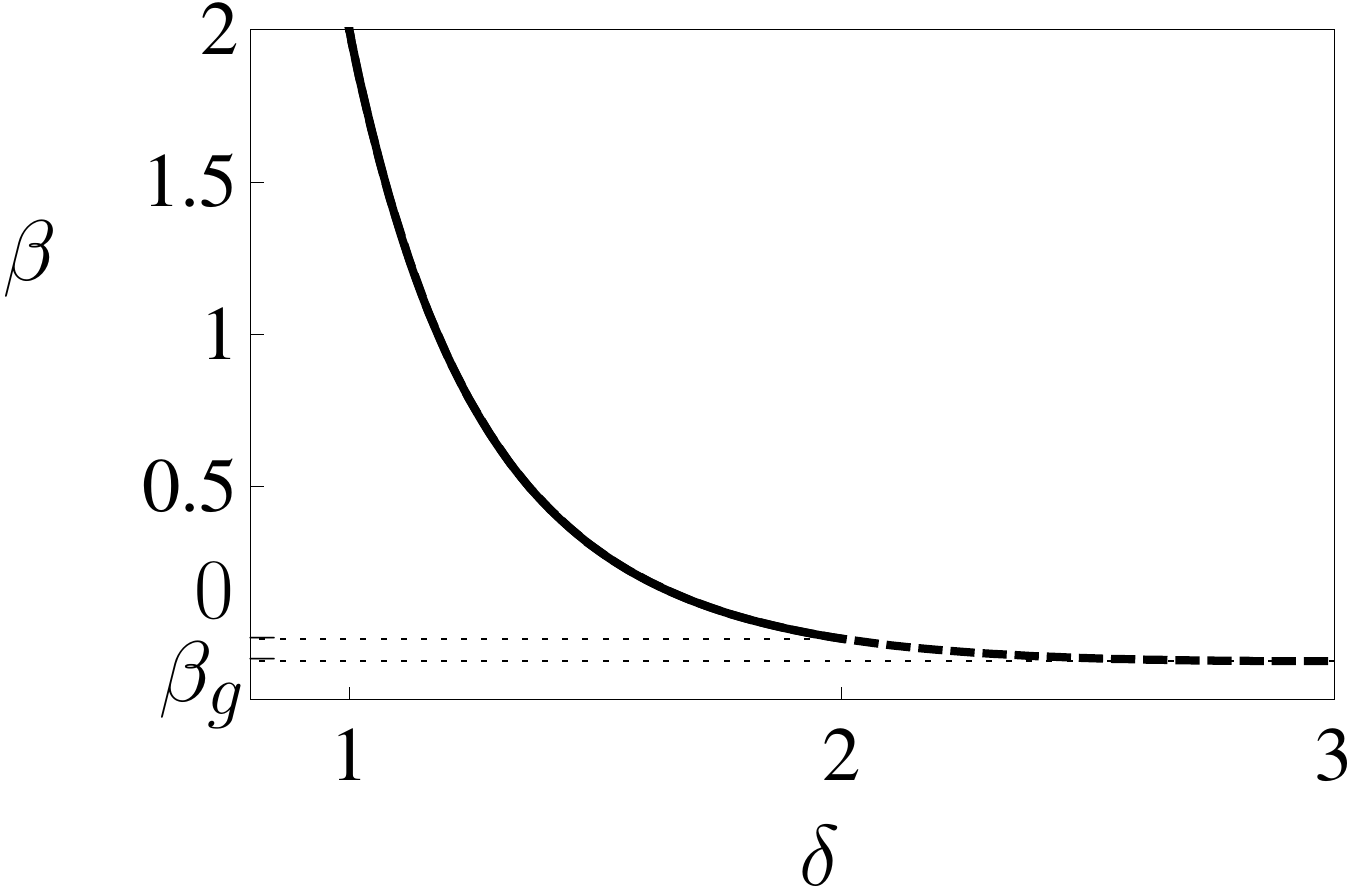}
\caption{Plot of Eq.\ (\ref{eq:beta<2}) for positive  (solid line) and negative (dashed line) temperatures. The minimum $\beta_g=-2/27$ is attained at $\delta=3$.}
\label{fig:betadeltapos}
\end{figure}

\begin{figure}[h]
\centering
\includegraphics[width=0.5\columnwidth]{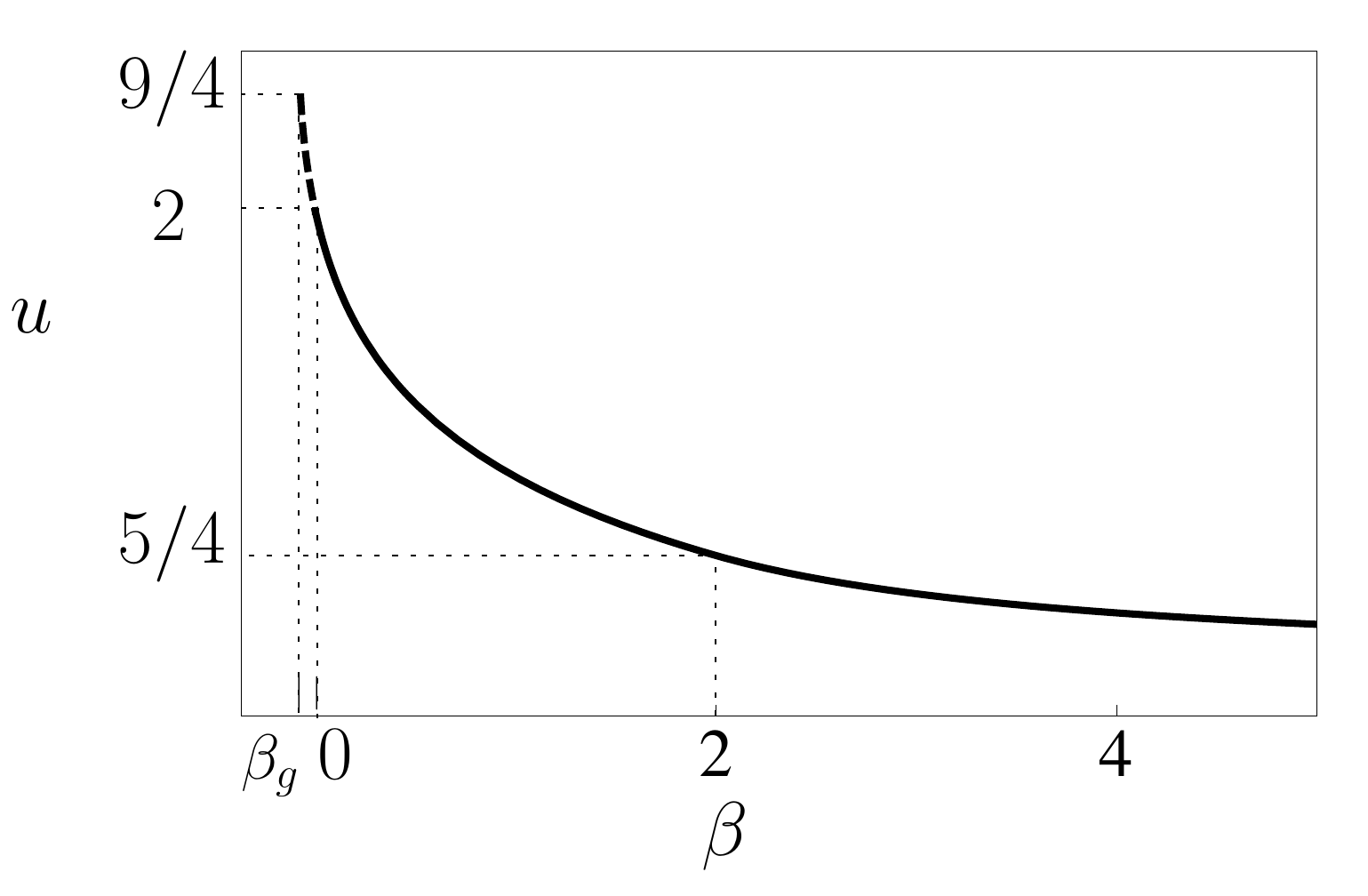}
\caption{$u$ as a function of the (inverse)
temperature. Notice that $u=2$ at $\beta=0$ (typical states).
In the $\beta\to\infty$ limit we find the minimum $u=1$. We will see that $\beta_g=-2/27,u=9/4$
(left point) and $\beta_+=2,u=5/4$ (right point) are critical points for the system. However the phase for $\beta_g \leq \beta < 0$
is unstable (dashed line) towards another phase as soon as $\beta<0$ (in this
scaling of $\beta$).}\label{fig:uvsbetapos}
\end{figure}

At higher temperatures $0\leq \beta < \beta_+$ the solution 
acquires a different shape, becoming asymmetric with respect to the center of its support. Analogously to the previous case, by plugging the relations (\ref{eq:beta<2}) between $m$, $\delta$ and $\beta$ at the minimum of $\beta f$ into the physical solution of the saddle point equations (\ref{eq:phi(x)}), we get the 
distribution of the Schmidt coefficients:
\begin{equation}
\phi(x)=\frac{2}{\pi\delta}\sqrt{\frac{1-x}{1+x}}\big(1+(2-\delta)x\big) .
\label{eq:ansatz1phi}
\end{equation}
By (\ref{eq:new variable}) it yields
\begin{equation}
\label{eq:ansatz1}
\rho(\lambda)= \frac{4}{\pi b^2} \sqrt{\frac{b-\lambda}{\lambda}} \left(b-2+\frac{2(4-b)}{b}\lambda\right),
\end{equation}
with $b=2\delta$.  See Fig.~\ref{fig:Wisharts}. This is the Wishart distribution.
\\
The change from the semicircle to the Wishart distribution is accompanied by a phase transition (the first of a series), as we shall see in Sec.~\ref{secondOrderPhTR}.
In Fig.~\ref{fig:betadeltapos} we plot the relation (\ref{eq:beta<2}) between the half width of the distribution $\delta=b/2$ and $\beta$.
It runs monotonically from $\beta =2$ when $\delta=1$ to $\beta=0$ when $\delta=2$.
\\Moreover, the inverse temperature $\beta$ reaches a minimum equal to 
\begin{equation}
\label{eq:betagdef}
\beta_g=-\frac{2}{27} 
\end{equation}
at $\delta=3$.
Therefore, the above solution can
be analytically continued down to $\beta_g$, which is slightly negative, but not
below.
We will study the solution for negative
temperatures in the next chapter.
Let us now compute the thermodynamic quantities for the Wishart distribution.
The internal energy $u$ (proportional to the average purity) is obtained by plugging
(\ref{eq:beta<2}) into the expression (\ref{eq:energy
density1})
\begin{equation}
\label{eq:puri1}
u=\frac{3}{2}\delta -\frac{\delta^2}{4} .
\end{equation}
In particular, at $\beta=0$ ($\delta=2$) we get $u=2$,
at $\beta_+=2$ ($\delta=1$) we get  $u=5/4$, and at $\beta_g=-2/27$ ($\delta=3$) we get $u=9/4$. See Fig.~\ref{fig:uvsbetapos}. Observe that $u(\beta)$ is continuous at $\beta_{+}$.
From Eqs. (\ref{eq:beta<2}) and (\ref{eq:entrophy
density1}) we can also compute the entropy and the free energy for the Wishart distribution
\begin{eqnarray}
\label{eq:entropydensityWishart}
s&=& -\frac{9}{4} + \frac{5}{\delta} - \frac{3}{\delta^2}
+\ln\frac{\delta}{2},\\
\beta f&=& \frac{9}{\delta^2} - \frac{9}{\delta} +\frac{11}{4} -\ln\frac{\delta}{2} ,
\label{eq:Fbeta0}
\end{eqnarray}
in terms of  $\delta=\delta(\beta) \in ]1,3]$.

Summarizing, for $\beta \to+ \infty$ all eigenvalues of $\rho_A$ are equal to $1/N$, that is subsystem $A$ is in a maximally mixed state. As the inverse temperature is decreased,  for $\beta \geq \beta_{+}$, all eigenvalues remain of $\Ord{1/N}$ their distribution being characterized by the Wigner semicircle law, while for $0\leq \beta < \beta_{+}$ the eigenvalues, still of $\Ord{1/N}$, follow the Wishart distribution which diverges at the origin, that is some eigenvalues of $\rho_A$ vanish.
Let us remark that the distributions of eigenvalues we have found are consistent with the scaling coefficient $\alpha=3$. Indeed, for any value of the inverse temperature $\beta$, the domain of $\rho(\lambda)$ involves the rescaled eigenvalues $\lambda$ which are of $\mathcal{O}(1)$, that is the eigenvalues $\lambda_i$ are of $\mathcal{O}(1/N)$. In other words, the range of positive temperatures refers to those submanifolds of the total Hilbert space $\Ham$ given by pure states $\ket{\psi}$ showing a high level of entanglement between subsystems $A$ and $B$. Observe that the internal energy is of $\mathcal{O}(1)$, that is the purity is of $\mathcal{O}(1/N)$, as will be discussed at the end of this chapter. 
\subsection{Analysis of moments and cumulants}\label{sec:momentipureCaseCap3}
Notice that from the relation
\begin{equation}
\cZ_{AB}(\beta)=e^{-\beta N^2 f}
\end{equation}
we have that the free energy  is proportional to the generating function of the connected
correlations of $\pi_{AB}$. More precisely, the derivatives  of $\cZ_{AB}(\beta)$ evaluated for $\beta=0$, yield
 the moments of $\pi_{AB}$ with respect to the measure $d \mu(\rho_A)$ introduced in Sec.~\ref{sec:Induced measure on subsystemsPURE}, i.e.
 \begin{eqnarray} \label{moma}
 {\cal M}_n= \left\langle{\pi_{AB}}^n \right\rangle=
 \int d\mu(\rho_A) \; \pi_{AB}(\rho_A)^n = 
 \frac{(-1)^n}{N^{3n}} \; \left. \frac{\partial^n  \cZ_{AB}(\beta)}{\partial \beta^n }\right|_{\beta=0} \;.
 \end{eqnarray} 
These functions fully determine the statistical distribution of $\pi_{AB}$ on the set of pure bipartite states and
 in the high temperature regime provide an expansion of $\cZ_{AB}(\beta)$. 
In~\cite{paper1} we can find the expression for the $n-$th cumulant ${\cal K}_n$ of the purity $\pi_{AB}$ (in one to one correspondence with the first $n$ moments) in the limit of large $N$ and $\beta \to 0$, for the case of balanced bipartitions, $\dim \cH_A = \dim \cH_B = N$:
  \begin{equation}\label{eq:mompureLargeN}
  {\cal K}_n= - \frac{(-1)^n}{N^{3n}}\frac{\partial^n (\beta f)}{\partial \beta^n}\Bigg|_{\beta \to 0}=\frac{2^{n+1}}{N^{3n-2}}\frac{(3n-3)!}{(2 n)!}
  \end{equation}
 and also the cumulants for the more general case of unbalanced bipartitions. For instance, the first and second cumulants are given by 
  \begin{eqnarray}
 {\cal K}_1&=& \frac{N+M}{N M }\label{eq:cum1GirNM}\\ \nonumber\\ 
  {\cal K}_2&=& \frac{2}{N^2 M^2},\label{eq:cum2GirNM}
 \end{eqnarray}
 with $N=\dim \cH_A$ and $M=\dim \cH_B$. They reduce to (\ref{eq:mompureLargeN}) for $N=M$ and $n=1,2$.
The exact expression (that is valid also for small $N$)  at $\beta=0$ of the first $5$ cumulants can be found in~\cite{Giraud}. In particular
for the case of balanced bipartitions, $\dim \cH_A = \dim \cH_B =N$, the first two cumulants are given by
\begin{eqnarray}
 {\cal K}_1&=& \frac{2N}{1+N^2}\\ \nonumber\\ 
  {\cal K}_2&=& \frac{2 (N^2-1)^2}{(1 +N^2 )^2(2 +N^2 )(3 +N^2)},\label{eq:cum2pureGiraudNN}
 \end{eqnarray}
while for more general case of $\dim \cH_A \neq \dim \cH_B$ we have
\begin{eqnarray}
 {\cal K}_1&=& \frac{N+M}{1 + NM }\label{eq:cum1pureGiraudNM}\\ \nonumber\\ 
   {\cal K}_2&=& \frac{2(N^2-1)(M^2 -1)}{(1 +N M)^2 (2 +N M)(3 + N M)}.\label{eq:cum2pureGiraudNM}\nonumber\\ 
 \end{eqnarray} 
 We recall the relationship between the cumulants and moments of a generic probability distribution
 \begin{equation}
 {\cal K}_n= {\cal M}_n- \sum_{1\leq m \leq n-1} \begin{pmatrix}
      n-1   \\
      m-1
\end{pmatrix} {\cal K}_m {\cal M}_{n-m},
 \end{equation}
and in particular
 \begin{eqnarray}
 {\cal K}_1&=&{\cal M}_1\\
 {\cal K}_2&=&{\cal M}_2-{{\cal M}_1^2}.
 \end{eqnarray}
 More generally, 
 we can also define 
 the moments of $\pi_{AB}$ for $\beta\neq 0$ as  
 \begin{eqnarray}
 {\cal M}_n(\beta) =   \left\langle{\pi_{AB}}^n \right\rangle_{\beta}  = 
 \label{ffd1pure} 
  \int d\mu_\beta(\rho_A) \; {\pi_{AB}(\rho_A)}^n = 
 \frac{(-1)^n}{ N^{3n} \cZ_{AB}(\beta) } \;  \frac{\partial^n  \cZ_{AB}(\beta)}{\partial \beta^n } \;,
 \end{eqnarray} 
where $d \mu_\beta(\psi)$ is the canonical measure
\begin{eqnarray}\label{ffhh1}
d \mu_\beta(\rho_A) =d \mu(\rho_A) \; \frac{  \e ^{-\beta \;  \pi_{AB}(\rho_A)} }{\cZ_{AB}(\beta)} \;.
\end{eqnarray} 
The latter is a deformation of the Haar measure $d \mu(\rho_A)$ obtained by including a non uniform weight 
which explicitly depends upon the purity $\pi_{AB}$, thru $\beta$. In particular, as $\beta$ increases, 
$d \mu_\beta(\rho_A)$ enhances the role of the states with lower values of $\pi_{AB}$ (i.e. larger values of bipartite entanglement between the two subsystems $A$ and $B$), to the extent that
for $\beta \rightarrow +\infty$  only the maximally entangled states contribute to~(\ref{ffd1pure}).


\section{Second order phase transitions}\label{secondOrderPhTR}
We are now ready to unveil the presence of the first critical point for the system
at $\beta_{+}=2$. In the previous section, we have found that the densities of eigenvalues corresponding to the two contiguous intervals of temperature $0 \leq\beta < \beta_{+}$ and $\beta\geq\beta_{+}$ show a different symmetry, which correspond to the Wishart and the Wigner semicircle distributions. They are associated to two distinct entanglement phases for the system: the \textit{maximally entangled} and the \textit{typical} phases. In the former all the eigenvalues are of $\Ord{1/N}$, whereas in the second regime some eigenvalues go to zero. A second order phase transition at $\beta_{+}$ separates the two phases. 
It is due to
the restoration of the ${\mathbb Z}_2$ symmetry $P$ (``parity"), present in the thermodynamic quantities, see 
 Eqs.\  (\ref{eq:energy density1}), (\ref{eq:entrophy density1}) and (\ref{eq:freeenergy}).
It corresponds to the reflection symmetry of the semicircle distribution 
$\rho(\lambda)$ around the center of its support ($m=\delta=b/2$ for
$\beta < \beta_+$ and $m=1$ for $\beta\geq \beta_+$), that is not present in the Wishart distribution.

\begin{figure}[h]
\centering
\includegraphics[height=0.33\columnwidth]{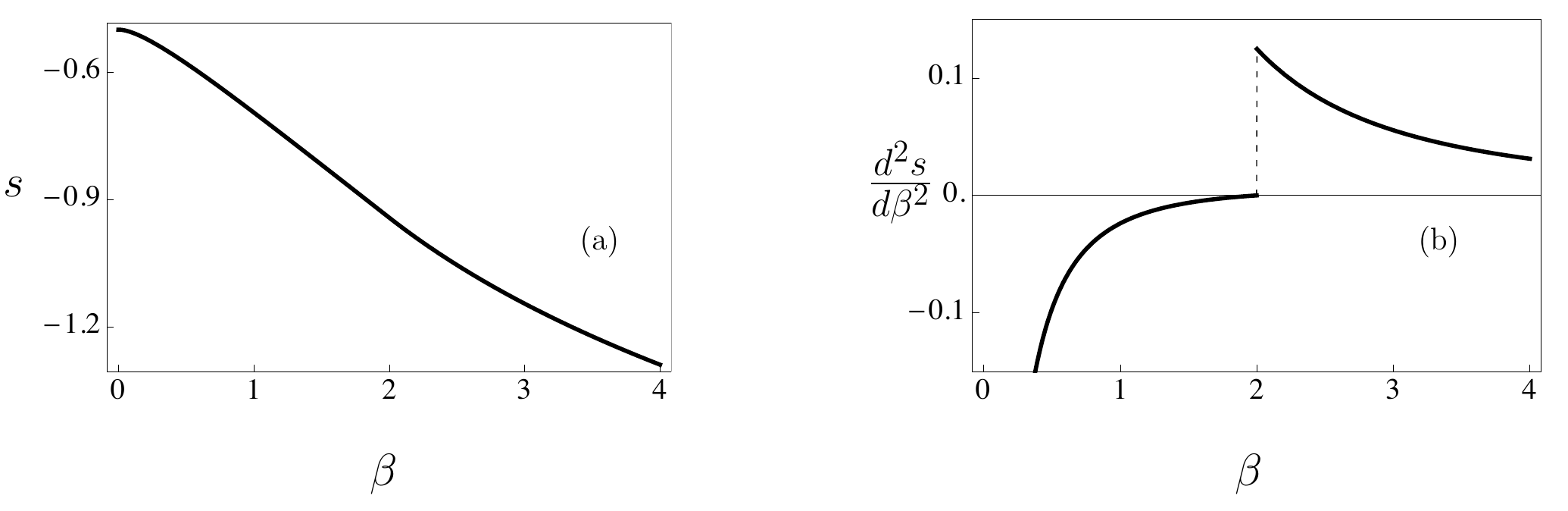} 
\caption{Entropy (a) and its second derivative with respect to $\beta$ (b).
The entropy is continuous in $\beta_{+}$ while its second derivative
presents a finite discontinuity. }
\label{fig:entrophy at beta+}
\end{figure}

The direct proof of the presence of a second
order phase transition in the system at $\beta=\beta_{+}$ can be obtained from the expression of the entropy density $s=\beta(u-f)$. See Eq. (\ref{eq:entropy density for semicircle}) for $\beta\geq \beta_+$ and  Eq. (\ref {eq:entropydensityWishart}) for $\beta<\beta_+$.
From both equations we get the same values of the entropy at $\beta=\beta_+$ $\;(\delta=1)$
\begin{equation}
s=-1/4-\ln 2.
\end{equation}
On the other hand
the first derivative of $s$ with respect to $\delta$ is discontinuous at $\delta=1$. However, also $\beta$ as a function of $\delta$ has a discontinuous first derivative at $\delta=1$, see Eqs. (\ref{eq:mdeltasemicircle}) and (\ref{eq:beta<2}). By reminding that
\begin{eqnarray}
\frac{ds}{d\beta} &=& \frac{ds}{d\delta}\Big/\frac{d\beta}{d\delta},\nonumber \\
\frac{d^2s}{d\beta^2}&=& 
\frac{d^2s}{d\delta^2} \Big/\left(\frac{d\beta}{d\delta}\right)^2  - \frac{ds}{d\delta}\, \frac{d^2\beta}{d\delta^2} \Big/ \left(\frac{d\beta}{d\delta}\right)^3 ,
\end{eqnarray}
one easily obtains that the discontinuities compensate. For $\beta \to \beta_+$ we have
\begin{eqnarray}
s \sim -\frac{1}{4}-\ln 2 -\frac{\beta-\beta_+}{4} +
\theta(\beta-\beta_+) \frac{(\beta-\beta_+)^2}{16},\;
\end{eqnarray}
where $\theta$ is the step function. Thus the system undergoes a second order phase transition at $\beta_{+}$ that, however, is fairly mild. Indeed, the entropy $s$, together with its first
derivative, is
continuous, although the second derivative shows a finite discontinuity of $1/8$ at $\beta=\beta_{+}$. See Fig.~\ref{fig:entrophy at beta+}. In other words, the heat capacity of the system, corresponding to $d s / d \beta$, does not diverge, although its first derivative is discontinuous. 

Observe, also, that the entropy is unbounded from below when $\beta \to
+\infty$, see Eq. (\ref{eq:entropy density for semicircle}). The interpretation of this result is quite
straightforward: the minimum value of $\langle\pi_{AB}\rangle$ is reached on a
submanifold (isomorphic to $SU(N)/\mathbb{Z}_{N}$~\cite{Kus01}) of dimension
$N^2-1$, as opposed to the typical vectors which form a
manifold of dimension $2N^2-N-1$ in the Hilbert space $\mathcal{H}$.
Since this manifold has zero volume in the original Hilbert space,
the entropy, being the logarithm of this volume, diverges.
\section{Entropy density  vs  average entanglement}\label{EntropyVSentanglement}
\begin{figure}[h]
\centering
\includegraphics[width=0.5\columnwidth]{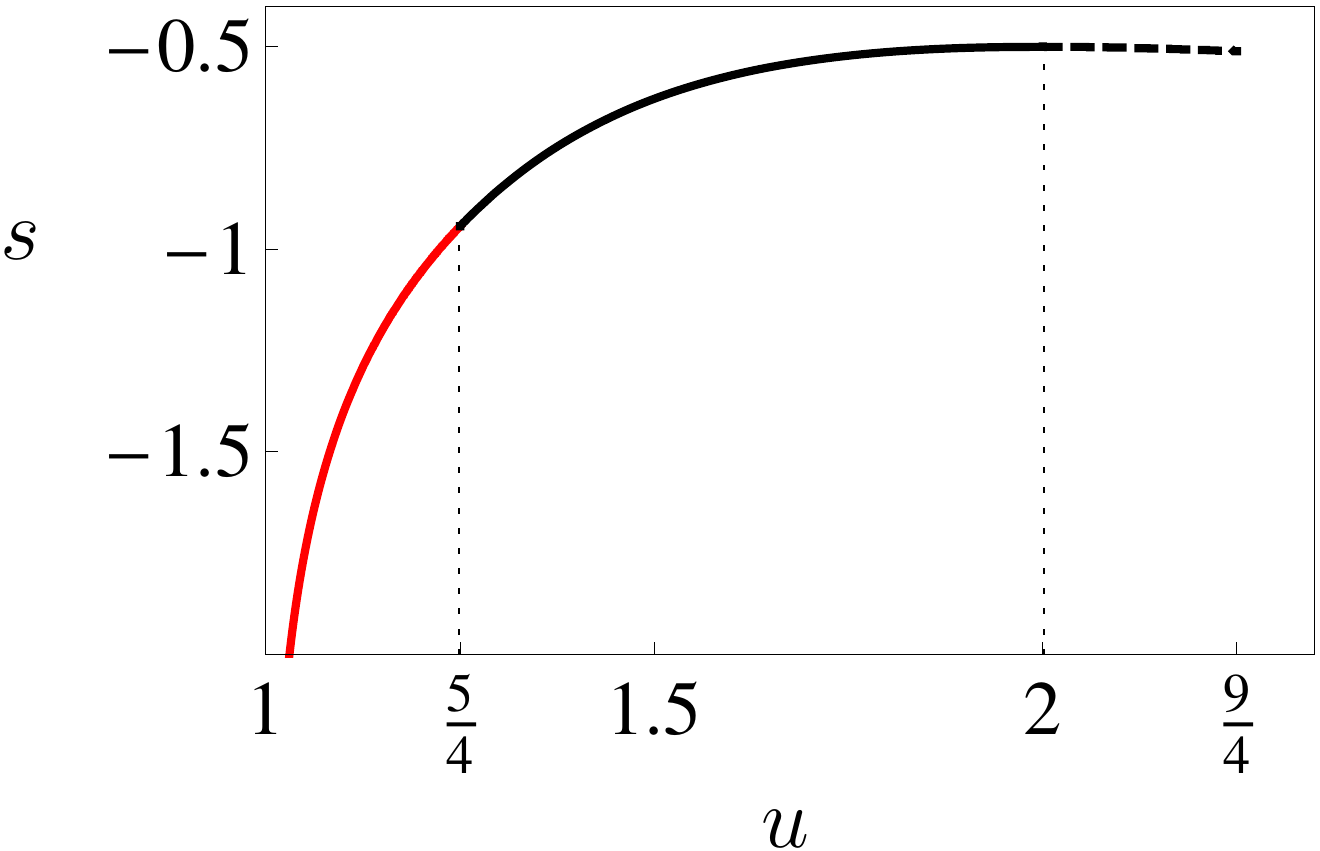}
\caption{Entropy density $s$ versus internal energy density $u=N\langle\pi_{AB}\rangle$. See Eq.\ (\ref{eq:112}).}
\label{fig:svsuN}
\end{figure}

\begin{figure}[h]
\centering
\includegraphics[width=1\columnwidth]{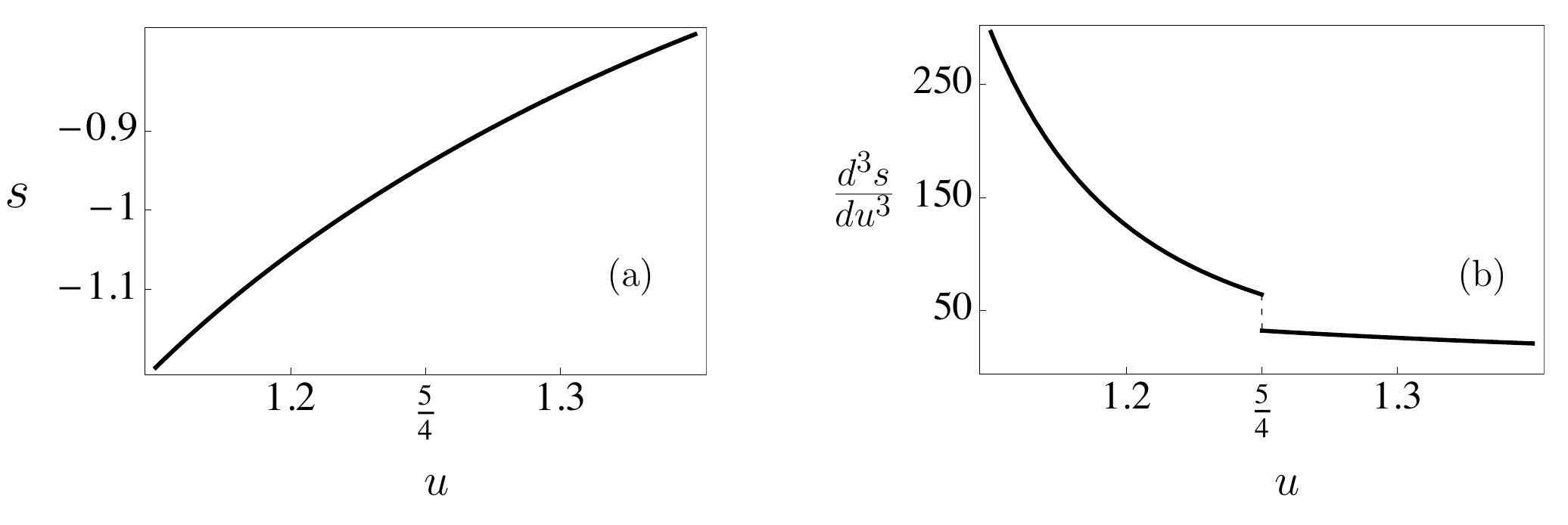}
\caption{Entropy (a) and its third derivative with respect to $u$ (b) in the critical region $u \to 5/4$.
The entropy is continuous at $u=5/4$ while its third derivative displays a finite discontinuity.}
\label{fig:svsUbetapiur}
\end{figure}

We conclude the analysis of positive temperatures, reinterpretating our results from the point of view of quantum information. To this end we will write the entropy density $s$ as a function of the internal energy density $u$, or average entanglement $\langle \pi_{AB} \rangle$. In Sec.~\ref{SemicircleEWishart} we computed these thermodynamic quantities as functions of the parameter $\delta$, both for the semicircle law and for the Wishart distribution. If we invert the expressions for the internal energy, Eqs. (\ref{eq:puri2}) and  (\ref{eq:puri1}), we get
\begin{equation}
\delta  = \begin{cases}
2\sqrt{u-1},  & 1 < u \leq  \frac{5}{4} ,  \\
\\
3-\sqrt{9-4u},  & \frac{5}{4} < u \leq 2
\end{cases}
\label{eq:deltavsu}
\end{equation}
which plugged into Eqs. (\ref{eq:entropy density for semicircle}) and (\ref{eq:entropydensityWishart}) give 
\begin{eqnarray}\label{eq:svsupositive}
s(u)  = \begin{cases}
\frac{1}{2} \ln (u-1) -\frac{1}{4},  & 1 < u \leq  \frac{5}{4} ,  \\
\\
 \ln \left( \frac{3}{2}  - \sqrt{\frac{9}{4} -u}
\right) - \frac{9}{4} + \frac{5}{2\left( \frac{3}{2}  - \sqrt{\frac{9}{4} -u}
\right)} -
\frac{3}{4\left( \frac{3}{2}  - \sqrt{\frac{9}{4} -u}
\right)^2},
& \frac{5}{4} < u \leq 2.
\end{cases}
\label{eq:112}
\end{eqnarray}
This function is plotted in Fig.~\ref{fig:svsuN}.  It counts the number of vectors with a given (average) degree of bipartite entanglement between subsystems $A$ and $B$: $s = (\ln V)/N^2$ where $V$ is the volume occupied by each submanifold. \\
\\
In particular by expanding (\ref{eq:svsupositive}) in the critical region, that is for $u\to 5/4$, we get\\
\begin{eqnarray}\label{eq:sAt54}
s(u)&=&-\frac{1}{2}\left(\frac{1}{2}+\ln 4\right)+ 2 \left(u -\frac{5}{4}\right)-4 \left(u -\frac{5}{4}\right)^2 \nonumber\\&&+ \frac{32}{3}\left(1-\frac{1}{2}\theta \left(u -\frac{5}{4}\right)\right) \left(u -\frac{5}{4}\right)^3 +\Ord{\left(u -\frac{5}{4}\right)^4}.
\end{eqnarray}\\
Therefore, the second order phase transition at $u=5/4$ is signaled  by a discontinuity in the third derivative of $s(u)$, see Fig.~\ref{fig:svsUbetapiur}, whereas at this point the entropy shows a discontinuity of the second derivative with respect to $\beta$, see Fig.~\ref{fig:entrophy at beta+}. In fact, in general, if $s, u$ and $T$ are entropy, energy and temperature, respectively, and $C=du/dT$ is the specific heat, one gets $ds/du =\beta=1/T$ and $d^2s/du^2= -1/(T^2 C)= -(1/T^3)(ds/dT)^{-1}.$ Discontinuities of the $n$-th derivative of $s(T)$ translate, therefore, into discontinuities of the $(n+1)$-th derivative of $s(u)$.
Let us finally discuss the significance of these results.
\\In the range of positive temperatures $T=1/\beta$ the eigenvalues of the reduced density matrix of our 
$N^2$ dimensional system
are always of $\Ord{1/N}$. 
As a consequence, the value of energy (average purity) in Eq.\ (\ref{eq:purityN}) 
\begin{equation}
\langle\pi_{AB}\rangle= \Bigg \langle \sum_{1 \leq j \leq N} \lambda_j^2 \Bigg\rangle = \frac{1}{N}\int \lambda^2 \rho(\lambda) d\lambda = \Ord {\frac{1}{N}}
\label{eq:puritysmall}
\end{equation}
is always small: there is therefore a lot of entanglement in our system. Notice that the purity becomes an intensive quantity if multiplied by $N$: $u=N\langle\pi_{AB}\rangle$.
There are however, important differences as purity changes (it is important to keep in mind that in the statistical mechanical approach pursued here, the Lagrange multiplier $\beta$ fixes the value of the energy/purity). As pointed out in Sec. ~\ref{SemicircleEWishart}, when $1/N < \langle\pi_{AB}\rangle \leq 5/4N$ (where $\beta \geq \beta_{+}$) the eigenvalues are distributed according to the semicircle law (Fig.~\ref{fig:semicircle}), while for $5/4N < \langle\pi_{AB}\rangle \leq 2/N$ (where $0\leq \beta< \beta_{-}$) they follow the Wishart distribution (Fig.~\ref{fig:Wisharts}). The two regimes are separated by a second order phase transition. Notice that the value $\langle\pi_{AB}\rangle = 2/N$, which corresponds to infinite temperature $\beta=0$, refers to vectors in the Hilbert space that are typical according to the Haar measure. One is therefore tempted to extend these results to negative temperatures~\cite{paper1} and indeed this can be done up to $\langle\pi_{AB}\rangle = 9/4N$, corresponding to the slightly negative temperature $\beta_g = -2/27$. However, as we have seen, a mathematical difficulty emerges and no smooth continuation of this solution seems possible beyond $\beta_g$.

In the next chapter we will see that two branches exist for negative $\beta$: one containing a phase transition at $\beta=\beta_g$ and in which purity is always of $\Ord{1/N}$ and one in which purity is of $\Ord{1}$. The latter becomes stable as soon $\beta$ becomes negative through a first order phase transition.
We remind that larger values of purity, towards the regime $\pi_{AB} = \Ord{1}$ yield separable (factorized) states. We are therefore going to look at the behavior of our quantum system towards separability (regime of small entanglement).

\chapter{Negative temperatures: metastable branch}\label{chap4}
\markboth{Negative temperatures: metastable branch}
{Negative temperatures: metastable branch}
By analytic continuation, the solution at positive temperatures of the previous chapter can be turned into a solution for negative $\beta$, satisfying the constraints of positivity and normalization. As this is an analytic continuation of the solution obtained for $\beta \geq 0$, we are not assured that this is indeed a stable branch. In this chaper we will study this analytic continuation, but we anticipate that this is metastable as soon as $\beta$ becomes negative (namely for $\beta<-2.455/N$). We call this solution the metastable branch, postponing the proof of its metastability to Chap.~\ref{chap5}. 
The metastable branch is of interest in itself because, as we shall see, it entails two phase transitions. The first one is at $\beta=-2/27\equiv \beta_g$ and corresponds to the so-called 2D quantum gravity free energy (see~\cite{Di Francesco1993nw}), provided an appropriate double-scaling limit (jointly $\beta\to\beta_g$ and $N\to\infty$) is performed. The second phase transition, at $\beta=-\beta_{+}=-2$, is due to the restoration of the $\mathbb{Z}_2$ symmetry that was broken at the phase transition at $\beta=\beta_{+}=2$, described in the previous chapter.

This chapter is organized as follows. In Secs.~\ref{metastabledomain} and~\ref{metastabledistributions} we will determine the physical solution of the Tricomi equation for the range of negative temperatures and select the one which maximizes the potential energy density of the system. 
Thus, in Sec.~\ref{metastablePhaseTrans}, we will unveil the presence of the two mentioned second order phase transitions for the system, and discuss them by studying the behavior of the entropy density. Finally, we will devote Sec.~\ref{metastableentropyVSentanglement} to the overview and the reinterpretation of our results in terms of the average entanglement.  
\section{Physical domain and maximization of the free energy}\label{metastabledomain}
\begin{figure}[h]
\centering
\includegraphics[width=0.78\columnwidth]{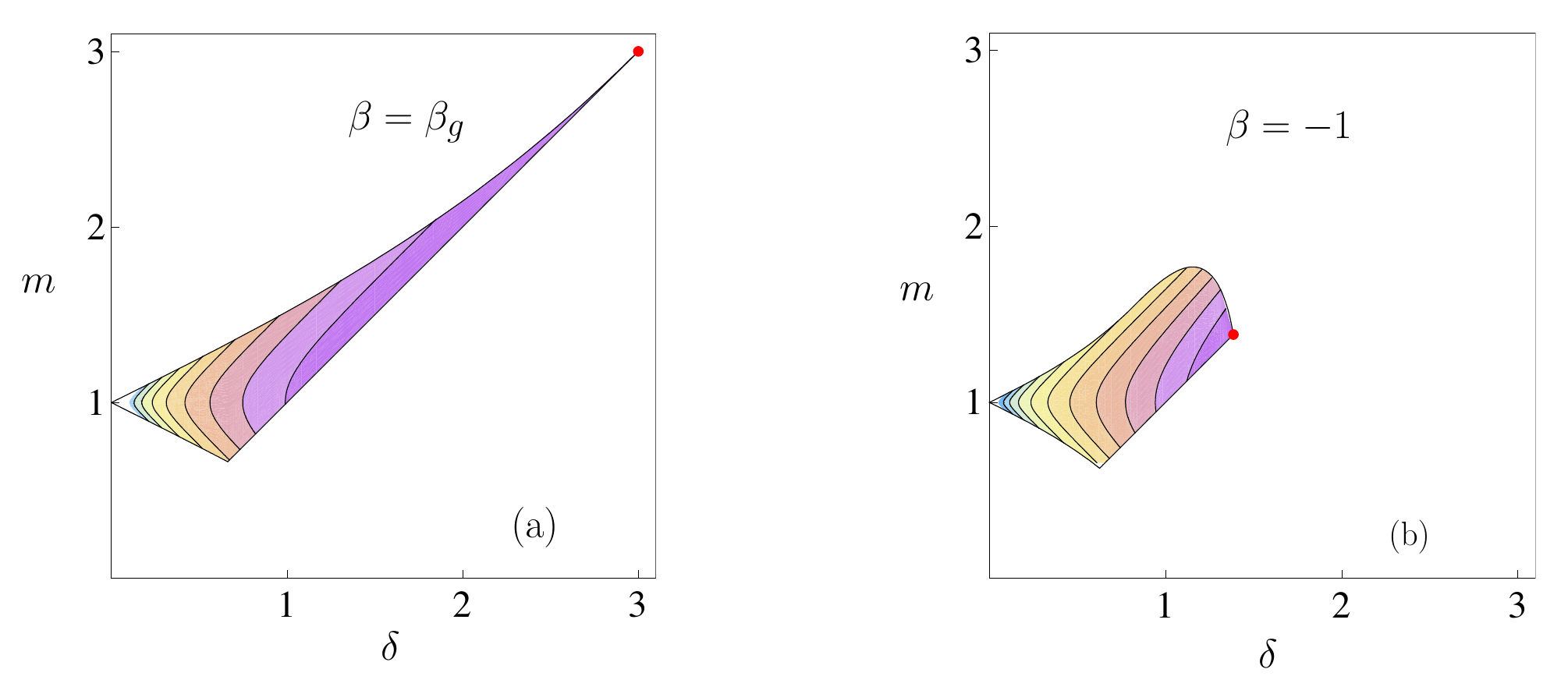}\\
\includegraphics[width=0.35\columnwidth]{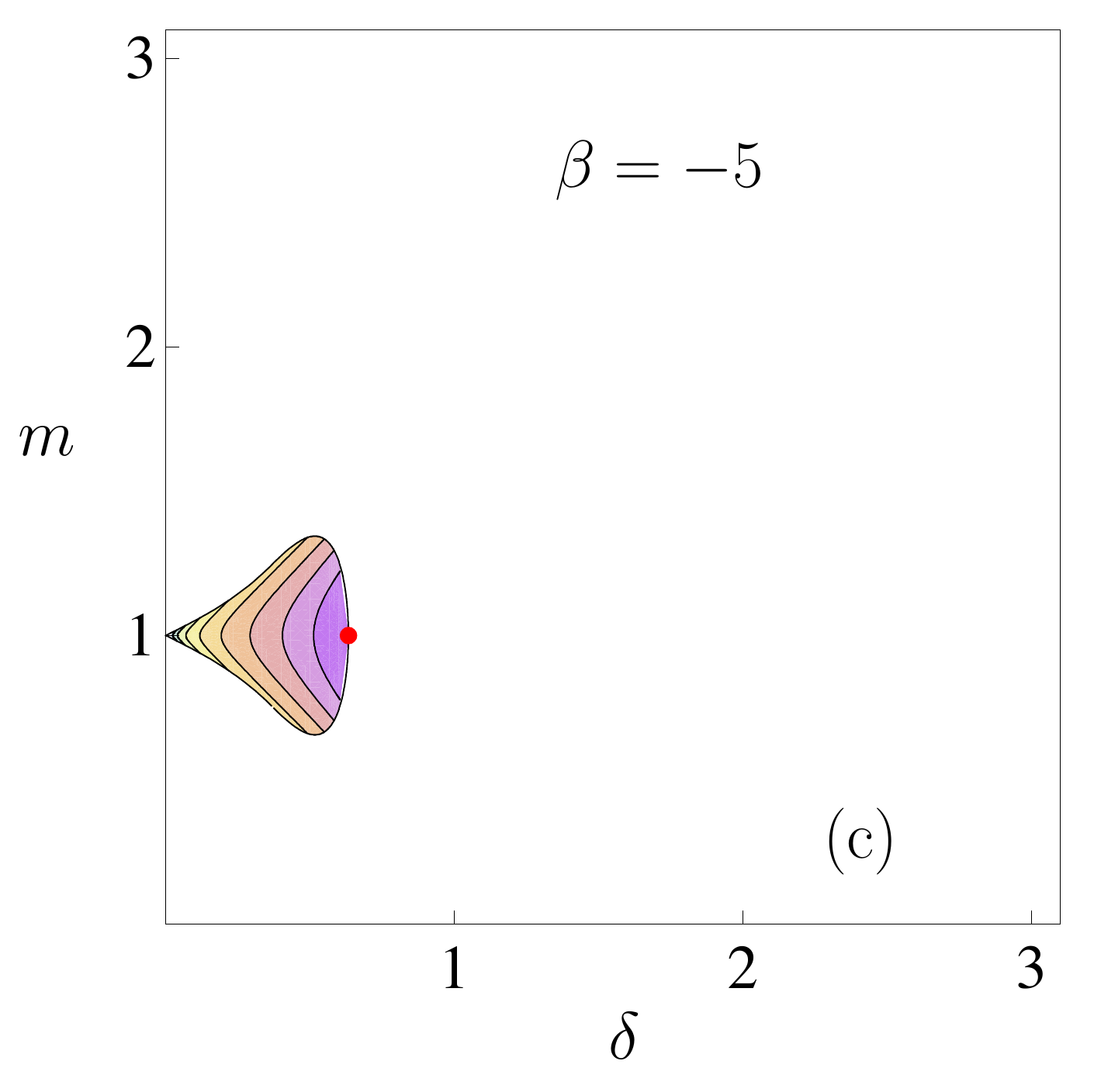}\quad
\caption{Metastable branch. Contour plots of $-f$ in regions of the
parameter space such that $\phi(x)\geq0$, at (a) $\beta=\beta_g$,
(b) $\beta=-1$ and (c) $\beta=-5$. Darker regions have lower free
energy.} \label{fig:contoursneg}
\end{figure}

By keeping the same value of the scaling coefficient $\alpha=3$ used for the positive temperatures, the saddle point equations for the potential energy of the Coulomb gas, lead in the continuous limit to the Tricomi equation (\ref{eq:Tricomi})-(\ref{gx}), now with $\beta<0$.
We recall that its solution is given by Eq. (\ref{eq:phi(x)}) that is 
\begin{equation}\label{eq:phi(x)met}
\phi(x)=
\frac{1}{\pi \sqrt{1-x^2}}\left[ 1+\frac{\beta \delta^2}{2}
+\frac{2(1-m)}{\delta}x -\beta \delta^2 x^2\right],
\end{equation}
being defined for $x \in [-1,1]$. The density distribution is nonnegative for $x\notin (x_{-},x_{+})$, where $\phi(x_{\pm})=0$ and
\begin{equation}
x_{\pm}=\frac{1}{\beta\delta^2} \left(\frac{1-m}{\delta}\pm\sqrt{\Delta}\right),
\end{equation}
with
\begin{equation}
\Delta=\left(\frac{1-m}{\delta}\right)^2+\beta\delta^2 \left(1+\frac{\beta \delta^2}{2}\right). 
\end{equation}
See Eqs. (\ref{eq:roots}) and (\ref{eq:discriminant}). 
For $\beta <0$ we have $\Delta\geq 0$ when
\begin{equation}\label{Gamma2}
\Gamma_2^{-} (\delta,\beta) \leq m \leq \Gamma_2^{+} (\delta,\beta)
\end{equation}
where $m=\Gamma_2^{\pm} (\delta,\beta)$ are the level curves associated to $\Delta = 0$
\begin{equation}\label{levelcurves2}
\Gamma_2^{\pm} (\delta,\beta) = 1\pm\delta^2 \sqrt{-\beta\left(1+\frac{\beta\delta^2}{2}\right)}.
\end{equation}
\begin{figure}[h]
\centering
\includegraphics[width=1\columnwidth]{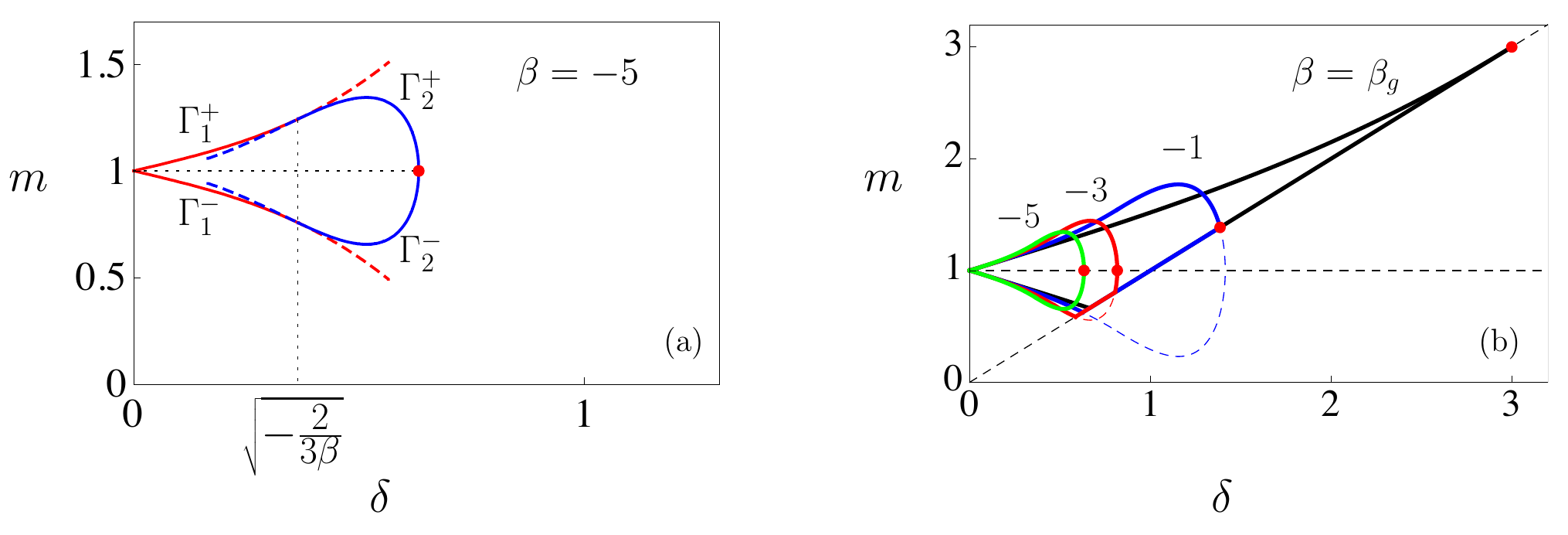}
\caption{Metastable branch. Domain of existence of the solution $(m,\delta)$ for negative temperatures. (a) Solution domain at $\beta=-5$. (b) Solution domains at different
temperatures: for each value of $\beta$ (indicated) the relative full line encloses the region of the parameter space such that the
eigenvalue density is positive, and the red dot indicates its right corner.}
\label{fig:domneg}
\end{figure}
As displayed in Fig.~\ref{fig:domneg}(a) at $\beta=-5$, they are symmetric with respect to the line $m=1$ and intersect at $\delta=0$ and $\delta=\sqrt{-2/\beta}$.
On the other hand the condition $(-1,1)\cap[x_{-},x_{+}]=\emptyset$, implies
\begin{equation}\label{levelcurves1}
\Gamma_1^{-} (\delta,\beta) \leq m \leq \Gamma_1^{+} (\delta,\beta)
\end{equation} 
being $m=\Gamma_1^{\pm} (\delta,\beta)$ the level curves (\ref{Gamma1}), $x_{\pm}=\pm 1$, already computed for $\beta \geq 0$. For $\beta<0$ they are still symmetric with respect to the line $m=1$ but intersect only in $\delta=0$.
The two couples of level curves, $\Gamma_1^{\pm}$ and $\Gamma_2^{\pm}$ intersect at points
\begin{equation}
(\delta,m)=\left(\sqrt{-\frac{2}{3\beta}},\; 1\pm\frac{2}{3}\sqrt{-\frac{2}{3\beta}}\right).
\end{equation}
Therefore, conditions (\ref{levelcurves1}) and $m\leq \delta$, expressing the positivity of the eigenvalues, determine the physical domain of the solution $\phi(x)$  of the Tricomi equation, namely the (possibly cut) eye-shaped domain
\begin{equation}\label{eq:negative_temp_eyesShapeDomain}
\max\left\{ \delta, \; h_{-}(\beta,\delta)\right\} \leq m \leq  h_{+}(\beta,\delta),
\end{equation}
where
\begin{equation}\label{eq:hpm}
h_{\pm}(\delta,\beta) =
\begin{cases}
\Gamma_1^{\pm} (\delta,\beta), &  0\leq \delta\leq \sqrt{-\frac{2}{3\beta}}, \\
\\
\Gamma_2^{\pm} (\delta,\beta), &  \delta > \sqrt{-\frac{2}{3\beta}} .
\end{cases}
\end{equation}
The right corner of the eye is given by
\begin{equation}
\label{eq:rightcornerneg}
(\delta, m)=\left(\sqrt{-\frac{2}{\beta}},  1\right)
\end{equation}
and belongs to the boundary as long as $\beta\leq -\beta_{+}=-2$. This equation is the analogue of Eq.~(\ref{eq:beta>2}) for positive temperatures.
For $\beta\geq-4$ the eye is cut by the line $m=\delta$. See Fig.~\ref{fig:domneg}(b). Notice that, similarly to what found for the case of positive temperatures, the physical domain does not display the same symmetries for the entire range of negative temperatures.
We have already noticed that all points inside the physical domain correspond to solutions of the saddle point equations, for every fixed temperature. The solution corresponding to the most probable distribution of the Schmidt coefficients can be found by maximizing the free energy density $f$ (or $-\beta f$, $\beta<0$) of our Coulomb gas. For the explicit expression of the free energy we still refer to Eq. (\ref{eq:freeenergy}). 
The contour plots of the free energy  density are shown in Fig.~\ref{fig:contoursneg}.
\\
The behavior of $\beta f$ at the boundaries of the allowed domain is shown in Fig.~\ref{fig:3freeneg} for different temperatures.  Notice that the free energy density displays a $\mathbb{Z}_2$ symmetry with respect to the line $m=1$, as already found for the range of positive $\beta$. See Fig.~\ref{fig:contourspos}. The symmetries displayed by the physical domains for different ranges of $\beta$ will naturally emerge in the distribution of the Schmidt coefficients. 
Apart from $\beta=\beta_g$, the free energy density $f$ has no stationary points for $\beta<0$. Thus, the most probable distribution will correspond to an absolute minimum of $\beta f$ on the boundary.
\begin{figure}[h]
\centering
\includegraphics[width=0.91\columnwidth]{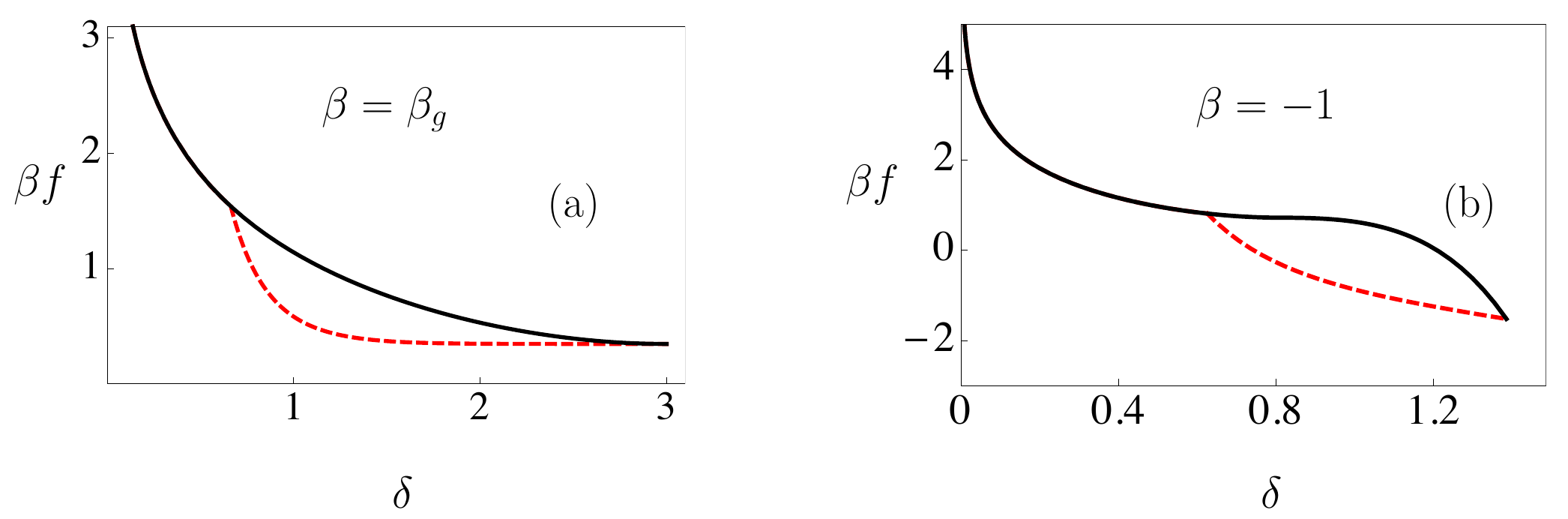}
\\  \vspace{0.2cm}
\includegraphics[width=0.45\columnwidth]{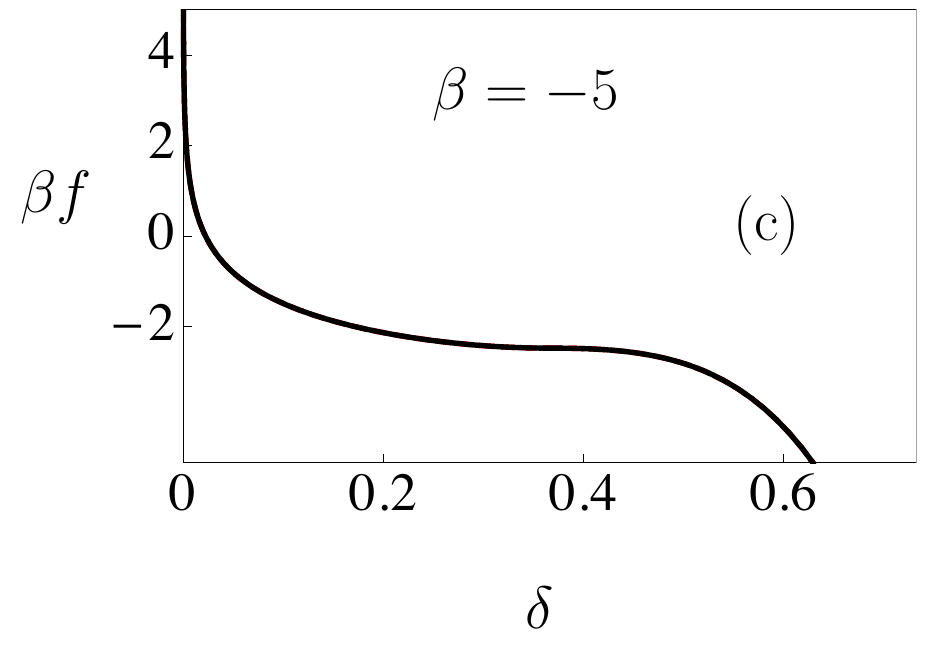}
\caption{Metastable branch. Free energy on the boundary of the region of the domain
where $\phi(x)\geq0$, for different temperatures (indicated).
Dashed lines: free energy $\beta f$ on the lower boundary of the eye-shaped domain;
full lines: free energy on the upper boundary. The sought minima of the free energy can be inferred from the graphs and coincide with the dots in Figs.~\ref{fig:domneg} and~\ref{fig:contoursneg}.}
\label{fig:3freeneg}
\end{figure}\\
In particular we have that:
\begin{itemize}
\item for $\beta\leq -\beta_{+}=-2$ the right corner of the eye (\ref {eq:rightcornerneg})  is the global minimum,
\item
for $-\beta_{+}<\beta<0$ the absolute minimum of $-f$ is at the right upper corner of the allowed region, namely at
\begin{equation}
m=\delta,  \quad \delta= h_{+}(\beta,\delta).
\label{eq:beta-2<0}
\end{equation}
\end{itemize}
The division of the range of negative temperatures in two parts is the first signature that the point $\beta=-\beta_{+}$ will correspond to a change of symmetry in the distribution of the Schmidt coefficients, as can be expected from what seen for $\beta=\beta_{+}$ in the previous chapter. Condition (\ref{eq:beta-2<0}) implies a further arrangement in subintervals of the range $-\beta_{+}<\beta<0$. Recall that the boundary $h_{+}(\delta, \beta)$ consists of two level curves, $\Gamma_2^{+}(\delta,\beta)$ and $\Gamma_1^{+}(\delta,\beta)$, corresponding to nearby intervals of inverse temperature.  See Eq. (\ref{eq:hpm}). Therefore Eq. (\ref{eq:beta-2<0}) entails
\begin{itemize}
\item for $\beta_g\leq \beta < 0$
\begin{equation}
m=\delta ,  \quad \beta = \frac{4}{\delta^3}-\frac{2}{\delta^2} 
\label{eq:prolong}
\end{equation}
\item for $-\beta_{+} < \beta < \beta_g$
\begin{equation}
m=\delta ,  \quad \delta -1 = \delta^2\sqrt{-\beta\left(1+\beta\frac{\delta^2}{2}\right)}.
 \label{eq:prolong1}
\end{equation}
\end{itemize}
Notice that Eq. (\ref{eq:prolong}) coincides with the condition (\ref{eq:beta<2}) found for $0\leq \beta < \beta_{+}$, and leads to the Wishart distribution. We conclude that Eq. (\ref{eq:prolong}) is the analytic continuation of the curve (\ref{eq:beta<2})
which runs monotonically from $\beta=0$ when $\delta=2$ to its minimum  $\beta_g=-2/27$ at $\delta=3$. See Fig.~\ref{fig:betadeltapos}: the continuous and the dashed lines refer to $0\leq \beta < \beta_{+}$ and $\beta_g\leq \beta< 0$, respectively. On the other hand, Eq.(\ref{eq:prolong1})
\begin{equation}
\qquad \delta -1 = \delta^2\sqrt{-\beta\left(1+\beta\frac{\delta^2}{2}\right)},
 \label{eq:prolong1bis}
\end{equation}
admits two real solutions:
\begin{equation}\label{eq:2rami}
\beta =
\begin{cases}
 -\frac{1}{\delta^2}+\frac{1}{\delta^3}\sqrt{(2+\sqrt{2}-\delta)(\delta-2+\sqrt{2})}, &   3 < \delta \leq 2+\sqrt{2}\\
\\
 -\frac{1}{\delta^2}-\frac{1}{\delta^3}\sqrt{(2+\sqrt{2}-\delta)(\delta-2+\sqrt{2})}, & 1<\delta < 2+\sqrt{2} .
\end{cases}
\end{equation}
They run from $\beta= \beta_g$, when $\delta=3$, (with derivative zero) up to $\beta=-3/2 +\sqrt{2}$ when $\delta=2+\sqrt{2}$ (with derivative $-\infty$) and then  from $\beta=-3/2 +\sqrt{2}$, when $\delta=2+\sqrt{2}$, (with derivative $+\infty$) up to $\beta=-\beta_{+}$, when $\delta=1$.
\begin{figure}[h]
\centering
\includegraphics[width=0.47\columnwidth]{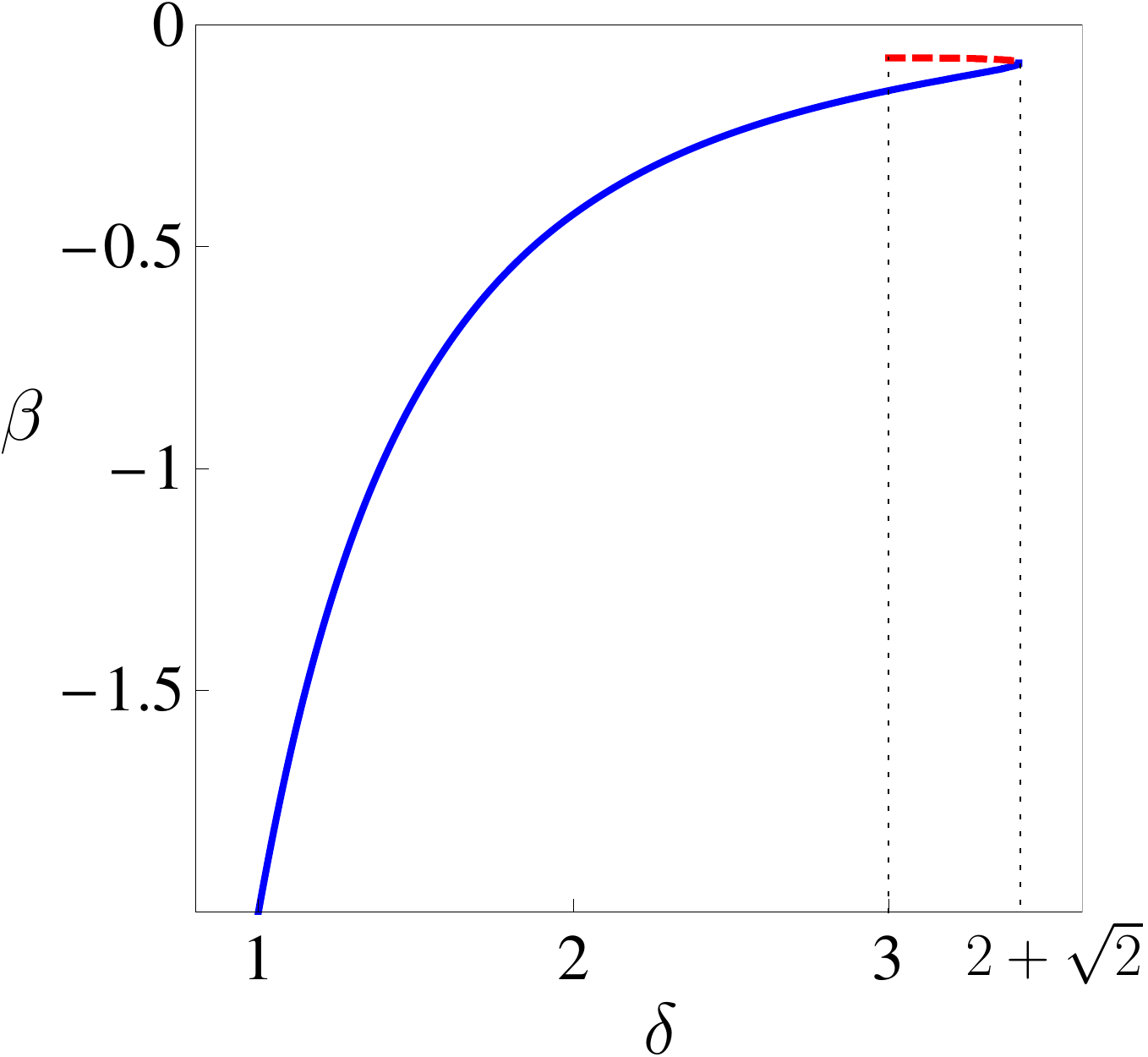}
\caption{Metastable branch. $\beta$ vs $\delta$. See Eqs.(\ref{eq:prolong}) and (\ref{eq:2rami}).}
\label{fig: betadeltaneg}
\end{figure}
See Fig.~\ref{fig: betadeltaneg}.
\section{Distribution of the Schmidt coefficients and thermodynamics of the metastable branch}\label{metastabledistributions}
In this section we will compute the density function for the Schmidt coefficients. They will display different symmetries for different ranges of $\beta$ (or $\delta$), reflecting the symmetry of their domains discussed in the previous section. We will explore the whole range of negative temperatures, from $\beta=0$ ($\delta=2$) to $\beta \to -\infty$ ($\delta=0$). We will then compute the entropy, the internal energy and the free energy of the Coulomb gas of eigenvalues. 
\subsubsection{Wishart distribution: $\beta_g \leq \beta < 0$ - typical states and quantum gravity}
\begin{figure}
\centering
\includegraphics[width=1\columnwidth]{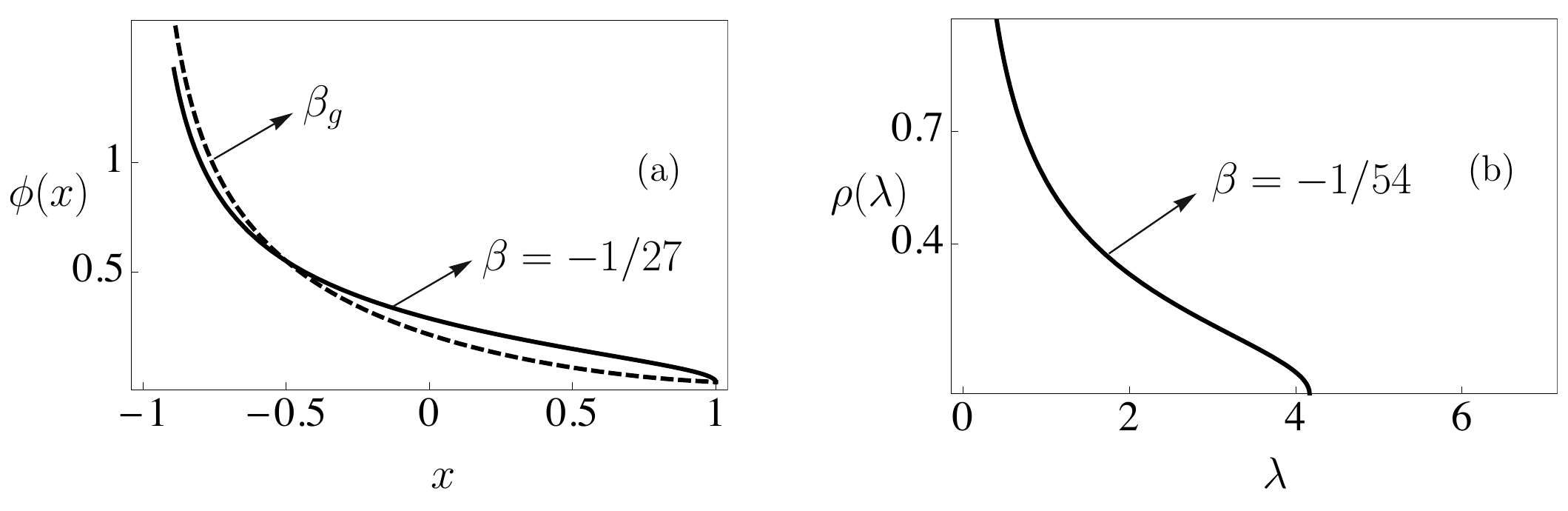}
\caption{Metastable branch. (a) Density of the eigenvalues at $\beta=-1/27$ (solid) and $\beta_g$ (dashed). (b) Density of eigenvalues at $\beta=-1/54$, the solution is given by the Wishart distribution.}
\label{fig:wishartnegr}
\end{figure}

When   $\beta_g\leq \beta < 0$ the probability distribution of the Schmidt coefficients is obtained by plugging (\ref{eq:prolong}) into the generic expression  (\ref{eq:phi(x)met}) for the solution of the Tricomi equation
\begin{equation}
\phi(x)=\frac{2}{\pi\delta}\sqrt{\frac{1-x}{1+x}}\big(1+(2-\delta)x\big) ,\quad 2 < \delta\leq3.
\label{eq:Whishprolong}
\end{equation}
It is the Wishart distribution (\ref{eq:ansatz1phi}) we have found for $\beta<\beta_{+}$ in Chap.~\ref{chap4}. 
At $\beta_g$, where the inverse temperature 
\begin{equation}\label{eq:betadeltaWishartmet}
\beta = \frac{4}{\delta^3}-\frac{2}{\delta^2} 
\end{equation}
is minimum, for $\delta=3$, (see Fig.~\ref{fig:betadeltapos}) the density function becomes
\begin{equation}\label{eq:phibetag}
\phi(x)=\frac{2}{3 \pi}\sqrt{\frac{(1-x)^3}{1+x}} ,
\end{equation}
or equivalently
\begin{equation}\label{eq:Whish0}
\rho(\lambda)=\frac{2}{27\pi} \sqrt{\frac{(6-\lambda)^{3}}{\lambda}}.
\end{equation}
It displays a different shape with respect to the other distributions in this range of temperatures. Indeed for $\beta_g<\beta < \beta_{+}$ the derivative at the right edge of the distribution diverges, whereas at $\beta_g$ it vanishes. See Fig. \ref{fig:wishartnegr}. In Sec.~\ref{metastablePhaseTrans} we will show that the Coulomb gas of eigenvalues undergoes a second order phase transition at $\beta_g$. 
The internal energy, entropy and free energy for the system corresponding to this family of solutions can be determined by inserting (\ref{eq:betadeltaWishartmet}) with $m=\delta$
in Eqs. (\ref{eq:energy density1})-(\ref{eq:freeenergy})
\begin{eqnarray}
u &=& \frac{3}{2}\delta -\frac{\delta^2}{4}, \\
s &=&  -\frac{9}{4} + \frac{5}{\delta} - \frac{3}{\delta^2}
+\ln\frac{\delta}{2} , \label{eq:svsdeltawhishartmetastable}\\
\beta f &=& \frac{11}{4} -\frac{9}{\delta} + \frac{9}{\delta^2} - \ln \frac{\delta}{2} .
\label{eq:gravitysol}
\end{eqnarray}
These expressions determine the thermodynamic properties of the Coulomb gas, and will be employed in the analysis of the phase transition for the system.
\subsubsection{Asymmetric arcsine distribution: $-2 = -\beta_{+}< \beta<\beta_g$ - towards maximally entangled states}
\begin{figure}[h]
\centering
\includegraphics[width=1\columnwidth]{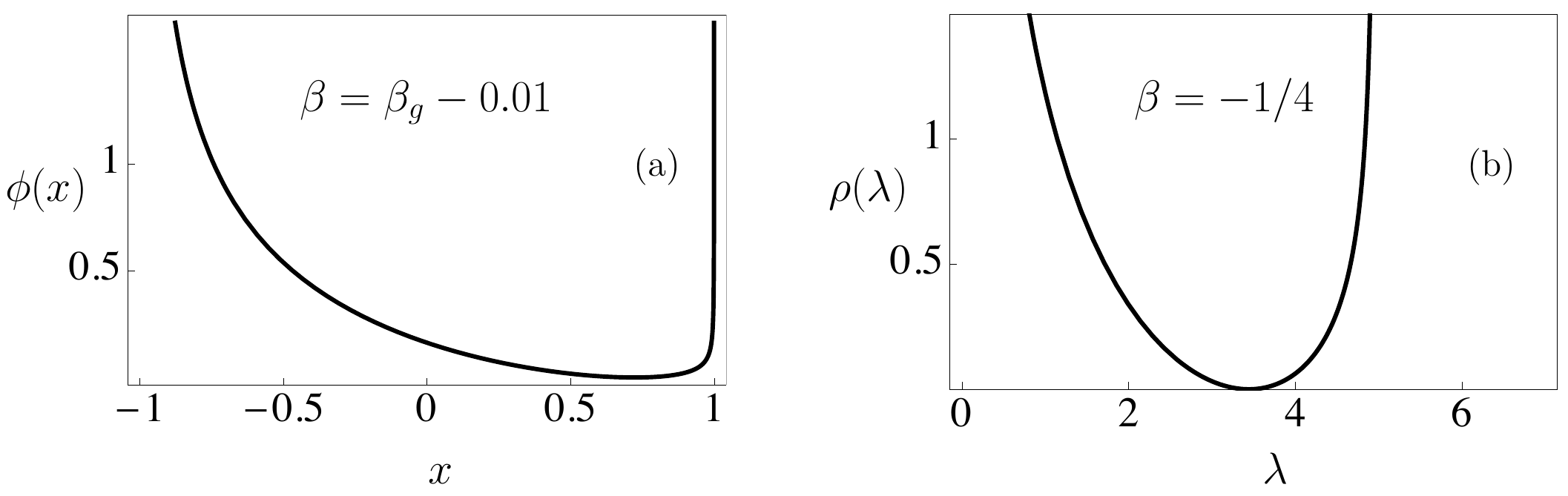}
\caption{Metastable branch. (a) Eigenvalue density for $\beta= \beta_g -0.01 \in [-3/2+\sqrt{2}\leq\beta < \beta_g]$ (see Eq. (\ref{eq:primo})). (b) Eigenvalue density for $\beta=-1/4 \in [-2 < \beta < -3/2+\sqrt{2}]$ (see Eq. (\ref{eq:secondo}), with 
$\lambda= m+ x \delta$, Eq. (\ref{eq:new variable}). In both cases, the distribution of the Schmidt coefficients diverges at the edges and is not symmetric with respect to the center of its domain.} 
\label{fig:m025Embgm001}
\end{figure}

For this range of temperatures the relation between $\beta$ and $\delta$ is more involved. Indeed we find two families of solutions.
By (\ref{eq:2rami}) and (\ref{eq:phi(x)met}) we have that: 
when   $-3/2+\sqrt{2}\leq\beta < \beta_g$ the distribution of the Schmidt coefficients is
\begin{eqnarray}\label{eq:primo}
\phi(x)&=&\frac{1}{\pi\delta\sqrt{1-x^2}}\Big[\frac{1}{2}\left(\delta + \sqrt{-\delta^2+ 4 \delta-2} \right)\nonumber\\ &&\qquad+2(1-\delta)x  + \left(\delta - \sqrt{-\delta^2+ 4 \delta-2} \right) x^2 \Big] ,
\end{eqnarray}
with
\begin{equation}\label{eq:betamestastsub1}
\beta= -\frac{1}{\delta^2}+\frac{1}{\delta^3}\sqrt{(2+\sqrt{2}-\delta)(\delta-2+\sqrt{2})}, \quad   3 < \delta=m \leq 2+\sqrt{2},
\end{equation}
while for   $-2 < \beta< -3/2+\sqrt{2}$ we have
\begin{eqnarray}\label{eq:secondo}
\phi(x)&=&\frac{1}{\pi\delta\sqrt{1-x^2}}\Big[\frac{1}{2}\left(\delta - \sqrt{-\delta^2+ 4 \delta-2} \right)\nonumber\\ &&\qquad+2(1-\delta)x  + \left(\delta + \sqrt{-\delta^2+ 4 \delta-2} \right) x^2 \Big] ,
\end{eqnarray}
with
\begin{equation}\label{eq:betamestastsub2}
\beta= -\frac{1}{\delta^2}-\frac{1}{\delta^3}\sqrt{(2+\sqrt{2}-\delta)(\delta-2+\sqrt{2})}, \quad 1 < \delta < 2+\sqrt{2} .
\end{equation}
The density functions corresponding the these two nearby subintervals of inverse temperatures are shown in Fig.~\ref{fig:m025Embgm001}. Notice that this eigenvalue density diverges \emph{both} at the left edge $x=-1$ \emph{and} at the right edge $x=+1$. It can be seen as an asymmetric arcsine distribution, as it is not symmetric with respect to the center of its domain, $x=0$. 
\\
By plugging (\ref{eq:betamestastsub1}) and (\ref{eq:betamestastsub2}), with $m=\delta$ in the Eqs. (\ref{eq:energy density1})-(\ref{eq:freeenergy}) we compute the internal energy, entropy and free energy for the system. Beyond  $\beta_g$ we get 
for $-3/2+\sqrt{2}\leq\beta < \beta_g$ ($3 < \delta\leq2+\sqrt{2}$) 
\begin{eqnarray}
u &=& 2\delta -\frac{3}{8} \delta^2 -\frac{\delta}{8} \sqrt{-\delta^2+ 4 \delta-2}, \\
s &=&-2 + \frac{15}{4\delta} -\frac{15}{8\delta^2} +\frac{1}{8\delta} \sqrt{-\delta^2+ 4 \delta-2} + \ln \frac{\delta}{2},
\label{eq:svsdeltaasymarcine1}\\
\beta f &=&\frac{5}{2} -\frac{25}{4\delta} +\frac{17}{8\delta^2} -\left(\frac{3}{8\delta} -\frac{2}{\delta^2} \right) \sqrt{-\delta^2+ 4 \delta-2}
- \ln \frac{\delta}{2},
\end{eqnarray}
and 
for $-2 < \beta < -3/2+\sqrt{2}$ ($1< \delta < 2+\sqrt{2}$),
\begin{eqnarray}
u &=& 2\delta -\frac{3}{8} \delta^2 +\frac{\delta}{8} \sqrt{-\delta^2+ 4 \delta-2}, \\
s &=& -2 + \frac{15}{4\delta} -\frac{15}{8\delta^2} -\frac{1}{8\delta} \sqrt{-\delta^2+ 4 \delta-2} + \ln \frac{\delta}{2},  \\
\beta f &=&  \frac{5}{2} -\frac{25}{4\delta} +\frac{17}{8\delta^2} +\left(\frac{3}{8\delta} -\frac{2}{\delta^2} \right) \sqrt{-\delta^2+ 4 \delta-2}
- \ln \frac{\delta}{2}.
\end{eqnarray}
\subsubsection{Arcsine distribution: $\beta \leq -\beta_{+}= -2$ - almost maximally entangled states}
At $\beta=-\beta{+}$ ($\delta=1$) the $\mathbb{Z}_2$ symmetry is restored. We have seen that for $\beta\leq-\beta_{+}$, the global minimum of the physical domain is given by Eq. (\ref{eq:rightcornerneg})
\begin{equation}\label{eq:mEdeltaarcsine}
m=1, \quad \delta=\sqrt{-\frac{2}{\beta}}.
\end{equation}
Observe the analogy with Eq. (\ref{eq:mdeltasemicircle}) for the Wigner distribution, $\beta \geq \beta_{+}$.
We immediately have that for all inverse temperatures lower than $-\beta_{+}$ the solution of the Tricomi equation is given by
\begin{equation}\label{eq:revsemicirclephi}
\phi(x)=\frac{2 x^2}{\pi\sqrt{1-x^2}}, \qquad 0 < \delta \leq 1
\end{equation}
that is
\begin{equation}\label{eq:revsemicirclerho}
\rho(\lambda)=-\frac{\beta(1 - \lambda)^2}{\pi\sqrt{1-\frac{\beta}{2} (1-\lambda)^2}}, \qquad  \beta \leq -2,
\end{equation}
being $\lambda=m+\delta x$ (Eq. (\ref{eq:new variable})).
\begin{figure}[h]
\centering
\includegraphics[width=1\columnwidth]{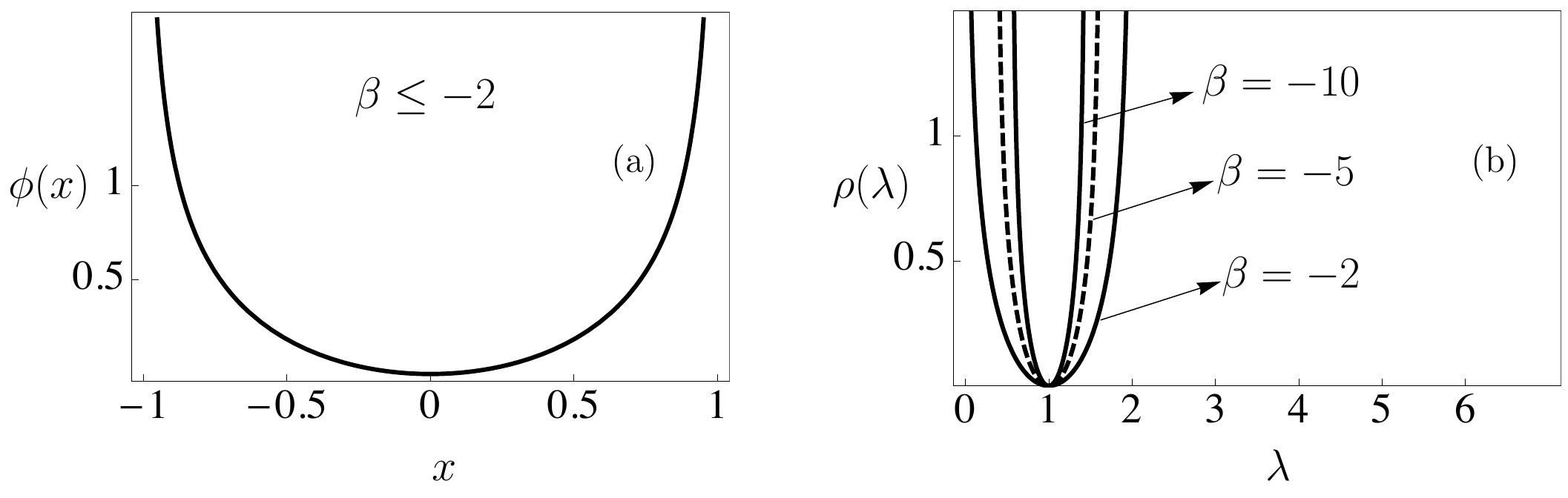}
\caption{Metastable branch. (a) Eigenvalue density for $\beta\leq-\beta_{+}=-2$. (b) Eigenvalue density for $\beta=-2,-5,-10$. For $\beta\leq-\beta_{+}$ the $\mathbb{Z}_2$ symmetry of the eigenvalues distribution with respect to $m=1$ is restored.} 
\label{fig:revsemicircle}
\end{figure}
See Fig.~\ref{fig:revsemicircle}.
The eigenvalues distribution is symmetric with respect to the center of its domain. In the next section, we will show that at $\beta=-\beta_{+}=-2$ the system undergoes a second order phase transition related to the $\mathbb{Z}_2$ symmetry, that mirrors the critical point at $\beta_+=2$. The function (\ref{eq:revsemicirclephi}) or (\ref{eq:revsemicirclerho}) is the arcsine distribution. Furthermore, while $\phi(x)$ does not depend on the temperature, $\rho(\lambda)$ shrinks around $\lambda=1$ as $\beta$ lowers, i.e., recalling the natural scaling (\ref{eq:scalinglambda}), $\lambda_1=\lambda_2=\ldots=\lambda_N= 1/N$ for $\beta \to -\infty$. In other words, the metastable branch for the system leads to maximally entangled states, for very high $-\beta$'s. However, as already anticipated at the beginning of this chapter, we will show that this solution, characterized by the same scaling coefficient $\alpha=3$ of positive temperatures, belongs a metastable branch  for the system. On the contrary, the stable branch will approach the region of states showing a significant separability between systems $A$ and $B$. Furthermore, we will find that the volume occupied by  states with $\langle\pi_{AB}\rangle=1$ ($\beta \to -\infty$) diverges.    

The  thermodynamic functions $u$ (internal energy density) and $s$ (entropy density) computed for the arcsine distribution are
\begin{eqnarray}
u &=& 1+\frac{3}{4} \delta^2 =1-\frac{3}{2\beta},\\
s &=& \ln \frac{\delta}{2} - \frac{1}{4} =  -\frac{1}{2} \ln \left(- 2 \beta \right) - \frac{1}{4} ,\label{eq:svsdeltaminoremeno2}\\
\beta f &=& - \frac{5}{4} -\frac{2}{\delta^2} - \ln \frac{\delta}{2}= -\frac{5}{4}  +\beta + \frac{1}{2} \ln \left(- 2 \beta \right).
\end{eqnarray}
The next section will be devoted to the analysis of the two critical points for the metastable branch, $\beta_g$ and $-\beta_{+}$.

\section{Second order phase transitions for the\\metastable branch}\label{metastablePhaseTrans}
The metastable branch for $\beta<0$, is characterized by the presence of two second order phase transitions for the system. Similarly to what seen in the previous chapter for the range of positive temperatures, they correspond to a change in the symmetry of the distribution of eigenvalues. The thermodynamic quantity we will consider in order to characterize these critical points is the density entropy.

\begin{figure}[h]
\centering
\includegraphics[width=1\columnwidth]{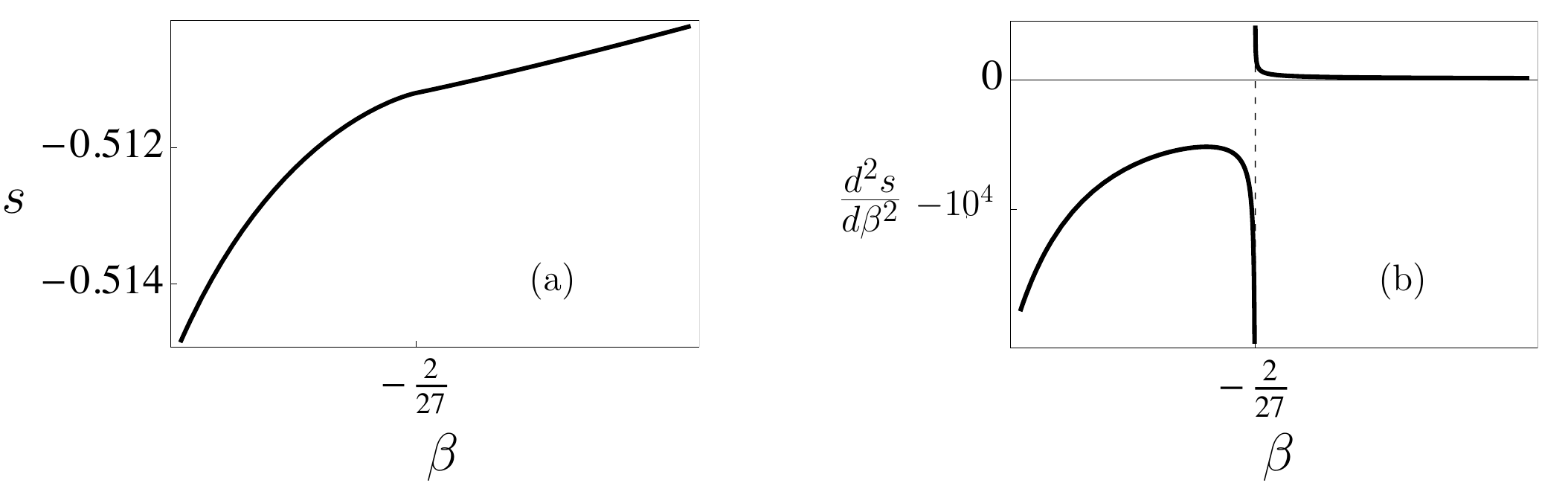}
\caption{Entropy density (a) and its second derivative (b) versus $\beta$.
The entropy is continuous at $\beta_g$ while its second derivative
diverges.}
\label{fig:svsbetag}
\end{figure}

\begin{figure}[h]
\centering
\includegraphics[width=0.52\columnwidth]{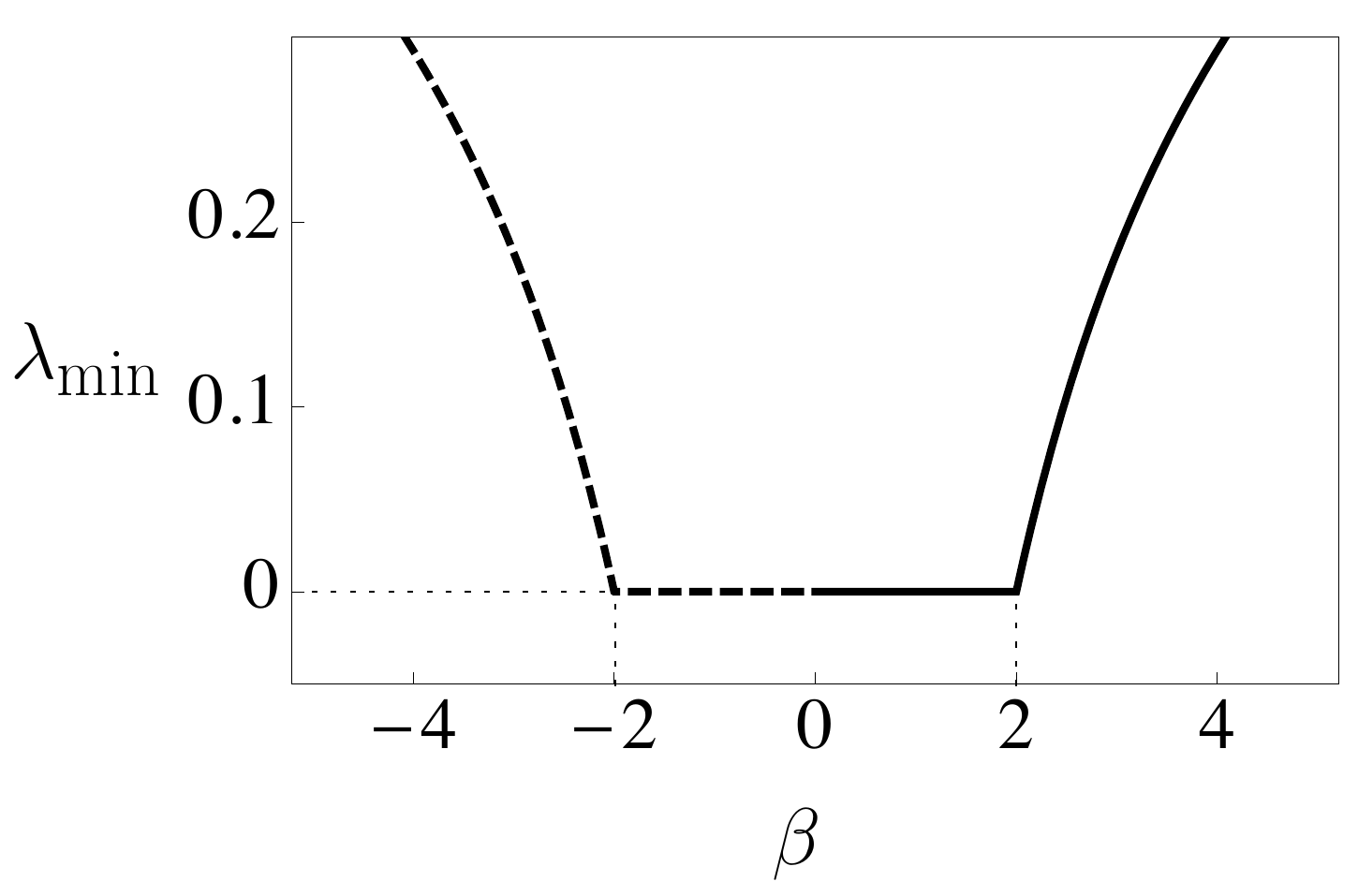}
\caption{Minimum eigenvalue $a=m-\delta$ versus $\beta$. Solid line: stable branch. Dotted line: metastable branch.}
\label{fig:a}
\end{figure}

\begin{figure}[h]
\centering
\includegraphics[width=1\columnwidth]{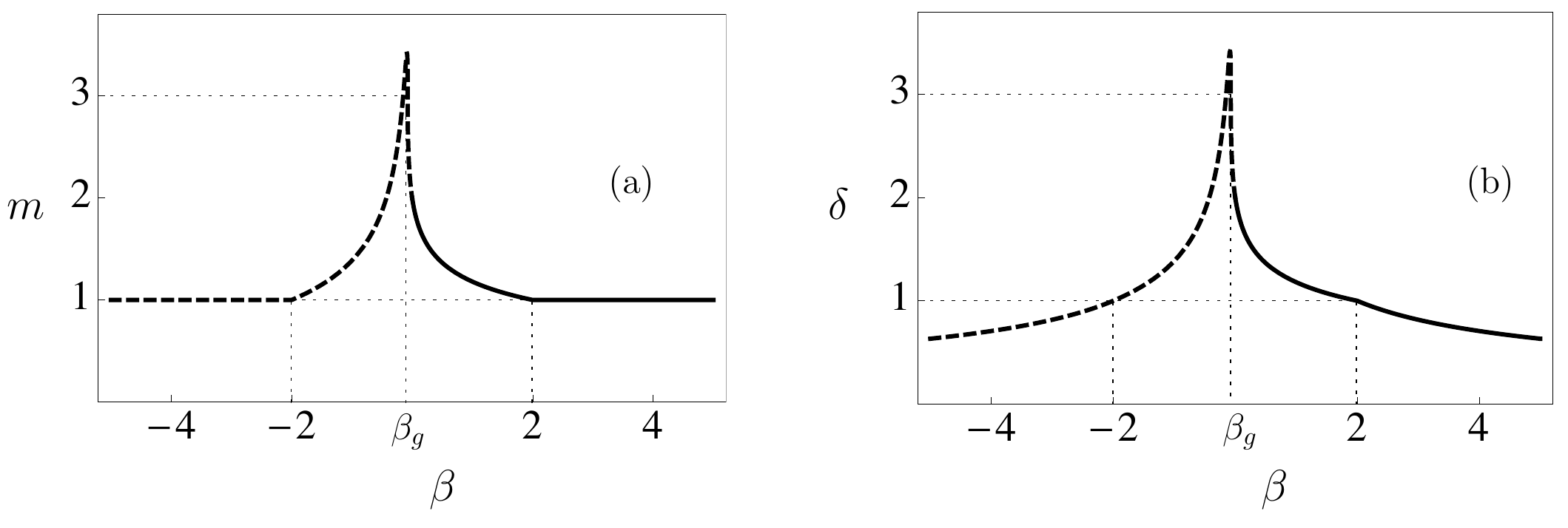}
\caption{Average $m$ (a) and width of the solution domain $\delta$ (b) (Eq.\ (\ref{eq:mdeltadef})) as a function of $\beta$. Solid line: stable branch. Dotted line: metastable branch.}
\label{fig:mEdeltavsbeta}
\end{figure}

\begin{figure}[h]
\centering
\includegraphics[width=1\columnwidth]{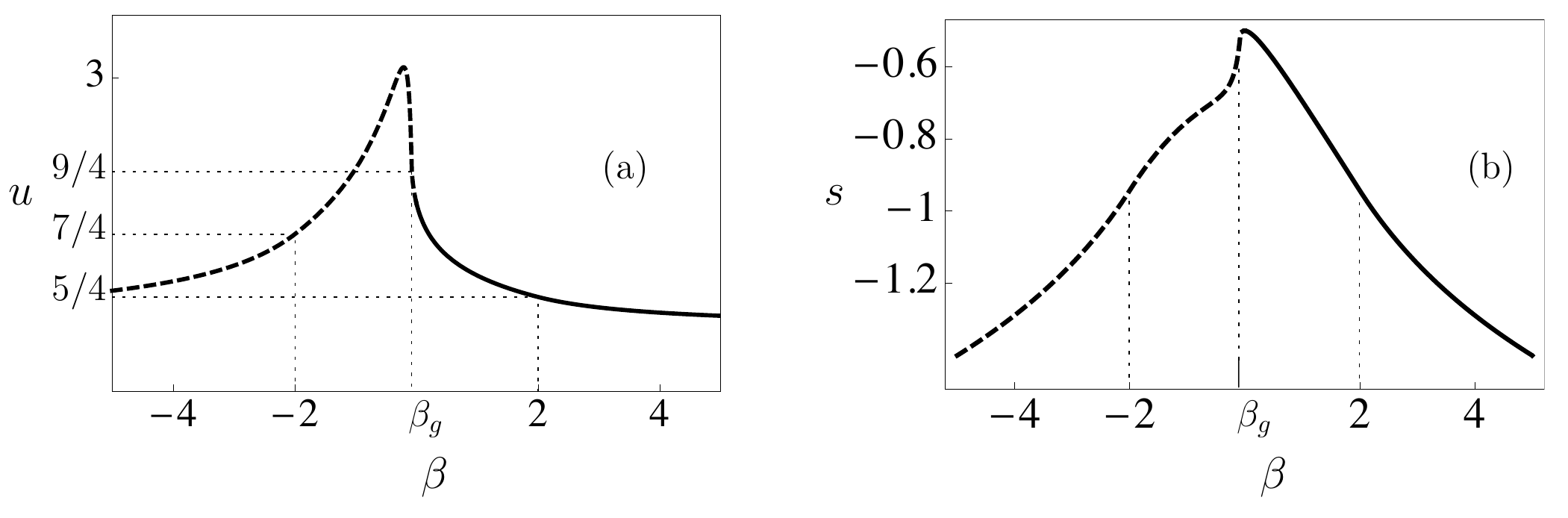}
\caption{Internal energy density $u$ (a) 
and entropy density $s$ (b)
versus $\beta$. Solid line: stable branch. Dotted line: metastable branch.}
\label{fig:uvsbeta}
\end{figure}

The first phase transition of the system, as $\beta$ lowers from $0$ to $-\infty$, appears at the \textit{gravity temperaure}, $\beta_g$, that is $\delta=3$. This point is at the convergence of two solutions, the Wishart and the asymmetric arcsine distribution, given by Eqs. (\ref{eq:Whishprolong}) and (\ref{eq:primo}) respectively. We have also observed that the Wishart distribution at $\beta_g$ shows a vanishing derivative at the right edge. See Figs. ~\ref{fig:Wisharts}(b) and~\ref{fig:wishartnegr}(a). 
In order to characterize this phase transition we expand  Eqs. (\ref{eq:betadeltaWishartmet}) and (\ref{eq:betamestastsub1})  for $\delta \to 3$ and invert the series:
\begin{eqnarray}
\beta&=&-\frac{2}{27} + \frac{1}{81}\left(2-3\ \theta(\delta-3)\right)(\delta-3)^2 -\frac{16}{729}(\delta-3)^3\nonumber\\
&&+\frac{1}{81}\left(\frac{10}{9}-\frac{7}{2}\theta(\delta-3)\right)(\delta-3)^4 +\Ord{(\delta-3)^5},
\end{eqnarray}
which, by setting $x=\sqrt{9|\beta-\beta_g|/2}$, returns
\begin{eqnarray}
\delta&=&3+3x \left(\sqrt{2} +(1-\sqrt{2}) \theta(\beta-\beta_g)\right)+4 x^2\left(4 -3 \theta(\beta-\beta_g)\right)\nonumber\\
&&+\frac{x^3}{\sqrt{18}}\left(- 253 + \left(\frac{35}{\sqrt{2}}+253\right) \theta(\beta-\beta_g)\right) + \Ord{x^4}.
\end{eqnarray}
Therefore the entropy in the critical region $\beta \to \beta_g$ is given by
\begin{eqnarray}
s&=&-\frac{11}{12}+ \ln\left(\frac{3}{2}\right)+\frac{3}{4}(\beta-\beta_g) + 3\left( 5 + (\sqrt{2}-5)\theta(\beta-\beta_g)\right)|\beta-\beta_g|^{3/2}\nonumber\\
&& +\Ord{(\beta-\beta_g)^2}.
\end{eqnarray}
The entropy and its first derivative are continuous at $\beta=\beta_g$, while its second derivative diverges at this point. See Fig.~\ref{fig:svsbetag}.
Thus the critical exponent for the first derivative of the heat capacity is $-1/2$.
\\ By expanding the free energy for $\beta \to \beta_g^+$  we have
\begin{eqnarray}
\beta f &=& \frac{3}{4}- \log \frac{3}{2} + \frac{9}{4} (\beta-\beta_g) - \frac{81}{16} (\beta-\beta_g)^2 \nonumber \\ 
& & 
- \frac{81 \sqrt{2}}{5} (\beta-\beta_g)^{5/2}+\Ord{(\beta-\beta_g)^3}.
\end{eqnarray}
If one relaxes the unit trace condition~(\ref{eq:normeigv}), our model coincides with one that has been studied in the context of
random matrix theories~\cite{Morris91}. The objects generated
in this way, thru an appropriate double scaling limit $\beta \to \beta_g$ and $N \to +\infty$ correspond to chequered polygonations of 2D surfaces, a theory of pure gravity. Therefore, the constraint $\Tr\ \rho_A=1$ is irrelevant for the critical exponents in this region. This is why we call this solution the \textit{gravity distribution} for the gas of eigenvalues.
The second critical point for the system is given by $\beta=-\beta_{+}$ ($\delta=1$), and corresponds to the restoration of the $\mathbb{Z}_2$ symmetry of the density function of the eigenvalues, namely from the asymmetric to the symmetric arcsine distribution. We already noticed that it represents the specular counterpart of the phase transition at $\beta_{+}=2$. Similarly to the gravity point, we expand Eqs. (\ref{eq:betamestastsub2}) and (\ref{eq:mEdeltaarcsine}) for $\delta \to 1$
\begin{eqnarray}
\beta=-2 +4(\delta-1)-(6-\theta(\delta-1))(\delta-1)^2 +\Ord{(\delta-1)^3},
\end{eqnarray}
and for $\beta \to -\beta_{+}=-2$ we have
\begin{eqnarray}
\delta=1+\frac{\beta+2}{4}+\frac{1}{32}\left(3-\frac{1}{2}\theta(\beta+2)\right) (\beta+2)^2+\Ord{(\beta+2)^3}.
\end{eqnarray}
We finally expand the entropy for $\beta \to -\beta_{+}$
\begin{eqnarray}
s=-\frac{1}{4}-\ln2 +\frac{\beta+2}{2} + \frac{1}{8}\left(\frac{1}{2}- \theta(\beta+2)\right)(\beta+2)^2+\Ord{(\beta+2)^3}.
\end{eqnarray}
The entropy together with its first derivative is continuous,
while its second derivative displays a finite jump ($5/32$) at $\beta=-\beta_{+}=-2$.
\begin{figure}[h]
\centering
\quad \includegraphics[width=1.05\columnwidth]{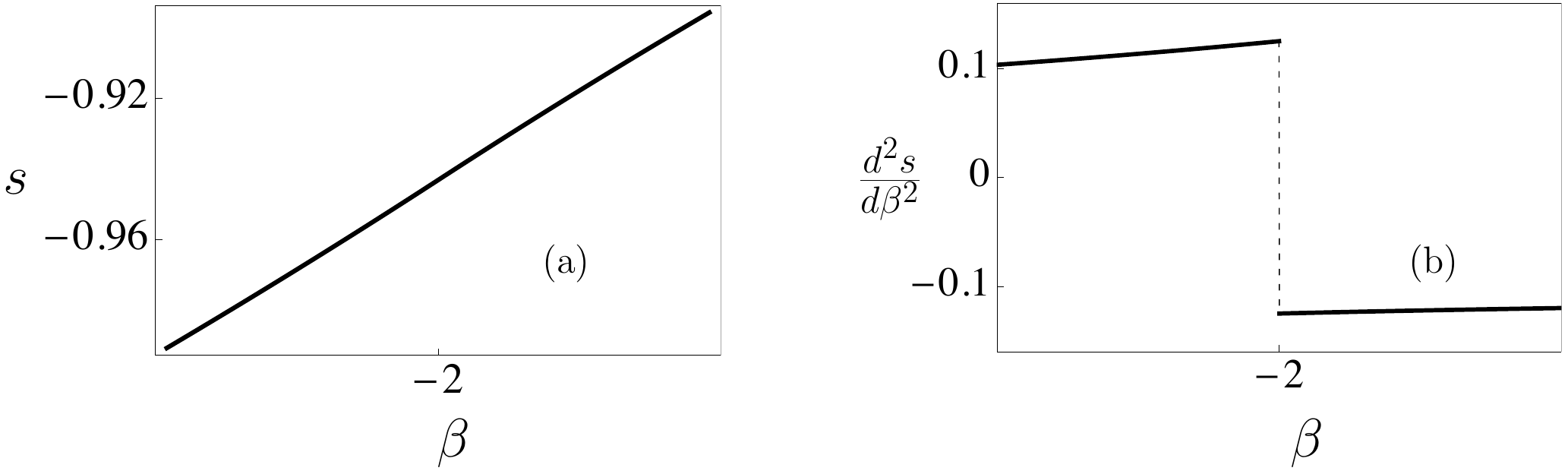}
\caption{Entropy density (a) and its second derivative with respect to $\delta$ (b).
The entropy is continuous at $\beta=-\beta_{+}=-2$ while its second derivative
presents a finite discontinuity.}
\label{fig:svsbetam2}
\end{figure}
See Fig.~\ref{fig:svsbetam2}.\\
Let us briefly overview the behavior of the system for the scaling $\alpha=3$.
We plot the interesting behavior of the minimum eigenvalue $\lambda_{\mathrm{min}}=a=m-\delta$ (see Eqs. (\ref{eq:new variable}) and (\ref{eq:mdeltadef})) plotted in Fig.~\ref{fig:a}.
\\
For $-\beta_{+}<\beta<\beta_{+}$, $a$ coincides with the origin (left border of the solution domain).
This variable signals both second order phase transitions at $-2=-\beta_{+}$ and at 
$2=\beta_+$. 
The $\mathbb{Z}_2$ symmetry is broken for
$-\beta_{+}<\beta<\beta_{+}$. Notice, however, that the gravity critical point at $\beta_g=-2/27$ remains undetected by $a$, while it is detected by the vertical derivative displayed by the average and the width of the solution domain at $\beta_g$. See Fig.~\ref{fig:mEdeltavsbeta}.
However, since $\beta_g$ and $-\beta_{+}$
lie on an analytic continuation of the solution obtained for $\beta \geq 0$, we are not assured that this is a stable branch. In the next chapter we will show that a first order phase transition occurs  at $\beta \simeq -2.455/N$ in this scaling. 
In Fig.~\ref{fig:uvsbeta} we overview the results presented for the scaling $\alpha=3$ by plotting the internal energy $u$ and the entropy density $s$ as a function of the inverse temperature, $-\infty < \beta < \infty$. For $\beta\geq 0$ we have a stable branch for the system (solid line), whereas for $\beta<0$ we will see that the solution is metastable. For the sake of future convenience let us record that at $\beta_{+}=2$  the internal energy density reads $u=5/4$, at $\beta_g=-2/27$ $u=9/4$,  and at $-\beta_{+}=-2$ we have $u=7/4$. 
\section{Entropy density as a function of average entanglement}\label{metastableentropyVSentanglement}
Let us now reinterpret the behavior of the system directly from the point of view of quantum information. In this section, we will study the behavior of the entropy as a function of the internal energy, that is the average purity (see Eq. (\ref{eq:puritysmall})). 
Its expression  for $3 < \delta\leq2+\sqrt{2}$ and $1 <  \delta <  2+\sqrt{2}$ is extremely involved. On the contrary for the other two ranges of $\delta$ we have:
\begin{equation}
\delta  = \begin{cases}
2\sqrt{(u-1)/3},  &  1 < u \leq \frac{7}{4}\\ \\ 
3+\sqrt{9 -4 u},  & 2 < u \leq  \frac{9}{4} , 
\end{cases}
\label{eq:deltavsumetastble}
\end{equation}
which plugged into Eqs. (\ref{eq:svsdeltawhishartmetastable}) and (\ref{eq:svsdeltaminoremeno2}) give 
\begin{eqnarray}\label{eq:svsumetastable}
s(u)  = \begin{cases}
\frac{1}{2} \ln (u-1) -\frac{1}{2} \ln3-\frac{1}{4},  &  1  < u \leq  \frac{7}{4} ,  \\
\\
 \ln \left( \frac{3}{2}  - \sqrt{\frac{9}{4} -u}
\right) - \frac{9}{4} + \frac{5}{2\left( \frac{3}{2}  - \sqrt{\frac{9}{4} -u}
\right)} -
\frac{3}{4\left( \frac{3}{2}  - \sqrt{\frac{9}{4} -u}
\right)^2},
& 2 < u \leq \frac{9}{4}.
\end{cases}
\end{eqnarray}
As can be expected, by analytic continuation we find that $s(u)$ for  $2 < u \leq 9/4$  has the same expression obtained for the interval $5/4 < u \leq 2$ (see Eq. (\ref{eq:112})). Let us now compute $s(u)$ in the critical region unveiled in the last section.
At $\beta_g$ ($\delta=3$) the average purity is $\langle \pi_{AB} \rangle =u/N=9/4N$. By expanding $u$ for $\delta \to 3$ we have
\begin{equation}
u = \frac{9}{4}-\frac{1}{4} \left(1-\frac{3}{2}\theta(\delta-3)\right)(\delta-3)^2+\Ord{(\delta-3)^3}
\end{equation}
and inverting the series we find
\begin{eqnarray}
\delta&=& 3 + \Bigg( 2 + (2 \sqrt{2}-2)\theta\!\!\left(u-\frac{9}{4}\right)\Bigg)x \nonumber \\ &&+ \theta\!\!\left(u-\frac{9}{4}\right) (- 16 x^2 + 116 \sqrt{2} x^3-2272 x^4 ) + \Ord{x^5},
\end{eqnarray}
where $x= \sqrt{|u-9/4|}$.
From Eqs. (\ref{eq:svsdeltawhishartmetastable}) and (\ref{eq:svsdeltaasymarcine1}) we get that for $u \to 9/4$ the entropy is given by
\begin{eqnarray}
s &=&-\frac{11}{12}+ \ln \frac{3}{2} - \frac{2}{27}x^2 -\frac{4}{81}x^4\nonumber \\&&+\frac{1}{3645}\Bigg(256+(1504\sqrt{2}-256) \theta\!\!\left(u-\frac{9}{4}\right)\Bigg) x^5 +\Ord{x^6}.
\end{eqnarray}
\begin{figure}[h]
\centering
\includegraphics[width=1\columnwidth]{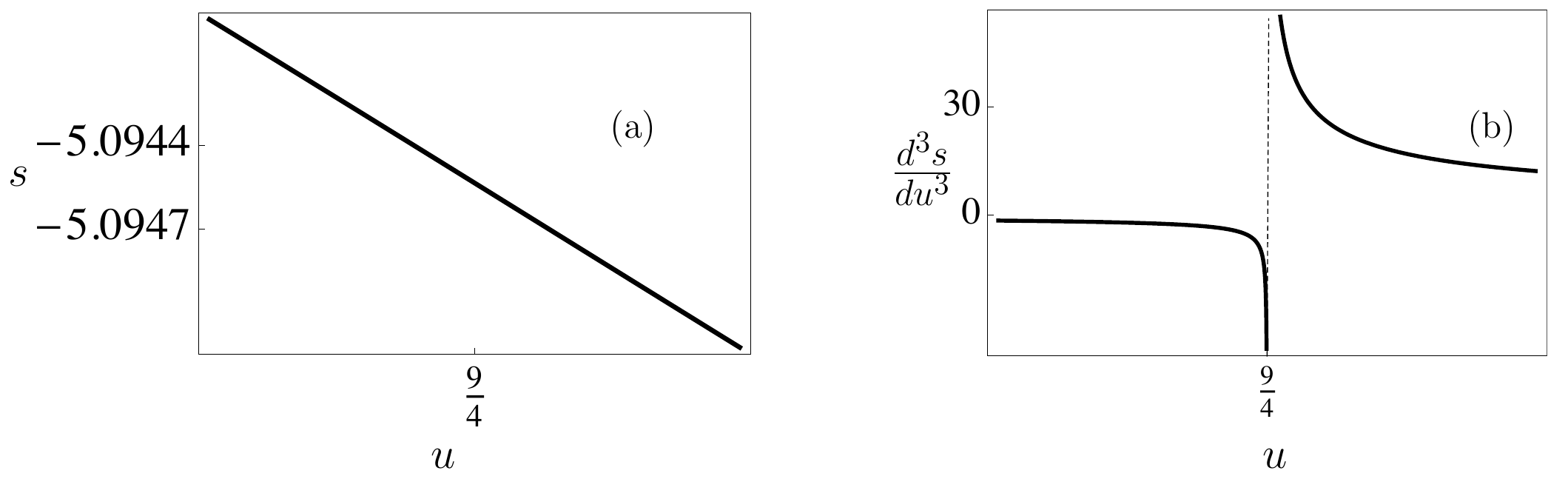}
\caption{Entropy (a) and its second derivative (b) versus $\beta$.}
\label{fig:SvsUbetag}
\end{figure}
Notice that even if we are exploring a metastable solution for the system, and thus the Maxwell relations may not be satisfied, at $\beta=\beta_g$ we have $d s / d u =\beta_g$. Furthermore, the second order phase transition at $u=9/4$ is unveiled by a divergence for $u\to (9/4)^{+}$ of the third derivative of the entropy, $d^3 s / d u^3$ (see Fig.~\ref{fig:SvsUbetag}), while we have seen that its second derivative with respect to $\beta$
diverges with different rates on both sides of $\beta_g$.
This behavior can be justified by observing that the solution $m=\delta=3$ found by minimizing the free energy in the physical domain, is also a stationary point for the system. This can be inferred from the plot in  Fig.~\ref{fig:3freeneg}(a) of the free energy on the boundary of the domain, as $m$ and $\delta$ tend to $3$ (see Fig.~\ref{fig:contoursneg}(a)).

\begin{figure}[h]
\centering
\includegraphics[width=1\columnwidth]{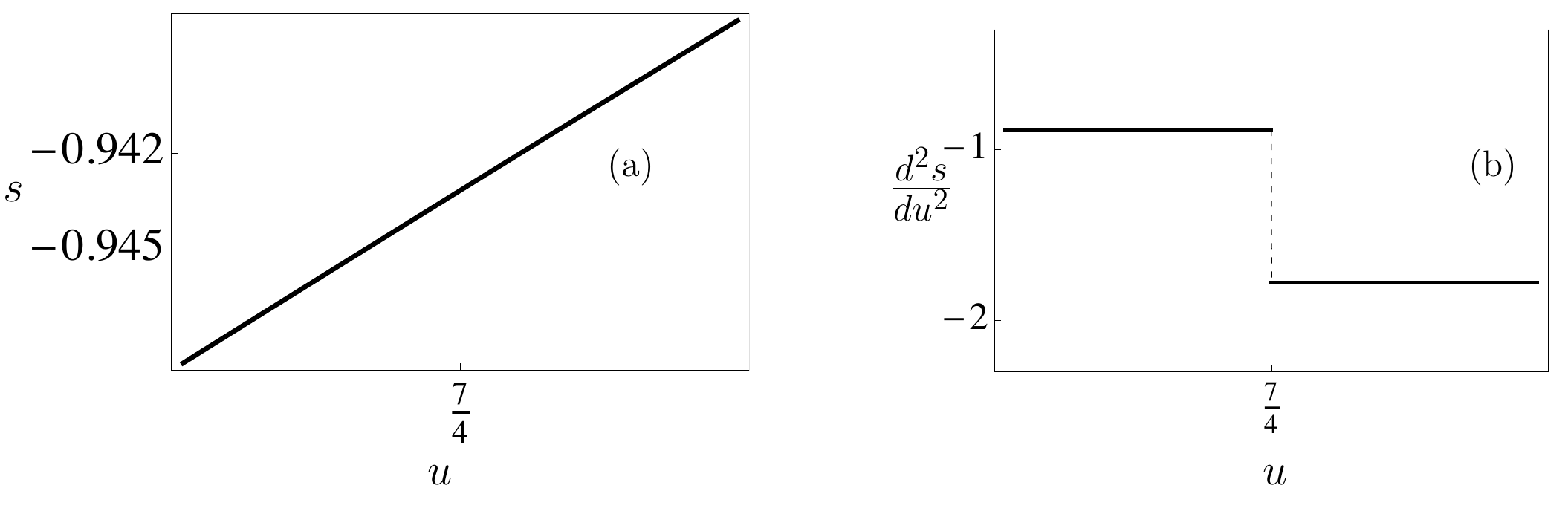}
\caption{Entropy (a) and its second derivative (b) versus $\beta$.}
\label{fig:SvsUmeno2}
\end{figure}

Let us now consider the other critical point at $\beta=-\beta_{+}$ ($\delta=1$). We have 
\begin{eqnarray}
u&=& \frac{7}{4} +\frac{3}{2} (\delta-1)+ \frac{3}{4}\left( 1- \frac{3}{2} \theta(\delta-1)\right)(\delta-1)^2-\frac{1}{16}\theta(\delta-1)(\delta-1)^4\nonumber\\&&+\Ord{(\delta-1)^5}\\
\delta&=&1+\frac{2}{3}\left(u-\frac{7}{4}\right)-\frac{2}{9}\left(1+\frac{5}{9}\theta\left(u-\frac{7}{4}\right)\right)\left(u-\frac{7}{4}\right)^2 \nonumber\\\qquad&&+\frac{1}{27}\left(4-3\theta\left(u-\frac{7}{4}\right)\right)\left(u-\frac{7}{4}\right)^3+\Ord{\left(u-\frac{7}{4}\right)^4}\\ 
s&=&-\frac{1}{4}-\ln2+\frac{2}{3}\left(u-\frac{7}{4}\right)-\frac{4}{9}\left(1+\theta\left(u-\frac{7}{4}\right)\right)\left(u-\frac{7}{4}\right)^2\nonumber\\&&+\frac{4}{81}\left(8+9\theta\left(u-\frac{7}{4}\right)\right)\left(u-\frac{7}{4}\right)^3+\Ord{\left(u-\frac{7}{4}\right)^4}.\label{eq:svilSvsBetaAm2}\end{eqnarray}
See Fig.~\ref{fig:SvsUmeno2}. Observe that in this case, the second derivative of the entropy with respect to the internal energy is discontinuous, as found by deriving $s$ with respect to $\beta$.

\begin{figure}[h]
\centering
\includegraphics[width=0.58\columnwidth]{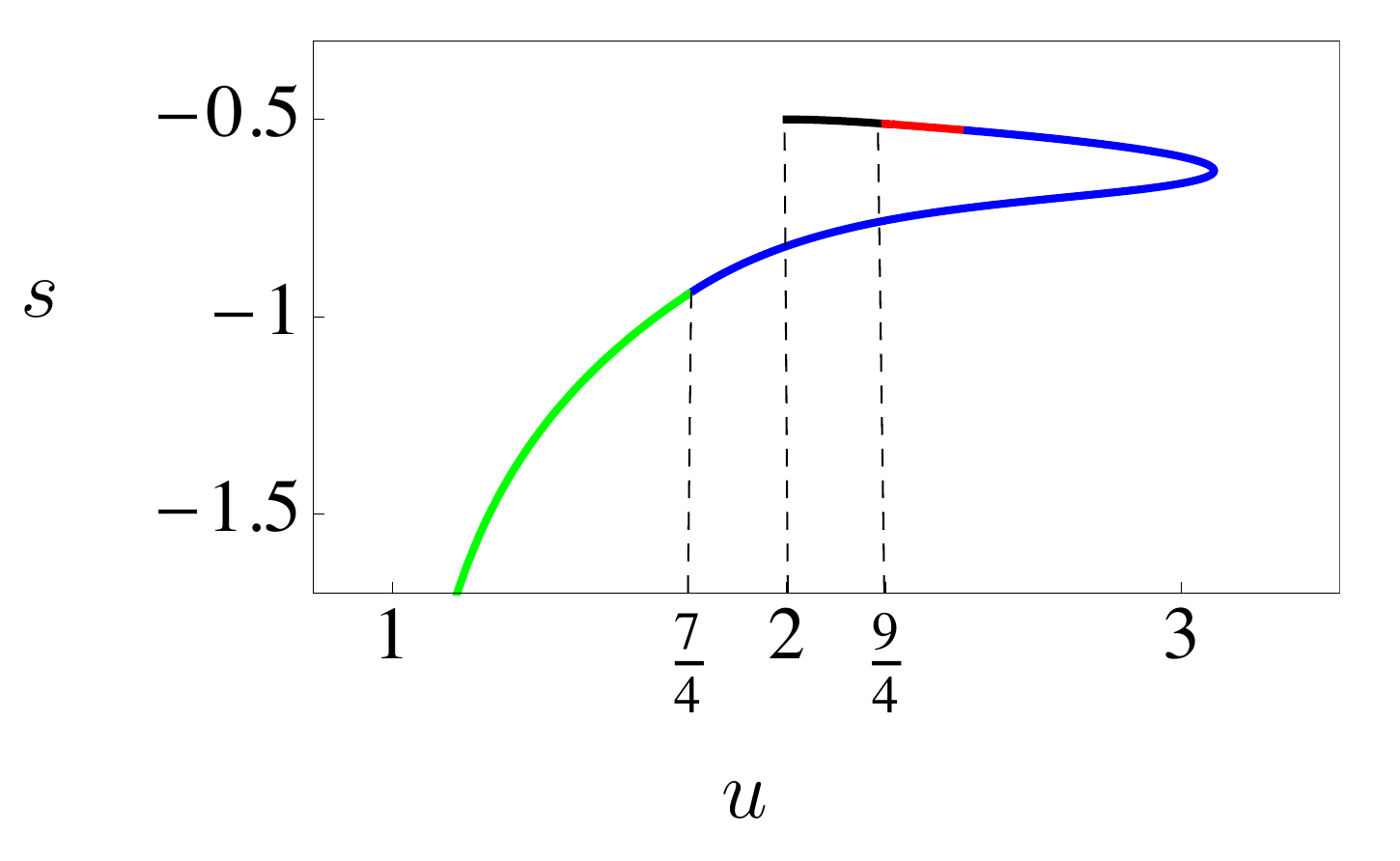}
\caption{Entropy density versus average entanglement $u=N \langle \pi_{AB}\rangle$, for the metastable branch. The four solutions are plotted with different colors. In particular both the red and the blue lines refer to the asymmetric arcsine distribution, and correspond to the real solutions of Eq. (\ref{eq:prolong1bis}).}
\label{fig:SvsUr}
\end{figure}

\begin{figure}[h]
\centering
\includegraphics[width=0.58\columnwidth]{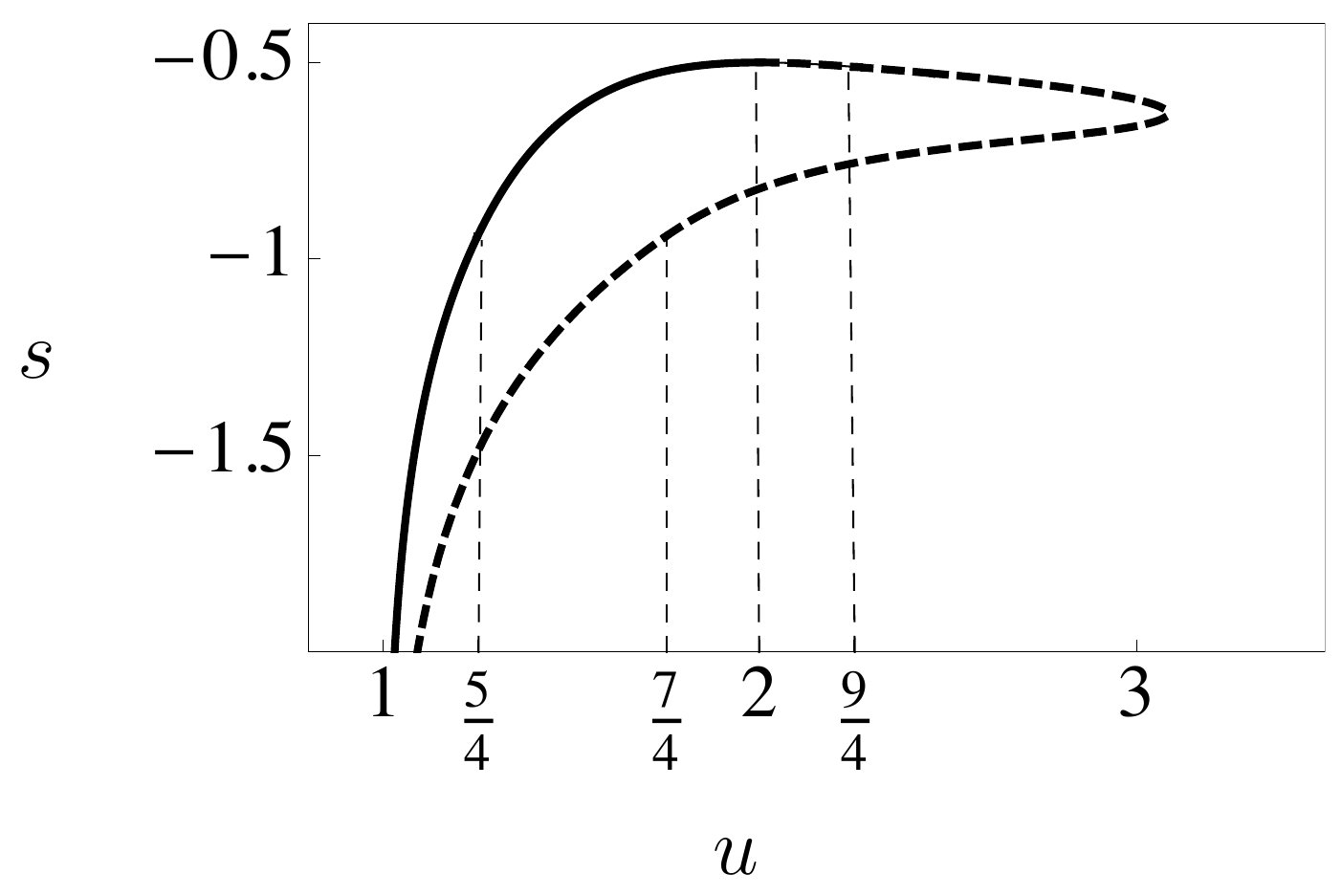}
\caption{Entropy density versus average entanglement $u=N \langle \pi_{AB}\rangle$ for the scaling 
$\alpha=3$: positive temperatures and metastable branch of negative temperatures. See Figs.~\ref{fig:svsuN} and~\ref{fig:SvsUr}.}
\label{fig:SvsUalpha3r}
\end{figure}

Note incidentally that the point $\beta=-3/2+\sqrt{2}$ ($\delta=2+\sqrt{2}$), connecting real solutions of Eq. (\ref{eq:prolong1}), is not a critical point for the system. Indeed for $\delta \to 2+\sqrt{2}$ we get
\begin{eqnarray}
\left(\beta+\frac{3}{2}-\sqrt{2}\right)^{2}&=& \frac{2\sqrt{2}(2+\sqrt{2}-\delta)}{(2+\sqrt{2})^6}+\mathcal{O}((2+\sqrt{2}-\delta)^2)\\
\left(s-s_0\right)^{2}&=&\frac{1}{64}(2+\sqrt{2}-\delta)(3\sqrt{2}-4)+\Ord{(2+\sqrt{2}-\delta)^2}\!,
\end{eqnarray}
where
\begin{equation}
s_0=-\frac{17}{16} + \ln\left(1 + \frac{1}{\sqrt{2}}\right).
\end{equation}
It then follows that the entropy has a linear behavior for $\beta=-3/2+\sqrt{2}$, that is
\begin{equation}
(s-s_0) =\left(\frac{3}{4}+\frac{1}{\sqrt{2}}\right)\left(\beta+\frac{3}{2}-\sqrt{2}\right) + \Ord{(2+\sqrt{2}-\delta)^2}.
\end{equation}
Indeed the divergences in the derivative of the entropy and in the derivative of the inverse temperature with respect to $\delta$ at this point compensate. In Fig.~\ref{fig:SvsUr} we show the behavior of the entropy for the metastable branch.
 \\
Let us briefly comment on the fact that  the metastable branch which emanates from the analytic continuation of the solution at positive $\beta$ described in the previous chapter has \emph{not} led us towards separable states. See also Fig.~\ref{fig:SvsUalpha3r} where we plot the entropy as a function of the internal energy. We recall that in this scaling ($\alpha=3$) the average purity is $\langle \pi_{AB} \rangle=u/N$, namely of $\mathcal{O}(1/N)$. The eigenvalues remain of $\Ord{1/N}$ (and so does purity) even though the temperature can be (very) negative (as $\beta$ crosses $0$). In order to find separable states we will have to look at the stable branch in the next chapter.
We conclude that until now we have studied the region of the Hilbert space given by highly entangled states.

\chapter{Negative temperatures: stable branch of separable states}\label{chap5}
\markboth{Negative temperatures: stable branch of separable states}
{Negative temperatures: stable branch of separable states}

In this chapter we will search the stable solution of the system at negative temperatures. We will find that for $\beta \to -\infty$ the most probable solution for the Coulomb gas of eigenvalues involves the evaporation of one Schmidt coefficient from $\Ord{1/N}$ to $1$. In other words, it will appear a new phase for the system related to the set of separable states. This solution will be associated to the appearance of a new critical point for the system, whose nature will be different from the phase transitions discussed in the previous chapters. 
In particular, by maximizing the potential energy of the Coulomb gas, we will find that at $\beta=-2/N$ a competition starts between two solutions: one with all the eigenvalues of order $\Ord{1/N}$, and one with an isolated eigenvalue of $\Ord{1}$ and all the others of $\Ord{1/N}$, namely the \textit{typical} and \textit{separable} phases. At $\beta_{-}/N\simeq -2.455/N$ we will unveil the coexistence of these two phases, and the system undergoes a first order phase transition. Finally, for $\beta \leq \beta_{-}/N$ the stable phase for Coulomb gas will be given by separable states. However, in order to yield the correct thermodynamic limit in this statistical analysis, we will need to properly modify the scaling coefficient with respect to the stable solution of positive temperatures (and the metastable branch for negative temperatures). This will introduce a new scale in the inverse temperatures and, in particular, the phase transition in the new scaling will be at $\beta_{-}\simeq -2.445$.

In Sec.~\ref{natural scaling neg} we will discuss the new scaling for the system, which will affect the saddle point equations, whose physical solution will be obtained in Sec.~\ref{saddle point neg}. 
In Sec.~\ref{thermodynamicSepStates} we will look for the most probable distribution of the Schmidt coefficients by maximizing the free energy of the Coulomb gas, and then unveil the presence of the new separable phase. In Sec.~\ref{firstOrderphTrans} we will discuss the thermodynamics of the system and identify the first order phase transition from typical to separable states. We will conclude this chapter with a brief overview about the finite size corrections, in Sec.~\ref{sec:finitesize}: on the one hand we will explain our numerical results, on the other hand we will provide a mapping  between the results shown in Chaps.~\ref{chap3} and~\ref{chap4} and the stable solution for negative temperatures described in the next sections.

\section{Natural scaling for $\beta<0$}\label{natural scaling neg}
\begin{figure}[h]
\centering
\includegraphics[width=1\columnwidth]{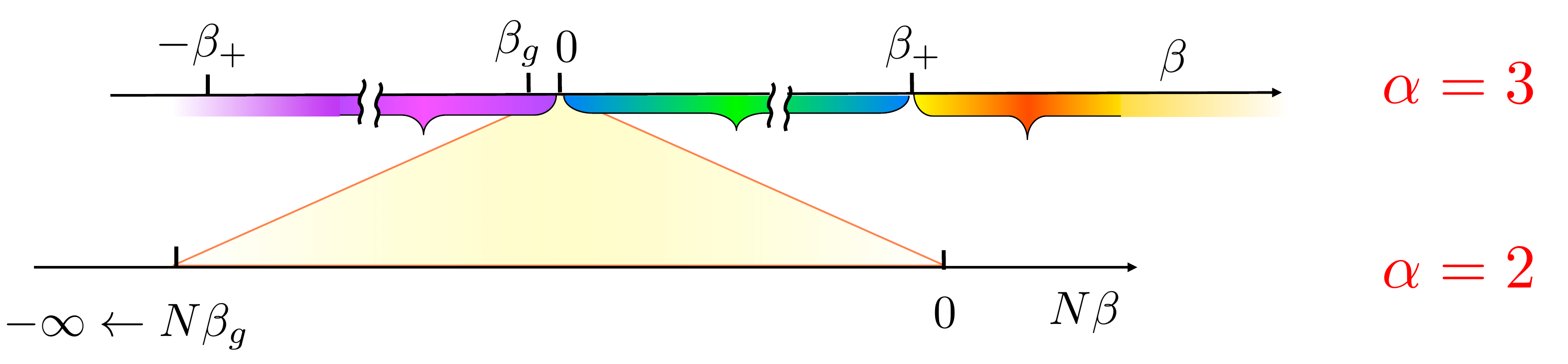}
\caption{By setting the scaling coefficient $\alpha=2$ for $\beta<0$ we zoom on the region $\beta=\Ord{1/N}$ of the scaling $N^3$  (i.e. $\alpha=3$)
introduced for positive temperatures. The critical point
$\beta_g=-2 / 27$ now reads
$\beta=-(2/27) N$ and escapes to $-\infty$.}
\label{fig:zooming}
\end{figure}

From definition (\ref{eq:partitionfunction}) of the partition
function, one expects that for any $N$, as $\beta \rightarrow
-\infty$, the system approaches the region of the phase space
associated to separable states: here the purity is $\Ord{1}$
and thus the scaling coefficient in Eqs.\ (\ref{eq:partitionfunction})-(\ref{eq:piscale})  is $\alpha = 2$.
From the point of view of the parameter $\beta$, this can be interpreted as follows: by adopting the scaling $N^2$ for the exponent of the partition function, we will
explore the region $\beta=\Ord{1/N}$ of the scaling $N^3$
introduced for positive temperatures. Notice that the critical point
$\beta_g=-2 / 27$ for the solution at negative temperatures now reads
$\beta=-(2/27) N$ and escapes to $-\infty$ in the thermodynamic
limit, as pictorially shown in Fig.~\ref{fig:zooming}.
\\
We will show that the solution (\ref{eq:ansatz1}), according to which all the eigenvalues are $\Ord{1/N}$, becomes metastable in the region of negative
temperatures, and the most probable distribution of the eigenvalues is such that one eigenvalue is $\Ord{1}$: this
solution in the limit $\beta \rightarrow -\infty$ will correspond to the
case of separable states. By following an approach similar to that adopted for positive temperatures, we will first look for
the set of eigenvalues $\{\lambda_1, \ldots, \lambda_N\}$ satisfying
the saddle point equations (\ref{eq:stat1})-(\ref{eq:normal}) with
$\alpha=2$, and get, as in Sec.~\ref{solution}, a continuous
family of solutions. We will select among them the set maximizing
($\beta<0$) the free energy (\ref{eq:freeF}), with $\alpha =2$:
\begin{eqnarray}\label{eq: free energy}
f_{N}= \sum_{1\leq i \leq N} \lambda_i^2-\frac{2}{N^2 \beta} \sum_{1\leq
i<j\leq N} \ln |\lambda_j-\lambda_i|.
\end{eqnarray}
Since we are
approaching the limit $\beta \rightarrow -\infty$ the states
occupying the largest volume in phase space are separable. We define $\lambda_N= \mu$ as the maximum eigenvalue of order of unity, whereas the other eigenvalues
are $\Ord{1/N}$:
\begin{equation}\label{eq:eigenvalues for negative temperatures}
\lambda_N=\mu= \Ord{1}, \quad \sum_{1 \leq i \leq N-1}\lambda_i=1-\mu .
\end{equation}
From this it follows that we need to introduce the natural scaling
only for the \textit{sea} of the first $N-1$ eigenvalues:
\begin{eqnarray}
\label{eq:naturalRescaling for Negative temeperatures}
\lambda_i &=& (1-\mu)\frac{\lambda(t_i)}{N-1},  \quad 0<t_i=\frac{i}{N-1}\leq1, \quad \forall \ i \in \{1, \ldots, N-1\}. 
\end{eqnarray}
In the limit $N \to \infty$ we will describe the behavior of the rescaled eigenvalues in terms of a continuous, non negative and normalized density function as discussed in section Sec.~\ref{natural scaling} (see Eqs. (\ref{eq:rhodef}) and (\ref{eq:norm})).
\section{Saddle point equations}\label{saddle point neg}
\begin{figure}[h]
\centering
\includegraphics[width=0.55\columnwidth]{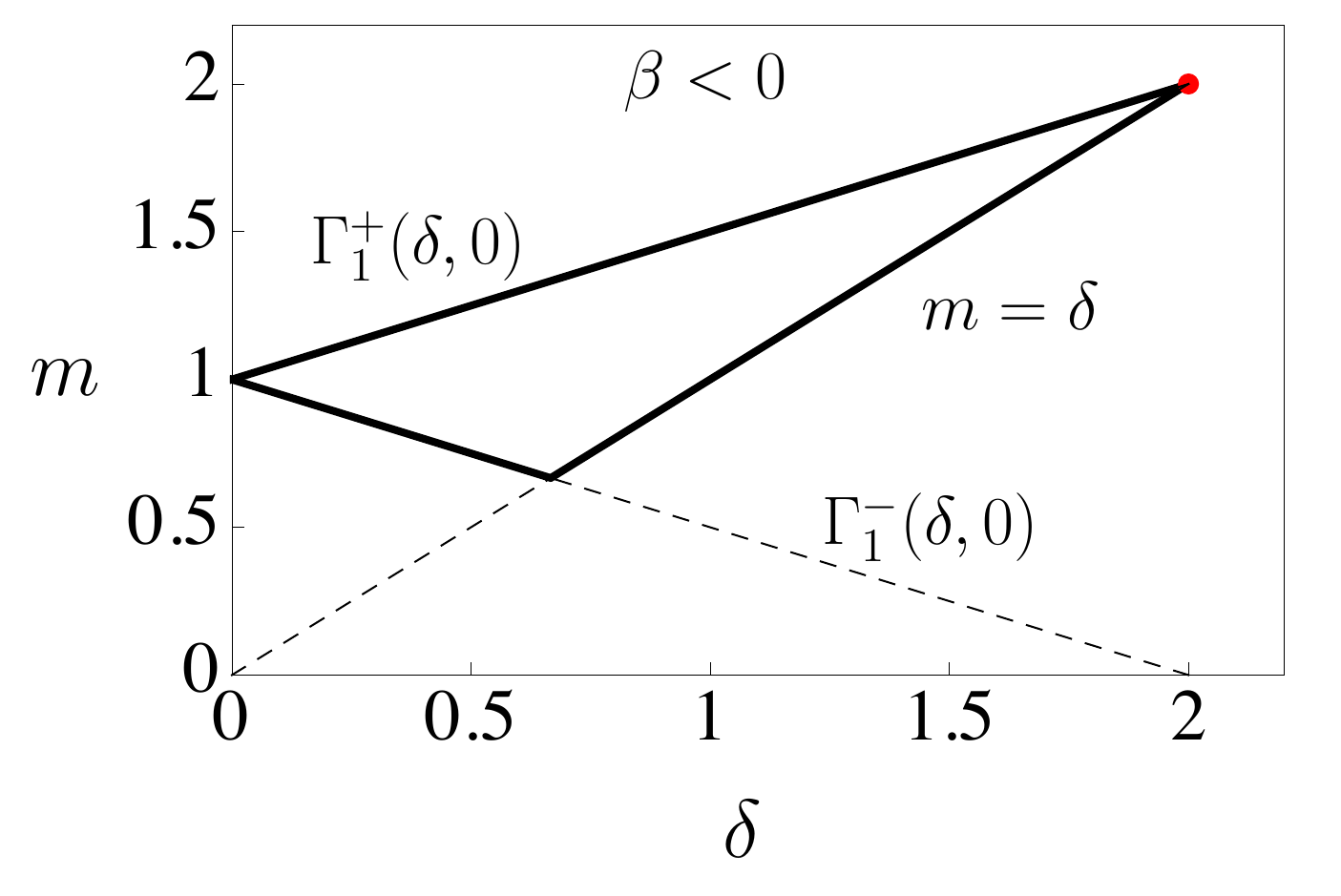}
\caption{Physical domain for the sea of the first $N-1$ eigenvalues, for $\beta<0$.}
\label{fig:domsea}
\end{figure}

In order to solve the saddle point equations for the expression (\ref{eq:normal0})-(\ref{eq:normal}), we will separately consider the contribution of the first $N-1$ eigenvalues and of the isolated eigenvalue:
\begin{eqnarray}
&&\lambda_i - \frac{1}{\beta N^2} \frac{1}{\lambda_i-\mu}- \frac{1}{\beta N^2} \sum_{1\leq j \leq N-1, j\neq i}\frac{1}{\lambda_i-\lambda_j}+ i\frac{\xi}{2\beta}=0,\quad \forall \ i \in \{1, \ldots, N-1\} \nonumber \\ \\
&&\mu- \frac{1}{\beta N^2} \sum_{1\leq j \leq N-1}\frac{1}{\mu- \lambda_j}+ i\frac{\xi}{2\beta}=0\label{eq:saddle point eq discr for mu}\\
&&\mu + \sum_{1\leq j \leq N-1} \lambda_j = 1. 
\end{eqnarray}
In the $N \to \infty$  limit, by neglecting contributions of $\Ord{1/N}$, we get:
\begin{eqnarray}
P \int_0^\infty \frac{\bar{\rho}(\lambda')d
\lambda'}{\lambda-\lambda'} - i \frac{\xi}{2}(1-\mu)&=0\label{eq:neg_temp_saddle point equations
for the sea} \\
 2\mu \beta + i\xi&= 0,\label{eq:saddle point eq for mu}\\
 \int_0^\infty \lambda \bar{\rho}(\lambda)d
\lambda&=1\label{eq:temp neg normcontr} 
\end{eqnarray}
where $\bar{\rho}$ 
is the density of the first $N-1$ eigenvalues
and (\ref{eq:saddle point eq for mu}) is the condition deriving from the saddle
point equation (\ref{eq:saddle point eq discr for mu}) associated to $\mu$. By the same change of variables
introduced in Sec.~\ref{tricomi}, see Eqs.~(\ref{eq:new variable}) and (\ref{eq:mdeltadef})
\begin{equation}
\label{eq:new variablebis}
\lambda= m+ x \delta,
\end{equation}
with $m=(a+b)/2$, $\delta=(b-a)/2$ and $\lambda \in [a,b]$, the solution
of the integral equation (\ref{eq:neg_temp_saddle point equations
for the sea}) can be expressed in terms of $\bar{\phi}(x)=
\bar{\rho}(\lambda)\delta$. It can be easily shown that the solution of this Tricomi equation is  
\begin{equation}
\bar{\phi}(x)=\frac{1}{\pi
\sqrt{1-x^2}}\left(1-\frac{2x(m-1)}{\delta}\right),
\end{equation}
where the normalization constraint (\ref{eq:temp neg normcontr}) on the rescaled eigenvalues fixes the  Lagrange multiplier 
\begin{equation}\label{eq:xinegativestab}
\xi=-i \frac{4 (m-1)}{(\delta^2(1-\mu))}.
\end{equation}
The region of the parameter space $(m,\delta)$ such that the density
of eigenvalues $\bar{\phi}$ is nonnegative reads
\begin{equation}\label{eq:negative_temp_DomainForPhi}
\max\left\{ \delta, \;1-\frac{\delta}{2}\right\} \leq m \leq 1+
\frac{\delta}{2}.
\end{equation}
Notice that the physical domain for the sea of the first $N-1$ eigenvalues has the same expression of the domain found for the range of
positive temperatures (\ref{eq:positive_temp_eyesShapeDomain}) at
$\beta=0$, namely $\Gamma_1^\pm(\delta,0)=1\pm\delta/2$
(see Fig.~\ref{fig:domsea}, which is the analog of Fig.~\ref{fig:positive_eye}). 
\\
This is consistent with the change in the temperature scaling from $N^3$ to $N^2$: we are
``zooming'' into the region near $\beta \rightarrow 0^-$ of the range of
temperatures analyzed in Chap.~\ref{chap3}. Furthermore, as could be expected from what
seen for positive temperatures, the solution of the
saddle point equations is a two parameter continuous family of
solutions. In the next section we will determine the distribution of eigenvalues that maximizes  the free energy of the Coulomb gas of eigenvalues. 
\section{Thermodynamics of the Coulomb gas}\label{thermodynamicSepStates}
Will now set the basis for the thermodynamic characterization of the Coulomb gas of the Schmidt coefficient, when the scaling coefficient is $\alpha=2$.
From Eqs.~(\ref{eq: free energy}) and (\ref{eq:eigenvalues for negative temperatures}) we get
\begin{eqnarray}\label{eq:vn_stacc1}
f_N&=&\mu^2- \frac{2}{N^2 \beta} \sum_{1\leq i<j\leq N} \ln|\lambda_i-\lambda_j|+\Ord{\frac{1}{N}}
\end{eqnarray}
By applying the natural scaling we have
(\ref{eq:naturalRescaling for Negative temeperatures}) 
\begin{eqnarray}\label{eq:vn_negtemp}
f_N &=& \mu^2 -\frac{1}{\beta}
\ln{(1-\mu)} - \frac{2}{N^2\beta}\sum_{1\leq i<j \leq N-1}\ln |\lambda(t_i)-\lambda(t_j)|
+\frac{1}{\beta}\ln N
+\Ord{\frac{\ln N}{N}}
\nonumber\\
&=& u -\frac{1}{\beta} s + \frac{1}{\beta}\ln N+\Ord{\frac{\ln N}{N}}
\nonumber\\
&=& f +\frac{1}{\beta}\ln N+\Ord{\frac{\ln N}{N}},
\end{eqnarray}
where
\begin{eqnarray}
u&=&\mu^2\\ \nonumber\\
s&=&\ln(1-\mu)+ \frac{2}{N^2}\sum_{1\leq i<j \leq N-1}\ln |\lambda(t_i)-\lambda(t_j)| \\
f&=&\lim_{N \to \infty}\left(f_N-\frac{1}{\beta}\ln N\right).
\end{eqnarray}
Finally, in the limit $N\to\infty$ 
\begin{eqnarray}
u&=&\mu^2\\
s&=&\ln(1-\mu)-\beta f_{\mathrm{red}}(m,\delta,\beta)\\
\beta f&=& \beta u-s=\beta \mu^2 -\ln(1-\mu)+ \beta f_{\mathrm{red}}(\delta,m,\beta), \label{eq.potentialEnergy}
\end{eqnarray}
being $f_{\mathrm{red}}$ the reduced free energy of the sea of eigenvalues
\begin{eqnarray}
 f_{\mathrm{red}}(\delta,m,\beta)&=&-\frac{1}{\beta}\int_{-1}^{1} \ dx \bar \phi(x) \int_{-1}^{1} \ dy  \bar \phi(y) \ln ( \delta |x-y| ) \nonumber \\
&=& \frac{2(m-1)^2}{\beta \delta^2} -\frac{1}{\beta} \ln\left(\frac{\delta}{2}\right).
\end{eqnarray}
As shown for the range of positive temperatures and for the metastable branch, $f$, $u$ and $s$ will be the free energy, the internal energy and the entropy densities of our Coulomb gas when computed at its most probable distribution, i.e. on the solution $\{\mu, \bar{\phi}(x)\}$ which maximizes the exponent of the partition function or equivalently minimizes $\beta f$ ($\beta<0$).
\subsection{Minimization problem: separable states}\label{separablestates}
\begin{figure}[h]
\centering
\includegraphics[width=0.9\columnwidth]{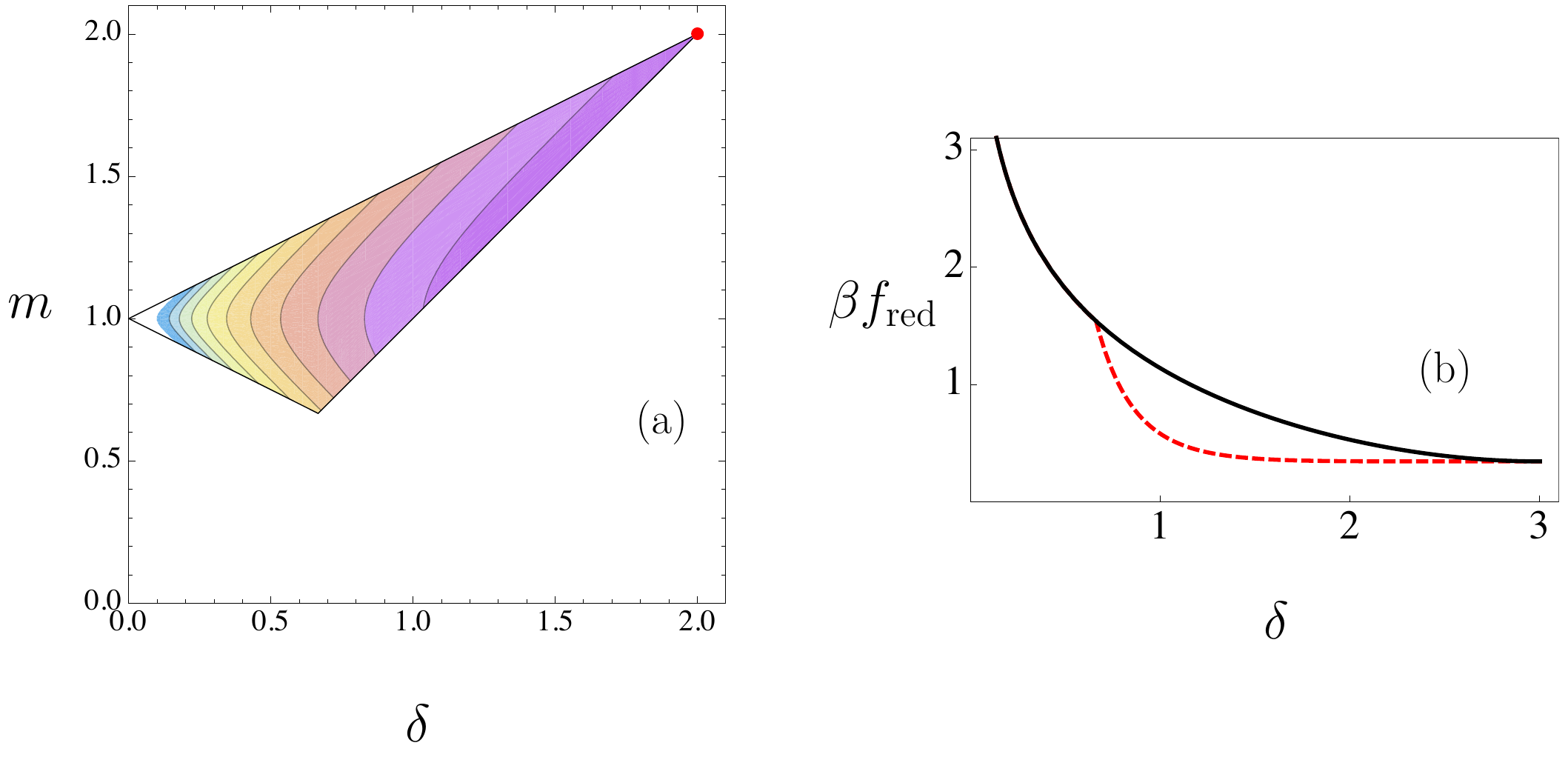}
\caption{(a) Contour plot of the reduced free energy density $\beta f_{\mathrm{red}}(\delta,m)$ of the sea for negative
temperatures. (b) Reduced free energy $\beta f_{\mathrm{red}}$ on the boundary of the triangular domain in Fig.~\ref{fig:contourPlotForNegativeTemp}(a) for
the case of negative temperatures. Solid line: upper boundary; dashed line: lower boundary.}
\label{fig:contourPlotForNegativeTemp}
\end{figure}

\begin{figure}[h]
\centering
\includegraphics[width=0.53\columnwidth]{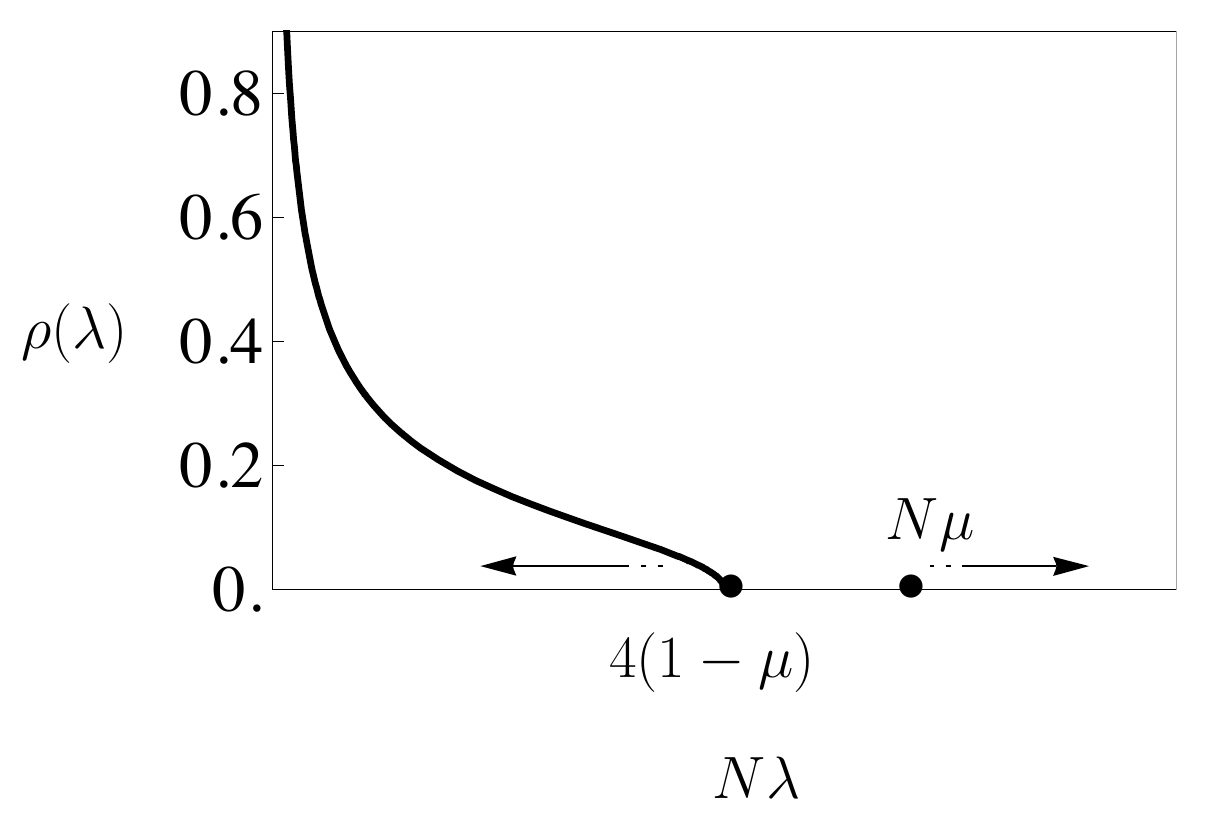}
\caption{Evaporation of the eigenvalue $\mu = \Ord{1}$ from the sea of eigenvalues $\Ord{1/N}$.}
\label{fig:evaporation}
\end{figure}

\begin{figure}[h]
\centering
\includegraphics[width=0.53\columnwidth]{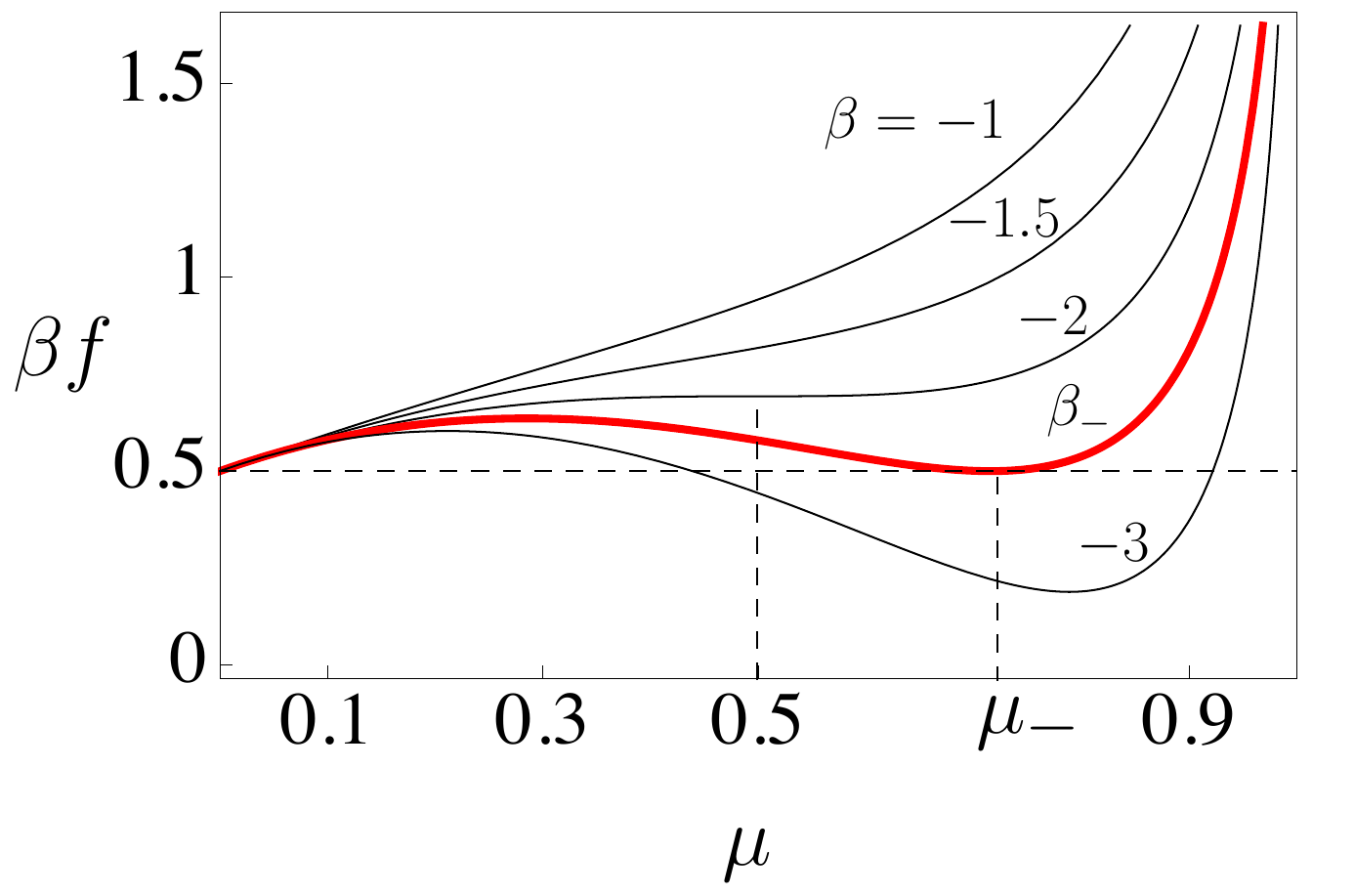}
\caption{Reduced free energy as a function of $\mu$ for different
values of $\beta(<0)$. Notice the birth of a stationary point for $\beta=-2$ ($\mu=1/2$) which becomes the global minimum for $\beta < \beta_-$ ($\mu > \mu_{-}$).}
\label{fig:neg_temp_free_energy}
\end{figure}

\begin{figure}[h]
\centering
\includegraphics[width=0.53\columnwidth]{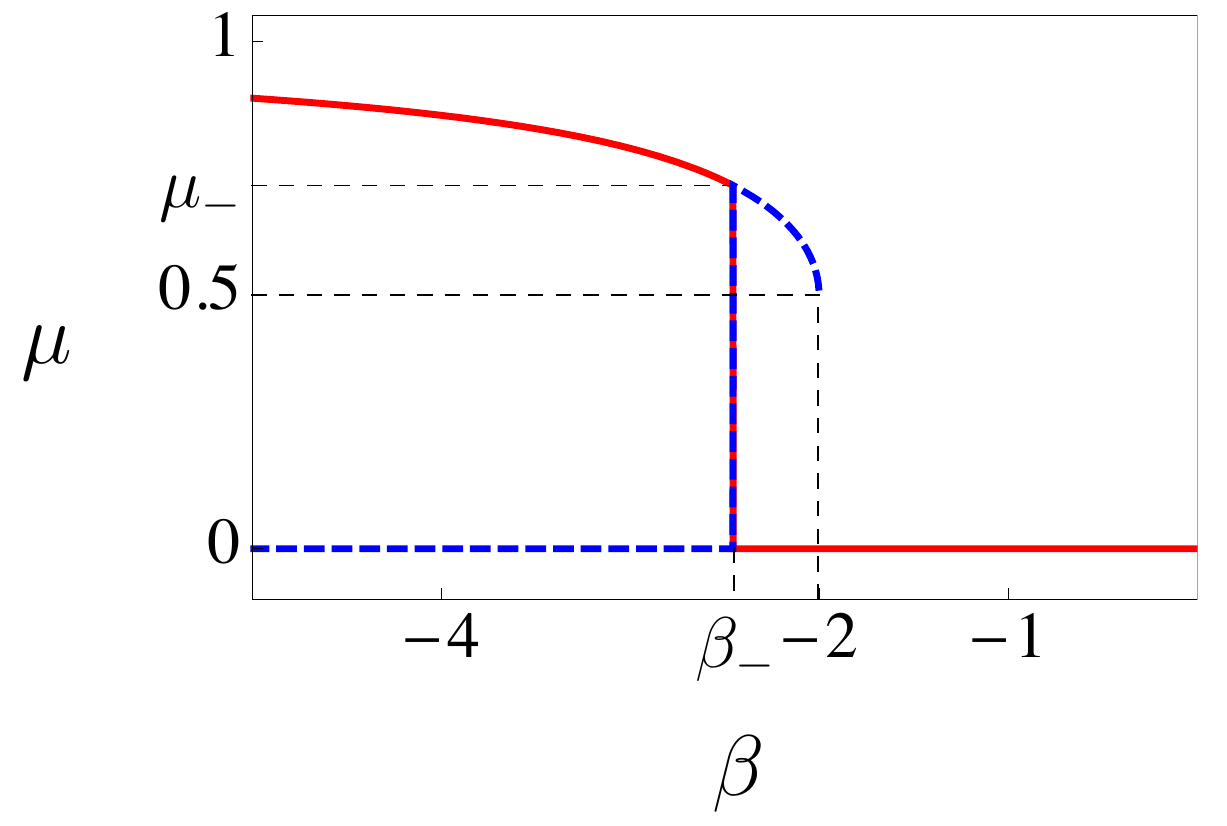}
\caption{Maximum eigenvalue at negative
temperatures. The two solutions are exchanged at $\beta_- \simeq
-2.45541$. The solution with $\mu = \Ord{1}$ appears at $\beta=-2$ ( $\mu = 0.5$) but is not the stable solution. Full line:
solution of minimal $\beta f$, i.e. maximal free energy; dashed thick line: solution of higher
$\beta f$.}
\label{fig:mustacc}
\end{figure}

It is easy to see that $\beta  f_{\mathrm{red}} (m,\delta)$,
has no stationary
points, but only a global minimum $\beta  f_{\mathrm{red}}=1/2$ at $(\delta,m, \beta)=(2, 2)$, indicated by the red dot in Fig.\
\ref{fig:contourPlotForNegativeTemp}(a). 
This point yields the Wishart distribution (\ref{eq:ansatz1phi}) or (\ref{eq:ansatz1})
found at $\beta=0$ for the case of positive temperatures:
\begin{equation}\label{eq:whistart in zero}
\bar \phi(x)=\frac{1}{\pi}\sqrt{\frac{1-x}{1+x}} , \quad
\bar \rho(\lambda)=\frac{1}{2\pi} \sqrt{\frac{4-\lambda}{\lambda}},
\end{equation}
where one should remember that now the $\lambda$'s are also scaled by $1-\mu$, see Eq.\ (\ref{eq:naturalRescaling for Negative temeperatures}).
We stress that this result is valid for all $\beta <0$.
In order
to check this solution we have to compute the free energy on the
boundary of this domain, see Fig.~\ref{fig:contourPlotForNegativeTemp}(b)
(which is the analog of Fig.~\ref{fig:3free}(a)).
\\
Thus by minimizing the potential energy for the sea of eigenvalues we get
\begin{eqnarray}
u&=&\mu^2\label{eq:vu1}\\
s&=&\ln(1-\mu)-\frac{1}{2}\label{eq:vs1}\\
\beta f&=&\beta \mu^2 -
\ln{(1-\mu)}+\frac{1}{2}.\label{eq:v1}
\end{eqnarray}
A new stationary solution, in which the largest isolated eigenvalue
$\mu$ becomes $\Ord{1}$, can be found by minimizing $\beta f$ and yields
\begin{equation}
\label{eq:isolated eigenvalue}
\mu(\beta)=\frac{1}{2}+\frac{1}{2}\sqrt{1+\frac{2}{\beta}},
\end{equation}
which is defined for $\beta\leq -2$. This expression can be also obtained directly by plugging Eq. (\ref{eq:xinegativestab}) into the saddle point equation (\ref{eq:saddle point eq for mu}) corresponding to the isolated eigenvalue $\mu$.
This eigenvalue, of $\Ord{1}$, evaporates from the sea of eigenvalues of $\Ord{1/N}$, as pictorially represented in Fig.~\ref{fig:evaporation}.
\\
The isolated eigenvalue moves at a speed $-d \mu /d\beta = 1/(2\sqrt{\beta^4+2\beta^3})$, which diverges at $\beta=-2$: another symptom of criticality. 
However, this new solution,  when it appears at $\beta=-2$, is not the global minimum of $\beta f$: it  becomes stable at a lower value of $\beta$. This can be immediately seen from the plot, in Fig.~\ref{fig:neg_temp_free_energy},  of the free energy (\ref{eq:v1}) as a function of the detached eigenvalue $\mu$ for different values of $\beta$. 
When $\beta>-2$ there are no stationary points, but only a global minimum at
$\mu=0$. According to the solution corresponding to this point, $\mu$ is still in the sea of the eigenvalues $\Ord{1/N}$ and the stable density function is given by the Wishart distribution (\ref{eq:Whishprolong}) with the free energy (\ref {eq:gravitysol}) (remember that, in the zoomed scale considered in this chapter, $\beta_g$ corresponds to the very large inverse temperature $N \beta_g$). At
$\beta=-2$ it appears a stationary point for the free energy density
corresponding to $\mu = 1/2=\Ord{1}$, see (\ref{eq:isolated
eigenvalue}). Notice, however, that $\beta f$ at this point remains larger than its
value at the global minimum, until $\beta$
reaches $\beta_-$. Finally, for $\beta<\beta_-$ the global
minimum of $\beta f$ moves to the right, to the solution containing $\mu = \Ord{1}$. 
The inverse temperature $\beta_{-}$ represents a critical point for the system, which we will soon see, undergoes a first order phase transition. The critical temperature $\beta_- $
is the solution of the transcendental
equation $f(\beta_-,0)=f(\beta_-,\mu_-)$, that is
\begin{equation}
\frac{\mu_-}{2(1-\mu_-)} =-\ln(1-\mu_-),
\label{eq:mu-}
\end{equation}
which yields
\begin{equation}
\mu_-\simeq 0.71533, \qquad \beta_- = -\frac{1}{2\mu_-
(1-\mu_-)} \simeq -2.45541.
\label{eq:beta-}
\end{equation}
The maximum eigenvalue is then a discontinuous function of the temperature at $\beta=\beta_-$
\begin{equation}\label{eq:mubeta}
\mu(\beta) =
\begin{cases}
0, &  0 < \beta < \beta_{-}, \\
\\
\frac{1}{2}+\frac{1}{2}\sqrt{1+\frac{2}{\beta}}, &  \beta \leq \beta_{-},
\end{cases}
\end{equation}
and in the limit $\beta \rightarrow -\infty$, $\mu$
approaches $1$: the state becomes separable.
\\ \\See Fig.~\ref{fig:mustacc}.
We stress that Eqs.~(\ref{eq:whistart in zero}) and (\ref{eq:isolated eigenvalue}) are the distributions of eigenvalues that gives the largest contribution to the partition function, that is the the most probable distribution for the isolated eigenvalue and for the sea of eigenvalues, respectively. 
\\
Thus, by inserting Eq.~(\ref{eq:isolated eigenvalue}) in Eqs.~(\ref{eq:vu1})-(\ref{eq:v1}), we determine the internal energy $u$, the entropy $s$ and the free energy $f$ densities for the system:
\begin{itemize}
\item
for $ \beta_{-} \leq \beta< 0$
\begin{eqnarray}
u&=& 0\label{eq:u2}\\
s&=& \frac{1}{2}\label{eq:s2}\\
\beta f &=& \frac{1}{2}\label{eq:bf2}
\end{eqnarray}
\item
for  $\beta < \beta_{-}$
\begin{eqnarray}
\label{eq:unegmu}
u &=&\frac{1}{2}+\frac{1}{2\beta}+\frac{1}{2}\sqrt{1+\frac{2}{\beta}},
\\
\label{eq:snegmu}
s &=& \ln \left(\frac{1}{2}-\frac{1}{2}\sqrt{1+\frac{2}{\beta}}\right) -\frac{1}{2}, \\
\label{eq:fnegmu}
\beta f &=&1+ \frac{\beta}{2}+\frac{\beta}{2}\sqrt{1+\frac{2}{\beta}}-\ln \left(\frac{1}{2}-\frac{1}{2}\sqrt{1+\frac{2}{\beta}}\right).
\quad
\end{eqnarray}
\end{itemize}
Summarizing, for
$\beta_-\leq\beta<0$ the solution of saddle point equations maximizing the free energy of the system is such that all eigenvalues are $\Ord{1/N}$, and through an appropriate scaling reduces to the Wishart distribution described in Sec.~\ref{metastabledistributions}. 
We remark that, since the scaling is $\alpha=2$, we are neglecting the contributions deriving from the sea of eigenvalues, since they scale as $\Ord{\ln N/N}$ (see Eq. (\ref{eq:vn_negtemp})).
At $\beta=-2$ a new metastable solution appears with one eigenvalue of $\Ord{1}$, and for $\beta <
\beta_-$ it becomes the stable solution, whereas the distribution of the eigenvalues described in Sec.~\ref{metastabledistributions} becomes metastable. It is important to emphasize that in the scaling $\alpha=3$, adopted in Chaps.~\ref{chap3} and~\ref{chap4}, the interval of negative temperatures where the Wishart distribution refers to all the eigenvalues, reads $\beta_{-}/N \leq \beta <0$, and shrinks to zero in the thermodynamic limit (that is why the solution found in Chap.~\ref{chap4} is metastable).
\section{First order phase transition}\label{firstOrderphTrans}
\begin{figure}[h]
\centering
\includegraphics[width=1.0\columnwidth]{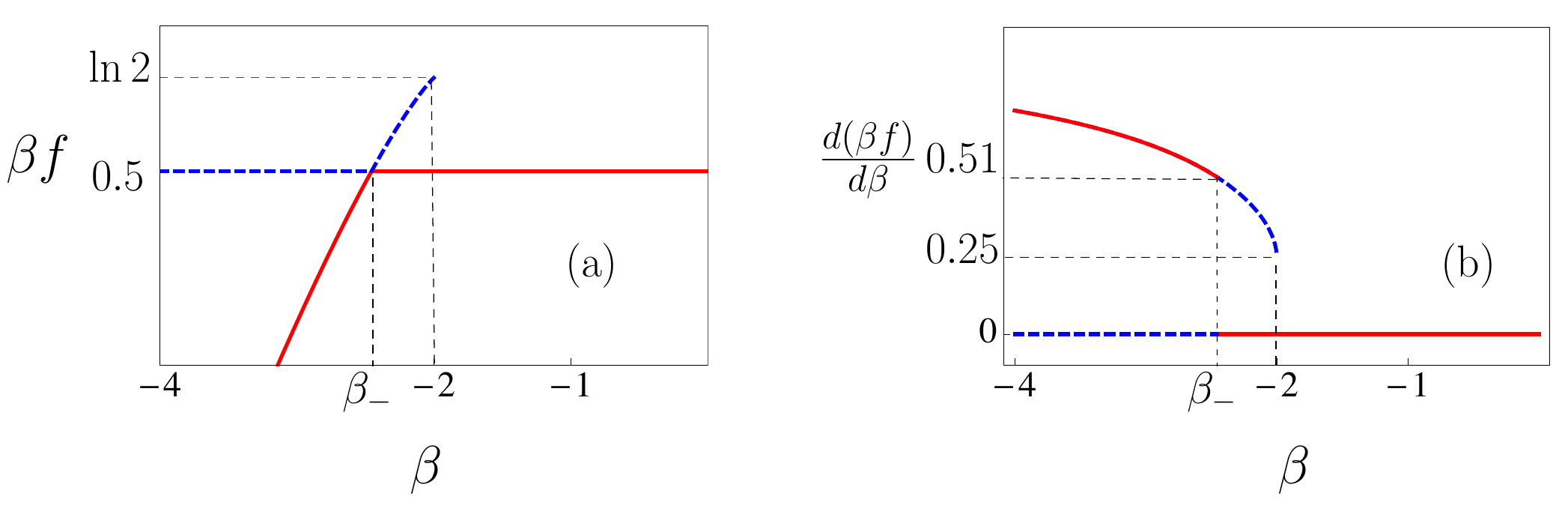}
\caption{Free energy (a) and first derivative (b) with respect to $\beta$. The two solutions are exchanged at $\beta_- \simeq
-2.45541$, where there is a first order phase transition. Full line:
solution of maximal free energy; dashed line: solution of lower
free energy.} \label{fig:BETAfreeEnergyTeoricaNegTemp1}
\end{figure}

\begin{figure}[h]
\centering
\includegraphics[width=1.0\columnwidth]{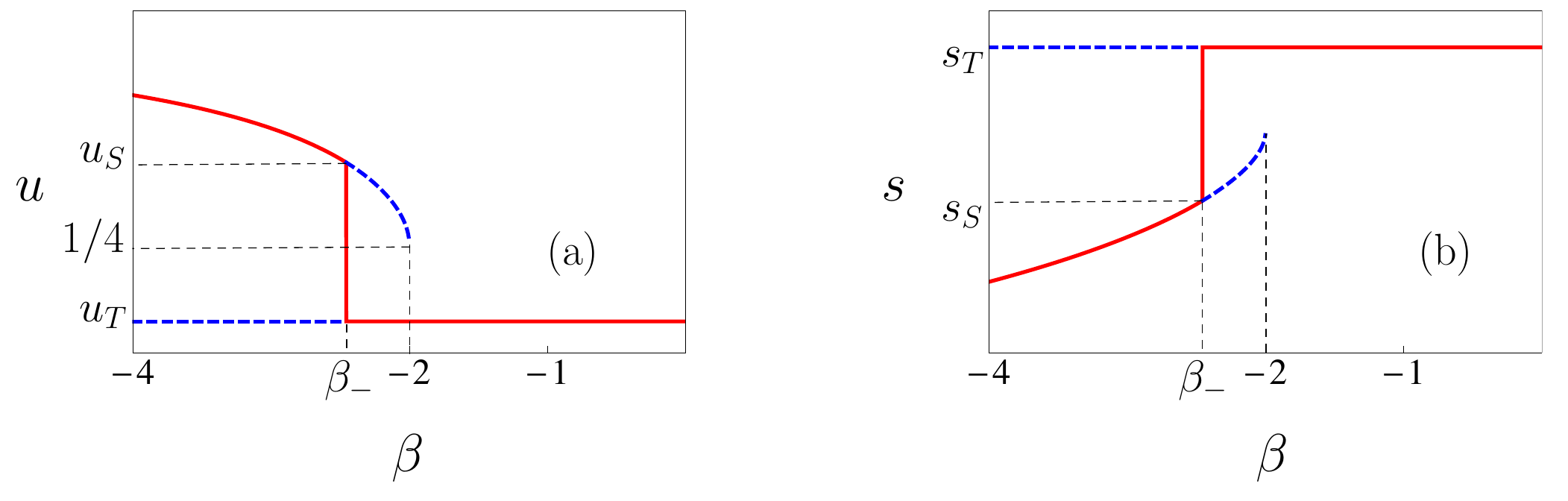}
\caption{Internal energy (a) and entropy density derivative (b). At fixed inverse temperature 
$\beta_{-}$ the internal energy goes up from  $u_T=0$ to $u_S \simeq 0.5117$ whereas the entropy goes down from $s_T=-0.5$ to $s_S \simeq -1.756$.} \label{fig:ues}
\end{figure}

\begin{figure}[h]
\centering
\includegraphics[width=1\columnwidth]{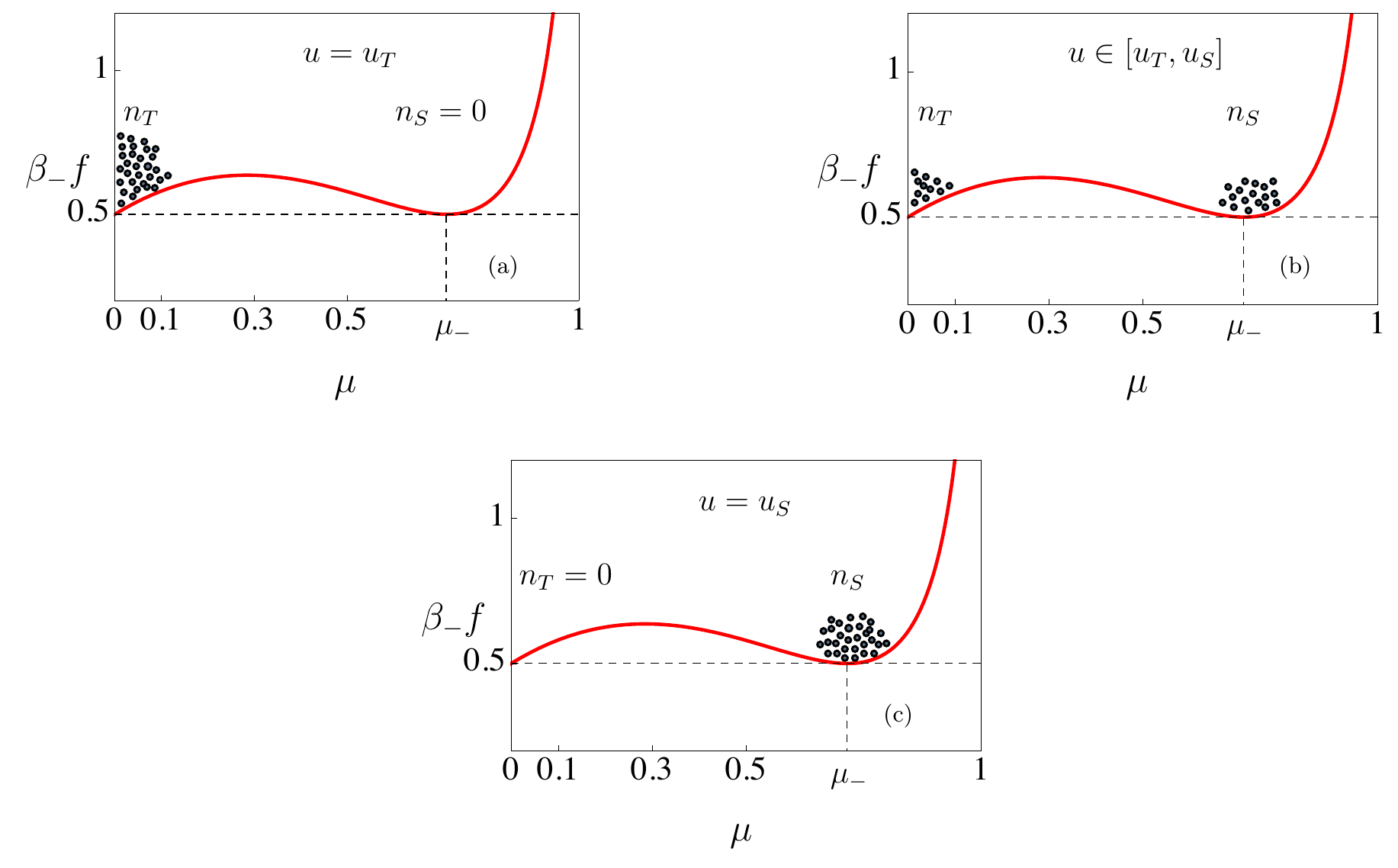}
\caption{First order phase transition, at fixed temperature $\beta=\beta_{-}$. In (a) and (c) the system is the typical and separable phase, respectively. In (b) the two phases coexist.} \label{fig:Liquidgasr}
\end{figure}

\begin{figure}[h]
\centering
\includegraphics[width=0.53\columnwidth]{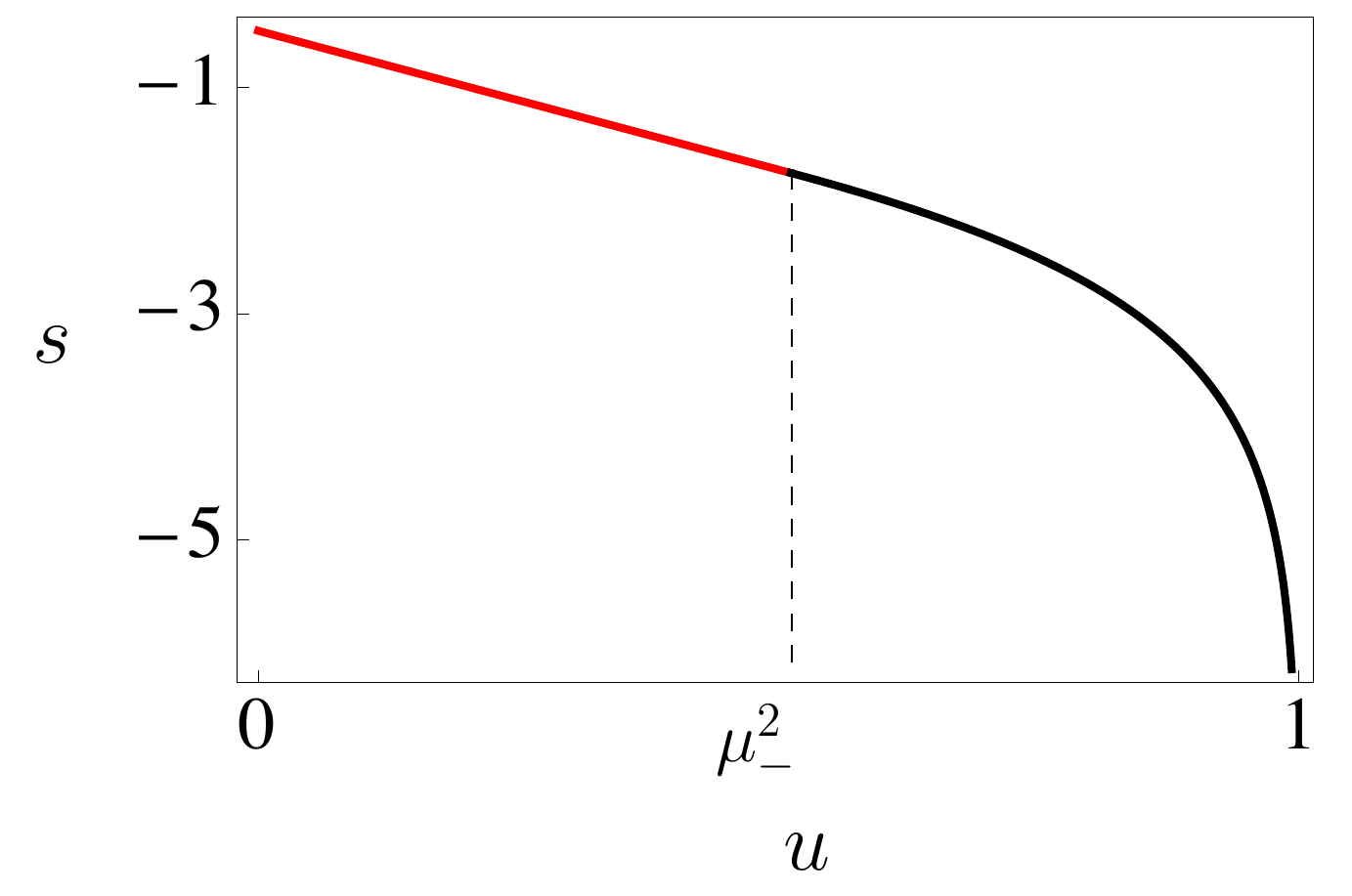}
\caption{Entropy density $s$ versus internal energy density $u=\langle\pi_{AB}\rangle$. See Eq.\ (\ref{eq:puritylarge}).} \label{fig:svsustacc}
\end{figure}

\begin{figure}[h]
\centering
\includegraphics[width=1\columnwidth]{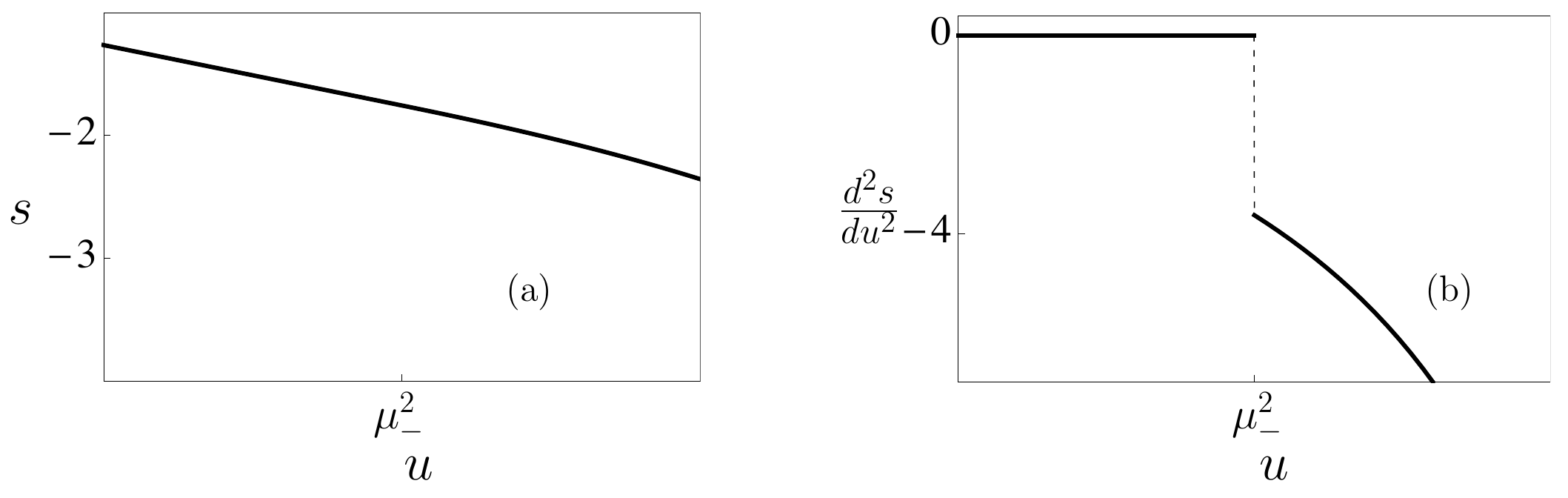}
\caption{Entropy density (a) and its second derivative (b) versus $u$.} \label{fig:svsu}
\end{figure}

We are now ready to unveil the presence of a first order phase transition in the system. 
The branch (\ref{eq:unegmu})-(\ref{eq:fnegmu})
is stable for $\beta < \beta_-$ while it becomes metastable for
$\beta_-\leq \beta<-2$. 
On the other hand, the solution $\mu=0$,
corresponding to (\ref{eq:u2})-(\ref{eq:bf2}) has a lower value of $\beta f$ for $\beta_- \leq \beta<0$, and a higher one for $\beta < \beta_-$.
See Fig.~\ref{fig:BETAfreeEnergyTeoricaNegTemp1}(a). At $\beta_{-}$ there is a first order phase transition, signaled by a discontinuity in the first derivative of the free energy, as shown in Fig.~\ref{fig:BETAfreeEnergyTeoricaNegTemp1}(b). 
\\Indeed for $\beta \to \beta_{-}$ we have
\begin{eqnarray}
\beta f &=& \frac{1}{2}+ \mu_{-}^2 (\beta-\beta_{-}) \theta(\beta_{-}-\beta)\nonumber\\ && -\frac{1}{4\beta_{-}^2}\left(1+\sqrt{\frac{\beta_{-}}{2+\beta_{-}}}\right) (\beta-\beta_{-})^2 \ \theta(\beta_{-}-\beta) + \Ord{(\beta-\beta_{-})^3},\nonumber\\
\end{eqnarray} 
with $u(\beta_{-})=\mu_{-}^2\simeq0.5117$ and 
\begin{equation}
\left(\frac{d u}{d \beta}\right)_{\beta_{-}} = -\frac{1}{2\beta_{-}^2}\left(1+\sqrt{\frac{\beta_{-}}{2+\beta_{-}}}\right)\simeq-0.2755,
\end{equation}
see Eq.~(\ref{eq:unegmu}).
We have seen that at $\beta_{-}$ the free energy of the global minimum at $\mu=0$ is equal to the free energy  of the stationary point at $\mu=\mu_{-}$. Thus at this temperature our statistical model is characterized by the coexistence of two phases, corresponding to typical and separable states. In Fig.~\ref{fig:ues} we plot the entropy and the internal energy densities versus $\beta$. 
\\
Observe that at fixed temperature, $\beta=\beta_{-}$, the internal energy of the system goes from $u_T=0$ up to $u_S = \mu_-^2\simeq0.5117$, while the entropy goes from $s_T=-1/2$ down to $s_S =  -1/2 + \ln(1-\mu_-)\simeq -1.75643$: the system moves from typical to separable states. We have that $u \in [u_T, u_S]$ and $s\in[s_S,s_T]$ are given by convex combinations of the boundaries of the associated intervals and describe the coexistence of the two phases. More precisely, if our statistical ensemble consists of $n$ identical copies of the system, we say that during this first order phase transition a certain amount $n_T$ of them is in the \textit{typical phase}, that is with the largest Schmidt coefficient null on average (i.e. of $\Ord{1/N}$ in the rescaling $\alpha=3$), and a complementary amount $n_S= n - n_T$ is in the \textit{seperable phase}, the average of $\mu$  being $\mu_{-}$.  Notice that there can be some (few) copies of the system whose free energy is not maximal, in this sense we should have written $n \simeq n_T + n_S$. However, these fluctuation can be considered negligible for our scopes. Thus, as the energy of the system increases we have a gradual transition from the typical to the separable phase. In particular for a generic state $X$, when we have both phases, the internal energy $u$ and the entropy $s$ are 
\begin{eqnarray}
u= \frac{u_T n_T + u_S n_S}{n_T+n_S} \qquad \mbox{and}\qquad s= \frac{s_T n_T + s_S n_S}{n_T+n_S}. 
\end{eqnarray}
See Fig.~\ref{fig:Liquidgasr}.
\\
Since this first order phase transition refers to a reversible isothermal process,  we have
\begin{equation}
\int_{s_T}^{s} ds = \beta_{-}\int_{u_T}^{u} du 
\end{equation}
thus
\begin{equation}
s= s_{T} + \beta_{-} (u_T - u) = - \frac{1}{2} + \beta_{-} u.
\end{equation}
We also have
\begin{equation}\label{eq:firstorderphasetr}
\frac{\Delta s} {\Delta u} = \beta_{-},
\end{equation}
with $\Delta s= s_T -s_S$ and $\Delta u =u_T - u_S$. The variation $\Delta u$ of the internal energy can be interpreted as the specific latent heat of the evaporation of the largest eigenvalue from the sea of the eigenvalues, from $\Ord{1/N}$ up to $\Ord{1}$.
Summarizing, the entropy density as a function of the internal energy density for the stable solution at negative temperatures reads
\begin{equation}
s(u) =
\begin{cases}
\beta_- u -\frac{1}{2}, &  0<u\leq\mu_-^2, \\
\\
\ln(1-\sqrt{u})-\frac{1}{2}, &  \mu_-^2 < u <1 .
\end{cases}
\label{eq:sumu}
\end{equation}
It is continuous together with its first derivative at $u=\mu_-^2$, while its second derivative shows a finite gap. See Figs.~\ref{fig:svsustacc} and~\ref{fig:svsu}.
\\
At the critical point, for $u\to\mu_-^2$, we have
\begin{eqnarray}\label{eq:SvsUmu-}
s(u) &=&  s_S + \beta_{-} (u-u_S)\nonumber\\&&+ \frac{\beta_{-}^2}{2}\left(-2 + \beta_{-}\left(-1+\sqrt{\frac{2+\beta_{-}}{\beta_{-}}}\right)\right) (u-u_S)^2 \theta(u-u_S)+\Ord{(u-u_S)^3},\nonumber\\
\end{eqnarray}
where 
\begin{eqnarray}
{\left(\frac{d \beta}{d u}\right)}_{\beta_{-}}=\left(\frac{d u}{d \beta}\right)^{-1}_{\beta_{-}}= \beta_{-}^2\left(-2 + \beta_{-}\left(-1+\sqrt{\frac{2+\beta_{-}}{\beta_{-}}}\right)\right)\simeq -3.629,
\end{eqnarray}
see Eq.~(\ref{eq:unegmu}).
\\
A few words of interpretation are necessary. As we have seen, it is the stable branch of the solution that lead us to separable states at negative temperatures. The analytic continuation of the stable solution for positive temperatures emanates a metastable branch in which all eigenvalues remain $\Ord{1/N}$. By contrast, the new stable solution consists in a sea of $N-1$ eigenvalues $\Ord{1/N}$ plus one isolated eigenvalue $\Ord{1}$. Let us now translate these results in terms of purity (we stress again that $\beta$ is a Lagrange multiplier that fixes the value of the purity of the reduced density matrix of our  $N$ dimensional
system). The average entanglement of the states described by the separable phase is given by
\begin{equation}
\langle\pi_{AB}\rangle= \Bigg \langle \sum_{1 \leq j \leq N} \lambda_j^2 \Bigg\rangle = \mu^2 + \frac{(1-\mu)^2}{N}\int \lambda^2 \bar{\rho}(\lambda) d\lambda = \mu^2 + \Ord {\frac{1}{N}}
\label{eq:puritylarge}
\end{equation}
Recall the scaling $(1-\mu)/(N-1)$ for the sea of eigenvalues.  
It is the analog of (\ref{eq:puritysmall}) for the scaling $\alpha=3$. Assume that we pick a given isopurity manifold in the original Hilbert space, defined by a given \emph{finite} value of the average purity $\langle\pi_{AB}\rangle$. If we randomly select a vector belonging to this isopurity manifold, its reduced density matrix (for the fixed bipartition) will have one finite eigenvalue $\mu \simeq \sqrt{\pi_{AB}}$ and many small eigenvalues $\Ord{1/N}$ (yielding a correction $\Ord{1/N}$ to purity). In this sense, the quantum state is largely separable. 
The probability of finding in the above mentioned manifold a vector whose reduced density matrix has, say, \emph{two} (or more) finite eigenvalues $\mu_1$ and $\mu_2$ (such that $\mu_1^2+\mu_2^2 \simeq \langle\pi_{AB}\rangle$, modulo corrections $\Ord{1/N}$) is vanishingly small. By contrast, remember (from the results of Chap.~\ref{chap3}) that if the isopurity manifold is characterized by a \emph{very small} value $\Ord{1/N}$ of purity, the eigenvalues of a randomly chosen vector on the manifold are \emph{all} $\Ord{1/N}$ (being distributed according to the semicircle or Wishart, depending on the precise value of purity).
This is the significance of the statistical mechanical approach adopted in this thesis. We will come back to this point in Chap.~\ref{chap6}.

\section{Finite-size corrections}
\label{sec:finitesize}
\begin{figure*}
\centering
\includegraphics[width=1\textwidth]{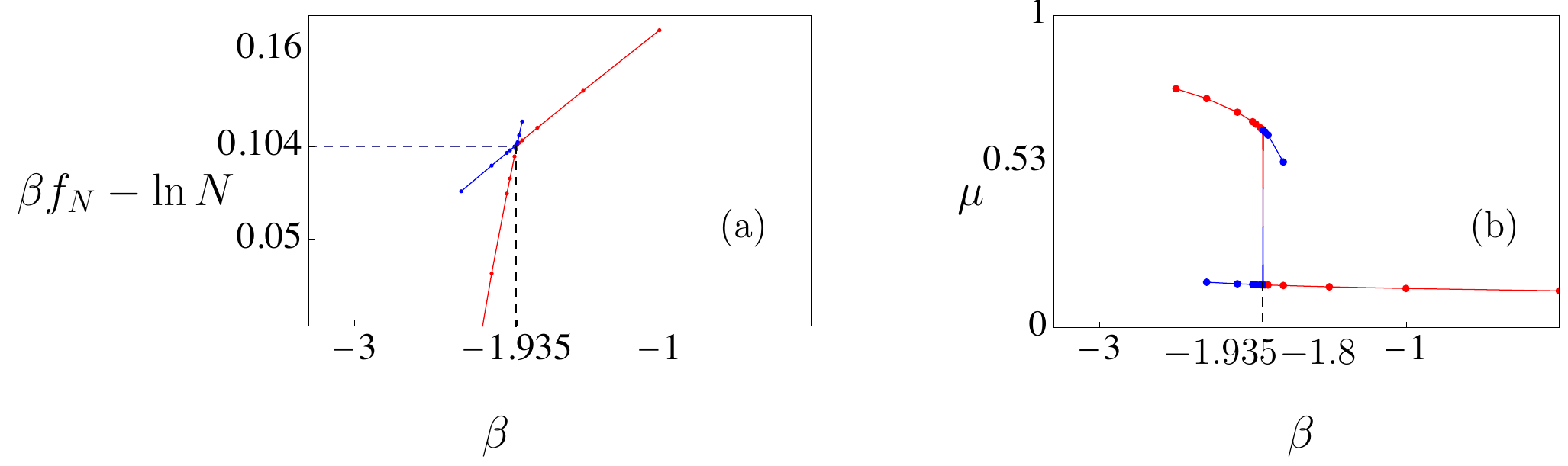}
\caption{Finite-$N$ version of Fig.~\ref{fig:BETAfreeEnergyTeoricaNegTemp1}. Free energy and maximum eigenvalue in the saddle point
approximation as function of $\beta$ at $N=30$. The local minimum is
in blue, the global one in red. The two minima swap stability at $\beta=-1.935$.
Notice the birth of the new local minimum at $\beta=-1.8$ (for $N=\infty$ this takes place at $\beta=-2$) and the exchange of stability at $\beta=-1.93$ (for $N=\infty$, $\beta=-2.45$). } 
\label{fig:betaF}
\end{figure*}

\begin{figure}[h]
\centering
\includegraphics[width=0.53\columnwidth]{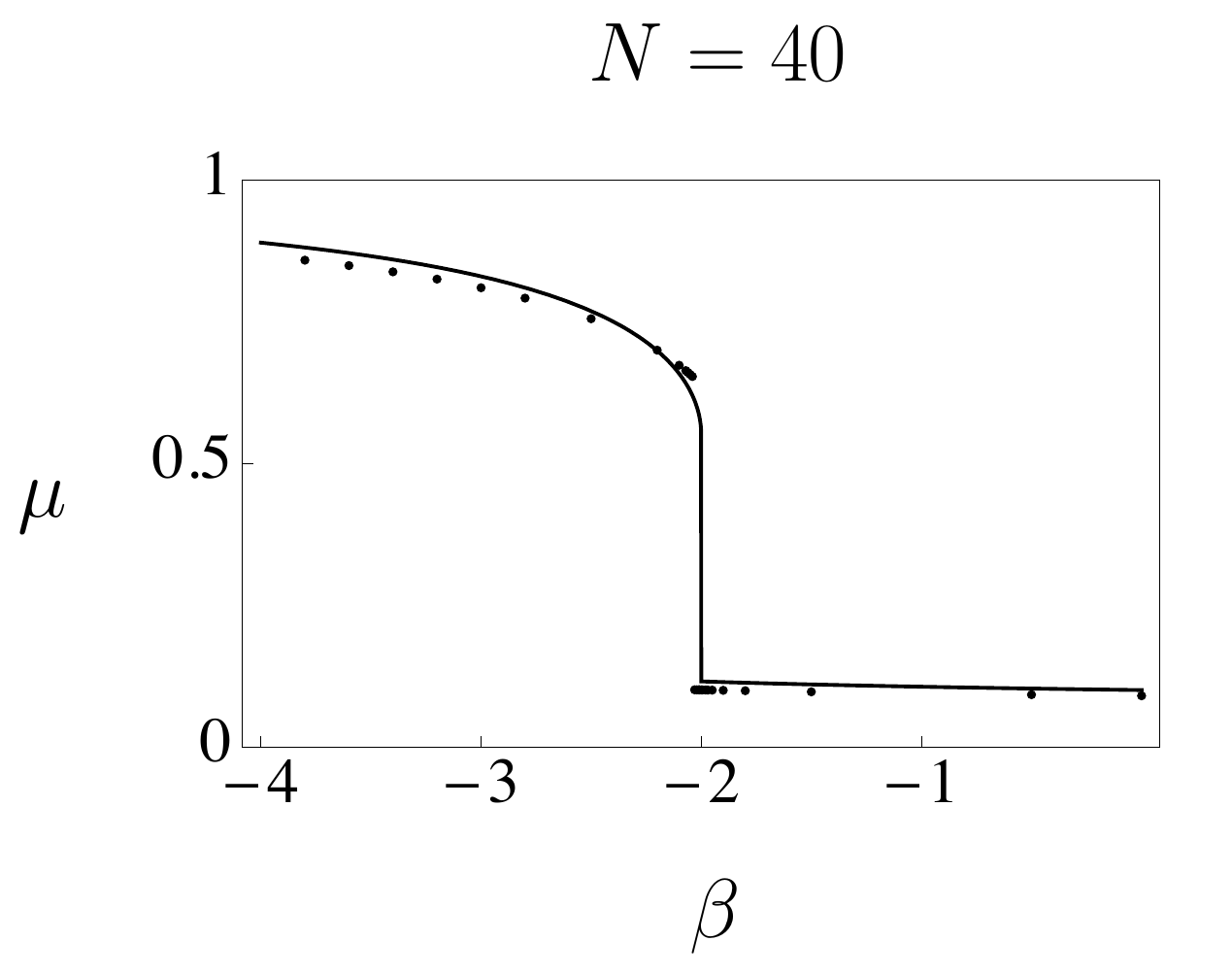}
\caption{Maximal eigenvalue in the saddle point approximation as function of $\beta$. The points are the result of a numerical evaluation for $N=40$, while the full line is the expression in Eq.\ (\ref{eq:Nfinfin}).}
\label{fig:betaF40}
\end{figure}

The results of the previous section refer to the $N\to\infty$ limit. In order to understand how finite-$N$ corrections affect our conclusions we have numerically maximized the free energy for various temperatures. The two phases of the system discussed in the previous sections correspond to the two solutions
obtained by minimizing $\beta f_N$  (\ref{eq: free energy}) on the $N$
dimensional simplex of the normalized eigenvalues. Indeed, we have
numerically proved that $\beta f_N(\beta)$ presents two local minima at negative temperatures: for $\beta_-^{(N)}  \leq \beta<0$ the minimum
giving the lower value of $\beta f_N(\beta)$ corresponds to the
distribution of eigenvalues (\ref{eq:Whishprolong}), found in the last chapter. The other minimum is reached when the highest eigenvalue is $\Ord{1}$. The point $\beta=\beta_-^{(N)}$ is a crossing point for these two solutions, and for $\beta < \beta_-^{(N)}$ these two solutions are inverted, see Figs.~\ref{fig:mustacc} and~\ref{fig:BETAfreeEnergyTeoricaNegTemp1}. Summarizing, there exists a negative temperature at which the system undergoes a first order phase transition, from typical to
separable states.
The first thing to notice is that qualitatively the phase transition remains of first order even for finite $N$. The second is that the finite-$N$ corrections are quite relevant for the location of the phase transition and the value of the maximum eigenvalue as a function of $\beta$.
For example, for $N=30$, the negative critical temperature $\beta_{-}^{(30)}= -1.935$ instead of $-2.455$.
This is evinced from Fig.~\ref{fig:betaF}, which is the finite-size version of Fig.~\ref{fig:BETAfreeEnergyTeoricaNegTemp1}.
This can be understood, as the corrections to $f(\mu)$ around $\mu=0$ are quite large. In the limit $\mu=1/N$ there is a hard wall for the maximum eigenvalue $\mu$, as the condition $\sum_i\lambda_i=1$ cannot be satisfied if $\mu<1/N$. It is therefore likely that all sorts of large corrections occur as $\mu$ tends to $1/N$, probably yielding an effective size to the corrections which is a lower power of $1/N$ (or even possibly $1/\ln N$). The limits $\mu\to 0$ and $N\to\infty$ do not commute.
\\ 
\begin{figure*}
\centering
\includegraphics[width=0.865\columnwidth]{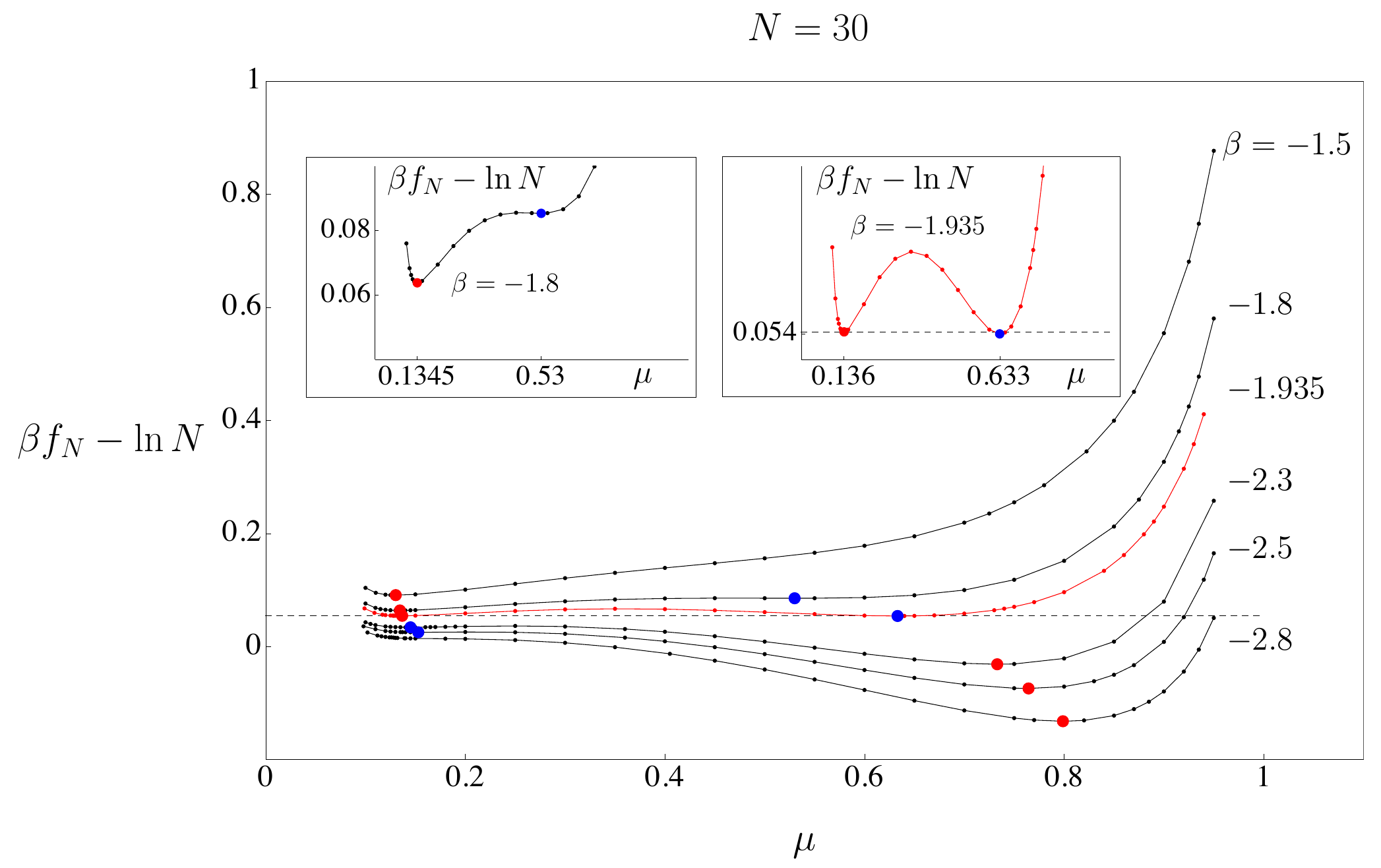}
\caption{Finite-$N$ version of Fig.~\ref{fig:neg_temp_free_energy}. $\beta f_N-\ln N$ as a function of the maximum eigenvalue $\mu$, obtained by numerical minimization over the remaining $N-1$ eigenvalues for various $\beta$. Observe the formation of a new minimum and the exchange of stability, although the critical values of $\beta$ at which these phenomena occur differ from the theoretical ones, due to large finite $N$ corrections. However, it is clear that at small $\mu$, $1/N$ corrections tend to increase the value of $\beta f_N$, making the critical value $\beta_-$ move towards 0, as observed in the numerics.}\label{fig:FreeEnergyAtFixedX}
\end{figure*}
To further explore this effect we have minimized $\beta f_N$ with respect to $\lambda_1, \ldots, \lambda_{N-1}$,
for fixed values of the largest eigenvalue $\lambda_N=\mu$ and for different temperatures. The results for $N=30$ are shown in Fig.~\ref{fig:FreeEnergyAtFixedX}. One can see that between $\beta=-1.8$ and $\beta=-2.5$ there is a competition between two well defined local minima, corresponding to the two solutions discussed above.
\\At
$\beta=\beta_{-}^{(30)}= -1.935$ their free energies are equal. For higher $\beta$ the global minimum corresponds to the solution (\ref{eq:Whishprolong}), whereas on the other side of $\beta_{-}^{(30)}$ the solution with $\mu = \Ord{1}$ minimizes $\beta f_N$.
We have seen that for $\beta\geq \beta_{-}$ the stable solution has no detached eigenvalues. By taking into account the scaling $\beta\to\beta/N$ we get that the solution is given by the very first part of the gravity branch (\ref{eq:gravitysol}). In particular, recalling the natural scaling (\ref{eq:naturalRescaling for Negative temeperatures}) for eigenvalues, the maximum eigenvalue is given by $b/N=(m+\delta)/N=2\delta/N$. On the other hand for $\beta<\beta_{-}$ the maximum eigenvalue is given by (\ref{eq:isolated eigenvalue}). Therefore, we get
\begin{eqnarray}
\mu = 
\begin{cases}
\frac{2}{N} \delta(\beta/N), &   \beta_- \leq \beta<0, \\
\\
\frac{1}{2}+\frac{1}{2}\sqrt{1+\frac{2}{\beta}}  , &  \beta < \beta_-,
\end{cases}
\label{eq:Nfinfin}
\end{eqnarray}
where $\delta(\beta)$ can be retrieved from (\ref{eq:betadeltaWishartmet}). The numerical results for $N=40$ are compared with the expressions in Eq.\ 
(\ref{eq:Nfinfin}) in Fig.~\ref{fig:betaF40}.
The agreement is excellent. The corresponding free energy follows from (\ref{eq:fnegmu}) and (\ref{eq:gravitysol}) with the appropriate scaling 
\begin{eqnarray}
\beta f = 
\begin{cases}
\frac{11}{4} -\frac{9}{\delta(\beta/N)} + \frac{9}{\delta(\beta/N)^2} - \ln \frac{\delta(\beta/N)}{2} , &  \beta_- \leq \beta<0 ,\\
1+ \frac{\beta}{2}+\frac{\beta}{2}\sqrt{1+\frac{2}{\beta}}-\ln \left(\frac{1}{2}-\frac{1}{2}\sqrt{1+\frac{2}{\beta}}\right), &  \beta < \beta_- .\\
\end{cases} 
\end{eqnarray}
In the limit $N \to \infty$ we get $\beta f = 1/2$, as expected from Eq. (\ref{eq:bf2}).
Notice that in order to have a correct finite size scaling of the critical temperature $\beta_{-}^{(N)}$ one should take into account $\Ord{1/N}$ corrections  to the expression of $\beta f$, see Eq. (\ref{eq:vn_stacc1}). This would enable a more exhaustive mapping between the two branches.  This analysis goes beyond our scopes.
In the next chapter we will overview the results of our statistical approach to bipartite entanglement, trying to connect what found in the two scaling regimes $\alpha=3$ and $\alpha=2$.

\chapter{Overview}\label{chap6}
\markboth{Overview}
{Overview}

In this chapter we will summarize the main results obtained in the first part of this thesis, dealing with the analysis of bipartite large quantum systems in pure states. We will focus on those quantities that are more directly related to physical intuition. In the classical statistical mechanics approach we have adopted, the temperature plays the usual role of a Lagrange multiplier, whose task is to fix the value of energy (purity in our case). A given value of $\beta$ determines a set of vectors in the projective Hilbert space whose reduced density matrices have a given purity (isopurity manifold of quantum states). The distribution of the eigenvalues of (the reduced density matrices associated to) these vectors has been investigated in the last three chapters and yields information on the separability (entanglement) of these quantum states. The distribution of eigenvalues is the most probable one \cite{Mehta2004}, in the same way as the Maxwell distribution of molecular velocities is the most probable one at a given temperature. In this chapter we will abandon the temperature and fully adopt purity as our physical parameter.
In Sec.~\ref{sec:overviewvolumeofstates} we will determine the probability distribution of the purity which is proportional to the volume of the isopurity manifolds. Then in Sec.~\ref{sec:OverviewdistributionOfEigenvalues} we will overview the different shapes of the distribution of eigenvalues for different phases of the system, corresponding to different ranges of $\pi_{AB}$ from $1/N$ to $1$. 
\section{Entropy and volume of isopurity manifolds}\label{sec:overviewvolumeofstates}
\begin{figure}[h]
\centering
\vspace{0.03cm}
\includegraphics[width=0.55\columnwidth]{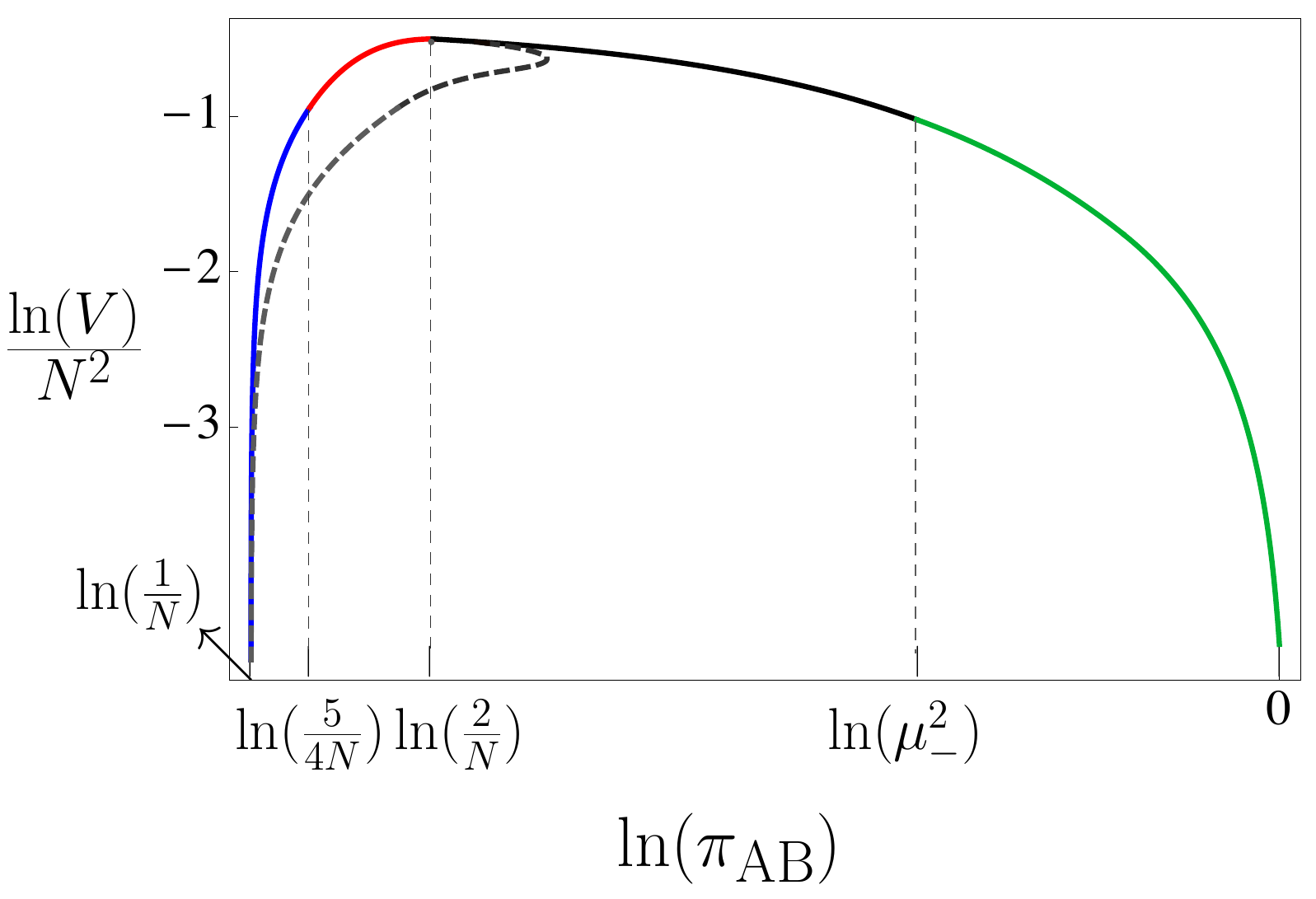}
\caption{Volume $V=\exp(N^2s)$ of the isopurity manifolds versus their purity $ \pi_{AB}$ for $N=50$. We indicate in blue, green and red maximally entangled,  typical and separable phases, respectively. The black line corresponds to the first order phase transition from the typical to the separable phase. The dashed line refers to the metastable branch, see Figs.~\ref{fig:SvsUr} and~\ref{fig:SvsUalpha3r}.}
\label{fig:volumestabEmet}
\end{figure}

Entropy counts the number of states with a given value of purity and is proportional to the logarithm of the volume in the projective Hilbert space. The explicit expressions of the entropy density $s$ as a function of the purity $\pi_{AB}$ of the state vectors in that volume, can be read directly from  Eqs.\ (\ref{eq:112}) and (\ref{eq:sumu}) for the stable solution, by taking into account the correct scaling when the system moves across different regions of the Hilbert space. 
In order to discuss this delicate and important point let briefly recall the expression of the partition function (\ref{eq:partitionfunction}):
\begin{equation}
\cZ_{AB}=\int d\mu(\rho_A)  \exp\left(-\beta {{N}^{\alpha}}
\pi_{AB}\right),
\end{equation}
where $N$ is the dimension of subsystem $A$ in the state $\rho_{A}$. The power $N^\alpha$ can be read as $N^\alpha=N^2 N^{\alpha-2}$, where $N^2$ is the number of degrees of freedom of $\rho_A$. The scaling coefficient $\alpha$ depends on the degree of entanglement of the global system, or equivalently on the temperature $\beta$. In the region of positive temperatures, the purity of the system scales as $\Ord{1/N}$, and becomes an intensive quantity if multiplied by $N^{\alpha-2}=N$, that is $\alpha=3$. The internal energy density is then given by $u = N \pi_{AB}$. On the other hand, when we approach the region of separable states, $\beta\rightarrow -\infty$,  the purity is $\Ord{1}$, from which it follows that $\alpha=2$ and $u= \pi_{AB}$. 
Thus, the entropy density (Eqs.\ (\ref{eq:112}) and (\ref{eq:sumu})) reads
\begin{eqnarray}
s(\pi_{AB})  = \begin{cases}
\frac{1}{2} \ln ( N \pi_{AB} -1) -\frac{1}{4}, \qquad\qquad\qquad\qquad    \frac{1}{N} < \pi_{AB} \leq  \frac{5}{4 N} ,  \\
\\
 \ln \left( \frac{3}{2}  - \sqrt{\frac{9}{4} - N \pi_{AB}}
\right) - \frac{9}{4} + \frac{5}{2\left( \frac{3}{2}  - \sqrt{\frac{9}{4} - N \pi_{AB}}
\right)} -
\frac{3}{4\left( \frac{3}{2}  - \sqrt{\frac{9}{4} - N \pi_{AB}}
\right)^2},\\ \qquad\qquad\quad\qquad\qquad\qquad\qquad\qquad\qquad \frac{5}{4 N} < \pi_{AB} \leq \frac{2}{N},
\\
\\
\beta_- 
\pi_{AB}
-\frac{1}{2},  \qquad\qquad\qquad\qquad\qquad\qquad \frac{2}{N} <\pi_{AB} \leq \mu_-^2 
, \\
\\
\ln\left(1-\sqrt{ 
\pi_{AB} 
}\right)-\frac{1}{2}, \qquad\qquad\quad\quad\qquad  ~\mu_-^2 
< \pi_{AB} <1 ,\qquad
\end{cases}
\label{eq:1121}
\end{eqnarray}
with $\mu_{-}^2\simeq 0.5117$ and $\beta_{-} \simeq -2.455$, given by Eqs. (\ref{eq:mu-}) and (\ref{eq:beta-}). 
By exponentiating expression (\ref{eq:1121}) we get the volume $V=\exp{(N^2 s)}$ (i.e.\ the probability) of the isopurity manifolds
\begin{eqnarray}
V(\pi_{AB})  \propto \begin{cases}
e^{-\frac{N^2}{4}} (N \pi_{AB} -1)^{N^2/2},\qquad \qquad\qquad\qquad  \frac{1}{N} < \pi_{AB} \leq  \frac{5}{4 N} ,  \\
\\
 \left( \frac{3}{2}  - \sqrt{\frac{9}{4} - N \pi_{AB}}
\right)^{N^2} \\ \qquad \times \exp \left[ N^2\left(- \frac{9}{4} + \frac{5}{2\left( \frac{3}{2}  - \sqrt{\frac{9}{4} - N \pi_{AB}}
\right)} -
\frac{3}{4\left( \frac{3}{2}  - \sqrt{\frac{9}{4} - N \pi_{AB}}
\right)^2}\right)\right] , \\
\qquad\quad \qquad\qquad\qquad \qquad\qquad\qquad\qquad~ \frac{5}{4 N} < \pi_{AB} \leq \frac{2}{N},
 \\
\\
\exp \left[ N^2 \left(\beta_- \pi_{AB}
-\frac{1}{2}\right) \right],\qquad\qquad \quad\qquad~\frac{2}{N} <\pi_{AB} \leq \mu_-^2 
, \\
\\
e^{-\frac{N^2}{2}} \left(1-\sqrt{\pi_{AB} }\right)^{N^2},  \qquad\qquad\qquad\qquad  \mu_-^2 
< \pi_{AB} <1 .
\end{cases}
\label{eq:1122}
\end{eqnarray}
We have three main regions of entanglement, corresponding to the maximally entangled phase ($1/N < \pi_{AB}\leq 5/4$), the typical phase ($5/4N < \pi_{AB}\leq \mu_{-}^2$) and the separable phase ($\mu_{-}^2 < \pi_{AB} < 1$). 
Notice that $\pi_{AB}=2/N$ refers to typical states, $\beta=0$, at the boundary between the solution for the positive and the stable solution at negative temperatures. We have already noticed in Sec.~\ref{sec:finitesize} that the thermodynamic potentials show large corrections in the region $\mu=\Ord{1/N}$. In particular, if on the one hand we have neglected contributions of $\Ord{1/N}$ to the potential energy density (\ref{eq:vn_stacc1}) in the region of negative temperatures ($\alpha=2$), on the other hand the average purity of the system in the region of positive temperatures ($\alpha=3$) is also of $\Ord{1/N}$ (see Eq. (\ref{eq:puritysmall})). In other words when we derive the expression of the entropy as a function of the internal energy we deal with the same order of fluctuations in $u$ and $s$. It follows that in order to yield a uniform connection between the two branches at $\pi_{AB}=2/N$, we should have considered higher order corrections in $1/N$. However, this goes beyond our scopes. 
In order to plot $s(u)$ for finite $N$, notwithstanding the presence of these fluctuations, we have artificially modified the entropy as 
\begin{equation}
\bar{s}(x)= \begin{cases}
s(N \exp(x)), &  x\leq \ln(2/N)\\ 
\\
s(\left(\exp(x)-2/N\right),& x > \ln (2/N).
\end{cases}
\end{equation}
This function is continuous in $x=\ln{2/N}$, for finite $N$, see Fig.~\ref{fig:volumestabEmet}. 
The presence of discontinuities in some derivatives of the entropy detects the two phase transitions for the stable solution. 
Notice that, as already pointed out, discontinuities of the $n$-th derivative of $s(T)$ translate in discontinuities of the $(n+1)$-th derivative of $s(u)$. 
At $\pi_{AB}=5/4N$ there is a second order phase transition signaled  by a discontinuity in the third derivative, see Sec.~\ref{EntropyVSentanglement} (Eq. (\ref{eq:sAt54})).
The first order phase transition, which takes place between $\pi_{AB}=2/N$ and $\pi_{AB}=\mu_{-}^2\simeq 0.512$ is signaled by discontinuities in the second derivative of the entropy at those points. Actually, differently from $\pi_{AB}=\mu_{-}^2$, analyzed in Sec.~\ref{firstOrderphTrans} (Eq. (\ref{eq:SvsUmu-})), the same check can not be explicitly performed for $\pi_{AB}=2/N$,  due to the large fluctuations in this region, mentioned above.

Observe that entropy is unbounded from below: 
at both endpoints of the range of purity, $\pi_{AB}=1/N$ (maximally entangled states) and $\pi_{AB}=1$ (separable states), when the isopurity manifold shrinks to a vanishing volume in the original Hilbert space, the entropy, being the logarithm of this volume, diverges, and the number of vector states goes to zero (compared to the number of typical vector states)

\section{Distribution of eigenvalues}\label{sec:OverviewdistributionOfEigenvalues}
\begin{figure}[h]
\centering
\includegraphics[width=0.5\columnwidth]{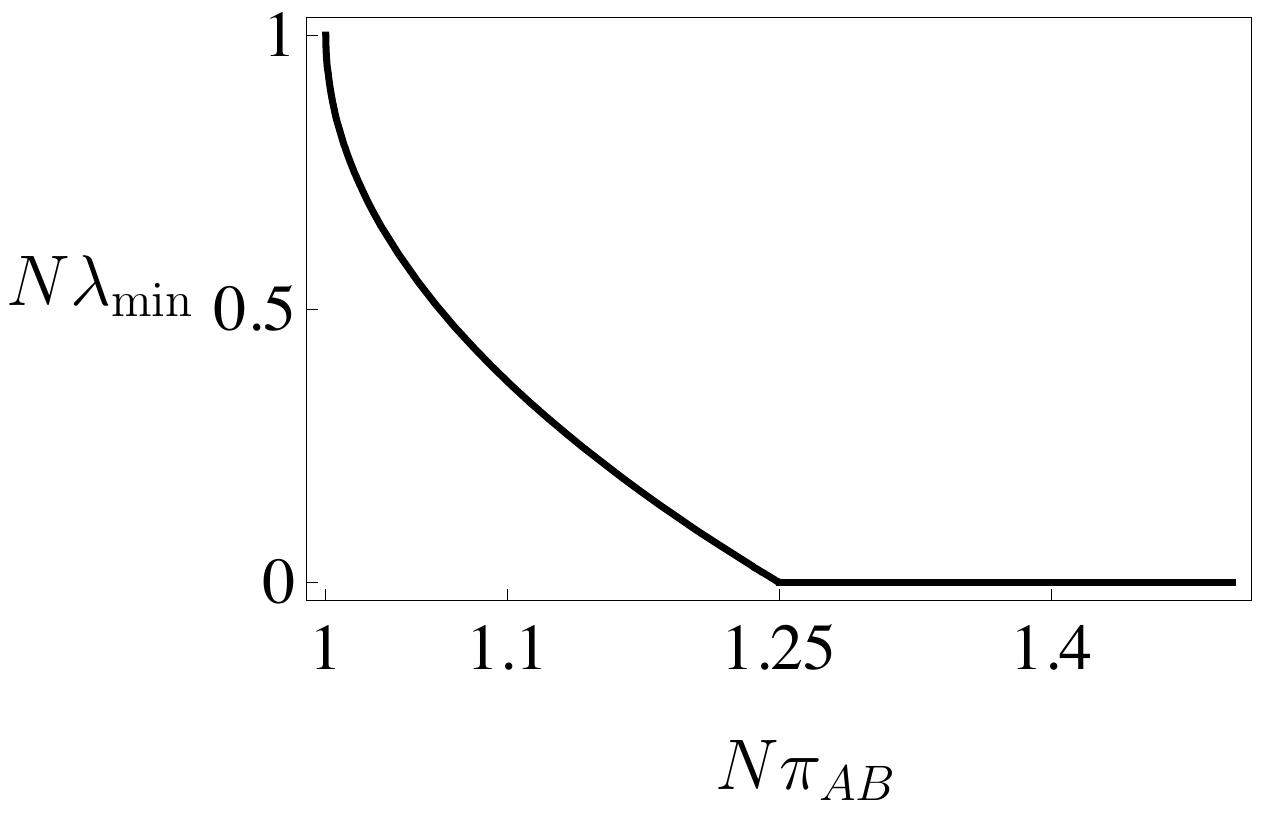}
\caption{Minimum eigenvalue as a function of $\pi_{AB}$.  At $\pi_{AB}=5/4N$ the gap vanishes.}
\label{fig:minmax}
\end{figure}

In the previous chapters we have stressed that the presence of phase transitions in the system can be easily read out from the behavior of the distribution of the Schmidt coefficients, i.e.\ the eigenvalues of the reduced density matrix $\rho_{A}$ of one subsystem. From Eq.~(\ref{eq:deltavsu}) we get the expression of the minimum eigenvalue  $\lambda_{\mathrm{min}}=a = m-\delta$
as a function of $\pi_{AB}$
\begin{eqnarray}
\lambda_{\mathrm{min}} = \begin{cases}
\frac{1}{N}\left(1-2 \sqrt{N \pi_{AB} -1 }\right),  & \frac{1}{N} < \pi_{AB} \leq  \frac{5}{4 N} ,  \\
\\
 0 ,
& \frac{5}{4 N} < \pi_{AB} < 1.
\end{cases}
\label{eq:lambdamin}
\end{eqnarray}
See Fig.~\ref{fig:minmax}.  We recall that $m=1$ for $1/N < \pi_{AB} \leq  5/4N$ and $m=\delta$ for $5/4 N < \pi_{AB} < 1$, Eqs. (\ref{eq:mdeltasemicircle}) and (\ref{eq:beta<2}). 
The second order phase transition at $\pi_{AB}=5/4N$, associated to a ${\mathbb Z}_2$ symmetry breaking, is detected by a vanishing gap.
\begin{figure}[h]
\centering
\includegraphics[width=0.9\columnwidth]{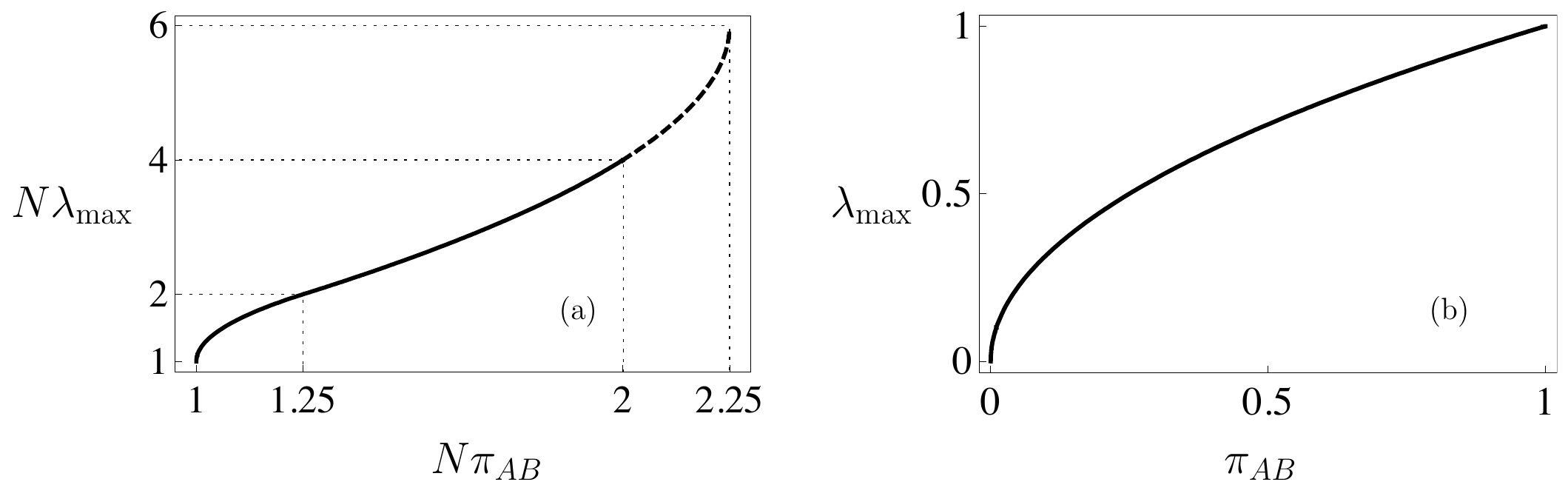}
\caption{Maximum eigenvalue as a function of $\pi_{AB}$. The dashed line in (a) represents the first part of the metastable branch, from $\pi_{AB}=2/N$ to $\pi_{AB}=9/4N$ ($\beta=\beta_{g}$).}
\label{fig:maximum}
\end{figure}\\ \\ \\
On the other hand, the maximum eigenvalue $\lambda_{\mathrm{max}}$ coincides with the upper edge of the sea of eigenvalues $b = m+\delta$, as given by (\ref{eq:deltavsu}), until it evaporates according to Eq.~(\ref{eq:unegmu}). Thus,
\begin{eqnarray}
\lambda_{\mathrm{max}} =  \begin{cases}
\frac{1}{N}\left(1+2 \sqrt{N \pi_{AB} -1 }\right),  & \frac{1}{N} < \pi_{AB} \leq  \frac{5}{4 N} ,  \\
\\
\frac{2}{N}\left( 3  - 2 \sqrt{\frac{9}{4} - N \pi_{AB}} \right),
 & \frac{5}{4 N} < \pi_{AB} \leq \frac{2}{N}\qquad\,
 \\
\\
\sqrt{\pi_{AB} }, &  \frac{2}{N} < \pi_{AB} < 1 ,
\end{cases}
\label{eq:lambdamax}
\end{eqnarray}
as shown in Fig.~\ref{fig:maximum}.
\begin{figure}[h]
\centering
\includegraphics[width=0.8 \columnwidth] {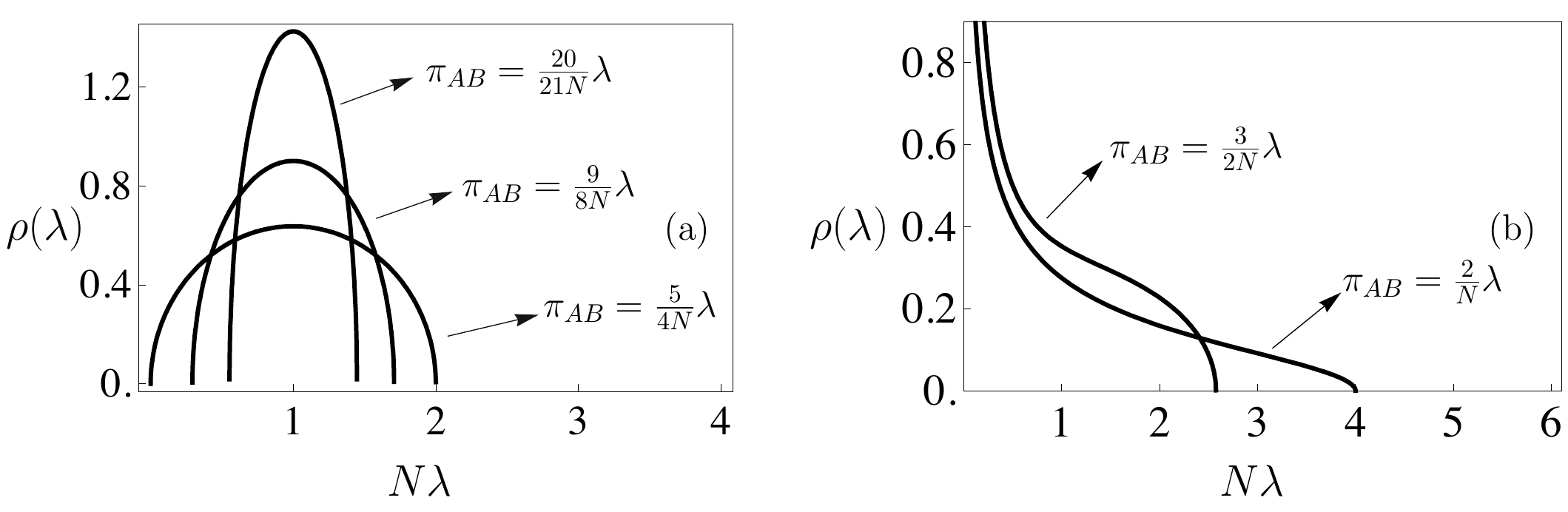}  \includegraphics[width=0.35\columnwidth] {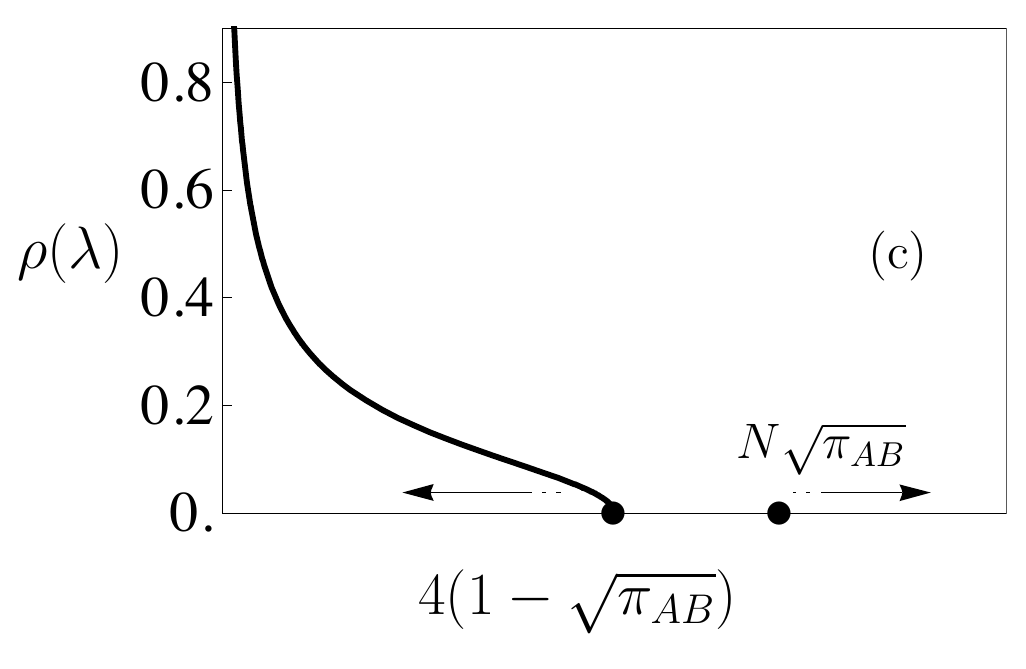}
\caption{Summary (stable branch): profiles of the eigenvalue density for $1/N < \pi_{AB} \leq \frac{5}{4N}$, $\frac{5}{4N} < \pi_{AB} \leq \frac{2}{N}$, and $\frac{2}{N} < \pi_{AB} < 1$.}
\label{fig:distributions}
\end{figure}\\
Notice that due to the double scaling in the maximum eigenvalue and the purity in the two nearby intervals, $1/N < \pi_{AB}\leq2/N$ and $2/N < \pi_{AB} < 1$, we preferred to separate the plot of $\lambda_{\mathrm{max}}$ vs $\pi_{AB}$ in two distinct figures, Fig.~\ref{fig:maximum}(a) and (b). 
In Chaps.~\ref{chap3} and~\ref{chap4} we have seen that in the different phases the distribution of the eigenvalues of $\rho_A$ has very different profiles. 
See Fig.~\ref{fig:distributions}.
While for $1/N < \pi_{AB} \leq \frac{5}{4N}$ the eigenvalues (all $\Ord{1/N}$) follow Wigner's semicircle law (maximally entangled phase), they become distributed according to Wishart for larger purities, $\frac{5}{4N} < \pi_{AB} \leq \frac{2}{N}$ (typical phase), across the second order phase transition.
This is a first signature of separability: some eigenvalues vanish and the Schmidt rank decreases. For even larger  values of purity, $\frac{2}{N} < \pi_{AB} < 1$, across the first order phase transition, one
eigenvalue evaporates, leaving the sea of the other eigenvalues $\Ord{1/N}$ and becoming $\Ord{1}$ (separable phase). This is the signature of factorization, fully attained when the eigenvalue becomes $1$ at $\pi_{AB} = 1$. 

\begin{figure}[h]
\centering
\includegraphics[width=0.9\columnwidth]{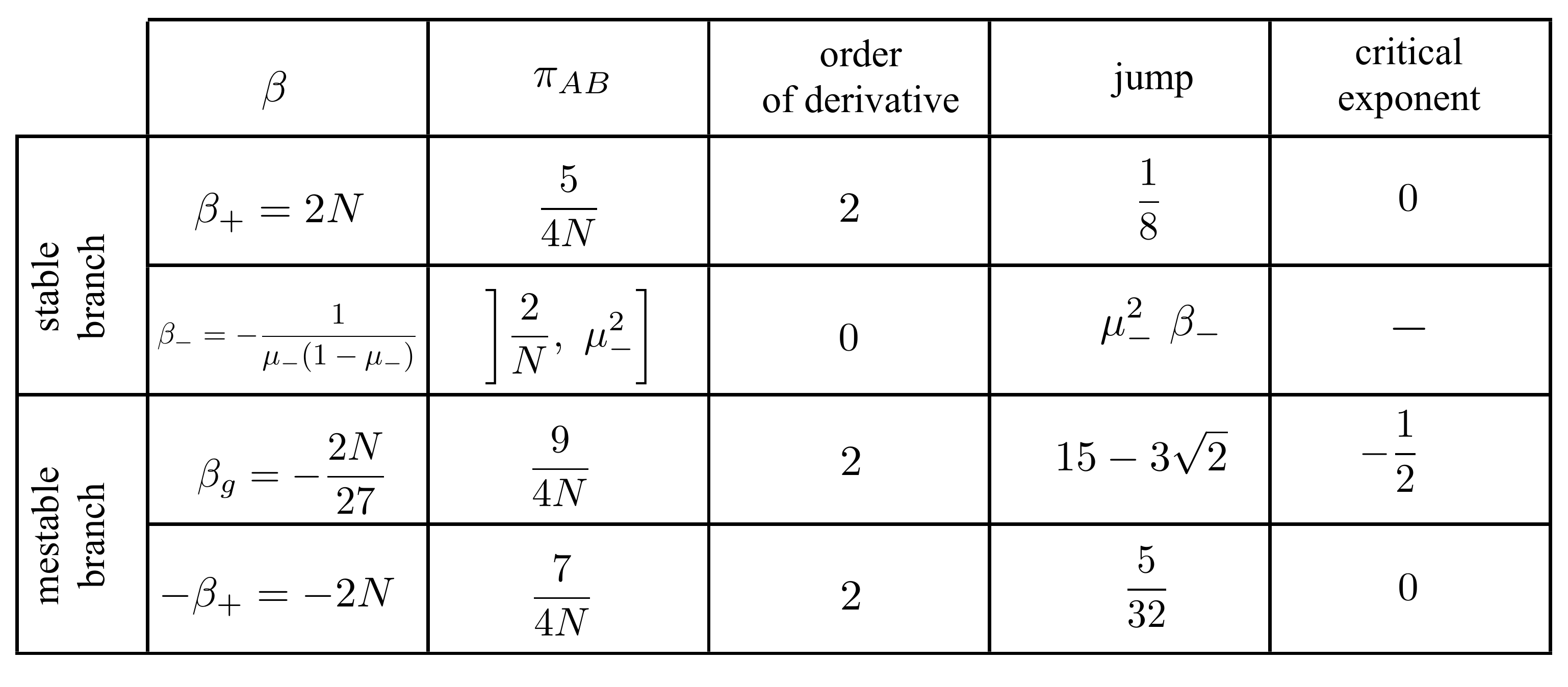}
\caption{Table of the critical points for the system, for the stable and metastable branches. In particular the third and the forth columns refer to the behavior of the entropy density $s$ at the critical point while the last column refers to derivative with respect to $\beta$ of the heat capacity. Notice that the first order phase transition at $\beta_{-}\simeq -2.455$ involves a broad interval of purity, from $2/N$ to $\mu_{-}^2\simeq 0.5117$, associated to the coexistence of typical and separable phases.}
\label{fig:table}
\end{figure}

In the previous chapters, we have associated to each phase transition in the system, a critical exponent, related to the behavior of the entropy as a function of the inverse temperature $\beta$. See the table in Fig.~\ref{fig:table}.
\\ \\
For the first order phase transition, we have replaced the critical exponent with the thermodynamic relation (\ref{eq:firstorderphasetr}).  Note that in this case, at fixed temperature $\beta_{-}$ the purity, together with the entropy of the system, varies in a finite interval associated to the coexistence of the typical and separable phases.

\begin{figure}[h]
\centering
\includegraphics[width=1\columnwidth]{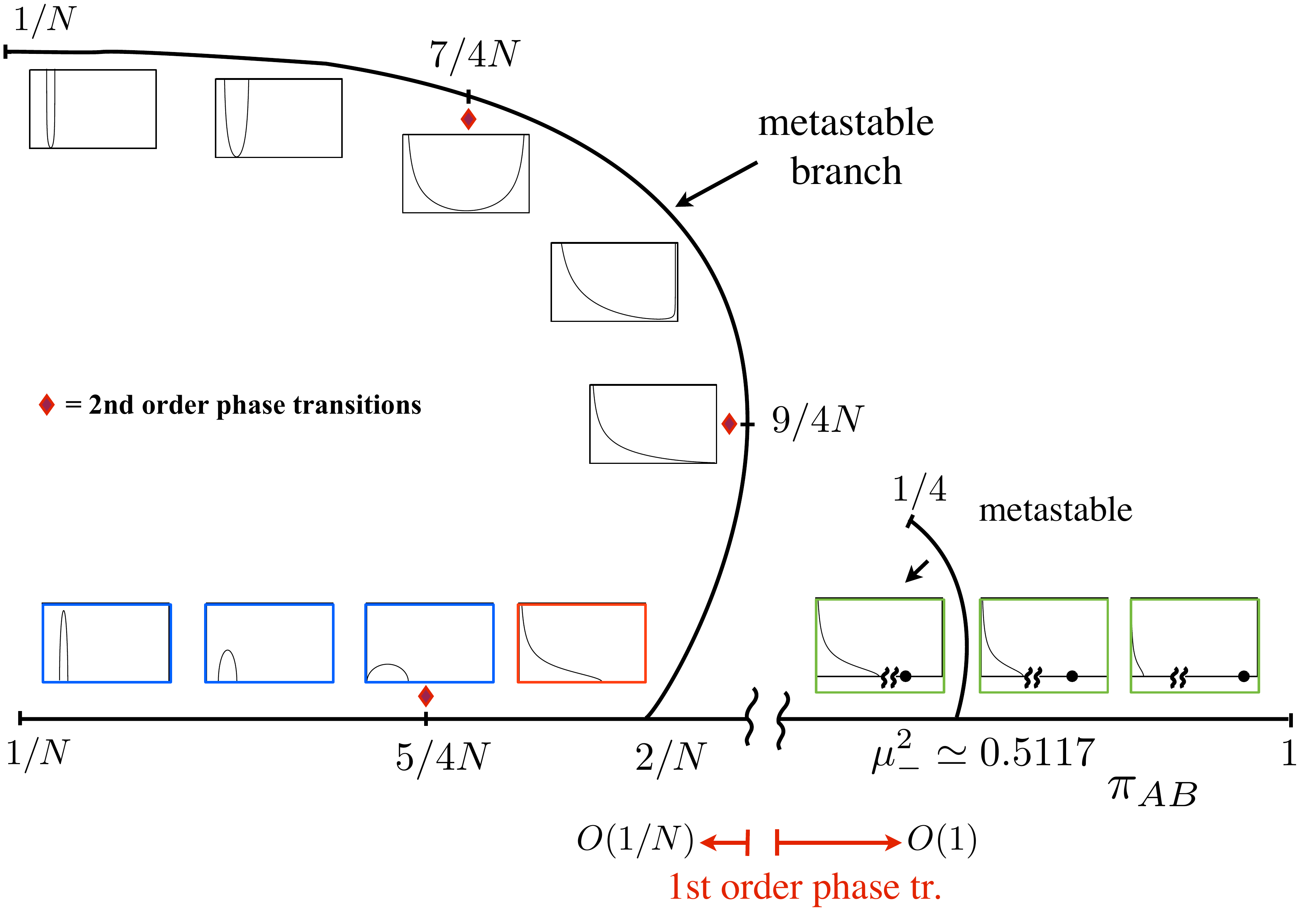}
\caption{Overview of the ``evolution" of the eigenvalue densities as a function of purity $(1/N < \pi_{AB} < 1)$.
The straight segment represents the stable branch.
The three phases of maximally entangled, typical and separable states are indicated in blue, red and green, respectively, as done in Fig.~\ref{fig:volumestabEmet} for the volumes of states. The curved lines refer to the two metastable branches. The diamonds indicate the three second order phase transitions.}
\label{fig:masterpiece}
\end{figure}

We conclude the first part of this Thesis with an overview in Fig.~\ref{fig:masterpiece} of the phenomenology of the phase transitions in bipartite entanglement. 
Observe that while the second order phase transitions correspond to a given value of the purity, the first order phase transition from typical to separable states lies on a broad interval of $\Ord{1}$, namely $2/N < \pi_{AB} \leq \mu_{-}^2 \simeq 0.5117$. We have also included the two metastable branches. A metastable branch is born at $\pi_{AB}=2/N$: it starts as Wishart, undergoes a second order phase transition at  $\pi_{AB}=9/4N$ (2D gravity), where a singularity is developed at its right edge, then its support starts decreasing, undergoes a second order phase transition at $\pi_{AB}=7/4N$ (${\mathbb Z}_2$ symmetry restoration) and eventually becomes sharply peaked (with two singularities). The other metastable branch is born at $\pi_{AB}=1/4$ and corresponds to the separable phase: it becomes stable at $\pi_{AB}=\mu_{-}^2$. \\

 \newpage
\thispagestyle{empty}
\mbox{}
\part{MIXED STATES}
\label{intropart2}
\newpage
\thispagestyle{empty}
\mbox{}

\chapter*{Summary}
\addcontentsline{toc}{chapter}{Summary}

In the first part of the thesis we have introduced a statistical approach to the study of the behavior of bipartite entanglement of a large quantum system in a pure state. In particular we have chosen as entanglement measure the purity of one of the two subsystems, namely the local purity of the global state. Nevertheless, pure states represent only a special set of states as they refer to isolated quantum systems. In practice physical systems are described by a statistical mixture of pure states due to the interaction with the environment. In this second part of the thesis we intend to generalize our statistical analysis to the case of arbitrary  mixed global states. Let us first observe that in order to study the properties of bipartite entanglement for a large quantum system in a mixed state, we should substitute the purity of one of the two subsystems with its convex-roof counterpart, as seen in Chap.~\ref{chap1}. This considerably enhances the difficulty of the analytical treatment of our problem and will not be discussed in this thesis. On the other hand, a related interesting topic is the analysis of the statistical distribution of the local mixedness of the global density matrix, as an indicator of the average lower bound for the bipartite entanglement (see Eq. (\ref{eq:convexpurity})). This problem will be tackled by studying the moments of the local purity of the system.

This part of the thesis consists of three chapters. In Chap.~\ref{chap7}, by following a standard approach in statistical mechanics, we will determine the formal expression of the moments of the purity at a given temperature. In fact, for the case of arbitrary mixed global states, due to the nonfactorization of the integral over the unitary group, we cannot directly compute the partition function by applying the method of the steepest descent, as was done for the case of pure states. In Chap.~\ref{chap8} we will introduce a Gaussian approximation for the elements of the unitary group acting on the global and local systems, in order to determine the first moment of the purity at infinite temperature ($\beta=0$), at $\Ord{1/N}$, being $N$ the dimension of the subsystem whose purity distribution we aim to study. This will lead to a first generalization of the results achieved for the case of pure states in~\cite{page,Lloyd1988,paper1}. Finally in Chap.~\ref{chap9}, by exploiting the symmetry properties of the twirling maps~\cite{tw1} and thru the solution of some basic integrals over the unitary group due to Zuber~\cite{Zuber}, we will compute the exact expression for the first two cumulants of the local purity at $\beta=0$. Our results, when reduced to the case of pure states, will be in agreement with~\cite{Scott2003,Giraud}. We will conclude our analysis by determining the high temperature expansion of the first moment of the local purity.

\chapter{Statistical distribution of local purity}\label{chap7}
\markboth{Statistical distribution of local purity}{Statistical distribution of local purity}

We will devote this chapter to the introduction of the physical and mathematical setup that generalizes to the case of mixed quantum states the statistical approach introduced in Part 1 for the study of the local mixedness of pure states. In Sec.~\ref{sec:localmixedness} we will introduce the concept of local purity for a generic mixed bipartite state. It will play, in Sec.~\ref{sec:introductionstatisticalapproach},  the role of energy of the system in the definition of the partition function for a generic set of states with a given value of the global purity. Finally in Sec.~\ref{sec:Asymptotic behavior and moment analysis} we will write the formal expression for the moments  of the local purity, with respect to the canonical ensemble, in order to characterize its statistical distribution. 
\section{Entanglement and local mixedness}\label{sec:localmixedness}
Consider a bipartite system given by two subsystems $A$ and $B$,  described by the Hilbert space $\mathcal{H}=\mathcal{H}_A \otimes \mathcal{H}_B$, with $\dim \mathcal{H}_A=N$, $\dim \mathcal{H}_B=M$ and $\dim \cH=L=N M$. Without loss of generality we will assume that $N \leq M$. 
In this second part of the thesis we will slightly change the notation introduced in the first part. In particular we will define the total Hilbert space as $\mathcal{H}_X$, being $X=AB$ our bipartite system, and will denote with $\lambda_i$, $1 \leq i \leq L$, the spectrum of the global states in $\cH_X$, while the eigenvalues of each subsystems will be indicated by $\lambda_{A, i}$ and $\lambda_{B, j}$, for $1\leq i \leq N$ and $1\leq j \leq M$. (Recall that in the first part of the thesis the symbol $\lambda_i$, $1 \leq i \leq N$, referred to the eigenvalues of the states belonging to $\mathcal{H}_A$).

In the previous chapters we have explored the total Hilbert space $\cH_X$ from the point of view of bipartite entanglement between $A$ and $B$, that is in terms of manifolds with a fixed value of the purity of one of the two subsystems (isopurity manifolds). 
The main result of our analysis was the computation of the probability of finding a random pure state (of a large quantum system) with a given value of average entanglement between $A$ and $B$, if sampled according to the unique invariant Haar measure. 

As mentioned in Chap.~\ref{chap1}, in general the states of $X$  are represented by the set of nonnegative unit trace operators (density matrices) $\rho$  which  act on the Hilbert space $\mathcal{H}_X$, and reduce to (projections onto) vectors in the Hilbert space $\cH_X$ if and only if they have only one nonnull eigenvalue (equal to $1$). 
We indicate with $\mathfrak{S}({\cal H}_X)$ the set of all the states of $X$ and with
\begin{eqnarray}
\pi_{X}(\rho) = \mathrm{Tr}\rho^2 \ \in [1/L, 1]\;, \label{purityglobal}
\end{eqnarray} 
the purity of an arbitrary state $\rho$. The latter quantity provides  a characterization of the global mixedness of the system and induces a partition of $\mathfrak{S}({\cal H}_X)$ into a collection of distinct subsets 
\begin{equation}\label{eq: set with fixed mixedness}
\mathfrak{S}_{x}({\cal H}_X) = \{ \rho \in \mathfrak{S}({\cal H}_X): \mathrm{Tr}\rho^2=x\}.
\end{equation}
The minimum value of  $x=1/L$ is attained  when $X$ is in the completely mixed  state $I/L$,  
whereas the maximum $x=1$ is attained over the set $\mathfrak{S}_{1}({\cal H}_X)$ which includes all pure states
 $\ket{\psi}_{X}$.
   For each   $\rho \in \mathfrak{S}({\cal H}_X)$ the quantyties
\begin{eqnarray}
\pi_{A}(\rho)= \Tr \rho_A^2
 \quad \mbox{and} \quad  
 \pi_{B}(\rho) =\Tr \rho_B^2
\label{eq:purityN_Aloc}
\end{eqnarray}   
represent its $A$-{\em local} and $B$-{\em local} purity functions, $\rho_A=\Tr_B \rho$ and $\rho_B=\Tr_A \rho$  being the reduced density matrices of the subsystems $A$ and $B$, respectively.
Notice that the local purities of $\rho$ can be different in general, and vary in different ranges:
\begin{eqnarray}
\pi_{A}(\rho)\in [ 1/N,1] 
 \quad \mbox{and} \quad  
 \pi_{B}(\rho)   \in [ 1/M,1]. 
\end{eqnarray} 
Clarifying the connection between the global and local purities of a state is an important problem of quantum information theory~\cite{MIX2,MIX3,MIX1} which is closely   related to the characterization of bipartite entanglement~\cite{ENTMEAS,ENTMEAS1}. In particular we have seen that for the pure states $\rho=\ket{\psi}_{X} \bra{\psi}$ of $X$, belonging to the set  $\mathfrak{S}_{1}({\cal H}_X)$, due to the Schmidt decomposition (\ref{eq:pureSchmidtDec}), the $A$ and the $B$-local purities coincide
\begin{equation}
\pi_{AB}(\psi)= \pi_{A}(\ket{\psi}_{X}\bra{\psi})=\pi_B(\ket{\psi}_{X}\bra{\psi})\;
\label{eq:purityN_A}
\end{equation}
and provide a  measure of the bipartite entanglement between $A$ and $B$: the smaller 
$\pi_{AB}(\psi)$ is, the larger is the entanglement contained in $\ket{\psi} _{X}\bra{\psi}$, see Sec.~\ref{pure state measures}. The double index of $\pi_{AB}$ refers to the equality between the local purities when the global state is pure (see  Eq. (\ref{eq:purityN_A})).

\section{The statistical approach: partition function} \label{sec:introductionstatisticalapproach}
In the first part of the thesis, we have characterized the bipartite entanglement between $A$ and $B$ by  studying the statistical distribution of $\pi_{AB}(\psi)$ on $\mathfrak{S}_{1}({\cal H}_X)$.
In this section we will generalize this statistical approach, based on the canonical ensemble, to the case of generic states, being the pure states only a special class of them. 

Let us briefly recall what seen for the $x=1$. We have introduced the partition function 
\begin{equation}
\mathcal{Z}_{AB}(\beta)=
\int
 d\mu (\psi)\; \mathrm{e}^{- \beta\; \pi_{AB}(\psi)} \;,
\label{eq:pure}
\end{equation}
where the local purity $\pi_{AB}(\psi)$ of  $|\psi\rangle_{X}$  plays the role of the internal energy density of the system and $\beta$ is a Lagrange multiplier which fixes the value of the purity/energy selecting an isopurity manifold~\cite{Kus01}. See Eq. (\ref{eq:partitionfunction}) for comparison: here the coefficient $N^{\alpha}$, related to the scaling properties of the system, has been reabsorbed in the definition of the fictitious temperature. We also have that, $d \mu (\psi)$ is a (normalized) measure on the space of pure states $\mathfrak{S}_{1}({\cal H}_X)$.
From the structure of complex vector space of $\cH_X$, we have that the natural choice  for $d \mu (\psi)$
has been induced by the Haar measure $d\mu_H(U)$ on the unitary group  $\mathcal{U}(\mathcal{H}_X)\simeq\mathcal{U}(L)$, through the mapping 
$\ket{\psi}_{X}=U_{X}\ket{\psi_0}_{X}\;$
with $\ket{\psi_0}_{X}$ an arbitrary reference unit vector of $\cH_X$. 
Noticing then that for every $U_X \in \mathcal{U}(L)$ the reduced density matrix $\rho_A=\Tr_B(\ket{\psi}_{X}\bra{\psi})$
can be written as $\rho_A=U_A \Lambda_A U_A^{\dagger}$, with $U_A \in \mathcal{U}(N)$ and $\Lambda_A= \mathrm{diag} \{ \lambda_{A, 1}, \lambda_{A, 2}, \ldots \lambda_{A, N} \}$, the  measure induced over the density matrices $\rho_A$ by 
$d \mu_H(U_X)$, 
factorizes into the product of a measure over the unitary group $d \mu_{H}(U_A)$ (related to the eigenvectors of $\rho_A$) and  a measure  $d \sigma(\Lambda_A)$ over the $(N -1)$ dimensional simplex of its eigenvalues $\lambda_{A,j}$, $1\leq j \leq N$~\cite{Zyczkowski98,Zyczkowski99}. Thus the partition function becomes
 \begin{equation}\label{eq:pure1}
 \cZ_{AB}(\beta)
=\int d \mu_{H}(U_A) \int d \sigma(\Lambda_A)\;  \e ^{-\beta\Tr (\Lambda_A^2)}
=\int d \sigma(\Lambda_A)\;  \e ^{-\beta\Tr (\Lambda_A^2)} ,
\end{equation}
where
\begin{equation}\label{eq:mis_autov_A}
 d \sigma(\Lambda_A)=   C_{N,M}\ \delta\Big(1-\sum_{1 \leq i \leq N} \lambda_{A,i}\Big) \prod_{1 \leq j \leq N} \theta(\lambda_{A,j})\lambda_{A,j}^{M-N} 
\prod_{1 \leq l<m \leq N}(\lambda_{A,l}-\lambda_{A,m})^2 d ^ {N} \lambda_A\;,
\end{equation}
is the measure on the spectrum, with $\theta(x)$ the unit step function and $C_{N,M}$  a normalization coefficient, see Eqs. (\ref{eq:measure}) and (\ref{eq:normcoeffpure}).

A natural question is what happens when the 
system $X$ is in a \emph{mixed} state $\rho$ of purity  $x\neq 1$, rather than in a pure state. 
A generalization of Eq.~(\ref{eq:pure}) is obtained by replacing $\pi_{AB}(\psi)$ with  (say)
the $A$-local purity 
$\pi_A(\rho)$ of Eq.~(\ref{eq:purityN_Aloc})  and 
the measure $d \mu(\psi)$ with a proper measure $d\mu_{x}(\rho)$ on the set $\mathfrak{S}_{x}({\cal H}_X)$.
This yields the following definition of the partition function of the $A$-local purity 
\begin{equation}
\mathcal{Z}_A(x, \beta)=\int d \mu_{x} (\rho) \;
\mathrm{e}^{-\beta\; {\pi}_A(\rho)}
 = C_{x} \int d\mu
(\rho) \; \delta(\Tr\rho^2 -x) \;
\mathrm{e}^{-\beta\; {\pi}_A(\rho)}\;,
 \label{eq:mixed}
\end{equation}
where $d\mu (\rho)$ is a probability measure on the set of mixed states (see below),
and
\begin{equation} 
C_{x} = \left[ \int d\mu
(\rho) \; \delta(\Tr\rho^2 -x)\right]^{-1}
\end{equation}
is a normalization factor. An analogous expression 
for the $B$-local purity partition function $\mathcal{Z}_B(x, \beta)$  is obtained  by simply replacing $\pi_A(\rho)$ with $\pi_B(\rho)$ in Eq.~(\ref{eq:mixed}). Notice however that for $x\neq 1$, $\mathcal{Z}_B(x, \beta)$  will in general differ from $\mathcal{Z}_A(x, \beta)$. 
Moreover, it is worth stressing that the  function $\mathcal{Z}_A(x, \beta)$  provides only statistical information on the  local mixedness  of $X$, but not directly on its  bipartite entanglement properties: this is  due to the fact that
for generic mixed states $\rho$ of $X$ the local purities $\pi_A(\rho)$ and $\pi_B(\rho)$ are not entanglement measures, see Sec.~\ref{generic state measures}. We have also seen that a generalization of Eq.~(\ref{eq:pure}) that retains the ability of characterizing the statistical properties of the
bipartite entanglement of $X$ for $x\neq1$  could in principle be constructed by
replacing $\pi_{A}(\psi)$ with 
 its convex-roof
 \begin{equation}\label{eq:pibarroof}
\bar{\pi}_{A}(\rho) = \max_{\rho=\sum_{j}p_j \ket{\psi_{j}}_{AB}\bra{\psi_j}} \; \sum_j p_j\;  \pi_{A} (\psi_j),
\end{equation}
where the maximum is taken over all ensembles $\{ p_j,  |\psi_j\rangle_{AB}\}_j$ which provide a convex decomposition of the mixed state $\rho$ (e.g. see~
\cite{entanglement2,CORNELIO}). Eq.~(\ref{eq:pibarroof}) coincides with Eq.~(\ref{eq:convexpurity}) being $E_{P}=\bar{\pi}_{A}$.
The quantity $\bar{\pi}_{A}(\rho)$ is a  proper measure of the bipartite entanglement, 
but  the resulting partition function 
does not allow for a simple analytic treatment and will not  be discussed in this thesis. The average local purity we will determine with our statistical approach can be seen as the average lower bound for the bipartite entanglement of the global system, with respect to the bipartition $AB$.

Finally, another delicate point is that, differently from what seen for the case of pure states, there is no unique measure  on mixed states~\cite{Zyczkowski}. Indeed, as previously discussed for $\rho_A$,  the Hermitian matrix $\rho$ can be always diagonalized by a unitary rotation, and as consequence we can write the measure $d \mu (\rho)$ as the product of a measure on the $(L-1)$ dimensional symplex of the eigenvalues and a measure of on the unitary group $\mathcal{U}(L)$ related to the eigenvectors. However, if on the one hand it would be natural to take the Haar measure on $\mathcal{U}(L)$ so that $d\mu (\rho)=d\mu (U_X \rho U_X^{\dagger})$, on the other hand the measure on the eigenvalues can be chosen in different possible ways~\cite{Slater}. Thus, we need to properly specify the choice of  $d \mu(\rho)$, since the consistency requirement 
that  for pure states Eq.~(\ref{eq:mixed}) should reduce back to Eq.~(\ref{eq:pure}), i.e.  $\mathcal{Z}_A(1, \beta)=\mathcal{Z}_{AB}(\beta)$, does not eliminate such ambiguity.     
In order to overcome this point we will use a \textit{balanced purification} strategy. In other words we introduce a composite Hilbert space
\begin{equation}
\mathcal{H}_{X X'}=\mathcal{H}_X\otimes \mathcal{H}_{X'}, \qquad \mathrm{with} \quad \mathcal{H}_{X} \simeq \mathcal{H}_{X'}
\end{equation}
that is $\mathcal{H}_{X}$ and $\mathcal{H}_{X'}$ are isomorphic Hilbert spaces. In this $L^2$ dimensional Hilbert space, each $\rho$ of $X$ can be represented by those pure states
 $\ket{\Psi}_{X X'}$  which  provide a \textit{purification} for such density matrix, i.e. which satisfy the identity 
\begin{equation} \rho=\mathrm{Tr}_{X'}
(\ket{\Psi}_{X X'}\bra{\Psi})\;.
\end{equation}
Thanks to this identification we can now induce a measure on  $\mathfrak{S}({\cal H}_X)$ by sampling 
the pure states on $\mathcal{H}_{X X'}$ according to the unique,
unitarily invariant Haar measure which, as usual,  is induced  by the Haar measure on the unitary group
$\mathcal{U}(L^2)$ through the mapping $\ket{\Psi}_{X X'}= U_{X X'}\ket{\Psi_0}_{X X'}$, where $\ket{\Psi_0}_{X X'}$ is an arbitrary reference vector and $U_{X X'}\in\mathcal{U}(L^2)$. Thus, with this choice the partition function becomes
%
\begin{eqnarray}\label{eq:ZUN2}
\mathcal{Z}_A(x, \beta)&=& C_x \int
d\mu_{H}(U_{X X'})\,
\delta\left(x-\Tr(\Tr_{X'}(\ket{\Psi}_{X X'}\bra{\Psi})^2\right)
\nonumber\\&&\qquad\qquad \mathrm{e}^{-\beta\; \Tr
\left(( \Tr_B \Tr_{X'} \ket{\Psi}_{X X'}\bra{\Psi})^2\right) },
\nonumber\\
\end{eqnarray}
where in computing $\pi_A(\rho)$ we have used the fact that 
the reduced density matrix  of subsystem $A$ is obtained from $\ket{\Psi}_{X X'}$ by tracing over $X'$ and $B$, i.e. 
\begin{equation}
\rho_A=\mathrm{Tr}_B(\rho)=\mathrm{Tr}_B(\mathrm{Tr}_{X'}
\ket{\Psi}_{X X'}\bra{\Psi}).
\end{equation}
Analogously to what seen for the pure case, $x=1$, by writing
$\rho=U_X \Lambda_X U_X^\dagger$ with 
$\Lambda_X=\mathrm{diag}(\lambda_1, \lambda_2, ..., \lambda_L)$,
we finally get
\begin{eqnarray}\label{eq:partition function unitary rotated}
\qquad \mathcal{Z}_A(x, \beta)= C_x \int
d\mu_H(U_X) \int d\sigma(\Lambda_{X})\; \delta\Big(x-\Tr \Lambda_X^2\Big)\; \mathrm{e}^{-\beta\;
\mathrm{Tr}
(\mathrm{Tr}_B(U_X\Lambda_X {U_X}^\dagger))^2}, \nonumber\\
\end{eqnarray}
where  $d\mu_H(U_X)$ is the Haar measure on $\mathcal{U}(L)$ 
and
\begin{equation}\label{eq:mixed1}
d\sigma(\Lambda_{X})=C_L\;  \delta\Big(1-\sum_{1\leq i \leq L}\lambda_i\Big)\; \prod_{1\leq i \leq L} \theta(\lambda_i)
\prod_{1\leq i<j \leq L}(\lambda_i-\lambda_j)^2 d^L \lambda,
\end{equation}
with
\begin{equation}
C_{L}=\frac{\Gamma(L^2)}{\Gamma(L+1) \prod_{1\leq k \leq L } \Gamma(k)^2}.
\end{equation}
Notice that, from the properties of the Euler gamma function $\Gamma$, the normalization constant $C_{L}$ concides with $C_{N, M}$ in Eq. (\ref{eq:normcoeffpure}) if we replace $N$ and $M$ with $L$, namely $C_L$ is equivalent to $C_{N,M}$ in Eq. (\ref{eq:mis_autov_A}) for $N=M=L$. The above definition is equivalent to have identified the measure $d \mu_x(\rho)$ of Eq.~(\ref{eq:mixed}) with  
\begin{eqnarray}\label{defdmu}
d \mu_x(\rho) = C_x d\mu_H(U_X) d\sigma(\Lambda_{X})\delta\Big(x-\Tr \Lambda_X^2\Big)\;.
\end{eqnarray} 
In the case of pure states, i.e.\ $x=1$, when the density operator of the system reduces to $\rho=\ket{\psi}_{X}\bra{\psi}$, the matrix $\Lambda_{X}$ becomes a rank one projection. Thus expression (\ref {eq:partition function unitary rotated}) reduces to  (\ref{eq:pure1}):
\begin{eqnarray}
\cZ_A(1, \beta)=\cZ_{AB}(\beta)\;. \label{identita}
\end{eqnarray}

\section{Asymptotic behavior and analysis of moments} \label{sec:Asymptotic behavior and moment analysis}
In the first part of this thesis we have applied the method of the steepest descent in order to determine the leading contribution to the partition function at $x=1$. This analysis could be performed because of the factorization of the integration over the unitary group.
It led to a complete characterization (in the thermodynamic limit) of the statistical distribution of the local purity of $X$, $\pi_{AB}(\psi)=\pi_{A}(\psi)=\pi_{B}(\psi)$  (and hence of the bipartite entanglement of the system) computed with respect to a {\em canonical} ensemble for different regions of the Hilbert space, selected by the inverse temperature $\beta$.
In other words, in the limit of large $N$, we have determined the most probable distribution of the Schmidt coefficients
which maximizes the distribution (\ref{ffhh1})
\begin{eqnarray}
d \mu_\beta(\rho_A) =d \mu(\rho_A) \; \frac{  \e ^{-\beta \;  \pi_{AB}(\rho_A)} }{\cZ_{AB}(\beta)} \;,
\end{eqnarray}
i.e.\ typical states with respect to the above canonical measure. In this context we have noticed that the system undergoes two main phase transitions (for the stable solution), related to different distributions of the eigenvalues $\Lambda_{A}$ of the typical states: a second order phase transition, mentioned above, associated to a ${\mathbb Z}_2$ symmetry breaking, and related to the vanishing of some eigenvalues of $\rho_A$, followed by a first order phase transition, associated to the evaporation of the largest eigenvalue from the sea of the others. By tuning the inverse temperature (or equivalently by deforming the Haar measure with the non uniform weight $e^{-\beta \pi_{AB}(\rho_A)}/\cZ_{AB}(\beta)$), we have determined the statistical behavior of the system at different levels of entanglement. As noticed in Sec.~\ref{sec:momentipureCaseCap3}, this is equivalent to determine all the moments of the purity at different values of $\beta$. 

For $x\neq 1$, the integration over the unitary group $\mathcal{U}(L)$ in Eq.~(\ref{eq:partition function unitary rotated})
 does not factorize, making the computation of the partition function far more complicated than for the case of pure states~(\ref{eq:pure1}). 
 The only notable exception is the case of maximally mixed states (i.e. $x = 1/L$), when the Dirac delta of 
 Eq.~(\ref{eq:partition function unitary rotated}) selects a unique diagonal matrix $\Lambda_X$ (the totally mixed state of $X$).
 This makes the exponent equal to  $\mathrm{e}^{-\beta/L}$ for all $U_X$ and yields the following exact expression 
\begin{eqnarray}\label{ffhh11}
\cZ_A(1/L,\beta) = \mathrm{e}^{-\beta/L}\;.
\end{eqnarray}
However, even if for $x\neq 1,1/L$ the situation is much more complicated, for small $\beta$, the evaluation of  the moments ${\cal M}^{A}_n(x, \beta)$ associated with $\cZ_A(x,\beta)$  admits exact analytical treatment, as we will show in the next two chapters. 
The moments of the purity for a generic set of states $\mathfrak{S}_{x}({\cal H}_X)$ are
formally defined as
\begin{equation}
 {\cal M}^{A}_n(x, \beta)=  \int d \mu_{x,\beta}(\rho) \; {\pi_A}^n(\rho) = \label{defma}
 \frac{(-1)^n}{\cZ_A(x,\beta) } \;  \frac{\partial^n  \cZ_A(x,\beta)}{\partial \beta^n } \;,
 \end{equation} 
and represent the average value of ${\pi_A}^n(\rho)$ over the canonical measure
\begin{eqnarray}\label{eq:mubeta}
d \mu_{x,\beta}(\rho) = d \mu_x(\rho)\; \frac{  \e ^{-\beta \;  \pi_A(\rho)} }{\cZ_A(x,\beta)} \;,
\end{eqnarray} 
with $d \mu_x(\rho)$ as in Eq.~(\ref{defdmu}). 
For pure states ($x=1$) they coincide with the moments ${\cal M}_n(\beta)$ defined in Eq.~(\ref{ffd1pure}). 
In the case of a
 totally mixed state  ($x=1/L$) Eq.~(\ref{ffhh11}) yields instead values which are independent from the
temperature $\beta$, namely 
\begin{eqnarray} \label{kkll}
 {\cal M}^{A}_n(1/L, \beta) =  {\cal M}^{A}_n(1/L, 0) = {N}^{-n}.
\end{eqnarray}
For intermediate values of $x$ and for sufficiently small $\beta$ we can write  
\begin{equation}
\quad {\cal M}_n^A(x,\beta)  \sim
 {\cal M}_n^A(x,0) - \beta \; [  {\cal M}^A_{n+1}(x,0) - {\cal M}^A_1(x,0)\; {\cal M}^A_{n}(x,0)] 
\;, \quad \beta\to 0 ,
\label{exp333}
\end{equation} 
which has been obtained by 
expanding Eq.~(\ref{defma}) up to the first order contribution in $\beta$. Incidentally, 
notice that in agreement with Eq.~(\ref{kkll}),  the $\beta$-corrections of Eq.~(\ref{exp333}) vanish when $x=1/L$. 
The above expression shows that, at least in the high  temperature regime, we can focus
on the unbiased moments  $ {\cal M}^{A}_n(x, 0)$. 
In the remaining part of the thesis we will compute the moments of the purity and infer, at the end of Chap. $\ref{chap9}$, the asymptotic behavior of the average local purity in the mean field approximation.

\chapter{Gaussian approximation}\label{chap8}
\markboth{Gaussian approximation}{Gaussian approximation}
In this chapter we will focus on the first moment of the local purity $\pi_{A}(\psi)$ for a generic state $\rho \in \mathfrak{S}_{x}({\cal H}_X)$. In particular, we will consider two distinct approaches based on the Gaussian approximation for the elements of the unitary matrices belonging to $\mathcal{U}(L)$ and $\mathcal{U}(L^2)$, where $L$ is the dimension of the global system.

In Sec.~\ref{sec:dc} we will use the fact that the integral over the eigenvalues of the density operator $\rho$, with purity $x$, and the integral over the unitary group associated to its eigenvectors can be evaluated independently, the former by introducing the empirical density distribution of the eigenvalues, the latter by approximating the elements of the unitary matrices acting on $\mathcal{H}_X$ as independent and identical complex Gaussian random variables. 
In Sec.~\ref{sec:mc}, we will follow a different strategy  which applies the same Gaussian approximation to the enlarged Hilbert space $\mathcal{H}_{X X'}$. Both approaches will lead to the evaluation of the first moment of the local purity at the order $\Ord{1/L}$.

\section{Divide and conquer}\label{sec:dc}
The approach we will discuss in this section for the computation of the first moment ${\cal M}_1^A(x,0)$ of the local purity $\pi_A(\psi)$ is based on the possibility to perform separately the integral over the unitary group and  the integral over the spectrum.
Indeed, let us consider the expression (\ref{eq:partition function unitary rotated}) of the partition function. It is worth introducing 
the spectral decomposition of $\Lambda_X$, i.e. 
\begin{equation}
\Lambda_X=\sum_{1\leq i \leq L} \lambda_i P_i, \quad \lambda_i \in [0,1]
\end{equation}
where  $P_i=|i\rangle_X\langle i|$ with $1 \leq i \leq L$  are the projections over the reference basis
$\{\ket{i}_X\}\subset \mathcal{H}_X$, that is $P_i=(\delta_{k,i} \delta_{i,l})$, $1\leq k, l \leq L$.  By defining the reduced density matrix $\rho_{Ai}=\rho_{Ai}(U_X)=\mathrm{Tr}_B(U_XP_i {U_X}^\dagger)$, associated to each eigenstate $U_X\ket{i}_X$ of $\rho$, the partition function becomes
\begin{equation}
 \mathcal{Z}_A(x,\beta)
 = C_x\int d\mu_H(U_X) \int d\sigma(\Lambda_{X})
 \delta\Big(x-\sum_{1 \leq i \leq L} \lambda_i^2\Big)
 \mathrm{e}^{-\beta\sum_{1\leq i,j \leq L}\lambda_i\lambda_j
 \pi_{ij}(U_X)}\;,
\label{eq:main}
 \end{equation}
 where $\pi_{ij}(U_X)= \mathrm{Tr}_A(\rho_{Ai}\rho_{Aj})$.
 As already observed the explicit computation of  $\mathcal{Z}_A(x, 0)$  of Eq.~(\ref{eq:main})
 involves a nontrivial integration over the unitary group~\cite{planar}. We will consider here the first
 order moment~(\ref{defma}). We get
  \begin{eqnarray}
{\cal M}_1^A(x,0)&=&\sum_{1 \leq i, j \leq L} \left[\int
d\sigma(\Lambda_{X})\delta\Big(x-\sum_{1 \leq k \leq L}
 \lambda_k^2\Big)\lambda_i\lambda_j\right]\left[\int
d\mu_H(U_X)\pi_{ij}(U_X)\right]\nonumber\\
 &=&\sum_{1 \leq i, j \leq L}\langle\lambda_i\lambda_j\rangle
 \langle \pi_{ij}(U_{X})\rangle\;,
 \label{eq:mean}
 \end{eqnarray}
 where $\langle\lambda_i\lambda_j\rangle$ and 
 $\langle \pi_{ij}(U_X)\rangle$ stands for two independent averages (the first over the eigenvalues and the second over the unitary matrices $U_X$). 
We will separately compute  these quantities 
  in Sec.~\ref{sec:Average purity_eigenvalues} and~\ref{sec:Average purity_unitaryGroup}, respectively, and then compute the first order moment of the purity ${\cal M}_1^A(x,0)$  up to order $\Ord{1/L}$.

\subsection{Average over the eigenvalues}\label{sec:Average purity_eigenvalues}
In order to evaluate the average over the eigenvalues $\left\langle \lambda_i \lambda_j\right\rangle$, ${1 \leq i, j \leq L}$, we will introduce the empirical density function of the eigenvalues $p(\lambda)$ and the joint density function $p(\lambda,\lambda' )$.
Given a typical realization of eigenvalues, the \textit{empirical density function} is defined as
\begin{equation}
p(\lambda)=\frac{1}{L} \sum_{1\leq i\leq L} \delta(\lambda-\lambda_i), \qquad  \lambda \in [0,1].
\label{eq:empiricalp}
\end{equation}
The above definition enables to encode the information on the eigenvalues $\lambda_{i}$, $1\leq i \leq L$, in the probability distribution $p(\lambda)$ on the interval $[0,1]$. The empirical density function is thus correctly normalized 
\begin{equation}
\int_0^1 d\lambda \, p(\lambda) =1.
\end{equation}
We also have 
\begin{equation}
\avg{\lambda^\alpha}\simeq\int_0^1 d\lambda \, \lambda^ \alpha \, p(\lambda) = \frac{1}{L} \sum_{1\leq i \leq L} \lambda_i^\alpha, 
\end{equation}
where, in fact, in order to compute $\avg{\lambda^\alpha}$ we should take the average of (\ref{eq:empiricalp}) over the distribution of the eigenvalues (in particular we get $\left\langle \lambda \right\rangle= 1/L$). We will show that the contribution of fluctuations is of higher order in $1/L$.
 If the global system is in a generic mixed state, with purity $\Tr \rho^2=x\in[1/L, 1]$, the distribution of the eigenvalues must satisfy the constraint
\begin{equation}\label{eq:purity}
\sum_{1\leq i \leq L} \lambda_i^2 = x.
\end{equation}
Since the purity is of $\Ord{1}$, the above condition can be satisfied only if one or more eigenvalues are of $\Ord{1}$.
In what follows we will first consider the simplest scenario with only one dominating eigenvalue, and then we will generalize the analysis to the case of more than one eigenvalue of order $\Ord{1}$, proving the equivalence of the two approaches.
\subsubsection{One eigenvalue of $\Ord{1}$}
Let assume to have one eigenvalue of $\Ord{1}$, detached from the sea of the first $L-1$ eigenvalues of $\Ord{1/L}$,
namely
\begin{equation}\label{eq:onedetachedEigenvalues}
\lambda_L=\sqrt{x}, \quad \sum_{1\leq i \leq L-1} \lambda_i=1-\sqrt{x},
\end{equation} 
with $\lambda_i=\Ord{1/L}$ for $i <L$. The constraint (\ref{eq:purity}) is satisfied up to order $\Ord{1/L}$
\begin{equation}
\sum_{1\leq i \leq L} \lambda_i^2 = x + \Ord{\frac{1}{L}}.
\end{equation}
In this hypothesis, the distribution of the eigenvalues (\ref{eq:empiricalp}) splits as
\begin{equation}\label{eq:OneEmpiricalp}
p(\lambda)= \frac{1}{L} \, \delta\left(\lambda-\sqrt{x}\right) + \left(1-\frac{1}{L}\right) p_{\mathrm{sea}} (\lambda) ,
\end{equation}
where $p_{\mathrm{sea}}(\lambda)$ is the normalized probability density of the first $L-1$ eigenvalues 
\begin{equation}
p_{\mathrm{sea}} (\lambda)= \frac{1}{L-1} \sum_{1 \leq i\leq L-1} \delta(\lambda-\lambda_i), \qquad \int_0^1 d\lambda \, p_{\mathrm{sea}}(\lambda) =1.
\label{eq:psea}
\end{equation}
Its first two moments are given by
\begin{eqnarray}
 \label{eq:lambdasea}\left\langle \lambda \right\rangle_{\mathrm{sea}}&=&\int_0^1 d\lambda \, \lambda\, p_{\mathrm{sea}}(\lambda) = \frac{1-\sqrt{x}}{L-1},
\\ 
\label{eq:lambda2sea}\left\langle\lambda^2  \right\rangle_{\mathrm{sea}}&=&\int_0^1 d\lambda\,  \lambda^2\, p_{\mathrm{sea}}(\lambda) = \frac{1}{L-1} \sum_{1 \leq i \leq L-1} \lambda_i^2 = \Ord{\frac{1}{L^2}}.
\end{eqnarray}
As before,  $\avg{\lambda^2}_{\mathrm{sea}}$ has been computed by neglecting the average over the distribution of the eigenvalues. This choice will be justified at the end of this subsection.
In order to evaluate the average $\left\langle \lambda_i \lambda_j\right\rangle$, $1\leq i,j \leq L$ we need to introduce the joint empirical density function, which similarly to Eq. (\ref{eq:empiricalp}) can be written as
\begin{equation}
p(\lambda,\lambda')= \frac{1}{L(L-1)} {\sum_{1\leq i,j\leq L}}^{\!\!\!\!\!\prime} \,\,\,\, \delta(\lambda-\lambda_i)\, \delta(\lambda-\lambda_j),
\end{equation}
the prime standing for $i\neq j$. Under the above hypothesis the joint density probability splits as
\begin{equation}
 p(\lambda,\lambda')= \frac{1}{L} \delta(\lambda-\sqrt{x})\,  p_{\mathrm{sea}} (\lambda') +
 \frac{1}{L} \delta(\lambda'-\sqrt{x})\, p_{\mathrm{sea}} (\lambda) +
 \left(1-\frac{2}{L}\right)\, p_{\mathrm{sea}} (\lambda,\lambda') ,
\end{equation}
where we have introduced the joint probability density of the sea: 
\begin{equation}\label{eq:jointProbabilityDensityOfTheSea}
p_{\mathrm{sea}} (\lambda,\lambda')= \frac{1}{(L-1)(L-2)} {\sum_{1\leq i,j\leq L-1}}^{\!\!\!\!\!\!\!\!\prime} \,\,\,\, \delta(\lambda-\lambda_i)\, \delta(\lambda'-\lambda_j).
\end{equation}
It easy  to show that the above definition can be written as
\begin{eqnarray}
p_{\mathrm{sea}} (\lambda,\lambda')&=& p_{\mathrm{sea}} (\lambda)\, p_{\mathrm{sea}} (\lambda') -\frac{1}{L-2}\, p_{\mathrm{sea}}\, (\lambda)\left[\delta(\lambda-\lambda')-p_{\mathrm{sea}} (\lambda')\right].
\end{eqnarray}
Notice that $p_{\mathrm{sea}}(\lambda)$ is the marginal probability of the above joint density function, that is
\begin{equation}
p_{\mathrm{sea}}(\lambda)=\int_0^1 d\lambda' \, p_{\mathrm{sea}}(\lambda,\lambda').\\
\end{equation}
Furthermore, given $\lambda$ and $\lambda'$ belonging to the sea of the first $L-1$ eigenvalues, we have
\begin{eqnarray}
\left\langle \lambda \lambda' \right\rangle_{\mathrm{sea}} &=& \left\langle \lambda \right\rangle_{\mathrm{sea}}^2 - \frac{1}{L-2} \left( \left\langle \lambda^2 \right\rangle_{\mathrm{sea}} - \left\langle \lambda \right\rangle_{\mathrm{sea}}^2\right)
\end{eqnarray}
and from Eqs. (\ref{eq:lambdasea}) and (\ref{eq:lambda2sea}) we get
\begin{eqnarray}
\left\langle \lambda \lambda' \right\rangle_{\mathrm{sea}}& = & \frac{(1-\sqrt{x})^2}{(L-1)(L-2)} -  \frac{1}{L-2} \left\langle \lambda^2 \right\rangle_{\mathrm{sea}}.
\label{eq:lambda1lambda2sea}
\end{eqnarray}
We are now ready to compute the explicit expression for $\avg{\lambda^2}$ and $\avg{\lambda\lambda'}$ (with $\lambda \neq \lambda'$):
\begin{eqnarray}
\langle\lambda^2\rangle&=&\frac{x}{L}
+\left(1-\frac{1}{L}\right)\langle\lambda^2\rangle_ {\mathrm{sea}},
\label{lambda^2}\\
\langle\lambda \lambda' \rangle&=& \frac{2}{L} \sqrt{x}
\langle\lambda\rangle_{\mathrm{sea}}+\left(1-\frac{2}{L}\right)\langle\lambda \lambda'  \rangle_ {\mathrm{sea}}.
\label{lambda_1lambda_2}
\end{eqnarray}
In the limit of large values of $L$, we get
\begin{eqnarray}
\langle\lambda^2\rangle&=&\frac{x}{L}  + \Ord{\frac{1}{L^2}},
\label{mediaAutovaloriuguali}\\
\langle\lambda \lambda' \rangle&=&\frac{\left(1-\sqrt{x} \right)^2}{L(L-1)}+ \Ord{\frac{1}{L^3}}\label{mediaAutovaloridiversi} .
\end{eqnarray}
Notice that the average values computed above show fluctuations of $\Ord{1/L^2}$ and $\Ord{1/L^3}$.
We will soon show that this result  does not depend on our hypothesis on the distribution of the eigenvalues, and can be generalized to the case of more eigenvalues of $\Ord{1}$.

Before considering this point, let us compute the density distribution of the sea of eigenvalues in the thermodynamic limit.
Similarly to what seen in Sec.~\ref{natural scaling}
we can introduce the natural scaling for the sea of eigenvalues 
\begin{equation}\label{eq:naturalscalingbig}
\lambda_i = (1-\sqrt{x}) \frac{\tilde{\lambda}(t_i )}{L-1}, \qquad 0<
t_i=\frac{i}{L-1}\leq 1, \quad \forall \ i=1, \ldots, L-1
\end{equation}
and define, in the thermodynamic limit, the density function
\begin{equation}
\rho_{\mathrm{sea}}(\tilde{\lambda})= \int_0^1 d t \;
\delta\left(\tilde{\lambda}-\tilde{\lambda}(t)\right). \label{eq:rhodefb}
\end{equation}
The distribution $\rho_{\mathrm{sea}}(\tilde{\lambda})$ is normalized and satisfies the constraint (\ref{eq:onedetachedEigenvalues}):
\begin{eqnarray}
\label{eq:normconstr} \int_0^{a} d\tilde{\lambda} \;
\rho_{\mathrm{sea}}(\tilde{\lambda})&=&1\label{eq:normalization}\\
\int_0^{a} d\tilde{\lambda} \; \tilde{\lambda}\,
\rho_{\mathrm{sea}}(\tilde{\lambda}) &=& 1-\sqrt{x},\label{eq:constraint}
\end{eqnarray}
being $[0, a]$  the domain of the scaled eigenvalues and $a>0$. In analogy to the Wishart distribution found at $\beta=0$ for the eigenvalues $\rho_A$ in Chap.~\ref{chap3}, we set
\begin{equation}\label{eq:varro1}
\rho_{\mathrm{sea}}(\tilde{\lambda})=C \sqrt{\frac{a-\tilde{\lambda}}{\tilde{\lambda}}},
\end{equation}
where $a$ and $C$ are determined by conditions (\ref{eq:normalization}) and (\ref{eq:constraint}):
\begin{eqnarray}\frac{a C}{2}\pi&=&1\\\frac{a^2 C}{8} \pi &=&1-\sqrt{x}.
\end{eqnarray}
Therefore, expression (\ref{eq:varro1}) becomes:
\begin{equation}
\rho_{\mathrm{sea}}(\tilde{\lambda})=\frac{1}{2\pi \left(1-\sqrt{x}\right)}
\sqrt{\frac{4\left(1-\sqrt{x}\right)-\tilde{\lambda}}{\tilde{\lambda}}}, \quad \tilde{\lambda} \in [0, 4(1-\sqrt{x})]
\end{equation}
being $a=4(1-\sqrt{x})$ and $C=1/2 \pi \left(1-\sqrt{x}\right)$. Notice that when $x=0$, $\rho_{\mathrm{sea}}$ reduces to the
unbiased average density at $\beta=0$ found for the distribution of the eigenvalues of the reduce density matrix $\rho_A$ in Chap.~\ref{chap3}.
By comparing (\ref{eq:psea}) and (\ref{eq:rhodefb}) we get
\begin{equation}
p_{\mathrm{sea}}(\lambda)= \frac{L-1}{1-\sqrt{x}} \; \rho_{\mathrm{sea}}( \tilde{\lambda})=\frac{L-1}{1-\sqrt{x}}\; \rho_{\mathrm{sea}}\left(\lambda \; \frac{L-1}{1-\sqrt{x}}\right),
\end{equation}
from which all moments can be easily evaluated
\begin{eqnarray}
\left\langle \lambda^n \right\rangle_{\mathrm{sea}} &=& \int_0^1 d\lambda \,
\lambda^n  p_{\mathrm{sea}}(\lambda) =\left(\frac{1-\sqrt{x}}{L-1}\right)^n \int_0^{4(1-\sqrt{x})} dy \, y^n \rho_{\mathrm{sea}}(y)
\nonumber\\
& = & \frac{\left(1-\sqrt{x}\right)^{2 n}}{(L-1)^n}\cdot \frac{ 4^n\Gamma\left(n+\frac{1}{2}\right)}{\sqrt{\pi}(n+1)!}.
\end{eqnarray}
In particular, $\left\langle \lambda^2 \right\rangle_{\mathrm{sea}}=
2\left(1-\sqrt{x}\right)^4 /(L-1)^2$, in agreement with (\ref{eq:lambda2sea}), that is $\avg{\lambda^2}_{\mathrm{sea}}=\Ord{1/L^2}$.
\subsubsection{$k$ eigenvalues of $\Ord{1}$}

In general, for large $L$, a density matrix $\rho$ with purity $x$ can have finite number $k$ of eigenvalues of order $\Ord{1}$, say $\lambda_i=\sqrt{x_i}$ for $L-k < i\leq L$, with
\begin{equation}\label{eq:puritybis}
\sum_{L-k< i\leq L} x_i=x,
\end{equation}
and the first $L-k$ eigenvalues 
$\lambda_i=\Ord{1/L}$ for $1\leq i\leq L-k$.
Thus the constraint on the purity of the global system is satisfied up to order $\Ord{1/L}$, that is
\begin{equation}
\sum_{1\leq i \leq L} \lambda_i^2 = 
x+ \Ord{\frac{1}{L}}.
\end{equation}
The empirical density (\ref{eq:OneEmpiricalp}) generalizes to 
\begin{equation}\label{eq:plambdabis}
p(\lambda)= \frac{1}{L} \sum_{L-k < i\leq L}\delta\left(\lambda-\sqrt{x_i}\right) + \left(1-\frac{k}{L}\right) p_{\mathrm{sea}} (\lambda) ,
\end{equation}
with the empirical density for the sea of the first $L-k$ eigenvalues of $\Ord{1/L}$ given by
\begin{equation}
p_{\mathrm{sea}} (\lambda)= \frac{1}{L-k} \sum_{1 \leq i \leq L-k} \delta(\lambda-\lambda_i).
\label{eq:pseak}
\end{equation}
Thus in this set up the joint probability density shows three main contributions 
\begin{eqnarray}\label{eq:lambda1lambda2}
p(\lambda,\lambda')&=& \frac{1}{L(L-1)} {\sum_{L-k < i,j\leq L}}^{\!\!\!\!\!\!\!\!\prime} \delta\left(\lambda-\sqrt{x_i}\right)\, \delta\left(\lambda'-\sqrt{x_j}\right)
\nonumber\\
& & +\frac{L-k}{L(L-1)} \sum_{L-k < i\leq L} \left[\delta\left(\lambda-\sqrt{x_i}\right)\,  p_{\mathrm{sea}} (\lambda') +
\delta\left(\lambda'-\sqrt{x_i}\right)\,  p_{\mathrm{sea}} (\lambda)\right]
\nonumber\\
& & + \frac{(L-k)(L-k-1)}{L(L-1)}\, p_{\mathrm{sea}} (\lambda,\lambda') ,
\end{eqnarray}
being $p_{\mathrm{sea}} (\lambda,\lambda')$ the joint probability density for the sea, analog  to (\ref{eq:jointProbabilityDensityOfTheSea}),
\begin{equation}
p_{\mathrm{sea}} (\lambda,\lambda')= \frac{1}{(L-k)(L-k-1)} {\sum_{1\leq i,j \leq L- k}}^{\!\!\!\!\!\!\!\!\prime} \delta(\lambda-\lambda_i)\, \delta(\lambda'-\lambda_j).
\end{equation}
We can now compute the average product of eigenvalues $\avg{\lambda_i \lambda_j}$, $1\leq i,j \leq L$. For $\lambda_i = \lambda_j$, from Eq. (\ref{eq:plambdabis}) we have
\begin{eqnarray}\label{eq:lambdaquadroBis}
\langle\lambda^2\rangle&=&\frac{1}{L-k} \sum_{1\leq i \leq L-k} \Ord{\frac{1}{L^2}}+\frac{1}{L}\ \sum_{L-k < i \leq L} x_i = \frac{x}{L} + \Ord{\frac{1}{L^2}}.
\end{eqnarray}
On the other hand for $\lambda_i \neq \lambda_j$, from Eq. (\ref{eq:lambda1lambda2}) we have
\begin{eqnarray}
\langle\lambda \lambda' \rangle&=&
\frac{1}{L(L-1)} {\sum_{L-k < i,j\leq L}}^{\!\!\!\!\!\!\!\!\prime} \sqrt{x_i x_j} +  \frac{2}{L(L-1)} \sum_{L-k <  i \leq L} \sqrt{x_i} 
\left(1-\sum_{L-k <  j \leq L}\sqrt{x_j} \right)
\nonumber\\
& & + \frac{(L-k-1)}{L(L-1)(L-k)} \left(1-\sum_{L-k < i\leq L} \sqrt{x_i} \right)^2
\nonumber\\
& \simeq & \frac{1}{L(L-1)} \Bigg[ \ {\sum_{L-k < i,j\leq L}}^{\!\!\!\!\!\!\!\!\prime} \sqrt{x_i x_j}+ 2 \sum_{L-k < i\leq L} \sqrt{x_i}  \left(1-\sum_{L-k < j\leq L} \sqrt{x_j} \right)\nonumber\\&&\qquad \qquad+\left(1-\sum_{L-k < i\leq L} \sqrt{x_i} \right)^2\Bigg]\nonumber\\
\nonumber\\
& =& \frac{1}{L(L-1)} \left( 1-\sum_{L-k < i\leq L} x_i\right)
\end{eqnarray}
that recalling condition (\ref{eq:puritybis}) becomes:
\begin{equation}\label{eq:lambdaplambdaBis}
\langle\lambda \lambda' \rangle= \frac{1-x}{L(L-1)}.
\end{equation}
By comparing Eqs.~(\ref{mediaAutovaloriuguali}) and (\ref{mediaAutovaloridiversi}) with  Eqs. (\ref{eq:lambdaquadroBis}) and (\ref{eq:lambdaplambdaBis}), we conclude that the average over the eigenvalues does not depend on our assumption on the (finite) number of eigenvalues of $\Ord{1}$. 
\subsection{Unitary group average}\label{sec:Average purity_unitaryGroup}
The next step, in order to compute the first moment of the purity (\ref{eq:mean}), is the evaluation of the average over the unitary group
\begin{equation}\label{eq:unitaryAverage}
\langle \pi_{i j}(U_{X})\rangle=\int d\mu_H(U_X)\pi_{i j}(U_X),
\end{equation}
where $\pi_{i j}(U_X)= \mathrm{Tr}(\rho_{Ai}\rho_{Aj})$, and $\rho_{Ak}=\mathrm{Tr}_B(U_X\ket{k}_X\bra{k}{U_X}^\dagger)$,
$\ket{k}_X$  being the $k$-th  vector of the reference basis, with $1 \leq i, j , k \leq L$.

Let us, first of all, choose a convenient representation of the tensor product structure of  $\mathcal{H}_X=\mathcal{H}_A\otimes\mathcal{H}_B$, i.e.\
$\mathbb{C}^L= \mathbb{C}^{N}\otimes \mathbb{C}^{M}$, with $L= N M$. Given $u_A=(u_n)_{1\leq n\leq N} \in  \mathbb{C}^{N}$ and $v_B=(v_m)_{1\leq m\leq M} \in  \mathbb{C}^{M}$, we define
the tensor product of the two vectors as
\begin{equation}
w_X= u_A\otimes v_B = (w_\ell)_{1\leq \ell\leq L}\in  \mathbb{C}^{L}
\end{equation}
where  
\begin{eqnarray}
w_\ell = u_n v_m, \quad \begin{cases} \ell=M(n-1) + m, \\ \\
n = \left[\frac{\ell-1}{M}\right]+1, \\ \\ m= \ell-M(n-1),\\ 
\end{cases}
\end{eqnarray}
$[x]$ being the integer part of $x$. Therefore, for any $w\in\mathbb{C}^L$, the representation of the partial trace 
\begin{equation}
\rho_A(w) = \mathrm{Tr}_B \left(|w\rangle_X\langle w |\right)=(\rho_{A,ij})_{1\leq i,j\leq N},
\end{equation}
reads
\begin{equation}
\left(\rho_A(w) \right)_{l,m}= \sum_{1\leq k\leq M} w_{M(l-1)+k}\; \bar{w}_{M (m-1) + k}\ , \quad \forall \  l, m     \in \{1, \ldots, N\}.
\end{equation}
In order to write the explicit expression for $\pi_{ij}(U_X)$, we have to choose the computational orthonormal basis for the system. Let $U=(u_{k l})_{1\leq k,l \leq L}$ be an element of the unitary group $\mathcal{U}(L)$.  Its columns $u_{\bullet l}=(u_{kl})_{1\leq k \leq L}$,
with $1\leq l\leq L$, form  an orthonormal basis of $\mathbb{C}^L$, i.e.\ $\langle u_{\bullet l} | u_{\bullet m}\rangle=\sum_{1\leq k \leq L}\bar{u}_{k l}  u_{k m} = \delta_{l,m}$, as follows from $U^\dagger U = 1$. Thus, we can write 
\begin{equation}
\pi_{ij}(U)= \mathrm{Tr}_A\left(\rho_A(u_{\bullet i})\rho_A(u_{\bullet j})\right), \quad   \forall \ i, j \in \{1, \ldots, L\}
\end{equation}
where the elements of the reduced density matrix $\rho_A(u_{\bullet i})$, $1 \leq i \leq N$, are given by 
\begin{equation}
(\rho_A(u_{\bullet i}))_{m n}=  \sum_{1\leq k\leq M} u_{M(m-1)+k,i}\; \bar{u}_{M (n-1) + k,i} \quad \forall \ m, n \in \{1, \ldots, N\}.
\end{equation}
From the last two relations we can finally write the explicit expression for $\pi_{ij}(U_X)$
\begin{eqnarray}
 \pi_{ij}(U)
=\sum_{1\leq k,l\leq M} \sum_{1\leq m,n \leq N} u_{M(m-1)+k,i}\; \bar{u}_{M (n-1) + k,i}
u_{M(n-1)+l,j}\; \bar{u}_{M (m-1) + l,j}\ , \nonumber \\
\qquad \qquad \quad \forall \ i, j \in \{1,\ldots N\}.
\end{eqnarray} 
Let us compute the average (\ref{eq:unitaryAverage}) over the unitary group. If $U \in \mathcal{U}(L)$ is sampled according to the Haar measure, we can parametrize its columns as $u_{\bullet j}=(r_k e^{i \phi_k})_{1\leq k\leq L}$ for $1 \leq j \leq L$: the phases $\phi_k$ are independent identically distributed (i.i.d.) random variables with a uniform probability density on $[-\pi,\pi]$, and the squared moduli $r_k^2$ are uniformly distributed on the simplex $\Lambda_L=\{(r_k^2)_{1\leq k\leq L}\,|\, 0\leq r_k^2\leq 1, \sum r_k^2 =1 \}$~\cite{Zyczkowski}. Since the volume of the simplex is $|\Lambda_L|=1/(L-1)!$, the joint probability density of the square moduli, on $\mathbb{R}_+^L$, reads 
\begin{equation}\label{eq:pn(r2)}
p_L(r_1^2,r_2^2,\dots,r_L^2)= (L-1)! \, \delta\left(1-\sum_{1\leq k \leq L} r_k^2\right),
\end{equation}
and therefore 
\begin{equation}
p_L(r_1,r_2,\dots,r_L)=  2^L (L-1)! \, \delta\left(1-\sum_{1\leq k \leq L} r_k^2\right) \prod_{1\leq k \leq L} r_k.
\label{eq:pn(r)}
\end{equation}
In appendix~\ref{sec:generate uniform random variables on the simplex} we show how $L$ random variables can be generated with a uniform distribution on the simplex $\Lambda_L$.
\\If we integrate out the last $L-K$ variables, $1\leq K < L$, by defining 
\begin{equation}
y_i = r_{K+i}^{2}, \quad\qquad \forall \ i \in \{1, \ldots, L-K \}
\end{equation}
we find
\begin{eqnarray}
 p_L(r_1,r_2,\dots,r_K)&=& 
 2^K
(L-1)! \prod_{1\leq j \leq K} r_j \int_0^1 d^{L-K} y \ \delta\Big(1-\sum_{\substack{1\leq i \leq K }}r_i^2-\sum_{\substack {1\leq j \leq L-K}} y_i\Big)\nonumber\\
&=& 2^K(L-1)! \prod_{1\leq j \leq K} r_j \int_0^1 d y_1 \int_0^{1-y_1} d y_2\int_0^{1-y_1-y_2} d y_3 \ldots  \nonumber\\
&& \times\int_0^{1-y_1\ldots -y_{L-K-3}} d y_{L-K-2} \left(1-\sum_{1\leq i \leq K }r_i^2-\sum_{\ 1\leq j \leq L-K-2} y_j\right)\nonumber\\
\end{eqnarray}
and finally 
\begin{equation}
 p_L(r_1,r_2,\dots,r_K)=  2^K
\frac{(L-1)!}{(L-K-1)!} \,
\left(1-\sum_ {1\leq i\leq K} r_i^2\right)^{L-K-1} \prod_{1\leq j \leq K} r_j.
\end{equation}
In particular for $K=1$ we get the probability density of the amplitude of an arbitrary element of $U$ 
\begin{equation}
p_L(r)=  2 (L-1) r (1-r^2)^{L-2}.
\label{eq:pn1}
\end{equation}
The $n$-th moments of this density function are given by:
\begin{equation}\label{eq:momnesimo}
\avg{r^{n}}=(L-1) B \left(\frac{n}{2}+1,L-1\right)=(L-1)! \frac{\frac{n}{2}\Gamma\left(\frac{n}{2}\right)}{\Gamma\left(\frac{n}{2}+L\right)}
\end{equation} 
where $B(x,y)$ is the Beta function,
\begin{equation}
B(x,y)=\int_{0}^{1} dt \ t^{x-1}(1-t)^{y-1} =\frac{\Gamma(x)\Gamma(y)}{\Gamma(x+y)}.
\end{equation}
For large $L$ ($L\to\infty$) the probability density $p_{L}(r)$ becomes
\begin{equation}
p_L(r)\sim   2 L r e^{-L r^2}.
\end{equation}
It follows that in the thermodynamic limit we can approximate $U_X=(u_{ij})$ with $u_{ij}=r_{ij} e^{i \phi_{ij}}$, i.i.d. random variables with probability density functions
\begin{equation}
p(r_{ij})= 2 L r_{ij} e^{-L r_{ij}^2}, \qquad
p(\phi_{ij})=\frac{1}{2\pi},
\end{equation}
that is
\begin{equation}\label{Gaussianity}
p(u_{ij})= \frac{L}{\pi}  \exp(-L |u_{ij}|^2), \quad \forall \ i,j \in \{1, \ldots L\}.
\end{equation}
We then have that $u_{ij}$'s are i.i.d.\ complex Gaussian random variables with zero mean and variance $\langle |u_{ij}|^2\rangle=1/L$.
Notice that, in the above approximation, unitarity is satisfied at order $1/\sqrt{L}$. Indeed,
\begin{equation}
\left\langle\sum_{1\leq k \leq L} \bar{u}_{k l}  u_{k m}\right\rangle = \sum_{1\leq k \leq L} \left\langle \bar{u}_{k l}  u_{k m}\right\rangle=   \sum_{1\leq k \leq L}  \frac{ \delta_{l,m}}{L}  = \delta_{l,m} \ ,
\end{equation}
while
\begin{equation}\label{eq:fluctuations}
\left\langle \left|\sum_{1\leq k \leq L} \bar{u}_{k l}  u_{k m}- \delta_{l,m}\right|^2\right\rangle= \sum_{1\leq k,j \leq L} \left\langle\bar{u}_{k l}  u_{k m} u_{j l} \bar{u}_{j m}  \right\rangle- \delta_{l,m} =     \frac{1}{L}\  \quad \forall \ l,m \in \{1, \ldots L\}.
\end{equation}
The above quantity measures the additional fluctuations introduced by the Gaussian approximation, which enables us to easily evaluate the average
\begin{eqnarray}
 \left\langle \pi_{ij}\right\rangle
=\sum_{1\leq k,l\leq M} \sum_{1\leq m,n \leq N} \left\langle u_{M(m-1)+k,i}\; \bar{u}_{M (n-1) + k,i}
u_{M(n-1)+l,j}\; \bar{u}_{M (m-1) + l,j} \right\rangle, \nonumber\\  \qquad\qquad\quad \forall \ i,j \in \{1, \ldots L\}.
\end{eqnarray}
Indeed, by $\langle |u_{ij}|^2\rangle_{X}=1/L$ and $\langle |u_{ij}|^4\rangle_{X}=2/L^2$
it follows that, when $1\leq i\neq j\leq L$,
\begin{eqnarray}
& &\left\langle u_{M(m-1)+k,i}\; \bar{u}_{M (n-1) + k,i} u_{M(n-1)+l,j}\; \bar{u}_{M (m-1) + l,j} \right\rangle
\nonumber\\
& & \quad = \delta_{m,n} \left\langle |u_{M(m-1)+k,i}|^2\right\rangle\left\langle |u_{M(m-1)+l,j}|^2
\right\rangle= \frac{\delta_{m,n}}{L^2}
\end{eqnarray}
while, for $1\leq i= j\leq L$,
\begin{eqnarray}
 \qquad & &\left\langle u_{M(m-1)+k,i}\; \bar{u}_{M (n-1) + k,i} u_{M(n-1)+l,i}\; \bar{u}_{M (m-1) + l,i}
\right\rangle
\nonumber\\
 \qquad & & \quad = (\delta_{m,n} + \delta_{k,l})  \left\langle |u_{M(m-1)+k,i}|^2\right\rangle \left\langle |u_{M(n-1)+l,i}|^2
\right\rangle- \delta_{m,n} \delta_{k,l} \left\langle |u_{M(m-1)+k,i}|^4\right\rangle
\nonumber\\
 \qquad & & \quad = \frac{\delta_{m,n}+ \delta_{k,l}-2 \delta_{m,n} \delta_{k,l}}{L^2},\nonumber\\\nonumber\\&&\qquad\qquad\qquad\qquad\qquad \forall \ m,n \in \{1, \ldots N\} \quad \mbox{and} \quad \forall \ k,l \in \{1, \ldots M\}.\nonumber\\
\label{eq:i=j}
\end{eqnarray}
Incidentally, notice that (\ref{eq:i=j}) can be computed exactly by using  (\ref{eq:momnesimo}) and reads $(\delta_{m,n}+ \delta_{k,l})/L^2 - 2 \delta_{m,n} \delta_{k,l}/L(L+1)$.
Therefore,
\begin{eqnarray}
\left\langle \pi_{ij}(U_X)\right\rangle
&=& \delta_{i,j} \frac{N M^2+ N^2 M - 2 N M }{L^2} + (1-\delta_{i,j}) \frac{N M^2}{L^2} \nonumber\\
&=& \frac{1}{N} +  \frac{\delta_{i,j}}{M} + \Ord{\frac{1}{L}},\qquad \forall \ i,j \in \{1, \ldots L\}.
\label{eq:pi_ijmean}
\end{eqnarray}
When $N=M=\sqrt{L}$,  one gets $\left\langle \pi_{ij}\right\rangle=\langle\mathrm{Tr}(\rho_{A_i}\rho_{A_j})\rangle \sim (1+ \delta_{i,j})/N$.

\subsection{Average purity}\label{sec:averagepurity_finale}

From the eigenvalue averages (\ref{mediaAutovaloriuguali}) and (\ref{mediaAutovaloridiversi}), and  from the average (\ref{eq:pi_ijmean})
over the unitary group we can evaluate the average purity  at order $\Ord{1/L}$
\begin{eqnarray}
{\cal M}_1^A(x,0)&\simeq&
\sum_{1\leq i, j \leq L}\langle\lambda_i\lambda_j\rangle
 \langle \pi_{ij}(U_X)\rangle= \sum_{1\leq i, j \leq L}\langle\lambda_i\lambda_j\rangle\left(\frac{1}{N}+\frac{\delta_{i,j}}{M}\right)
\nonumber\\
&=& \frac{1}{N}  {\sum_{1\leq i, j \leq L}}^{\!\!\!\! \prime} \langle\lambda_i\lambda_j\rangle+ \left(\frac{1}{N}+\frac{1}{M}\right) \sum_{1\leq i \leq L}\langle\lambda_i^2\rangle
\nonumber\\ \label{eq:firstmixmonNM}
&=& \frac{1}{N} (1-x) +  \left(\frac{1}{N}+\frac{1}{M}\right) x = \frac{1}{N} + \frac{x}{M},
\label{eq:meanresult}
\end{eqnarray}
which for  $N=M=\sqrt{L}$ yields
\begin{eqnarray} \label{COnF}
 \qquad {\cal M}_1^A(x,0) \simeq
\frac{1}{N}(1+x).
\end{eqnarray}
Notice that for $x=1$ Eq. (\ref{eq:firstmixmonNM}) is in agreement with the thermodynamic behavior of the pure states distribution computed in~\cite{Scott2003,paper1} for the general case of unbalanced bipartions. Furthermore, Eq. (\ref{COnF}) yields ${\cal M}_1^A(1,0)=\avg{\pi_{AB}} =2/N$, according to the results of Chap.~\ref{chap3} for $N=M$ (see Fig.~\ref{fig:uvsbetapos} where $u= N \avg{\pi_{AB}}$), which can be also inferred from Eq. (\ref{eq:mompureLargeN}) for $n=1$.

\section{Multiply and conquer}\label{sec:larger_unitary_group} \label{sec:mc}
In this section we introduce an alternative approach for the computation of the cumulants of the partition function $\cZ_A(x,\beta)$
of Eq.~(\ref{eq:main}) that will allow us to determine the leading terms of the average purity $\cM_1^{A}(x,0)$.
We will then compute the first moment of the distribution of the local purity  and compare the result with the one obtained in the previous section.
In particular, we go back to expression (\ref{eq:ZUN2}) of the partition function which involves the integration over the larger unitary group $\mathcal{U}(L^2)$ acting on the enlarged purification space $\mathcal{H}_{X X' }=\mathcal{H}_X\otimes\mathcal{H}_{X'}$
\begin{eqnarray}\label{eq:ZUN2bis}
\mathcal{Z}_A(x, \beta)&=&C_x\int
d\mu_{H}(U_{X X'})\,
\delta\left(x-\Tr(\Tr_{X'}(\ket{\Psi}_{X X'}\bra{\Psi})^2\right)
\nonumber\\
& & \qquad\qquad\qquad \times
\mathrm{e}^{-\beta\; \Tr
\left(\Tr_B
(\Tr_{X'}(\ket{\Psi}_{X X'}\bra{\Psi}))\right)^2 }.
\end{eqnarray}
Here, $\ket{\Psi}_{X X'}= U_{X X'}\ket{\Psi_{0}}_{X X'}$, where $\ket{\Psi_{0}}_{X X'}$ is a reference state and $U_{XX'}$ an element of the unitary group $\mathcal{U}(L^2)$. If we choose the vector $(\delta_{i,1})_{1\leq i\leq L^2}\in\mathbb{C}^{L^2}$ as reference state, the partition function (\ref{eq:ZUN2bis}), according to the tensor product representation introduced in Sec.~\ref{sec:Average purity_unitaryGroup},  reads
\begin{equation}
\quad \mathcal{Z}_A(x, \beta)=C_{x}\int
d\mu_{H}(\tilde{u}_{\bullet 1})\,
\delta\left(x-\mathrm{Tr}(\mathrm{Tr}_{X'} |\tilde{u}_{\bullet 1}\rangle_{X X'} \langle\tilde{u}_{\bullet 1}| )^2\right)
\mathrm{e}^{-\beta \mathrm{Tr}
\left(\mathrm{Tr}_B\mathrm{Tr}_{X'} |\tilde{u}_{\bullet 1}\rangle_{X X'}\langle\tilde{u}_{\bullet 1}|\right)^2},
\label{eq:ZUn21}
\end{equation}
where
\begin{equation}
d\mu_{H}(\tilde{u}_{\bullet 1}) = \frac{1}{(2\pi)^{L^2}}\,   p_{L^2}(r_1,r_2,\dots,r_{L^2})\,
d^{L^2}\! \phi\; d^{L^2}\! r
\end{equation}
is the explicit expression of the Haar measure over the unitary group for the representation $(\tilde{u}_{i 1})=(r_i \mathrm{e}^{i \phi_i})$, $1 \leq i \leq L^2$, with the joint probability density  $p_{L^2}$ given by Eq. (\ref{eq:pn(r)}).
By introducing the simplified notation, $z_i=\tilde{u}_{i 1}$,  we rewrite the partition in the more compact form
\begin{equation}
\mathcal{Z}_A(x, \beta)=\int
d\mu_{H}(z)\,
\delta\left(x-\pi_{X}(z) \right)
\mathrm{e}^{-\beta \pi_A(z) },
\label{eq:ZUn2z}
\end{equation}
and the Haar measure over the unitary group reads
\begin{equation}
d\mu_H (z) = \delta\left(1-\sum_{1 \leq k \leq L^2 } |z_k|^2\right)\; \prod_{1 \leq i \leq L^2 } \frac{ dz_i \,d\bar{z}_i  }{\pi}.
\end{equation}
With the same technique introduced in the first part of this chapter, let us determine the explicit expression for the purity for $\pi_X(z)$ and $\pi_A(z)$. The matrix elements of the  density matrix $\rho$ according to the notation introduced in Sec.~\ref{sec:Average purity_unitaryGroup} are
\begin{equation}
\rho_{mn}= (\mathrm{Tr}_{X'} |\tilde{u}_{\bullet 1}\rangle \langle\tilde{u}_{\bullet 1}|)_{mn}=
 \sum_{1\leq k\leq L} z_{L(m-1)+k}\; \bar{z}_{L (n-1) + k}, \quad  \forall \ m, n \in \{1, \ldots, L\}.
\end{equation}
We then get that the purity of the global system, whose value is constrained to be $x$, is
\begin{equation}\label{eq:constraintx}
\pi_X(z)=\sum_{1\leq k,l,m,n\leq L} z_{L(m-1)+k}\; \bar{z}_{L(n-1) + k}\;
z_{L(n-1)+l}\; \bar{z}_{L (m-1) + l} = x,
\end{equation}
whereas for the purity of subsystem $A$ we get
\begin{eqnarray}\label{piAz}
 \qquad \pi_A(z)&=&\!\!\!\! \sum_{1\leq k,k'\leq L}  \sum_{1\leq i,j\leq N} \sum_{1\leq l,l' \leq M} z_{L(M(i-1)+l -1)+k}\; \bar{z}_{L(M(j-1)+l-1) + k}
\nonumber\\
\qquad &  &\phantom{\!\!\!\! \sum_{1\leq k,k'\leq L}  \sum_{1\leq i,j\leq N} \sum_{1\leq l,l' \leq M}}
\times  z_{L(M(j-1)+l' -1)+k'}\; \bar{z}_{L(M(i-1)+l'-1) + k'} . \nonumber\\
\label{eq:pAz}
\end{eqnarray}
Analogously to what seen in Sec.~\ref{sec:Average purity_unitaryGroup}, for large $L$ we will consider the $z_i$'s as i.i.d. complex Gaussian random variables, i.e.\
\begin{equation}
d\mu_H(z) = \prod_{1\leq i\leq L^2} \frac{L^2}{\pi}\exp(-L^2 |z_i|^2)\, dz_i \,d\bar{z}_i.
\end{equation}
The fluctuations due to this approximation scale as $\Ord{1/L^2}$ (see Eqs. (\ref{Gaussianity}) - (\ref{eq:fluctuations})). However, will show that the fluctuations of the average purity are of order $\Ord{1/L}$.   
We now have all the elements for obtaining an alternative -- and more direct -- derivation of  (\ref{eq:meanresult}).
\subsection{Average purity}
The average purity (\ref{eq:mean}) reads
\begin{equation}
{\cal M}_1^A(x,0)=\int
d\mu_{H}(z)\,
\delta\left(x-\pi_X(z) \right) \pi_A(z),
\end{equation}
and involves 4-point correlation functions the form
\begin{equation}
\left\langle z_{L(M(i-1)+l -1)+k}\, \bar{z}_{L(M(j-1)+l-1) + k} \,
  z_{L(M(j-1)+l' -1)+k'}\, \bar{z}_{L (M(i-1)+l'-1) + k'} \right\rangle,
\end{equation}
with $1\leq i,j\leq N$, $1\leq k,k'\leq L$, and $1\leq l,l' \leq M$. See Eq.~(\ref{eq:pAz}).
The two dominant contributions come from the contractions $\delta_{i,j}$ and $\delta_{l,l'}$ that involve sums over $N$ or $M$ terms in (\ref{eq:pAz}). The remaining contractions yield higher order corrections.
The first contraction reads 
\begin{equation}
 \delta_{i,j} \left\langle |z_{L(M(i-1)+l -1)+k}|^2 \right\rangle
 \left\langle |\bar{z}_{L (M(i-1)+l'-1) + k'}|^2 \right\rangle =
 \frac{\delta_{i,j}}{L^4} ,\quad \forall \ i ,j \in \{1,\ldots N \}
\end{equation}
hence the first leading term is given by 
\begin{equation}
\left\langle \pi_{A} \right\rangle_1 = \sum_{1\leq k,k'\leq L}  \sum_{1\leq i,j\leq N} \sum_{1\leq l,l' \leq M} \frac{\delta_{i,j}}{L^4} =\frac{L^2 N M^2}{L^4} = \frac{1}{N} .
\end{equation}
The second contraction reads
\begin{equation}
 \delta_{l,l'} \left\langle z_{L(m -1)+k}\, \bar{z}_{L(n-1) + k} \,
  z_{L(n-1)+k'}\, \bar{z}_{L(m-1) + k'} \right\rangle, \quad \forall \ l ,l' \in \{1,\ldots M \}
\end{equation}
with
\begin{equation}
m= M(i-1)+l, \qquad n=M(j-1)+l' \ .
\end{equation}
Note that the condition $l=l'$ is equivalent to $m_B = n_B$, where
\begin{eqnarray}
m_B= m - M \left[\frac{m-1}{M}\right], \quad n_B= n - M \left[\frac{n-1}{M}\right] .
\end{eqnarray}
Thus the other dominant contribution reads
\begin{equation}
 \left\langle \pi_{A} \right\rangle_2 = \sum_{1\leq k,k',m,n\leq L}
\delta_{m_B,n_B} \left\langle z_{L(m -1)+k}\, \bar{z}_{L (n-1) + k} \,
  z_{L(n-1)+k'}\, \bar{z}_{L(m-1) + k'} \right\rangle_X ,
\end{equation}
which, by virtue of (\ref{eq:constraintx}), yields
\begin{equation}
\left\langle \pi_{A} \right\rangle_2= \frac{x}{M}.
\end{equation}
On the other hand, if we require both the conditions $\delta_{i,j}$ and $\delta_{l,l'}$ for $1\leq i,j \leq N$ and $1\leq l,l' \leq M$ to be satisfied, we get:
\begin{equation}
 \sum_{1\leq n, k,k'\leq L} \left\langle z_{L(n -1)+k}\, \bar{z}_{L (n-1) + k} \,
  z_{L(n-1)+k'}\, \bar{z}_{L(m-1) + k'} \right\rangle_X = \frac{x}{L}=\Ord{\frac{1}{L}}.
\end{equation}
Other correlations between the above indices lead to higher order corrections. 
Summing up the results we obtain
\begin{equation}
{\cal M}_1^A(x,0) = \left\langle \pi_{A} \right\rangle_1+ \left\langle \pi_{A} \right\rangle_2 + \Ord{\frac{1}{L}} = \frac{1}{N} + \frac{x}{M} + \Ord{\frac{1}{L}},
\label{eq:primoMomApprox}
\end{equation}
in agreement with (\ref{eq:meanresult}), and with the first cumulant computed in~\cite{paper1} for both balanced and unbalanced bipartitions, at $x=1$.

The techniques discussed in this chapter can be generalized to the computation of the leading terms of the higher order moments of the local purity. In the next chapter we will exactly compute the first two moments of the local purity for a generic set of states in $\mathfrak{S}_{x}({\cal H}_X)$, confirming the results found in the previous section thru the Gaussian approximation.

\chapter{Exact moments and twirling maps}\label{chap9}
\markboth{Exact moments and twirling maps}{Exact moment and twirling maps}
In this chapter we will determine the high temperature expansion of the first moment of the local purity, from the exact computation of its first two moments for $\beta=0$. This generalizes previous results achieved for the case of pure states both in the thermodynamic limit~\cite{Lubkin78,Lloyd1988,Zyczkowski,paper1} and for systems with finite dimension~\cite{Scott2003,Giraud}. 
In Sec.~\ref{sec:tw} we will establish a formal connection between our problem and the theory of quantum channels. We will exploit the symmetry properties of the so called twirling transformations in order to compute the exact expression of the first moment of the purity of one subsystem for the case of pure states and then, through a purification scheme apply the same strategy to the case of arbitrary mixed states. However the generalization of this procedure to the computation of higher moments is very hard to be performed, since the purification procedure introduces a higher number of copies of the system making the structure of the total Hilbert space more involved. In Sec.~\ref{sec:Zuber1} we will introduce an alternative approach based on the solution of some basic integrals over the unitary group. We will then compute the exact expression of the first two cumulants, and determine in Sec.~\ref{sec:highTempRegime} the high temperature expansion of the first moment of the local purity.

\section{Exact computation of the first moment by a twirling map}\label{sec:tw}
In this section we will determine the exact expression of the first moment of the local purity of an arbitrary mixed state. In particular we will exploit the properties of the so called twirling transformations~\cite{tw1,tw2,tw3,entanglement2,BENNETT1}. This approach will also establish a formal connection between our problem and the theory of quantum channels.
We start in Sec.~\ref{sec:twirlingpure} by deriving a general expression for 
the moments ${\cal M}^{A}_n(1, 0)$ in the case of pure $\rho$ (i.e. the quantities ${\cal M}_n$ of Eq.~(\ref{moma}))
and verify that it yields the exact value given by Page~\cite{page,Giraud,Scott2003} for $n=1$.
We will then generalize this technique to the case of mixed $\rho$ in Sec.~\ref{sec:Twirling_mixed}, and compute the exact expression of the first moment of the local purity.

\subsection{Pure initial states}
\label{sec:twirlingpure}

Let us consider a fixed (normalized) pure state $\ket{\psi}_{AB}$ of the global system $X=AB$ and parametrize the pure states of $\cH_X$ through the mapping $\ket{\psi_{U}}_{AB}=U_{AB}\ket{\psi}_{AB}$,  with $U_{AB} \in \mathcal{U}(\mathcal{H}_X)\simeq\mathcal{U}(L)$
distributed according to the unique invariant Haar measure $d \mu_H(U)$, as done in Chap.~\ref{chap2}. 
The reduced density matrix of subsystem $A$ associated to $\ket{\psi_{U}}_{AB}$ is given by
\begin{eqnarray}\label{Vdue}
\rho_A&=& 
\mbox{Tr}_B [U_{AB} |\psi\rangle_{AB}\langle \psi| U_{AB}^{\dagger}] 
= \sum_{1\leq \ell \leq M}    {_B\langle} \ell |U_{AB} |\psi\rangle_{AB} \langle \psi| U_{AB}^\dag | \ell\rangle_B\;
\end{eqnarray}
 where  $\{ |\ell\rangle_B\}_{1 \leq \ell \leq M}$ is an orthonormal basis of $\cH_B$, with $\dim \cH_B=M$. From Eq. (\ref{Vdue})
the local purity of $|\psi_{U}\rangle_{AB}$ is given by  
\begin{eqnarray}
\pi_{AB}(\psi) 
&=& \sum_{1\leq \ell, \ell' \leq M}
  \mbox{Tr}\left[  {_B\langle} \ell |U_{AB} |\psi\rangle_{AB} \langle \psi| U_{AB}^\dag | \ell\rangle_B\;
   {_B\langle} \ell' |U_{AB} |\psi\rangle_{AB} \langle \psi| U_{AB}^\dag | \ell'\rangle_B \right] \label{Vtre}
\nonumber \\   
 \quad   &=& 
   \sum_{1\leq \ell, \ell' \leq M}
 {_{AB} \langle} \psi| \left( U_{AB}^\dag | \ell\rangle{_B\langle} \ell' |U_{AB} \right) |\psi\rangle_{AB}\;
 {_{AB}\langle} \psi| \left( U_{AB}^\dag | \ell'\rangle_B \langle \ell |U_{AB} \right) |\psi\rangle_{AB}\;,\nonumber\\\label{eq:purityPureV}
   \end{eqnarray}
  where
   in the last term we used the property of the trace $\Tr(\ket{\psi}_{AB}\bra{\psi} O)={_{AB} \langle} \psi| O \ket{\psi}_{AB}$, for any operator $O$ on $\cH_{X}$, and write the resulting expression   
   as a product of two expectation values on $|\psi\rangle_{AB}$.
We can recast Eq. (\ref{eq:purityPureV}) into a more compact form by doubling the Hilbert space, i.e. adding the auxiliary  copies $A'$ and $B'$  of $A$ and $B$, respectively. Namely,
\begin{eqnarray}
\pi_{AB}(\psi) &=&
   \sum_{1\leq \ell, \ell' \leq M}
 {_{AB} \langle} \psi| \left( U_{AB}^\dag | \ell\rangle{_B\langle} \ell' |U_{AB} \right) |\psi\rangle_{AB}\nonumber\\&&\qquad\times
 {_{A'B'}\langle} \psi| \left( U_{A'B'}^\dag | \ell'\rangle_{B'} \langle \ell |U_{A'B'} \right) |\psi\rangle_{A'B'}\nonumber\\
 &=& \sum_{1\leq \ell, \ell' \leq M}
 \Big({_{AB} \langle}\psi| \otimes  {_{A'B'} \langle}\psi|\Big)\left( U_{AB}^\dag \otimes U_{A'B'}^\dag \right) \Big(| \ell\rangle{_B\langle} \ell' | \otimes | \ell'\rangle_{B'} \langle \ell | \Big)\nonumber\\&&\qquad\times \Big(U_{AB} \otimes U_{A'B'} \Big)\Big( |\psi\rangle_{A'B'} \otimes |\psi\rangle_{AB} \Big),\nonumber\\
 \end{eqnarray}
 that is   
   \begin{eqnarray}\nonumber
\pi_{AB}(\psi) &=&  \mbox{Tr} \Big[ \Big( U_{AB} \otimes U_{A'B'}\Big)
 \Big( | \psi \rangle_{AB}\langle \psi|  \otimes  |\psi \rangle_{A'B'}\langle \psi|  \Big)
 \left( U_{AB}^\dag \otimes U_{A'B'}^\dag\right)
 \nonumber\\
 &&\qquad \quad \times \left(S_{B|B'} \otimes I_{AA'}\right) \Big] \label{Vpurity}\;.
    \end{eqnarray}
Here, the trace is performed over all degrees of freedom (i.e. $AA'BB'$), $I_{AA'}$ is the identity operator on $AA'$, and 
\begin{eqnarray}
 S_{B|B'} = \left( \sum_{1\leq \ell, \ell' \leq M}
 | \ell\rangle{_B\langle} \ell' | \otimes  | \ell'\rangle{_{B'}\langle} \ell|\right)  \label{Vswap}\;
\end{eqnarray}
is the \textit{swap} operator on $BB'$. The swap operator is a unitary, self-adjoint transformation which, for all operators $\Theta_B$ and $\Upsilon_{B'}$ on $\cH_{B}$ and $\cH_{B'}$,
gives
\begin{equation}
S_{B|B'} (\Theta_B\otimes \Upsilon_{B'}) S_{B|B'} =  \Upsilon_{B} \otimes \Theta_{B'}.
\end{equation}
Recall that by definition of the Haar measure on $\mathcal{U}(L)$, the first moment ${\cal M}_1^A(x=1,0) = {\cal M}_1$ of Eq.~(\ref{moma}) is obtained by averaging over all possible $U_X \in \mathcal{U}(L)$ (see also Sec.~\ref{Haar measure}), that is 
\begin{eqnarray}
{\cal M}_1&=& 
 \mbox{Tr} \Bigg[ {\cal T}^{(2)}\Big( | \psi \rangle_{X}\langle \psi|  \otimes  |\psi \rangle_{X'}\langle \psi|  \Big) \;
 \Big(S_{B|B'} \otimes I_{AA'}\Big) \Bigg]\;, \label{Vmedia}
    \end{eqnarray}
    where $X=AB$, $X'=A'B'$,  and ${\cal T}^{(2)}$ is the Completely Positive Trace Preserving (CPTP) {\em twirling} channel~\cite{tw1,tw2,tw3,boundent2}
    which transforms the operators $\Theta_{XX'}$ of $XX'$ into
  \begin{eqnarray}\label{VVTW}
{\cal T}^{(2)} (\Theta_{XX'} ) =  \int d \mu_{H}(U) \; \left( U_{X} \otimes U_{X'}\right)\; \Theta_{XX'} \;
( U_{X}^\dag \otimes U_{X'}^\dag) \;,
    \end{eqnarray}
$d\mu_{H}(U)$ being the usual Haar measure. 
This map plays  an important role in quantum information theory where it was
first introduced as a tool for
characterizing the distillability of bipartite entanglement~\cite{BENNETT1,entanglement2}.
It  has several properties
which  allow us to simplify the calculation.  For instance, it is known
that ${\cal T}^{(2)}$  maps all the states of the system into (generalized) Werner states~\cite{tw1,tw2}. Furthermore it  is self-adjoint -- i.e.
its description in Heisenberg picture, when we extend the action of quantum channels from density matrices to linear operators on $\cH_{XX'}$, coincides with ${\cal T}^{(2)}$.
In particular, this last property can be used to rewrite~(\ref{Vmedia}) as
    \begin{eqnarray}\label{Vmedia11}
    {\cal M}_1
  &=& \Big( {_{X}\langle}  \psi|  \otimes  {_{X'} \langle}  \psi| \Big)\;
  {\cal T}^{(2)}\left(S_{B|B'} \otimes I_{AA'}\right) \;   \Big(| \psi \rangle_{X}\otimes |\psi\rangle_{X'} \Big)
 \;.  \end{eqnarray}
 One of the main reasons that make the twirling channels very useful for our scopes is that the
 explicit expression for the action of ${\cal T}^{(2)}$ can be obtained by exploiting the symmetry of $d\mu_{H}(U)$. In particular it is possible to show that  ${\cal T}^{(2)} (\Theta_{XX'} )$ can be decomposed as
a linear combination of projectors on the symmetric and antisymmetric subspaces of $XX'=A B A' B'$ (with respect to the bipartition $AB | A'B'$). More precisely, introducing the swap operator which exchanges $X$ with $X'$ (or equivalently $A$ with $A'$ and $B$ with $B'$)
\begin{equation}
S_{X|X'}=S_{AB|A'B'}=S_{B|B'} \otimes S_{A|A'},
\end{equation}
it can be shown that the action of ${\cal T}^{(2)}$ is given by
 \begin{eqnarray}
{\cal T}^{(2)} (\Theta_{XX'} ) &=&  \frac{L I_{XX'} - S_{X|X'}}{L(L^2-1)} \; \mbox{Tr} ( \Theta_{XX'} )
+ \frac{L  S_{X|X'} - I_{XX'} }{L(L^2-1)}\;  \mbox{Tr} ( S_{X|X'} \Theta_{XX'})
\label{Vsimply}
\nonumber\\\nonumber\\
 &=&  \frac{L \mbox{Tr} (\Theta_{XX'}) - \mbox{Tr} (S_{X|X'} \Theta_{XX'})  }{L(L^2-1)} \; I_{XX'}
\nonumber\\\nonumber\\&&+ \frac{L  \mbox{Tr} ( S_{X|X'} \Theta_{XX'}) -\mbox{Tr} (\Theta_{XX'})
}{L(L^2-1)}\;  S_{X|X'}  \;,
 \label{VF}
\end{eqnarray}
where  $L=NM$ is the dimension of  ${\cal H}_{X}$ and
 $I_{XX'}$ is the identity operator on $\cH_{X X'}$.
 This is the central formula of our approach with the twirling transformations, and will lead to the exact computation of the first moment of the local purity both for pure and mixed global states. This can be achieved for $x=1$, either using Eq.~(\ref{Vmedia}) or Eq.~(\ref{Vmedia11}).
We have first  to compute the quantities $\mbox{Tr} (\Theta_{XX'})$ and $\mbox{Tr}(S_{X|X'} \Theta_{XX'})$
with $\Theta_{XX'}=|\psi\rangle_{X}\langle \psi| \otimes |\psi\rangle_{X'}\langle \psi|$: 
 \begin{eqnarray}
 \mbox{Tr} (|\psi\rangle_{X}\langle \psi| \otimes |\psi\rangle_{X'}\langle \psi|) =1\;,
 \nonumber \\
 \mbox{Tr}[(S_{B|B'} \otimes S_{A|A'} )|\psi\rangle_{AB}\langle \psi| \otimes |\psi\rangle_{A'B'}\langle \psi|] = 1\;,
 \end{eqnarray}
 where in the second expression we have used  the fact that
 $|\psi\rangle_{AB}\otimes |\psi\rangle_{A'B'}$ is invariant uder $S_{X|X'}$, i.e.
 \begin{equation}
 (S_{B|B'} \otimes S_{A|A'} )(|\psi\rangle_{AB}\otimes |\psi\rangle_{A'B'} )= |\psi\rangle_{AB}\otimes |\psi\rangle_{A'B'}.
 \end{equation}
Replacing all this in Eq.~(\ref{Vsimply}) we get
   \begin{eqnarray}
{\cal T}^{(2)} (|\psi\rangle_{X}\langle \psi| \otimes |\psi\rangle_{X'}\langle \psi|) &=&  \frac{I_{XX'} + S_{X|X'}}{L(L+1)}  \;,
\end{eqnarray}
and thus
    \begin{eqnarray}
{\cal M}_1&=&\frac{1}{L(L+1)}
 \mbox{Tr} \Big[ ( I_{ABA'B'} + S_{B|B'}\otimes S_{A|A'})  \;
 \left(S_{B|B'} \otimes I_{AA'}\right) \Big] \nonumber \\ &=&
 \frac{1}{N(N+1)}  \Big[ \mbox{Tr} ( S_{B|B'} \otimes I_{AA'} ) +
\mbox{Tr} (    I_{BB'} \otimes S_{A|A'} ) \Big]
 \;. \label{Vmedia10}
    \end{eqnarray}
    (Here we exploited the fact that $ S_{B|B'}^2=I_{BB'}$).
Notice that 
\begin{equation}
\Tr (S_{B|B'})= 
 \sum_{1\leq i, j,\ell, \ell' \leq M}
{_B\langle} i | \ell\rangle{_B\langle} \ell' |i\rangle_{B} {_{B'}\langle} j | \ell'\rangle{_{B'}\langle} \ell|j\rangle_{B'}=M,
\end{equation}
while
\begin{equation}
\Tr\left(I_{B B'}\right)=M^2=\dim (\cH_{B} \otimes \cH_{B'}).
\end{equation}
Thus, by using the condition $L=N M$ and the identities
\begin{eqnarray}
\mbox{Tr} (  S_{B|B'} \otimes I_{AA'} )  = M N^2\;, \quad
\mbox{Tr} (  I_{BB'} \otimes S_{A|A'} )  = N M^2\;,
\end{eqnarray}
we get  
\begin{eqnarray}
{\cal M}_1=
 \frac{N + M}{N M+1}
 \;,\label{Vmedia20}
 \end{eqnarray}
 which coincides with the exact value found in~\cite{Giraud}, see Eq. (\ref{eq:cum1pureGiraudNM}).
 
We mention that the same technique can also be applied to higher moments ${\cal M}_n$. 
The extension  of Eq.~(\ref{Vmedia}) for $n\geqslant 2$ is obtained by
introducing $2n$ copies of $AB$ organized  in $n$ couples, i.e.  $ A_1B_1, A_1'B_1'$, $A_2 B_2, A_2'B_2' $,
$\cdots$, $A_nB_n, A_n' B_n'$.
We then introduce the following generalized twirling transformation acting on
$XX'=A_1B_1A_1'B_1'A_2 B_2 A_2'B_2'$
$\cdots$ $A_nB_n A_n' B_n'$:
  \begin{eqnarray}\label{Vgentw}
\qquad {\cal T}^{(2n)} (\Theta_{XX'}) =  \int d \mu_{H}(U) \; ( \underbrace{U\otimes U \otimes \cdots \otimes U}_{2n}) \; \Theta_{XX'} \;
( \underbrace{U^\dag\otimes U^\dag \otimes \cdots \otimes U^\dag}_{2n}),\nonumber \\
    \end{eqnarray}
where $X=A_1B_1\ldots A_n B_n$ and $\Theta_{XX'}$ is a generic operator on ${\cal H}_{XX'}= {\cal H}_{AB}^{\otimes 2 n}$.  
  This channel is  a proper generalization of the map ${\cal T}^{(2)}$ whose properties can be established along the lines of~\cite{Zuber}. 
  With this choice we can express the $n$-th moment of the purity~(\ref{moma}) as
\begin{eqnarray}
{\cal M}_n^A(1,0)= {\cal M}_n&=& \mbox{Tr}\Big[ {\cal T}^{(2n)} \Big( |\Psi^{\otimes 2}\rangle\langle \Psi^{\otimes 2}|^{\otimes n} \Big) (S_{\cal B}^{(2n)} \otimes I_{\cal A}^{(2n)}) \Big]
 \label{Vmediak}\;,
    \end{eqnarray}
   where $|\Psi^{\otimes 2} \rangle^{\otimes n}= \otimes_{j=1}^n\big( |\psi\rangle_{A_jB_j} \otimes  |\psi\rangle_{A_j'B_j'}\big)$, $I_{\cal A}$ is the
   identity on the $2n$ copies of $A$, i.e. ${\cal A}=A_1 A_1' \cdots A_nA_n'$, and $S_{\cal B}^{(2n)}$ is the  swap operator which exchanges subsystems $B_1B_2 \cdots B_n$ with $B_1' B_2' \cdots B_n'$ pairwisely, i.e.
   $S_{\cal B}^{(2n)} = \otimes_{j=1}^{n} S_{B_j|B_j'}$.

\subsection{Mixed initial states}\label{sec:Twirling_mixed}
Consider now the case $x<1$.  Following the parameterization introduced in Sec.~\ref{sec:introductionstatisticalapproach} we split the average
over the set $\mathfrak{S}_{x}({\cal H}_X)$ of the density matrices of global purity $x$, as an average over the unitaries acting on ${\cal H}_{X}$ followed
by an average over the  space of the eigenvalues, see Eq.~(\ref{eq:partition function unitary rotated}). Specifically this is accomplished by writing 
\begin{eqnarray}\label{eq:rhouuu}
\rho(U)= U_X \;\Lambda_{X}  \;U_{X}^\dag\;,
\end{eqnarray}
with $X=AB$, $U_{X}$ a generic unitary transformation on ${\cal H}_{X}$, and $\Lambda_{X}$ a diagonal density matrix which can be chosen as 
\begin{eqnarray}\label{VVpuri}
\Lambda_{X} &= &\sum_{1\leq i \leq N}\sum_{1 \leq \ell \leq M}  \; \lambda_{i\ell} \; |i\rangle_{A}\langle i | \otimes
 |\ell \rangle_{B}\langle \ell |\nonumber\\\nonumber\\ &=& \mathrm{diag} (\lambda_{M( i -1)+ j})_{1\leq i \leq N, 1\leq \ell \leq M}.
\end{eqnarray}
Here, by keeping the same notation of Chap.~\ref{chap8}, we have set $\lambda_{i \ell }=\lambda_{M (i-1)+\ell}$, where $\{ |i\rangle_{A}\}_{1\leq i \leq N}$ and  $\{ |\ell\rangle_{B}\}_{1 \leq \ell \leq M}$ are orthonormal sets
 of ${\cal H}_{A}$ and ${\cal H}_B$, respectively.  
Introducing then  the ancillary systems $a$ and $b$ isomorphic to $A$ and $B$
respectively, we  define the following purification of $\rho$,
\begin{eqnarray}
|\Psi\rangle_{ABab}= \sum_{1 \leq i \leq N} \sum_{1 \leq \ell \leq M} \sqrt{\lambda_{i\ell}}\; \; |i\rangle_A\otimes |\ell\rangle_B \otimes
|i\rangle_a \otimes | \ell\rangle_b\;.
\end{eqnarray}
Let us fix the spectrum $\Lambda_X$ of the global density matrix $\rho$ of the system.  The set of vectors in $\cH_{X x} = \cH_{X} \otimes \cH_{x}$ with the same Schmidt coefficients is given by 
\begin{equation}
\ket{\Psi_{U}}_{Xx}=U_{Xx}\ket{\Psi}_{Xx}, \qquad U_{Xx}={U}_{X} \otimes {U}_{x}\in \mathcal{U}(L^2), 
\end{equation}
with $U_X, U_{x} \in \mathcal{U}(L)$ and yields the set of density matrices (\ref{eq:rhouuu}) with the same spectrum $\Lambda_X$. Here, we have set  $X=AB$ and $x=ab$. By partial tracing over subsystem $B$ we obtain
the set of reduced density matrices
\begin{eqnarray}
\rho_A(U) &=& \mbox{Tr}_B (\rho_{AB}(U)) =
 \mbox{Tr}_{Bab} (U_{AB} |\Psi\rangle_{ABab} \langle \Psi| U_{AB}^\dag) \nonumber \\
 &=& \sum_{1 \leq q \leq N M^2} {_{Bab} \langle} q | U_{AB} |  \Psi\rangle_{ABab} \langle \Psi| U_{AB}^\dag
 | q \rangle_{Bab}\;,
 \end{eqnarray}
where $\{ |q\rangle_{Bab}\}_{1\leq q \leq N M^2}$ is an orthonormal basis of $\cH_B \otimes \cH_{ab}$, and $U_{AB} \in \mathcal{U}(L)$. Notice that $\rho_{A}(U)$ does not depend on $U_{x}$.
The local A-purity of $\rho_A(U)$ becomes
\begin{eqnarray}
\pi_A(\rho)= \mbox{Tr} ({\rho_A}^2(U)) &=& \sum_{1 \leq q, q' \leq N M^2}
  \mbox{Tr}\Big( {_{Bab}\langle} q |U_{AB} |\Psi\rangle_{ABab} \langle \Psi| U_{AB}^\dag | q\rangle_{Bab} \nonumber \\
  && \qquad \times
   {_{Bab}\langle} q' |U_{AB} |\Psi\rangle_{ABab} \langle \Psi| U_{AB}^\dag | q'\rangle_{Bab} \Big)\nonumber \\
  & = &\label{Vtre11}
   \sum_{1 \leq q, q' \leq N M^2}
{_{ABab}  \langle} \Psi| U_{AB}^\dag | q\rangle_{Bab} {\langle} q' |U_{AB} |\Psi\rangle_{ABab}
\nonumber \\
&& \qquad \times {_{ABab}  \langle} \Psi| U_{AB}^\dag | q' \rangle_{Bab} {\langle} q|U_{AB} |\Psi\rangle_{ABab}.
 \end{eqnarray}
In analogy to what seen for the pure case, it can be casted as an expectation value on $|\Psi\rangle^{\otimes 2}$ by doubling the Hilbert space
\begin{eqnarray}
\pi_A(\rho)& = &\Big({_{Xx}  \langle} \Psi| \otimes  {_{X'x'}  \langle} \Psi| \Big) \Big(U_{AB}^\dag \otimes  U_{A'B'}^\dag\Big) \Big(S_{Bab|B'a'b'} \otimes I_{A A'}\Big) 
\nonumber \\
&& \qquad \times \Big(U_{AB} \otimes U_{A'B'}\Big) \Big(|\Psi\rangle_{X x} \otimes  |\Psi\rangle_{X' x'}\Big),
\end{eqnarray}
where
\begin{equation}
S_{Bab|B'a'b'}= \sum_{1 \leq q, q' \leq N M^2} | q\rangle_{Bab} {\langle} q' |\otimes | q' \rangle_{Bab} {\langle} q|
\end{equation}
is the swap operator between $B a b$ and $B'a'b'$. Here, we have introduced the ancillary bipartite systems $X'=A'B'$, and $x'=a'b'$ (here $A',a',B',b'$ are the auxiliary copies
 of $A,a,B$, and $b$ respectively), thus $|\Psi\rangle_{X x} = |\Psi\rangle_{ABab}$ and $|\Psi\rangle_{X' x'} = |\Psi\rangle_{A'B'a'b'}$.
Therefore, by integrating over $U_{AB}$ we get,
\begin{eqnarray}
{\cal M}_1^A(\Lambda_{X})= \mbox{Tr}
\Big[ {\cal T}^{(2)}\Big( |\Psi\rangle_{Xx}\langle \Psi| \otimes |\Psi\rangle_{X'x'}\langle \Psi|\Big)
\;\; \Big( S_{Bab|B'a'b'} \otimes I_{AA'} \Big)\Big],
 \end{eqnarray}
 where ${\cal T}^{(2)}$ is the twirling transformation on $\cH_{XX'}$, Eq.~(\ref{VVTW}). As mentioned in the previous subsection, from the properties of the twirling maps ${\cal T}^{(2)}$, ${\cal M}_1^A(\Lambda_{X})$ can be written as
\begin{eqnarray}
{\cal M}_1^A(\Lambda_{X})= \Big( {_{Xx}\langle} \Psi| \otimes  {_{X'x'}\langle} \Psi| \Big)\label{VIMPO1}
 {\cal T}^{(2)}\Big( S_{Bab|B'a'b'} \otimes I_{AA'} \Big) \Big( |\Psi\rangle_{Xx} \otimes |\Psi\rangle_{X'x'}\Big).
 \end{eqnarray}  
See Eq. (\ref{Vmedia11})  for comparison. We will thus use the last identity.
According to Eq.~(\ref{VF}) we have to compute $\mbox{Tr}_{XX'} ( \Theta_{X X'} )$ and $\mbox{Tr}_{XX'} ( S_{X|X'} \Theta_{X X'} )$ where $\Theta_{XX'}$ is
\begin{equation}
S_{Bab|B'a'b'} \otimes I_{AA'}=S_{B|B'}
\otimes S_{b|b'}\otimes S_{a|a'} \otimes I_{AA'}.
\end{equation}
We get
\begin{eqnarray}
 & & \mbox{Tr}_{ABA'B'} (S_{Bab|B'a'b'} \otimes I_{AA'}) = \mbox{Tr}_{ABA'B'} ( S_{B|B'} \otimes S_{b|b'}
 \otimes S_{a|a'} \otimes I_{AA'})\nonumber \\ 
 \qquad & &\qquad = 
 \mbox{Tr}_{BB'} ( S_{B|B'}) \;
 \mbox{Tr}_{AA'}( I_{AA'} )  \; S_{b|b'} \otimes S_{a|a'}  =  N^2 M \; S_{b|b'} \otimes S_{a|a'} \;,\\ \nonumber \\ 
 & & \mbox{Tr}_{ABA'B'} [S_{AB|A'B'}(S_{Bab|B'a'b'} \otimes I_{AA'})] =
  \mbox{Tr}_{ABA'B'} (I_{BB'} \otimes S_{A|A'} \otimes S_{b|b'}
 \otimes S_{a|a'} )
 \nonumber\\ 
  & & \qquad = 
  \mbox{Tr}_{BB'} (I_{BB'})
 \; \mbox{Tr}_{AA'} (S_{A|A'}) \; S_{b|b'} \otimes S_{a|a'}
 =  M^2 N \; S_{b|b'} \otimes S_{a|a'}  \;.
 \end{eqnarray}
 Thus, from Eq.~(\ref{VF}) we obtain
 \begin{eqnarray}
\qquad  {\cal T}^{(2)}\Big( S_{Bab|B'a'b'} \otimes I_{AA'} \Big)
&=&  \frac{M (N^2 -1) }{N^2M^2-1} \; I_{BB'} \otimes
I_{AA'} \otimes S_{b|b'} \otimes S_{a|a'}  \nonumber \\
\qquad & &+  \frac{N(M^2 -1) }{N^2M^2-1}  \;  S_{B|B'} \otimes S_{A|A'} \otimes S_{b|b'} \otimes S_{a|a'}\nonumber \\ &=& \frac{M (N^2 -1) }{N^2M^2-1} \; I_{XX'} \otimes S_{x|x'}  \nonumber \\
\qquad & &+  \frac{N(M^2 -1) }{N^2M^2-1}  \;  S_{X|X'} \otimes S_{x|x'}, 
 \label{VF11}
\end{eqnarray}
with $L=N M$. Replacing this expression into Eq.~(\ref{VIMPO1}) and using the identities
\begin{eqnarray}
\Big( {_{Xx}\langle} \Psi| \otimes  {_{X'x'}\langle} \Psi| \Big)
 (I_{XX'} \otimes  S_{x|x'} )
 \Big( |\Psi\rangle_{Xx} \otimes |\Psi\rangle_{X'x'}\Big) &=& \sum_{1\leq i \leq N} \sum_{1\leq \ell \leq M} \lambda_{i\ell}^2 = \mbox{Tr}(\Lambda_X^2)\;,
 \nonumber \\
 \Big( {_{Xx}\langle} \Psi| \otimes  {_{X'x'}\langle} \Psi| \Big)
 (S_{X|X'} \otimes  S_{x|x'} )
 \Big( |\Psi\rangle_{Xx} \otimes |\Psi\rangle_{X'x'}\Big) &=& 1\;,
\end{eqnarray}
we finally get 
\begin{eqnarray}
{\cal M}_1^A(\Lambda_X)\label{Vfin}
&=& \frac{M (N^2 -1) }{N^2M^2-1}   \mbox{Tr}(\Lambda_X^2) + \frac{N(M^2 -1) }{N^2M^2-1}, \;
 \end{eqnarray}
which depends on the spectrum $\Lambda_X$ only through the purity of $\rho$, i.e.
$ \mbox{Tr}(\Lambda_X^2) =  \mbox{Tr}(\rho^2)$. 
This immediately tells us that  averaging upon $\Lambda_X$, while keeping fixed $x$,  in (\ref{eq:mixed}), 
will give 
\begin{eqnarray}
\label{VVFIN}
{\cal M}_1^A(x,0) = \frac{M (N^2 -1) }{N^2 M^2-1} \;x +  \frac{N(M^2 -1) }{N^2M^2-1} \; .
 \end{eqnarray}
\paragraph{Some special cases:--} It is worth noticing that for a balanced bipartition $N=M=\sqrt{L}\gg1$ Eq.~(\ref {VVFIN}) yields 
\begin{equation}
{\cal M}_1^A(x,0) = \frac{\sqrt{L} (1+x)}{L+1}\sim \frac{1+x}{\sqrt{L}}\;,
\label{eq:VVFIN}
 \end{equation}
 as found by using the Gaussian approximation in Chap.~\ref{chap8}, see Eqs. (\ref{COnF}) and (\ref{eq:primoMomApprox}).
 On the other hand for $x=1$ (i.e. pure global states), Eq.~(\ref{eq:VVFIN}) coincides with that obtained in~\cite{Giraud,paper1}. 
 
 Finally, consider the case in which $\rho$ is
 maximally mixed, i.e. is the density matrix $I_{AB}/L$. In this case $x=1/{L}$ and Eq.~(\ref {VVFIN})
 correctly gives
 \begin{eqnarray}
{\cal M}_1^A(1/L,0) =\frac{1}{N}\;,
 \end{eqnarray}
 in agreement with the general result~(\ref{kkll}).
In analogy to what seen for pure states, the above analysis can in principle be extended to the case of higher moments of $\pi_A({\psi})$
\begin{eqnarray}
 {\cal M}_n^A(\Lambda_X)=\mbox{Tr}\Big[ {\cal T}^{(2n)} \Big( |\Psi\rangle_{X x}\langle \Psi|^{\otimes (2n)} \Big) \Big(S_{{\cal B}{\mathbf{a}}{\mathbf{b}}}^{(2n)} \otimes I_{\cal A}^{(2n)}\Big) \Big]
 \label{Vmediakmixed}\;,
    \end{eqnarray}
   where $|\Psi \rangle_{X x}^{\otimes (2n)}= \otimes_{j=1}^n\big( |\psi\rangle_{A_jB_j a_j b_j} \otimes  |\psi\rangle_{A_j'B_j' a_j' b_j'}\big)$, $I_{\cal A}$ is the
   identity on the $2n$ copies of $A$, i.e. ${\cal A}= A_1A_1' \cdots A_n A_n'$, and  $S_{{\cal B}{\mathbf{a}}{\mathbf{b}}}^{(2n)}$ is the  swap operator which exchanges $B_ia_ib_i$ with $B'_ia'_ib'_i$ for $1\leq i \leq n$,
   $S_{{\cal B}{\mathbf{a}}{\mathbf{b}}}^{(2n)} = \otimes_{j=1}^{n} S_{B_j a_j b_j |B_j'a_j' b_j'}$.
However, for $n > 1$ Eq. (\ref{VF}) is no more valid and the application of the twirling transformation is far more complicated than for $n = 1$, since it cannot be reduced to a combination of projection operators on the symmetric and antisymmetric subspaces. In the next section we will introduce an alternative approach for the computation of moments of the local purity of an arbitrary mixed state, that will lead to a formula for the $n$-th moment of $\pi_{A}(\psi)$ and to the exact expression for its first two cumulants.  
\section{Moments of the local purity at $\beta=0$}\label{sec:Zuber1}
The technique we will introduce in this section in order to exactly compute the moments of the purity is based on the explicit solution due to Zuber of some basic integrals over the unitary group~\cite{Zuber}. Notice that, as mentioned above, also the symmetry properties of the twirling maps can be determined along the lines of these solutions. We will exactly compute the first two moments and determine the dependence on the spectrum of the higher order ones.
\subsection{First moment}\label{sec:firtsmomentExact}
Let us introduce a purification for $\rho$ on the doubled Hilbert space $\cH_{X X'}=\cH_X \otimes \cH_{X'}$. According to what seen in the previous section, after integrating over all states in $\cH_X$ with a given spectrum $\Lambda_X$, we will show that the only dependence on the latter is in terms of the purity $x$, fixed by the delta function in the partition function (\ref{eq:mixed}). If we fix the spectrum of the global density matrix $\rho$ of the system as $\Lambda_X = \mathrm{diag} (\lambda_{M(\alpha-1)+ \beta})$ (see Eq. (\ref{VVpuri})),
the purification of  $\Lambda_X$ in the space $\cH_{X X'}$, with $X=AB$ and $X'=A'B'$ is
\begin{eqnarray}
|\Psi\rangle_{X X'}  = \sum_{1 \leq \alpha \leq N} \sum_{1 \leq \beta \leq M} \sqrt{\lambda_{\alpha \beta}}\; \; \ket{\alpha \beta}_{AB}\otimes  \ket{\alpha \beta}_{A'B'} ,
\label{eq.ref_state}\end{eqnarray}
where  $\ket{\alpha \beta}_{A B}=\ket{\alpha}_A\otimes \ket{\beta}_B$,  $\{ \ket{\alpha}_{A}
\}_{1 \leq \alpha \leq N}$ and  $\{ \ket{\beta}_{B}\}_{1 \leq \beta \leq M}$  
($\{ \ket{\alpha}_{A'}\}_{1 \leq \alpha \leq N}$ and $\{ \ket{\beta}_{B'}\}_{1 \leq \beta \leq M}$) are the reference bases in $\cH_A$ and $\cH_B$ ($\cH_{A'}$ and $\cH_{B'}$), respectively.
We have seen that those vectors in $\cH_{X X'}$ with the same Schmidt coefficients are given by
$\ket{\Psi_{U}}_{XX'}=U_{X X'}\ket{\Psi}_{XX'}$, $U_{X X'}={U}_{X} \otimes {U}_{X'}\in \mathcal{U}(L^2)$, 
with $U_X, U_{X'} \in \mathcal{U}(L)$ and yields the set of density matrices with the same spectrum $\Lambda_X$, namely $\rho= U_X \Lambda_X U_X^\dag$. By partial tracing over subsystem $B$ one obtains
the set of reduced density matrices $\rho_A(U)=
 \mbox{Tr}_{B}\Tr_{X'}( \ket{\Psi_U}_{XX'}\bra{ \Psi_U} )$:
 \begin{eqnarray}\label{eq:rhoAperZuber}
 \rho_{A}&=& \sum_{1 \leq \alpha \leq N} \sum_{1 \leq \beta \leq M} \lambda_{\alpha \beta} \Tr_{B}(U_{AB} \ket{\alpha \beta}_{AB}\bra{\alpha \beta}U_{AB}^{\dagger})\nonumber\\
 &=& \sum_{1 \leq \alpha \leq N} \sum_{1 \leq \beta, j \leq M}  \lambda_{\alpha \beta} \, \bras{j}{B} U_{AB} \ket{\alpha \beta}_{AB}\bra{\alpha \beta}U_{AB}^{\dagger} \ket{j}_B.
 \end{eqnarray}
Thus we can write the local purity as
  \begin{eqnarray}
\pi_A(U_X \Lambda_X U^\dagger_X)=
\Tr({\rho_A}^2)=\sum_{1\leq  \alpha_1,\alpha_2\leq N}\sum_{1 \leq \beta_1,\beta_2 \leq N} \sum_{1 \leq  j_1, j_2 \leq  N}   \lambda_{\alpha_1 \beta_1} \lambda_{\alpha_2 \beta_2} \nonumber \\\qquad\qquad\qquad\qquad \times \bras{\alpha_2 \beta_2}{AB}U_{AB}^{\dagger} \ket{j_2}_B\bra{j_1} U_{AB} \ket{\alpha_1 \beta_1}_{AB}\nonumber \\ \qquad\qquad\qquad\qquad\times\, \bras{\alpha_1 \beta_1}{AB}U_{AB}^{\dagger} \ket{j_1}_B\bra{j_2} U_{AB} \ket{\alpha_2 \beta_2}_{AB},
 \end{eqnarray}
which, by inserting the completeness relation for subsystem $A$, becomes
  \begin{eqnarray}\label{Vtre1}
\pi_A(U_X \Lambda_X U^\dagger_X)=\sum_{1\leq  \alpha_1,\alpha_2\leq N}\sum_{1 \leq \beta_1,\beta_2 \leq M} \sum_{1 \leq  i_1, i_2 \leq  N}\sum_{1 \leq  j_1, j_2 \leq M}   \lambda_{\alpha_1 \beta_1} \lambda_{\alpha_2 \beta_2} \nonumber \\\qquad\qquad\qquad\qquad \times \bras{\alpha_2 \beta_2}{AB}U_{AB}^{\dagger} \ket{i_1 j_2}_{AB}\bra{i_1 j_1} U_{AB} \ket{\alpha_1 \beta_1}_{AB}\nonumber \\ \qquad\qquad\qquad\qquad\times\, \bras{\alpha_1 \beta_1}{AB}U_{AB}^{\dagger} \ket{i_2 j_1}_{AB}\bra{i_2 j_2} U_{AB} \ket{\alpha_2 \beta_2}_{AB}.
 \end{eqnarray}
We can now compute the first moment of the purity (\ref{defma})
at $\beta=0$.  More generally, by recalling that
\begin{eqnarray}
\cZ_A(x, 0)&=&1,\\
d \mu_{x,0}(\rho)&=& d \mu_x(\rho),\\
d \mu_x(\rho)&=&C_x d \mu_H(U_X) d \sigma(\Lambda_{X})\delta\Big(x-\Tr (\Lambda_X^2)\Big),
\end{eqnarray}
we get
\begin{eqnarray}
\qquad   {\cal M}^{A}_n(x, 0) &=&  \int d \mu_{x}(\rho) \; {\pi_A}^n(\rho) 
\nonumber\\
&=& C_x \int d\sigma(\Lambda_X) \delta(x-\Tr (\Lambda^2_X))\;
 {\cal M}^{A}_n(\Lambda_X)
= \langle {\cal M}^{A}_n(\Lambda_X)\rangle_x,\nonumber\\
\label{eq:MAnx0}
 \end{eqnarray} 
with
\begin{equation}
 \qquad \qquad {\cal M}^{A}_n(\Lambda_X)= \int d\mu _H(U_X)\; {\pi_A}^n(U_X \Lambda_X U_X^\dagger).
\label{eq:MAnLX}
\end{equation}
From Eq.~(\ref{Vtre1}) we have that the average over the unitary group of the first moment particularizes to
\begin{eqnarray}
 {\cal M}_1^A(\Lambda_X)
 &=&\sum_{1\leq  \alpha_1,\alpha_2\leq N}\sum_{1 \leq \beta_1,\beta_2 \leq M}   \lambda_{\alpha_1 \beta_1} \lambda_{\alpha_2 \beta_2}\nonumber\\
  &&\times \sum_{1 \leq  i_1, i_2 \leq  N}\sum_{1 \leq  j_1, j_2 \leq M} \int d \mu_H(U_X) U_{i_1 j_1, \alpha_1 \beta_1}U_{i_2 j_2, \alpha_2 \beta_2} 
U^\dag_{\alpha_2 \beta_2,i_1 j_2}U^\dag_{\alpha_1 \beta_1,i_2 j_1},\nonumber \\
\label{eq:unint}\end{eqnarray}
where 
\begin{equation}
U_{i j, \alpha \beta}=\bras{i j }{AB}U_{AB} \ket{\alpha \beta}_{AB} \qquad \forall \  \alpha,  i \in \{1,\ldots, N\}, \qquad \forall \  \beta,  j \in \{1,\ldots,M\} 
\end{equation}
are the matrix elements of $U_X \in \mathcal{U}(L)$.
This integral can be explicitly done by applying the solution for the basic integral over the unitary group $\mathcal{U}(L)$ given by Zuber~\cite{Zuber}:
\begin{equation}
\quad \int d \mu_{H}(U) U_{i_1 j_1} \ldots U_{i_n j_n}  U_{k_1 l_1}^{\dag} \ldots U_{k_n l_n}^{\dag}=\sum_{\tau, \sigma \in S_n} C[\sigma]\prod_{a=1}^{n}\delta(i_a, \ell_{\tau(a)})\delta(j_a,k_{\tau \sigma(a)}), 
\label{eq:tecnicaZuber}
\end{equation}
with
\begin{equation}
C[\sigma]=\sum_{|Y|=n} \frac{(\chi^{(k)}(1))^2\chi^{(k)}([\sigma])}{n!^2 s_k(I)},
\label{eq:tecnicaZuber1}
\end{equation}
where $C[\sigma]$ is the sum over the Young diagrams $Y$ where $\chi^{(k)}([\sigma])$ is the character of the symmetric group $S_n$ associated to $Y$, depending on the conjugacy class of the permutation $\sigma$, $s_k(I)$ is the dimension of the representation of $Y$ in terms of the linear group $\mbox{GL}(L)$, and $\delta(a,b)$ is the Kroneker delta function $\delta_{a,b}$. Applying this solution to (\ref{eq:unint}) we get
\begin{eqnarray}
 {\cal M}_1^A(\Lambda_X) &=&\sum_{1\leq  \alpha_1,\alpha_2\leq N}\sum_{1 \leq \beta_1,\beta_2 \leq M}   \lambda_{\alpha_1 \beta_1} \lambda_{\alpha_2 \beta_2}
\nonumber\\ 
 &&\times\sum_{\tau, \sigma \in S_2} C[\sigma] f_1(\tau)\delta(\alpha_1 \beta_1,\alpha_{\tau\sigma(2)}\beta_{\tau\sigma(2)})\delta(\alpha_2 \beta_2,\alpha_{\tau\sigma(1)}\beta_{\tau\sigma(1)})\nonumber\\ 
  &=&\sum_{1\leq  \alpha_1,\alpha_2\leq N}\sum_{1 \leq \beta_1,\beta_2 \leq M}   \sum_{\tau, \sigma \in S_2} \sum_{c\in \mathcal{C}(S_2)} C[\sigma] f_1(\tau) \delta([\tau\sigma s],c) \nonumber\\ 
 &&\qquad\qquad\qquad\qquad\qquad \times\lambda_{\alpha_{c(1)}\beta_{c(1)}} \lambda_{\alpha_{c(2)}\beta_{c(2)}},
\nonumber\\ 
 \label{eq:prim_momZuber}
\end{eqnarray}
where $ f_1(p)$ depends on the permutation $p \in S_2$
\begin{equation}
\qquad f_1(p)=\sum_{1 \leq i_1,i_2 \leq N}\delta(i_1 ,i_{p(1)} ) \delta(i_2 ,i_{p(2)})\sum_{ 1\leq j_1,j_2 \leq M}\delta(j_1, j_{p(2)}) \delta(j_2, j_{p(1)}),
\end{equation}
$s$  is the transposition (swapping) of pairs of nearby indices ($s=[2]\in S_2$)
\begin{eqnarray}\label{eq:s}
i_{s(1)} = i_{2}  \quad  \mbox{and} \quad i_{s(2)} = i_{1}
\end{eqnarray}
and $\mathcal{C}(S_2)$  is the set of the conjugacy classes of the symmetric group $S_2$, $\mathcal{C}(S_2)=\{[1^2],[2]\}$. From (\ref{eq:prim_momZuber}) it can be easily inferred that the only possible contributions of the spectrum are related to the conjugacy  classes of the symmetric group $S_2$:
\begin{eqnarray}\label{contrSpectrum1}
\left[\tau\sigma s \right]=\left[1^2\right]& \longrightarrow&\left(\sum_{1\leq \alpha \leq N}\sum_{1\leq \beta \leq M}\lambda_{\alpha\beta}\right)^2=1,\nonumber\\
\left[\tau\sigma s\right]=\left[2\right]&\longrightarrow& \left(\sum_{1\leq \alpha \leq N}\sum_{1\leq \beta \leq M}\lambda_{\alpha \beta}^2\right)=\Tr(\Lambda_X^2).
\end{eqnarray}
By summing and by using the explicit expressions of the coefficients~(\ref{eq:tecnicaZuber1})~\cite{Zuber}
\begin{equation}
\label{ zubercoeff}
C[1^2]=\frac{1}{(L-1)(L+1)}, \qquad C[2]=-\frac{1}{(L-1)L(L+1)},
\end{equation}
we get
\begin{equation}\label{eq:primMomU}
{\cal M}_1^A(\Lambda_X)=\frac{N(M^2-1)}{N^2M^2-1} + \frac{M(N^2-1)}{N^2M^2-1} \Tr(\Lambda_X^2).
\end{equation}
The first moment of the purity of subsystem $A$, is the average (\ref{eq:primMomU}) over the spectrum of the system. By plugging~(\ref{eq:primMomU}) into~(\ref{eq:MAnx0}), we finally get\begin{equation}\label{eq:firstMom}
{\cal M}_1^A(x,0)
= \frac{N(M^2-1)}{N^2M^2-1} + \frac{M(N^2-1)}{N^2M^2-1} x,
\end{equation} 
in total agreement with Eq. (\ref{VVFIN}), derived by using the properties of the twirling transformations. 
\subsection{k-th moment} \label{sec:kthMom}
The technique shown in the previous section can be easily generalized in order to compute from~(\ref{eq:MAnLX})
higher moments  at $\beta=0$. More precisely, we can write
\begin{eqnarray}
{\cal M}_k^A(\Lambda_X)&=&\int d \mu_H(U_X) \; {\pi_A}^k(U_X\Lambda_X U_X^\dagger)
\nonumber\\ 
 &=&\sum_{1\leq  \alpha_1,\ldots, \alpha_{2k}\leq N}\sum_{1 \leq \beta_1,\ldots,\beta_{2k} \leq M}  \prod_{1\leq i \leq 2k} \lambda_{\alpha_i \beta_i} \nonumber\\
 &&\times \sum_{1 \leq  i_1, \ldots, i_{2k} \leq  N}\sum_{1 \leq  j_1, \ldots j_{2k} \leq M} \int d \mu_H(U_X) \prod_{1\leq \ell \leq 2k}U_{i_\ell j_\ell, \alpha_\ell \beta_\ell} \nonumber\\
 &&\qquad \qquad \qquad \qquad \qquad \times \prod_{1\leq m\leq k}U^\dag_{\alpha_{2m} \beta_{2m},i_{2m-1} j_{2m}}U^\dag_{\alpha_{2m-1} \beta_{2m-1},i_{2m} j_{2m-1}}. \nonumber\\\label{eq:unintk}
 \end{eqnarray}
Thus, Eq. (\ref{eq:tecnicaZuber}) for $n= 2 k$ gives
\begin{eqnarray}
{\cal M}_k^A(\Lambda_X) &=& \sum_{1\leq  \alpha_1, \ldots,\alpha_{2k}\leq N}\sum_{1 \leq \beta_1,\ldots, \beta_{2k} \leq M}   \sum_{\tau, \sigma \in S_{2k}} \sum_{c\in \mathcal{C}(S_{2k})} C[\sigma] f_k(\tau) \delta([\tau\sigma s],c)\nonumber \\&&\qquad \qquad \qquad \qquad\qquad \qquad  \qquad \qquad \times
\prod_{1\leq i \leq 2k}\lambda_{\alpha_{c(i)}\beta_{c(i)}}, \nonumber \\
 \label{eq:kMomUnitIntl}
\end{eqnarray}
where  $ f_k(p)$ depends on the permutation $p \in S_{2k}$
\begin{eqnarray}
f_k(p)=\sum_{1 \leq i_1, \ldots, i_{2k} \leq N}\prod_{1\leq \ell \leq 2k}\delta(i_{\ell} ,i_{p(\ell)} )\sum_{1 \leq j_1, \ldots, j_{2k} \leq M}\prod_{1\leq m \leq k}\delta(j_{2m-1} ,j_{p(2m)}) \delta(j_{2m} ,j_{p(2m-1)}) \nonumber\\
\label{eq:fk}
\end{eqnarray}
and, analogously to Eq. (\ref{eq:s}), $s$ is the swapping of pairs of nearby indices, $1\leq i \leq N$ or $1\leq i \leq M$:
\begin{equation}\label{eq:sk}
 i_{s(2\ell-1)}=i_{2\ell} \qquad \mbox{and}  \qquad   i_{s(2\ell)}=i_{2\ell-1}  \quad \forall  \, \ell \in \{1,\ldots, k\}.
\end{equation}
Observe that when $k=1$ we retrieve 
${\cal M}_1^A(\Lambda_X)$ (see Eq. (\ref {eq:prim_momZuber})).
The different contributions from the spectrum can be classified in terms of the conjugacy classes of the symmetric group, as shown in Eq. (\ref{contrSpectrum1}). However, for $k>1$, they do not depend only on the identity and the purity $\Tr(\Lambda_X^2)$. They show a more complex dependence on the spectrum, thru its higher order invariants $\Tr (\Lambda_X^k)$, with $k>2$. Thus the integral on the spectrum~(\ref{eq:MAnx0}) is in general non trivial. 
\vspace{-1.2cm}
\subsection{Second moment}
\vspace{-0.2cm}
Let us now compute the second moment of the purity. In this section we will consider the case of arbitrary bipartite states, with purity $x \in [1/L,1]$,  generalizing some results found for the case of pure states $x=1$,~\cite{Lubkin78,Lloyd1988,Scott2003,Giraud,paper1}.
The second moment of the purity can be directly computed by setting $k=2$ in Eq. (\ref{eq:unintk}). The expression for the coefficients $C[p]$ in (\ref{eq:tecnicaZuber}), when $p \in \mathcal{C}(S_{4})$  is~\cite{Zuber}:
\begin{eqnarray}
C\left[1^4\right]&=&\frac{L^4-8L^2+6}{(L-3)(L-2)(L-1)L^2(L+1)(L+2)(L+3)},\nonumber\\
C\left[2,1^2\right]&=&-\frac{1}{(L-3)(L-1)L(L+1)(L+3)},\nonumber\\
C\left[2^2\right]&=&\frac{L^2+6}{(L-3)(L-2)(L-1)L^2(L+1)(L+2)(L+3)},\nonumber\\
C\left[3,1\right]&=&\frac{2L^2-3}{(L-3)(L-2)(L-1)L^2(L+1)(L+2)(L+3)},\nonumber\\
C\left[4\right]&=&-\frac{5}{(L-3)(L-2)(L-1)L(L+1)(L+2)(L+3)}.
\end{eqnarray}
The symmetric group $S_4$ consists of five conjugacy classes, giving the following contributions to the integral (\ref{eq:kMomUnitIntl}) in terms of the spectrum of $\rho$:
\begin{eqnarray}
 \left[\tau\sigma s\right]=\left[1^4\right]& \longrightarrow&\left(\sum_{\mu,\nu}\lambda_{\mu,\nu}\right)^4=1,\nonumber\\  
\left[\tau\sigma s\right]=\left[2,1^2\right]&\longrightarrow& \left(\sum_{\mu_1,\nu_1}\lambda_{\mu_1,\nu_1}^2\right)\left(\sum_{\mu_2,\nu_2}\lambda_{\mu_2,\nu_2}\right)^2=\Tr (\Lambda_X^2),\nonumber\\
\left[\tau\sigma s\right]=\left[2^2\right]&\longrightarrow& \left(\sum_{\mu,\nu}\lambda_{\mu,\nu}^2\right)^2=(\Tr( \Lambda_X^2))^2,\nonumber\\
\left[\tau\sigma s\right]=\left[3,1\right]&\longrightarrow& \left(\sum_{\mu_1,\nu_1}\lambda_{\mu_1,\nu_1}^3\right)\left(\sum_{\mu_2,\nu_2}\lambda_{\mu_2,\nu_2}\right)= \Tr (\Lambda_X^3),
\nonumber\\
\left[\tau\sigma s\right]=\left[4\right]&\longrightarrow& 
\sum_{\mu,\nu}\lambda_{\mu,\nu}^4=
\Tr (\Lambda_X^4) ,
\end{eqnarray}
with $\tau, \sigma \in S_{4}$ and $s \in S_{2}$ defined in (\ref{eq:sk}), where, in this case, we set $k=2$.\\
We can now compute the expression for  ${\cal M}_2^{A}(\Lambda_X)$:
\begin{eqnarray}\label{eq:secMomNANB}
 {\cal M}_2^{A}(\Lambda_X)= c_{N,M}&&\Bigg[(M^2-1)(N^4M^2 (M^2-1) - 2 N^2(6M^2-7)+22)
\nonumber\\
 &&+ \Tr(\Lambda_X^2) \; (2 N M (N^2-1)(M^2-1)(N^2M^2-14))\nonumber\\ 
&&+ (\Tr(\Lambda_X^2))^2
 \;(N^2-1)(M^4N^4+M^4N^2 - 14 N^2 M^2 + 6 M^2+30)\nonumber\\ 
  &&+  \Tr(\Lambda_X^3)
 \; 40 (N^2-1)(M^2-1)\nonumber\\ 
  &&+ \Tr(\Lambda_X^4)
  \; (-10 N M)(N^2-1)(M^2-1)\Bigg], \qquad 
 \end{eqnarray}
 where
 \begin{eqnarray}\label{eq:COEFFsecMomNANB}
c_{N,M}=\frac{1}{N^2 M^2 (N^2 M^2-7)^2-36}.
 \end{eqnarray}
In particular if $M=N=\sqrt{L}$ we get \begin{eqnarray}
\label{VVFIN2}
{\cal M}_2^{A}(\Lambda_X)= c_{L}&&\Bigg[(L^5 -2 L^4-11L^3+26L^2+8L-22)\nonumber\\
&&+ \Tr(\Lambda_X^2) \; (2 L^5-4L^4-26L^3+56L^2-28L)\nonumber\\ 
&&+ (\Tr(\Lambda_X^2))^2 \;(L^5-15L^3+20L^2+24 L -30)\nonumber\\ 
&&+ \Tr(\Lambda_X^3) 
\; 40 (L-1)^2\nonumber\\
&&+  \Tr(\Lambda_X^4) 
\; (-10 L )(L-1)^2\Bigg],\qquad 
 \end{eqnarray}
 with
 \begin{eqnarray}
c_{L}=\frac{1}{L^2 (L^2-7)^2-36}.
 \end{eqnarray}
From the exact expression for ${\cal M}_1^A(x,0)$ and ${\cal M}_2^A(x,0)$, Eqs. (\ref{eq:firstMom}) and (\ref{eq:secMomNANB}) respectively, we can now compute the second cumulant of the purity at $\beta=0$:
\begin{eqnarray}\label{eq:secCumInZero}
  \mathcal{K}^A_2(x,0)&=&  {\cal M}^A_2(x,0) -{\cal M}^A_1(x,0)^2\nonumber\\
 &=& \frac{2(N^2 -1) (M^2 -1)( N^2 M^2+11) }{(N^2M^2-1)^2(N^4M^4-13 N^2M^2+36)}
 \nonumber\\ 
&&- x\;\frac{4 N M (N^2 -1) (M^2 -1)(N^2 M^2+11)  }{(N^2 M^2-1)^2(N^4 M^4-13 N^2 M^2+36)}
\nonumber\\ 
&&+ x^2\;\frac{2(N^2 -1) (M^2 -1) (N^4 M^4 - 4 N^2 M^2+15) }{(N^2M^2-1)^2(N^4M^4-13 N^2 M^2+36)}\nonumber\\ 
 && +\langle \Tr(\Lambda_X^3) \rangle_x  
 \; \frac{40 (N^2-1)(M^2-1)}{N^2 M^2 (N^2 M^2-7)^2-36}\nonumber\\  
 &&  -\langle \Tr(\Lambda_X^4) \rangle_x 
 \; \frac{10 N M(N^2-1)(M^2-1)}{N^2 M^2 (N^2 M^2-7)^2-36}.
\end{eqnarray}
For the case of balanced bipartitions, $N=M=\sqrt{L}$, we get
\begin{eqnarray}
 \mathcal{K}^A_2(x,0) &=& \frac{2(L -1)^2 ( L^2+11) }{(L^2-1)^2(L^4-13 L^2+36)}
 \nonumber\\ 
&&- x\;\frac{4 L (L -1)^2 (L^2+11)  }{(L^2-1)^2(L^4-13 L^2+36)}
\nonumber\\ 
&&+ x^2\;\frac{2(L -1) ^2 (L^4 - 4 L^2+15) }{(L^2-1)^2(L^4-13 L^2+36)}\nonumber\\ 
 && +\langle \Tr(\Lambda_X^3) \rangle_x  
 \; \frac{40 (L-1)^2}{L^2 (L^2-7)^2-36}\nonumber\\  
 &&  -\langle \Tr(\Lambda_X^4) \rangle_x 
 \; \frac{10 L(L-1)^2}{L^2 (L^2-7)^2-36}.
\end{eqnarray}
 Notice that for the pure case, corresponding to $\Tr(\Lambda_X^k)=1$ for all $k$, we retrieve the results of Eqs. (\ref{eq:cum2pureGiraudNM}) and (\ref{eq:cum2pureGiraudNN}), as found in~\cite{Scott2003,Giraud}. Furthermore in the large $N$ limit we get
 \begin{eqnarray}
  \mathcal{K}^A_2(x,0) &\sim& \frac{2}{N^ 2 M^2}, \qquad N \neq M \nonumber\\
    \mathcal{K}^A_2(x,0) &\sim& \frac{2}{L^2}, \qquad L = M = \sqrt{L}
 \end{eqnarray}
 as found in~\cite{paper1}, see  Eq. (\ref{eq:cum2GirNM}).
\section{High temperature expansion of the first moment of the purity}\label{sec:highTempRegime}
\begin{figure}[h]
\centering
\includegraphics[width=0.6\columnwidth]{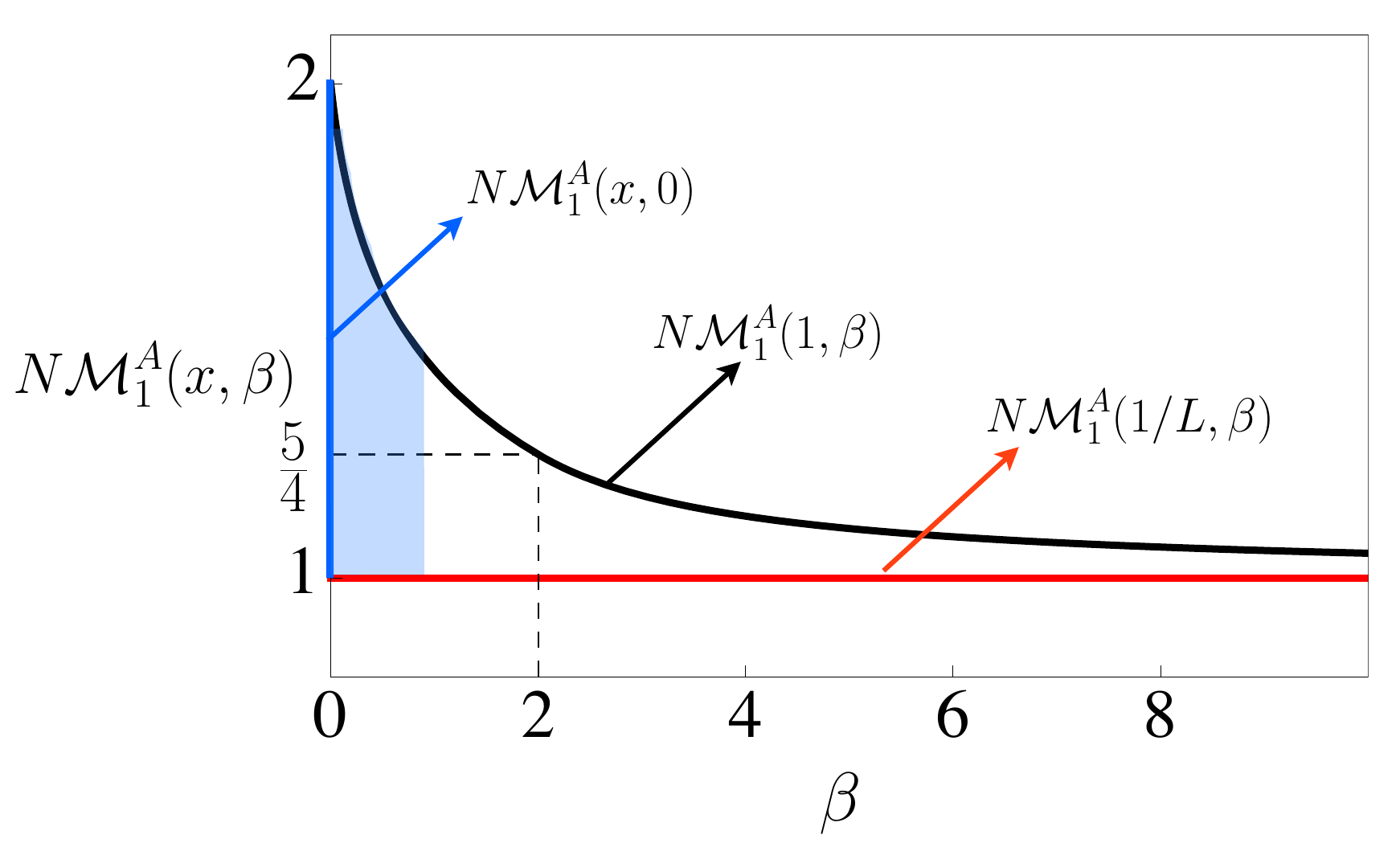}
\caption{First moment of the purity as a function of $\beta$. The red line refers to the set of totally mixed states, such that ${\cal M}_1^A(1/L,\beta)=1/N$ $\forall \ \beta$. The black one refers to pure states for $\beta>0$, see Fig.~\ref{fig:uvsbetapos}. Finally the blue line corresponds to the first moment of the local purity at $\beta=0$, for arbitrary mixed states, while the light blue region refers to the hight temperature region.}\label{fig:mixmom}
\end{figure}

In Chap.~\ref{chap7} we have seen that for $x\neq1$ due to the nonfactorization of the integral over the unitary group, the computation of the partition function is more involved than for pure states, with the only exception given by completely mixed states $\rho=I_X/L$. However, for sufficiently small $\beta$, that is in the high temperature regime, we can study the statistical distribution of the local purity by expanding its $n$-th moment around $\beta=0$, up to order $\beta^2$, as shown in Eq. (\ref{exp333}):
\begin{equation}
\quad {\cal M}_n^A(x,\beta)  \sim
 {\cal M}_n^A(x,0) - \beta \; [  {\cal M}^A_{n+1}(x,0) - {\cal M}^A_1(x,0)\; {\cal M}^A_{n}(x,0)] 
\;, \quad \beta\to 0.
\end{equation} 
In particular, by plugging in the above expression for $n=1$ Eqs. (\ref{eq:primMomU}) and (\ref{eq:secCumInZero}) for the first moment and the second cumulant in $\beta=0$ we get
\begin{eqnarray}
{\cal M}_1^A(x,\beta)  &\sim&
 {\cal M}_1^A(x,0) + \beta \; [ ({\cal M}^A_1(x,0))^2 - {\cal M}^A_2(x,0)] \nonumber \\
&=& 
\frac{M (N^2 -1) }{N^2M^2-1} \;x +  \frac{N(M^2 -1) }{N^2M^2-1}\nonumber \\ 
&&+ \beta \; \Bigg[ \;-\frac{2(N^2 -1) (M^2 -1)( N^2 M^2+11) }{(N^2M^2-1)^2(N^4M^4-13 N^2M^2+36)}\nonumber\\ 
&&\qquad\quad-x\;\frac{2(N^2 -1) (M^2 -1) (-2N M)(N^2 M^2+11)  }{(N^2M^2-1)^2(N^4M^4-13 N^2M^2+36)}\nonumber\\ 
 &&\qquad\quad- x^2\;\frac{2(N^2 -1) (M^2 -1) (N^4 M^4 - 4 N^2 M^2+15) }{(N^2M^2-1)^2(N^4M^4-13 N^2M^2+36)}\nonumber\\ 
&&\qquad\quad- \langle \Tr(\Lambda_X^3) \rangle_x 
\; \frac{40 (N^2-1)(M^2-1)}{N^2 M^2 (N^2 M^2-7)^2-36}\nonumber\\ 
&&\qquad\quad- \langle \Tr(\Lambda_X^4) \rangle_x 
\; \frac{(-10 N M)(N^2-1)(M^2-1)}{N^2 M^2 (N^2 M^2-7)^2-36} \Bigg], 
\end{eqnarray} 
which for the case of balance bipartitions, $N=M=\sqrt{L}$, particularizes to
\begin{eqnarray}
{\cal M}_1^A(x,\beta)  &\sim&
 {\cal M}_1^A(x,0) + \beta \; [ ({\cal M}^A_1(x,0))^2 - {\cal M}^A_2(x,0)] \nonumber \\
&=& 
\frac{\sqrt{L}(1+x) }{1+L} \nonumber \\ 
&&+ \beta \; \Bigg[ \;-\frac{2( L^2+11) }{(L+1)^2(L^4-13 L^2+36)}\nonumber\\ 
&&\qquad\quad+x\;\frac{4L( L^2+11)  }{(L+1)^2(L^4-13 L^2+36)}\nonumber\\ 
 &&\qquad\quad- x^2\;\frac{2(L^4 - 4 L^2 +15) }{(L+1)^2(L^4-13 L^2+36)}\nonumber\\ 
 &&\qquad\quad-  \langle \Tr(\Lambda_X^3) \rangle_x 
 \; \frac{40 (L-1)^2}{L^2 (L^2-7)^2-36}\nonumber\\ 
 &&\qquad\quad\langle \Tr(\Lambda_X^4) \rangle_x 
 \; \frac{10 L (L-1)^2}{L^2 (L^2-7)^2-36} \Bigg]. \label{exp3334}
\end{eqnarray} 
Eq. (\ref{exp3334}) for $x=1$ can be compared with the results we have found in Chap.~\ref{chap3}, where $\beta$ was scaled as $\beta' = \beta L^{3/2}=\beta N^3$. With this choice our expression yields 
\begin{eqnarray}
{\cal M}_1^A(1,\beta'L^{3/2})  &\sim&
 {\cal M}_1^A(1,0) + \beta' L^{3/2} \; [ ({\cal M}^A_1(1,0))^2 - {\cal M}^A_2(1,0)]
\nonumber \\ &\sim& (1  -  \beta') \;  \frac{2 }{\sqrt{L}}
\;, \label{exp3335}
\end{eqnarray} 
in exact agreement with the behavior of the average purity studied in Chap.~\ref{chap3}. Indeed, by expanding the inverse temperature in Eq. (\ref{eq:beta<2}) around $\delta=2$ (corresponding to $\beta=0$), we get
\begin{equation}
\beta= -\frac{\delta-2}{4}+\frac{3}{8} (\delta-2)^2 + \Ord{(\delta-2)^3}
\end{equation}
that is
\begin{equation}
\delta= 2-4 \beta +24 \beta^2 +\Ord{\beta^3}.
\end{equation}
Therefore, from Eq. (\ref{eq:puri1}) for the internal energy $u=N\avg{\pi_{AB}}=\sqrt{L} {\cal M}_1^A(1,\beta L^{3/2})$ we get
\begin{equation}
u= 2 (1-\beta)+ \Ord{\beta^2}.
\end{equation}
In Fig.~\ref{fig:mixmom} we show the behavior of the first moment of the local purity $\pi_{A}$ as a function of $\beta$, for different values of the global purity $\pi_{X}=x \in [1/L,1]$ and for balanced bipartions $N=M=\sqrt{L}$.
\\
Summarizing we have determined the exact analytical expression for ${\cal M}_1^A(x,\beta)$  for  the set $\mathfrak{S}_{x}({\cal H}_X)$ (\ref{eq: set with fixed mixedness}) of: 
\begin{itemize}
\item maximally mixed states, $x=1/L$ (see Eq. (\ref{kkll}))
\begin{equation}
{\cal M}_1^A(1/L,\beta)= 1/N, \qquad \forall \ \beta 
\end{equation}
\item arbitrarily mixed states $x \in [1/L,1]$ at $ \beta =0$ (see Eq. (\ref{eq:VVFIN}))
\begin{equation}
{\cal M}_1^A(x,0)= \frac{\sqrt{L} (1+x)}{L+1},
\end{equation}
\item pure states $x=1$ and for every $\beta$'s (see Eqs. (\ref{eq:puri1}), (\ref{eq:beta<2}) and (\ref{eq:puri2}))
\begin{eqnarray}
{\cal M}_1^A(x,\beta)&=&\frac{1}{\sqrt{L}}\left(1+\frac{1}{2 \beta}\right) , \quad \beta \geq \beta_{+}\\
{\cal M}_1^A(x,\beta)&=&\frac{1}{\sqrt{L}}\left(\frac{3}{2}\delta -\frac{\delta^2}{4}\right), \quad \beta=\frac{4}{\delta}- \frac{2}{\delta^2}, \quad 0\leq \beta \leq \beta_{+}, 1<\delta\leq2 \nonumber\\
\end{eqnarray}
for $0 \leq \beta<\beta_{+}=2$ and $ \beta_{+} \leq \beta$, respectively. 
\end{itemize}
Finally we have also found the high temperature expansion of the first moment, see Eq. (\ref{exp3334}).

 \newpage
\thispagestyle{empty}
\mbox{}
\newpage
\thispagestyle{empty}
\mbox{}
\chapter*{Conclusions and outlook}\label{conc}
\markboth{Conclusions and outlook}{Conclusions and outlook}
\addcontentsline{toc}{chapter}{Conclusions and outlook}

In this thesis, we have discussed the behavior of bipartite entanglement 
of a large quantum system. We have seen that given a bipartite system, the purity of one part can be considered a measure of quantum correlations between the two subsystems only for the case of pure states (that is for isolated quantum systems), while for the more general case of mixed states, this quantity is just a lower bound, and should be substituted with its convex roof. In particular, we have developed a canonical approach for the study of the distribution of the Schmidt coefficients for a fixed value of the average entanglement. We have thus introduced a partition function depending on a fictitious temperature, which localizes the measure on the set of states with higher and lower entanglement with respect to typical (random) states, with respect to the Haar measure. The role of the energy in the partition function is played by our entanglement measure/indicator, the purity of one subsystem.  

In the first part of the thesis, we have obtained a complete characterization of the distribution of the purity and of the eigenvalues for the case of pure states. The global picture is interesting as several locally stable solutions exchange stabilities. On the stable branch (solutions of minimal/maximal free energy for positive/negative temperatures) we have unveiled the presence of three main thermodynamic phases for the system, namely the maximally entangled, the typical and the separable phases, separated by critical points. In particular we have enlightened a second order phase transition, associated to a ${\mathbb Z}_2$ symmetry breaking and related to the vanishing of some Schmidt coefficients, followed by a first order phase transition, associated to the evaporation of the largest eigenvalue from the sea of the others.
In the different phases the distribution of the Schmidt coefficients
has very dissimilar profiles. While for large $\beta$ (small purity) the
eigenvalues, all $\Ord{1/N}$, follow the Wigner semicircle law,
they become distributed according to the Wishart law for smaller $\beta$ and larger purity,
across the second order transition. For even smaller (and eventually
negative) values of $\beta$, when purity becomes finite, across the first order phase transition, one eigenvalue evaporates, leaving the sea of the other eigenvalues
$\Ord{1/N}$ and becoming $\Ord{1}$. This is the signature of separability,
this eigenvalue being associated with the emergence of factorization
in the wave function (given the bipartition). This interpretation is
suggestive and hints at a profound modification of the distribution
of the eigenvalues as $\beta$, and therefore the purity, are changed. Summarizing, for the stable solution the purity goes asymptotically from $1/N$ to $1$. This analysis has been achieved by properly scaling $\beta$, in order to yield the correct thermodynamic limit in the partition function. By keeping the same scaling of positive temperatures also for negative $\beta$'s, we have determined a metastable solution for the system. It shows two phase transitions that are both of second order. The first one, called gravity branch, takes place when the Wishart distribution, characterizing typical states, evolves into an asymmetric arcsine distribution. Then, with a second order phase transition, the distribution of the Schmidt coefficients returns symmetric, leading for $\beta\to-\infty$ to maximally entangled states. At the gravity branch we can interpret the expansion of the free energy around the critical temperature as the partition function of random $2$D surfaces, yielding a theory of pure gravity~\cite{Morris91,Di Francesco1993nw}.

In the second part of the thesis we have considered the more involved case of mixed states and through the same statistical approach, we have determined the exact expression of the first two cumulants at $\beta=0$ and the high temperature expansion, for the first moment of the local purity. We have thus generalized the results of \cite{Giraud,Scott2003}, valid for the case of global pure states. Furthermore, we have also introduced an interesting connection between our problem and the theory of quantum channels, more precisely we have exploited the symmetry properties of the twirling transformations~\cite{tw1} in order to compute the exact expression for the first moment of the local purity.

Our characterization of bipartite entanglement of Haar-distributed states, where the least set of assumptions is made on their generation, can be used in an experiment, \emph{mutatis mutandis}, as a check of the lack of correlations. If one observes that the moments of the purity deviate from the expected values, one could argue for non-randomness (or additional available information) of the states. In turn, our fictitious inverse temperature $\beta$ acquires physical meaning, in that it measures deviations from typicality. This analysis also answers to another, more practical, question: since during the study of some features of entanglement one often relies on numerical simulations, it is important to know to what extent entangled quantum states may be considered  typical~\cite{Zyczkowski98}.

The analysis performed in this thesis for the study of bipartite entanglement of large quantum systems, offers several possible perspectives for future investigation.
From the point of view of quantum information, it would be of great interest to determine the evolution of the entanglement phase
transitions we have studied for the case of pure states, for the more general case of mixed states, when  a new degree of freedom, given by the purity of the global state, is introduced. In this context, another outstanding result would be to understand whether these phase transitions survive even in  the multipartite entanglement scenario \cite{multipartitoClassStat},
if one views the distribution of purity (over all balanced
bipartitions) as a characterization of the global entanglement of
the many-body wave function of quantum systems \cite{mmes}.
\\
Finally, we observe that the phase transitions investigated in this thesis, through a classical statistical mechanics approach are not quantum phase transitions~\cite{Sachdev99}. Nevertheless, entanglement is known to be a good indicator of  quantum phase transitions \cite{osterloh02}, not only for what concerns bipartite entanglement \cite{vidal03,VerstraetePopp2004} but also for the case of multipartite entaglement \cite{costantini2007,xxdiag}. Thus it would be interesting to investigate the link, if any, between these different phase transitions, and therefore apply the techniques shown in this work in order to explore new appealing features of many-body systems, such as spin chains and systems which show a critical behavior. 

We believe that, since the approach followed in this thesis is rooted on the intersection of many aspects of theoretical physics (from classical statistical mechanics, to quantum information, to random matrix theory) it can be regarded as a useful methodological framework for further and even more general topics of investigation. 

 \newpage
\thispagestyle{empty}
\mbox{}
\appendix
\addcontentsline{toc}{chapter}{A{\footnotesize PPENDIX}}

\chapter{Steepest descent method}\label{sec:Method of steepest descent}
\markboth{Steepest descent method}{Steepest descent method}
The steepest descent method is an approximation scheme to evaluate a class of contours integrals in a complex domain of the form
\begin{equation}\label{eq:int}
I_N = \int_{C} dz \ e^{N A(z)}
\end{equation}
in the large $N$ limit, being the contour $C$ contained within the domain of analyticity $D$ of $A$. The steepest descent method is an application of the Cauchy's integral theorem according to which the contour of integration can be deformed in $D$ without changing $I_N$. If we assume that the global maximum of $A(z)$ is not on the boundary of $D$, it corresponds either to a singularity of the function or to a regular point where the derivative of A vanishes,
\begin{equation}\label{eq:condition}
A'(z)=0.
\end{equation}
We will focus on the second case. A point $z_c$ where condition (\ref{eq:condition}) is satisfied is, in general, a \textit{saddle point} for the curves of constant $\mathrm{Re}A(z)$.  In order to avoid cancellations due to the oscillatory behavior of the phase of the integrand we need to choose as contour of integration a \textit{path of steepest descent}, that is a directed curve whose tangent at each point has a direction in which the rate of descent is maximal (\textit{direction of steepest descent}). Indeed it can be proved that such curves are defined by
\begin{equation}
\mathrm{Im}A(z)=\mathrm{Im}A(z_c).
\end{equation}
Thus along this path the phase of the integrand remains constant and the leading contribution to the integral comes from a neighbourhood of $z_c$. If we make the change of variables
\begin{equation}
t= A(z_c) - A(z)
\end{equation}
the path of steepest descent $C$ can be mapped onto the positive real axis:
\begin{equation}
I_N= e^{N A(z_c)} \int_{0}^{+\infty} dt \ \mathcal{A}(t) e^{N t}, \quad \mathcal{A}(t) = - \frac{d z}{d A(z)}\Bigg |_{z=A^{-1}(A(z_c)-t)}.
\end{equation}
We have assumed that the upper limit of integration can be extended up to $+\infty$ due to the fast exponential decay away from $A(z_c)$. By applying the Watson's lemma for $t \to 0_{+}$, in the limit $N\to \infty$, we find
\begin{equation}\label{eq:result1}
I_{N} \sim \sqrt{\frac{\pi}{ 2 N  |A''(z_c)|}}e^{N A(z_c)+i \theta_{1,2} }
\end{equation}
where $\theta_1$ and $\theta_2$ are the directions of steepest descent for the case $A''(z_c) \neq 0$, that is $(z-z_c)= |z-z_c|e^{i \theta_i}$ for $z \in C$ and $1\leq i \leq 2$.
For a more general form of the integral (\ref{eq:int}) 
\begin{equation}
I_N = \int_{C} dz \ g(z)\ e^{N A(z)},
\end{equation}
provided $\ln g(z)$ is an analytic function in a region of the complex plane that includes $C$, we have to multiply expression (\ref{eq:result1}) by the constant factor $g(z_c)$.
We remark that in most cases the leading term of this general expression can be retrieved by expanding $A$ around the maximum and approximate expression (\ref{eq:int}) with a Gaussian integral.

Another generalization of the above results is given by the $n$ dimensional integral on $\mathbb{R}^n$
\begin{equation}\label{eq:int3}
I_N = \int d^n z \ e^{N A(z_1, z_2, \ldots z_n)}.
\end{equation} 
In the limit $N\to \infty$ it can be shown that 
\begin{equation}\label{eq:int3}
I_N = \left(\frac{2 \pi}{N} \right)^{n/2}\left[\det \mathbf{A}^{(2)}\right]^{-1/2} e^{N A(\vec{z}_c) },
\end{equation}
being $\vec{z}_c=\{{z_c}_1, \ldots , {z_c}_n\} $ and $\mathbf{A}^{(2)}$ the Hessian matrix 
\begin{equation}
\left[ \mathbf{A}^{(2)}\right]_{i j}= \frac{\partial ^2 A(\vec{z}_c)}{\partial z_i z_j},\qquad \forall \ i, j \in  \{1, \ldots, n\}.
\end{equation}

\chapter{How to generate uniform random variables on the simplex}\label{sec:generate uniform random variables on the simplex}
\markboth{How to generate uniform random variables on the simplex}{How to generate uniform random variables on the simplex}
In this appendix we discuss a useful trick in order to generate random numbers with the probability density function (\ref{eq:pn(r2)}). 
\\Consider $L$ i.i.d.\ random variables $(y_j)_{1\leq j\leq L}$ with an exponential density function
\begin{equation}
p(y_j) = \exp(-y_j), \quad \forall \ j \in \{1, \ldots, L\}
\end{equation}
Let us define $x_j= y_j / \sum_{1\leq k \leq L} y_k$, for $1\leq j \leq L$. 
In what follows, in order to simplify the notation, if not specified the indices will run between $1$ and $L$.
The joint density function of the normalized random variables $x_j$ is
\begin{eqnarray}
p_L (x_1,\dots,x_L) &=&  \int_{\mathbb{R}_+^L} \mathrm{d}^L y\; \mathrm{e}^{-\sum_k y_k}
\prod_j \delta\left(x_j- \frac{y_j}{\sum_k y_k}\right) \nonumber \\
&  & = \int_{\mathbb{R}_+^L} \mathrm{d}^L y\; \mathrm{e}^{-\sum_k y_k}
\prod_j \delta\left(x_j- \frac{y_j}{\sum_k y_k}\right)
\int_0^\infty \mathrm{d} t\; \delta\Big(t-\sum_k y_k\Big)
\nonumber \\
&  & = \int_0^\infty \mathrm{d} t\; \mathrm{e}^{-t} \int \mathrm{d}^L y\; \delta \Big(t-\sum_k y_k \Big)\prod_j \delta\left(x_j- \frac{y_j}{t}\right)
\nonumber \\
&  & = \int_0^\infty \mathrm{d} t\; \mathrm{e}^{-t} \int \mathrm{d}^L z\; t^L \delta\left(t \Big(1-\sum_k z_k \Big)\right)\prod_j \delta\left(x_j- z_j\right)
\nonumber \\
&  & = \int_0^\infty \mathrm{d} t\; t^{L-1} \mathrm{e}^{-t} \int \mathrm{d}^L z\;  \delta\Big(1-\sum_k z_k \Big) \prod_j \delta\left(x_j- z_j\right)
\nonumber \\
&  & = (L-1)!\; \delta\Big(1-\sum_k x_k \Big),
\end{eqnarray}
where we have used the definition of the Gamma function
\begin{equation}
\Gamma(L) = \int_0^\infty \mathrm{d} t\; t^{L-1} \mathrm{e}^{-t} 
\end{equation}
and since $L$ is an integer number, we get $\Gamma(L)=(L-1)!$.
Therefore, the random variables
\begin{equation}
{r_j}^2 = \frac{y_j}{\sum_k y_k}\ , \qquad \forall \ j \in \{1,\ldots, L\}
\end{equation}
are distributed with joint probability density (\ref{eq:pn(r2)}).

 \newpage
\thispagestyle{empty}
\mbox{}
\newpage
\thispagestyle{empty}
\mbox{}

 \newpage
\thispagestyle{empty}
\mbox{}
 \newpage
\thispagestyle{empty}
\mbox{}
  \newpage
\thispagestyle{empty}
\mbox{}

\chapter*{Acknowledgments}\label{Acknowledgments}
\addcontentsline{toc}{chapter}{Acknowledgments}

I would like to express my gratitude to all those who supported and helped me through the realization of  this work, and, more in general, during my Ph.D years.

 I am deeply indebted to my thesis advisors Paolo Facchi and Saverio Pascazio, with whom I  collaborate since my Master thesis, in 2007. Paolo has been a precious guide  to me: he made me  explore many fields of theoretical and mathematical physics in such a pleasant way that  it made me almost not notice the  difficulties this subjects are naturally endowed with. Besides,  Paolo has also been a friend to me,  helping me through personal difficulties without even asking.  I have always considered Saverio as a reference point both during the Master and  the Ph.D thesis and  I would like to deeply thank him for many useful discussions on the most diverse subjects. He taught me how life can be hard, and how we can and must be even harder! 

Finishing this thesis has been challenging due to the amount of people I had the pleasure to meet and collaborate with during this years and that I would like to thank.  In particular, I spent one month at the University Federico II in Naples where I met three young researchers, B. Bellomo, G. Gualdi and U. Marzolino. Our collaboration, started under the supervision of G. Marmo, has represented a challenging and outstanding adventure which still goes on.  I would also like to address a heart-felt thank you to H. Nakazato and K. Yuasa, for their care  while I was in Japan. They both gave me  their precious scientific collaboration, and their human attentions, helping me to feel at home. Finally, I wish to mention  G. Florio, V. Giovannetti, G. Marmo,  G. Parisi and A. Scardicchio. Each of them has been for me a unique resource toward the comprehension of the physical phemomena discussed in this thesis and in the papers I published during my Ph.D. Finally, I would like to thank the referees of this thesis, R. Fazio and A. Marrone, for their helpfulness and for interesting discussions with them.

I conclude my acknowledgments with a personal thank to Davide. His patience and care have been of fundamental support during this last year and in particular the months involved with the preparation of this work. 
  \newpage
\thispagestyle{empty}
\mbox{}

  \newpage
\thispagestyle{empty}
\mbox{}
\newpage
\thispagestyle{empty}
\mbox{}
\end{document}